\documentclass[%
reprint,
 amsmath,amssymb,
 aps, pra,
longbibliography
]{revtex4-2}

\usepackage{natbib}
\usepackage{graphicx}% Include figure files
\usepackage{dcolumn}% Align table columns on decimal point
\usepackage{bm}% bold math
\usepackage[caption=false]{subfig}
\usepackage{physics} 
\usepackage{parskip} 
\usepackage{mathrsfs}
\usepackage{array}
\usepackage{ytableau}
\usepackage{color}
\definecolor{Blue}{rgb}{0.3,0.3,0.9}
\definecolor{Red}{rgb}{0.9,0.3,0.3}
\definecolor{Green}{rgb}{0.3,0.6,0.3}
\newcommand{\revision}[1]{#1}
\newif\ifNOSUP \NOSUPfalse

\newcommand{\figdir}{figure-sub}

\begin{document}

%\preprint{APS/123-QED}

\title{Spectral and Entanglement Properties of the Random Exchange Heisenberg Chain
%Study of random exchange Heisenberg chain
}% Force line breaks with \\

\author{Yilun Gao}
\author{Rudolf A. R\"{o}mer}%
\affiliation{%
 Department of Physics, University of Warwick, Gibbet Hill Road, Coventry, CV4 7AL, UK
}%


\date{\today}% It is always \today, today,

\begin{abstract}
We study the many-body localization problem in the non-abelian SU(2)-invariant random antiferromagnetic exchange model in 1D. Exact and sparse matrix diagonalization methods are used to calculate eigenvalues and eigenvectors of the Hamiltonian matrix. We investigate the behaviour of the energy level gap-ratio statistic, participation ratio, entanglement entropy and the entanglement spectral parameter as a function of disorder strength. Different distributions of random couplings are considered. 
We find, up to $L=24$, a clear distinction between our non-abelian model and the more often studied random field Heisenberg model: the regime of seemingly localized behaviour is much less pronounced in the random exchange model than in the field model case. 
\end{abstract}

%Use showkeys class option if keyword
\maketitle

%\tableofcontents

%%%%%%%%%%%%%%%%%%%%%%%%%%%%%%%%%%%%%%%%%%%%%%%%%%%%%%%%%%%%%%%%%%%%%%%%%%%%%%
\section{\label{sec:introduction} Introduction}
%%%%%%%%%%%%%%%%%%%%%%%%%%%%%%%%%%%%%%%%%%%%%%%%%%%%%%%%%%%%%%%%%%%%%%%%%%%%%%

An isolated system at thermal equilibrium is subject to equilibrium statistical mechanics as described by the eigenstate thermalization hypothesis (ETH) \cite{Deutsch1991QuantumSystem}: expectation values of physical observables for generic quantum many-body systems can be evaluated using standard ensembles of statistical mechanics \cite{DAlessio2016FromThermodynamics}. Such a system is traditionally called \emph{ergodic} since during its time evolution, it can explore all configurations in Hilbert space allowed by global constraints \cite{LudwigBoltzmann1896VorlesungenGastheorie}. 
On the other hand, in the presence of disorder, ergodicity can break down and thermalization avoided, due to a suppression of energy-exchange processes \cite{Gornyi2005InteractingTransport,Basko2006Metal-insulatorStates}. This many-body localization (MBL) phenomenon has been extensively studied in the last nearly two decades \cite{Abanin2017RecentLocalization,Imbrie2017Review:Systems,Abanin2019Colloquium:Entanglement}, both theoretically \cite{Oganesyan2007LocalizationTemperature,Znidaric2008Many-bodyField,Pal2010Many-bodyTransition, Serbyn2014QuantumPhase,DeLuca2013ErgodicityLocalization,Kjall2014Many-bodyChain,Luitz2015Many-bodyChain,Vasseur2015QuantumChains,Imbrie2016OnChains,Naldesi2016DetectingQuenches,Benini2020,Schliemann2021Many-bodyModels} and, somewhat less vigorous, also experimentally \cite{Schreiber2015ObservationLattice,Kondov2015Disorder-inducedGas,Smith2016Many-bodyDisorder}.
Various models have been employed, mostly numerically, to investigate the key features of MBL phases \cite{Oganesyan2007LocalizationTemperature,Znidaric2008Many-bodyField,Pal2010Many-bodyTransition,Imbrie2016OnChains}. Among them, the disordered Heisenberg spin chain has been discussed very often \cite{Pal2010Many-bodyTransition,DeLuca2013ErgodicityLocalization,Kjall2014Many-bodyChain,Vasseur2015QuantumChains,Luitz2015Many-bodyChain,Schliemann2021Many-bodyModels,Naldesi2016DetectingQuenches,Benini2020,Schliemann2021Many-bodyModels}.
The most commonly studied variant of the model has onsite disorder. Let us highlight a few of these studies:
Pal et al.\ \cite{Pal2010Many-bodyTransition} investigate the scaling of the probability distribution of long-distance spin correlations and propose that the MBL transition is driven by an infinite-randomness fixed point.
De Luca et al.\ \cite{DeLuca2013ErgodicityLocalization} study the scaling of participation ratios and identify the MBL transition by calculating wave function coefficients.
Also, Luitz et al.\ \cite{Luitz2015Many-bodyChain} suggest the existence of a many-body mobility edge in terms of an energy-resolved phase diagram while
Schliemann et al.\ \cite{Schliemann2021Many-bodyModels} find the critical disorder of transitions between ergodic and many-body localized phases by investigating the inflection point of the average of the consecutive-gap ratio. 
% give more on results with Heisenberg since this is "the most commonly studied variant"
%??? More !!! ???
%
Often, the random disorder is replaced with quasi-periodic ``disorder'' values, e.g., following the Aubry-Andre potential \cite{Naldesi2016DetectingQuenches,Benini2020,Falcao2024Many-bodyPotential,Hetenyi2024LocalizationModel}. Again, an energy-resolved MBL-type phase diagram can be found \cite{Naldesi2016DetectingQuenches}. 
% avalanches
However, recent work has begun to question whether a true MBL phase can really exist. Namely, it was argued that avalanche instabilities \cite{DeRoeck2017StabilitySystems}, due to rare regions of weak disorder, can destablize the MBL phase \cite{Leonard2023ProbingSystem,Morningstar2022AvalanchesSystems,Evers2023InternalDelocalization,Sierant2024Many-bodyComputing,Scocco2024ThermalizationSystems,Lu2024ExactChains} when $L\rightarrow\infty$. 
In many ways, this state of the field is astounding given the effort of the last decades. On the other hand, it seems to mimic earlier difficulties associated with formulating a consistent picture of ground state properties for such disordered interacting quantum systems \cite{Giamarchi1988AndersonMetals,Emery1993a,Voit1999,Giamarchi2003QuantumDimension,Schuster2002b}.

The study of the interplay of disorder and many-body interactions hence remains full of surprises.
Here, we study the SU(2)-invariant, disordered, anti-ferromagnetic Heisenberg spin chain with $L$ sites, the \emph{poster child} of the non-abelian ETH \cite{Murthy2023Non-AbelianHypothesis,Lasek2024NumericalHypothesis}. Often called the random exchange model, its ground state and low-temperature properties were first studied by Ma, Dasgupta and Hu \cite{Ma1979RandomChains,Dasgupta1980Low-temperatureChain}. They used their celebrated real-space renormalization group (RSRG) approach, based on successive spin-singlet formation starting from the largest $|J|$, and established power-law temperature dependencies for specific heat and magnetic susceptibility. Fisher \cite{Fisher1994RandomChains} discussed the transition from the ordered antiferromagnetic phase to the random singlet phase as a function of increasing disorder strength. In this case, the rare regions formed by singlets, composed of two spins separated by a long distance, dominate the mean correlation functions. 
% MBL results
The model was studied in the MBL context first by Vasseur et al.\  \cite{Vasseur2016Particle-holeModes} up to $L=16$. Breaking the full SU(2) invariance by choosing an XXZ-type coupling, they report a transition from ergodic states at weak disorders to MBL states at strong disorders \cite{Vasseur2016Particle-holeModes,Aramthottil2024PhenomenologyChains}.  
Protopopov et al.\  \cite{Protopopov2020Non-AbelianThermalization} studied the MBL aspects of the model in terms of the RSRG approach for the $s=0$ spin sector. By studying the $L$ scaling of the entanglement entropy, they find states intermediate between extended states and MBL states even at strong disorders.
Siegl and Schliemann focused on level statistics and argue that there exists a transition from the ergodic phase to a phase that is different from both ergodic and MBL phase \cite{Siegl2023ImperfectChains}.
Very recently, Saraidaris et al.\ \cite{Saraidaris2024Finite-sizeChains} suggest that eventual thermalization and delocalization appear at large system sizes, $L=48$, by looking at the distribution of entanglement entropy and correlation functions in the $s=0$ sector using tDMRG \cite{Saraidaris2022InfluenceChain}. 
Han et al.\ \cite{Han2024EntropyChain} found that there is no evidence of an MBL transition in the random exchange model by studying the time and disorder dependence of multifractal exponents.
In the 1D experimental realizations of the MBL situation \cite{Kondov2015Disorder-inducedGas,Smith2016Many-bodyDisorder}, the system sizes range from $L=10$ to $100$. Furthermore, a precise filling of spins per site is hard to achieve for all current such realizations \cite{Schreiber2015ObservationLattice,Kondov2015Disorder-inducedGas,Smith2016Many-bodyDisorder}. For an experimental realization of the random exchange model, it, therefore, seems useful to consider a range of spin sectors as well as to offer a range of possible distributions of random couplings. This is what we set out to do in this work: We characterize the behaviour of the disordered SU(2)-invariant Heisenberg model for three distributions of its random couplings and in various spin sectors.

%%%%%%%%%%%%%%%%%%%%%%%%%%%%%%%%%%%%%%%%%%%%%%%%%%%%%%%%%%%%%%%%%%%%%%%%%%%%%%
\section{\label{sec:modelsandmethods} Model and methods}
%%%%%%%%%%%%%%%%%%%%%%%%%%%%%%%%%%%%%%%%%%%%%%%%%%%%%%%%%%%%%%%%%%%%%%%%%%%%%%

%%%%%%%%%%%%%%%%%%%%%%%%%%%%%%%%%%%%%%%%%%%%%%%%%%%%%%%%%%%%%%%%%%%%%%%%%%%%%%
% model
%%%%%%%%%%%%%%%%%%%%%%%%%%%%%%%%%%%%%%%%%%%%%%%%%%%%%%%%%%%%%%%%%%%%%%%%%%%%%%
\subsection{\label{sec:model} The random exchange model}

\begin{figure*}[t]
(a)\includegraphics[width=0.4\textwidth]{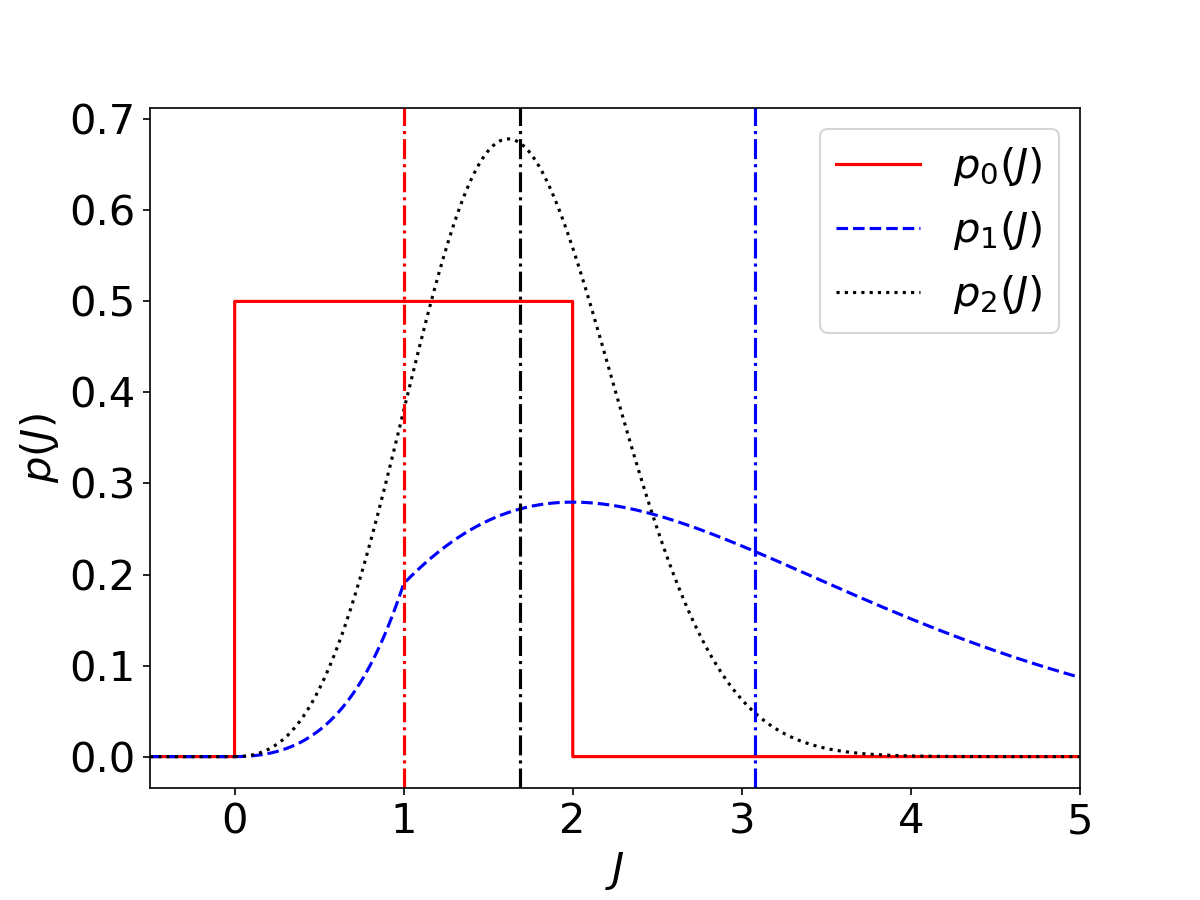}
(b)\includegraphics[width=0.4\textwidth]{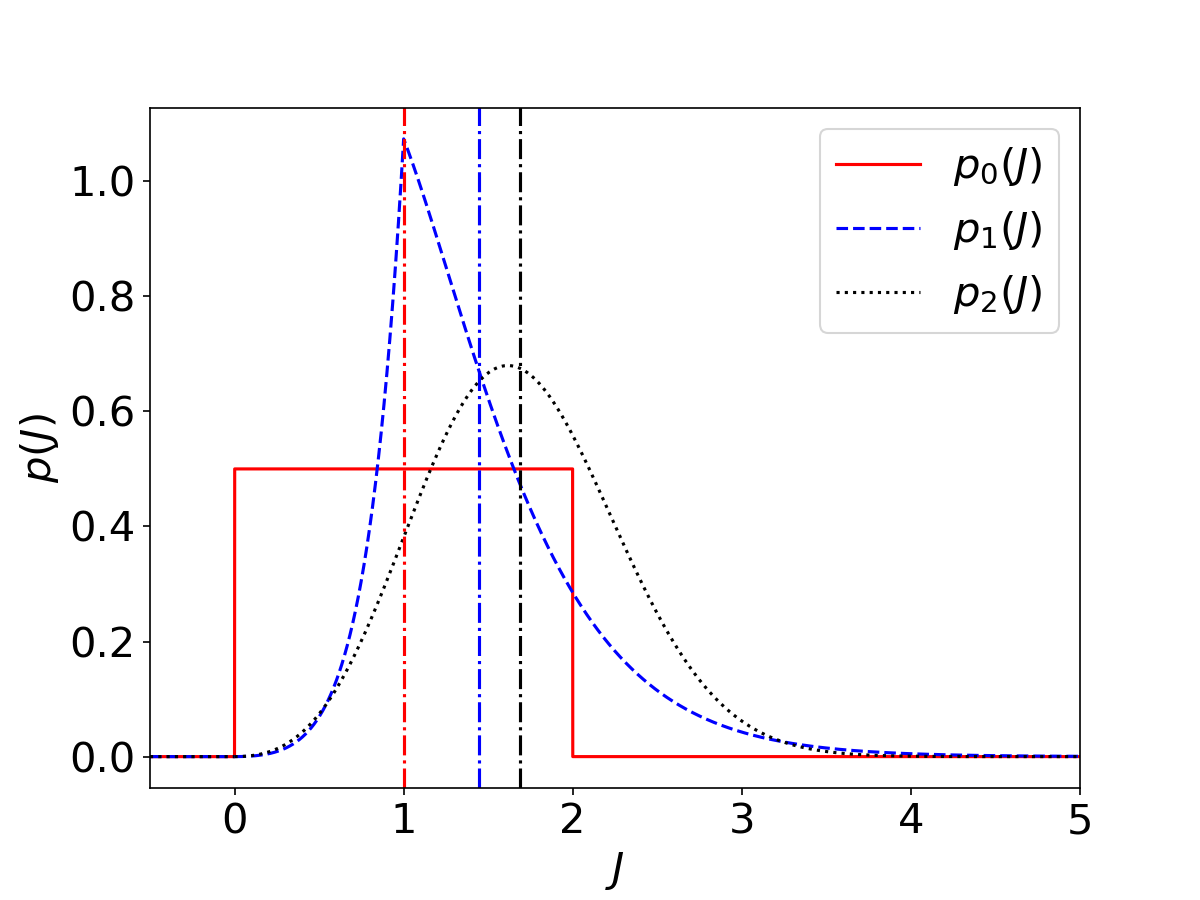}
\caption{The disorder distributions with (a) $\Delta/J_0=1$ for all three distributions and (b) with identical variances such that $\Delta/J_0=1, 0.37, 0.99$ for $p_0(J)$ (red solid line), $p_1(J)$ (blue dashed) and $p_2(J)$ (black dotted), respectively. 
The three vertical dash-dotted lines stand for the averages $\langle J \rangle_i = \int J p_i(J) dJ$, which are for (a) $\left\langle J\right\rangle_0=1$, $\left\langle J\right\rangle_1\approx 3.1$, $\left\langle J\right\rangle_2\approx 1.7$. \revision{Here, we set $J_0=1$.} 
For (b), we have $\left\langle J\right\rangle_0=1$, $\left\langle J\right\rangle_1\approx 1.4$, $\left\langle J\right\rangle_2\approx1.7$. 
With $100$ samples for $L=16$, we find that our numerical averages remain within $1.5\%$ of these estimates.
} 
\label{fig-disorder-distributions}
\end{figure*}

We study the antiferromagnetic 1D Heisenberg spin-$1/2$ chain with \emph{disordered} nearest neighbor exchange couplings. The Hamiltonian reads as
\begin{equation}
\begin{split}
H&=\sum_{i=1}^{L}J_i\mathbf{S}_i\cdot \mathbf{S}_{i+1}\\
&=\sum_{i=1}^{L}J_i
\left[ 
\frac{1}{2}(S^{+}_iS^{-}_{i+1}+S^{-}_iS^{+}_{i+1})+S^z_iS^z_{i+1}
\right],
\end{split}
\label{eq:ham}
\end{equation}
with $J_i$ corresponding to the exchange couplings between nearest neighbors and the $\mathbf{S}_i=\left( S^x_i, S^y_i, S^z_i \right)$ to the spin-$1/2$ operators at site $i=1, \ldots, L$. 
We take $J_i$ to obey three different distributions as shown in Fig.\ \ref{fig-disorder-distributions}. 
%%%%%%%%%%%%%%%%%%%%%%%%%%%%%%%%%%%%%%%%%%%%%%%%%%%%%%%%%%%%%%%%%%%%%%%%%%%%%%
%%%%%%%%%%%%%%%%%%%%%%%%%%%%%%%%%%%%%%%%%%%%%%%%%%%%%%%%%%%%%%%%%%%%%%%%%%%%%%
We start with (i) $J_i\in [-\Delta+J_0,\Delta+J_0]$, i.e.\ a uniform distribution $p_0(J)= 1/2 \Delta$ when $|J-J_0| \leq \Delta$ and $0$ otherwise. 
Furthermore, we choose (ii) $ p_1(J) \propto J^2e^{-|J-J_0|/\Delta}$ and (iii) $p_2(J)\propto J^2e^{-|J-J_0|^2/\Delta^2}$ with $J>0$ and all distributions normalized such that $\int_{\revision{-\infty}}^{\infty} p_j(J) \textrm{d}J = 1$ for $j=0,1,2$. Hence $\Delta/J_0$ parametrizes the strength of the disorder, i.e.\ the deviation from a constant $J_0$. In the following, we fix the energy scale by choosing $J_0=1$.
The reason for studying cases with disorder distributions $p_1(J)$ and $p_2(J)$ is because the simple uniform disorder distribution has possibly uncoupled spins for $\Delta \geq J_0$ ($J_i =0$ for some $i$'s) and also begins to mix antiferromagnetic and ferromagnetic sectors for $\Delta > J_0$. For $p_1(J)$ and $p_2(J)$, the leading $J^2$ term avoids ever running into the "zero coupling" problem and couplings always remain in the AFM regime. 
%Furthermore, the $J_0$ term continues to be of central importance for both $p_1$ and $p_2$.
%
We recall that for $p_0(J)$, the model is the standard random exchange model \cite{Ma1979RandomChains, Dasgupta1980Low-temperatureChain,Fisher1994RandomChains}.

We apply periodic boundary conditions such that $\mathbf{S}_{i+L}=\mathbf{S}_i$. Both the total spin $\mathbf{S_\text{tot}}=\sum_{i=1}^L\mathbf{S}_i$ and the $z$-projection of the total spin $S_{\text{tot}}^z=\sum_{i=1}^LS^z_i$ commute with the Hamiltonian, i.e.\ $[\mathbf{S_\text{tot}},H]=0$ and $[S_{\text{tot}}^z,H]=0$ \cite{Ma1979RandomChains, Dasgupta1980Low-temperatureChain,Fisher1994RandomChains}. The Hamiltonian satisfies SU(2) symmetry and we denote $s_\text{tot}= 0, 1, \ldots, L/2$ as the \emph{total spin} quantum number while $m= -L/2, -L/2+1, \ldots, 0, \ldots, L/2$ is the \emph{magnetic} quantum number.
We can diagonalize the Hamiltonian in each of the $(L/2+1)$ $s$ and $(L+1)$ $m$ subspaces. The eigenvectors $|{n,s,m}\rangle$ as usual obey $\mathbf{S_\text{tot}^2} |{n,s,m}\rangle = s(s+1) |{n,s,m}\rangle$ and $S_{\text{tot}}^z |{n,s,m}\rangle = m |{n,s,m}\rangle$ with $n$ labelling the energies $E_n$ in each $(s,m)$ sector.
In each such sector, the dimension of the Hilbert space, $\text{dim}\, \mathcal{H} (s,m)$, is given by a difference of binomials, i.e.\  
$\text{dim}\, \mathcal{H} (s,m) = \binom{L}{\lfloor L /2 \rfloor + m} - \binom{L}{\lfloor L /2 \rfloor + m +1}=\frac{2(\lfloor L /2 \rfloor + m)+1-L}{L+1}\binom{L+1}{\lfloor L /2 \rfloor + m}$ for $m \leq \lfloor L /2 \rfloor$\cite{Zachos1992AlteringSupersymmetry,Siegl2023ImperfectChains}.
For a given $m$ sector, we then have $\text{dim}\, \mathcal{H} (m) = \sum_{s} \text{dim}\, \mathcal{H} (s,m) = \binom{L}{\lfloor L /2 \rfloor  -m}$ with eigenstates indexed by $n=1, 2, \ldots, \text{dim}\, \mathcal{H} (m)$. 
We often shall simply write $\text{dim}\, \mathcal{H}$ for convenience if the context is clear.
We also note that the spin-flip operator $\mathcal{C}=\Pi_i^L S_i^x$ commutes with the Hamiltonian, i.e.\ $[\mathcal{C},H]=0$ \cite{Vasseur2016Particle-holeModes}. Therefore, if we apply the $\mathcal{C}$ operator and the Hamiltonian to an eigenstate $|n,s,m\rangle$, we get $H\mathcal{C}|n,s,m\rangle=\mathcal{C}H|n,s,m\rangle=E_n\mathcal{C}|n,s,m\rangle=E_n|n,s,-m\rangle$. This $Z_2$ symmetry is usually called the particle-hole symmetry. 
%%%%%%%%%%%%%%%%%%%%%%%%%%%%%%%%%%%%%%%%%%%%%%%%%%%%%%%%%%%%%%%%%%%%%%%%%%%%%%
% methods
%%%%%%%%%%%%%%%%%%%%%%%%%%%%%%%%%%%%%%%%%%%%%%%%%%%%%%%%%%%%%%%%%%%%%%%%%%%%%%
\subsection{\label{sec:methods} Spectral and entanglement measures}

We focus on four different measurements, namely (i) the consecutive-gap ratio, (ii) the normalized participation ratio, (iii) the entanglement entropy and (iv) the entanglement spectral parameter for different disorder strengths across the whole spectrum. 
The first measure provides a convenient characterization of the properties of the eigenspectra of our model. The \emph{consecutive-gap ratio} $r_n$ is defined as \cite{Oganesyan2007LocalizationTemperature}
\begin{equation}
r_n=\frac{\text{min}\{E_{n+1}-E_{n},E_{n+2}-E_{n+1}\}}{\text{max}\{E_{n+1}-E_{n},E_{n+2}-E_{n+1}\}}, n \leq \text{dim}\,\mathcal{H}-2,
\label{eq:rn}
\end{equation}
where $E_n$ is the $n^{\text{th}}$ energy eigenvalue of the system with $E_n<E_{n+1}$. 
According to Eq.\ \eqref{eq:rn} we can see that $r_n \in [0,1]$. When the state of a system is ergodic, the distribution $P(r)$ of the consecutive-gap ratio is expected to follow the predictions of the Gaussian orthogonal ensemble (GOE) \cite{Atas2013DistributionEnsembles}, namely,
\begin{equation}
P_{\text{GOE}}(r)=\frac{27}{4}\frac{r+r^2}{(1+r+r^2)^{\frac{5}{2}}}.
\end{equation}
The expectation value of \revision{$\left<r\right>_{\text{GOE}}$} is calculated to be $\left<r\right>_{\text{GOE}}=\int_0^1 P_{\text{GOE}}(r)rdr=4-2\sqrt{3}\approx0.5359$. There exists level repulsion in this case as \revision{$P_{\text{GOE}}(r\rightarrow 0)\rightarrow 0+$}.
In contrast, the $P(r)$ for (many-body) localized states is expected to satisfy the ensemble originating from a Poisson distribution \cite{Oganesyan2007LocalizationTemperature}, which can be written as
\begin{equation}
P_{\text{Poisson}}(r)=\frac{2}{1+r^2}.
\end{equation}
The expectation value in this case is given by $\langle r\rangle_{\text{Poisson}}=\int_0^1 P_{\text{Poisson}}(r)rdr=2\ln 2-1\approx0.3863$. We can see that there is no level repulsion here as \revision{$P_{\text{Poisson}}(0)=2$}.

% Rosenzweig-Porter approach
The above averages and distributions of the $r_n$'s are valid for spectra of irreducible random matrices \cite{Atas2013DistributionEnsembles}. For the model \eqref{eq:ham}, this approximation is valid in each $(s,m)$ sector. However, when taking into account a complete sector for $s$, the $2s+1$ subspaces with varying $m$ yield spectral degeneracies \cite{Grimm2021,ZHONG1998} which result in a peak for $P(0)$ \cite{Giraud2022ProbingStatistics}. Hence, the number $L+1$ of subspaces as well as the dimension of each subspace has to be taken into account for both GOE and Poisson cases as detailed in Appendix \ref{sec:rosenzweigporter}. In the following, we shall denote the $r$-distributions based on $k$ irreducible blocks by $P^{(k)}_\text{GOE}(r)$ and $P^{(k)}_\text{Poisson}(r)$ for purely Gaussian or purely Poissonian blocks, respectively.

The following three measures allow us to characterize the Hilbert-space localization properties via analysis of the eigenstates $|{n,s,m}\rangle$. With $f$ labelling a particular Fock state $|{n,s,m}; f\rangle$ in $\mathcal{H}$, we can write $|{n,s,m}\rangle = \sum_{f} \psi_{nsm}(f) |{n,s,m}; f\rangle$ such that $\psi_{nsm}(f)$ represents the expansion coefficients in the Fock basis. For convenience, we shall often simply write $\psi_{nsm}(f) = \psi_f$ when the chosen spin sector $(s,m)$ is clear.

A normalized \emph{participation ratio} $\mathcal{P}$ in Fock space for spin sector $(s,m)$ can be defined as 
\begin{equation}
\mathcal{P}(s,m) = \revision{\frac{1}{\text{dim}\,\mathcal{H}(s,m)}} \times \frac{1}{\sum_f|\psi_f|^4},
\end{equation}
with $\psi_f$ as given in the previous paragraph \cite{Krameri1993,Buijsman2018Many-bodyOrbitals}. For notational consistency, we shall also often abbreviate $\mathcal{P}=\mathcal{P}(s,m)$. 
When \revision{$\mathcal{P}\approx 0$}, this usually indicates the case when a state is localized in Fock space, while the most Fock-extended states give $\mathcal{P}\lesssim 1$ \cite{Buijsman2018Many-bodyOrbitals}. 
Furthermore, due to the aforementioned symmetries, we have $\psi_{n,s,m}(f) = \psi_{n,s,-m}(f)$.
This yields the same normalized participation ratio for eigenstates $|n,s,m\rangle$ and $|n,s,-m\rangle$.
% discuss that this really requires multifractal analysis

The \emph{entanglement entropy} $S_E$ can be written as 
\begin{equation}
S_E=-\text{Tr}\rho_A\text{ln}\rho_A,
\end{equation}
here $\rho_A$ stands for the reduced density matrix of subsystem $A$, when the lattice has been divided into two parts, namely, $A$ and $B$. We always \revision{make this bipartite division} at $\lfloor L/2 \rfloor$. In the following, we shall always assume that $L$ is even for convenience.
%
%limiting cases/expectations ????
Small values of $S_E$ correspond to localized states and are expected to follow an area-law scaling \cite{Abanin2017RecentLocalization}, i.e. $S_E$ $\sim$ $\ln L$ in 1D. In contrast, extended states give larger $S_E$ and yield a volume-law scaling, i.e. $S_E$ $\sim$ $L$ in 1D.

The \emph{entanglement spectral parameter} $\lambda$ reads 
\begin{equation}
\lambda=w_1-w_2+w_3-w_4,
\end{equation}
where $w_1$, $w_2$, $w_3$ and $w_4$ are the four largest eigenvalues of the reduced density matrix $\rho_A$ \cite{Deng2013BosonizationLattices}.
%discuss ???
Previous works have shown that $\lambda$ can be used to identify phases of many-body systems without the need to distinguish its $L$ scaling behaviour, at least for ground state properties \cite{Deng2013BosonizationLattices,Goldsborough2015UsingModel}. 

%%%%%%%%%%%%%%%%%%%%%%%%%%%%%%%%%%%%%%%%%%%%%%%%%%%%%%%%%%%%%%%%%%%%%%%%%%%%%%
\subsection{\label{sec:numerics} Diagonalization strategies}

We first diagonalize the Hamiltonian in each $m$ spin sector in terms of the exact diagonalization method based on the \textsc{QuSpin} package \cite{
Weinberg2019QuSpin:Spins}. The eigenvalues and eigenvectors that we compute are then used to calculate the quantitative localization measures mentioned above. We consider all the non-negative $m$ sectors as all the physical quantities related to eigenstates take identical values for $-m$ for a particular $s$ sector (see below).
We compute at least $100$ different random realizations at each disorder strength $\Delta/J_0$. With the \textsc{QuSpin} package, we can reach system sizes up to $L=16$, corresponding to $\text{dim}\,\mathcal{H} = 12870$ for the largest sector, i.e.\ $m=0$.
For the two non-uniform disorder distributions, $p_1(J)$ and $p_2(J)$, we build up a close correspondence between them and the $p_0(J)$ by equaling their variances.
E.g.\ $\Delta/J_0=1$ corresponds to $\Delta/J_0=0.37$ for $p_1(J)$ and $\Delta/J_0=0.99$ for $p_2(J)$ (cp.\ Fig.\ \ref{fig-disorder-distributions}). 
We also use the sparse matrix diagonalization method based on {\sc JaDaMILU} \cite{Bollhofer2007JADAMILU:Matrices} and \textsc{PETSc} \cite{Balay1997EfficientLibraries,S.Balayetal.2017PETScManual} and \textsc{SLEPc} \cite{Hernandez2005SLEPc,Roman2017SLEPcComputations} to diagonalize the Hamiltonian matrix \cite{Pietracaprina2018Shift-invertChains}. This enables us to deal with system size up to $L=18$, i.e.\ $\text{dim}\,\mathcal{H} = 48620$ for $m=0$ and $L=20$ with $\text{dim}\,\mathcal{H} =184756 $ for $m=0$ respectively. Here, 100 realizations are calculated for $p_0(J)$ via {\sc JaDaMILU} and for \emph{all} $p_i(J)$'s with \textsc{SLEPc}. As the method returns eigenvalues and eigenvectors close to a particular energy $E$, we do calculations on several $E$'s with more $E$'s in the center of the spectrum and less $E$'s on the spectral edges. More details will be provided when discussing our numerical results below.
(Sparse) diagonalization in each $s$ spin sector can also be implemented in terms of Young tableaux \cite{Nataf2014ExactModels,Dabholkar2024ErgodicChains,Dabholkar2024PrivateCommunication,Nataf2024DensityCoefficients}, which is explained further in the appendix. We can then solve -- and average over -- the Hamiltonian up to system size $L=24$ in the $s=0$ sector with $\text{dim}\,\mathcal{H}(s=0) = 208012$. In this case, we compute 100 realizations for $p_0(J)$ and 80 realizations for $p_1(J)$ and $p_2(J)$.
%%%%%%%%%%%%%%%%%%%%%%%%%%%%%%%%%%%%%%%%%%%%%%%%%%%%%%%%%%%%%%%%%%%%%%%%%%%%%%
\section{\label{sec:results} Results}
%%%%%%%%%%%%%%%%%%%%%%%%%%%%%%%%%%%%%%%%%%%%%%%%%%%%%%%%%%%%%%%%%%%%%%%%%%%%%%

%%%%%%%%%%%%%%%%%%%%%%%%%%%%%%%%%%%%%%%%%%%%%%%%%%%%%%%%%%%%%%%%%%%%%%%%%%%%%%
\subsection{\label{sec:results-degeneracies} Spectral degeneracies and simple localization measures}

%%%%%%%%%%%%%%%%%%%%%%%%%%%%%%%%%%%%%%%%%%%%%%%%%%%%%%%%%%%%%%%%%%%%%%%%%%%%%%
\begin{figure*}[tb]
  (a)\includegraphics[width=0.4\textwidth]{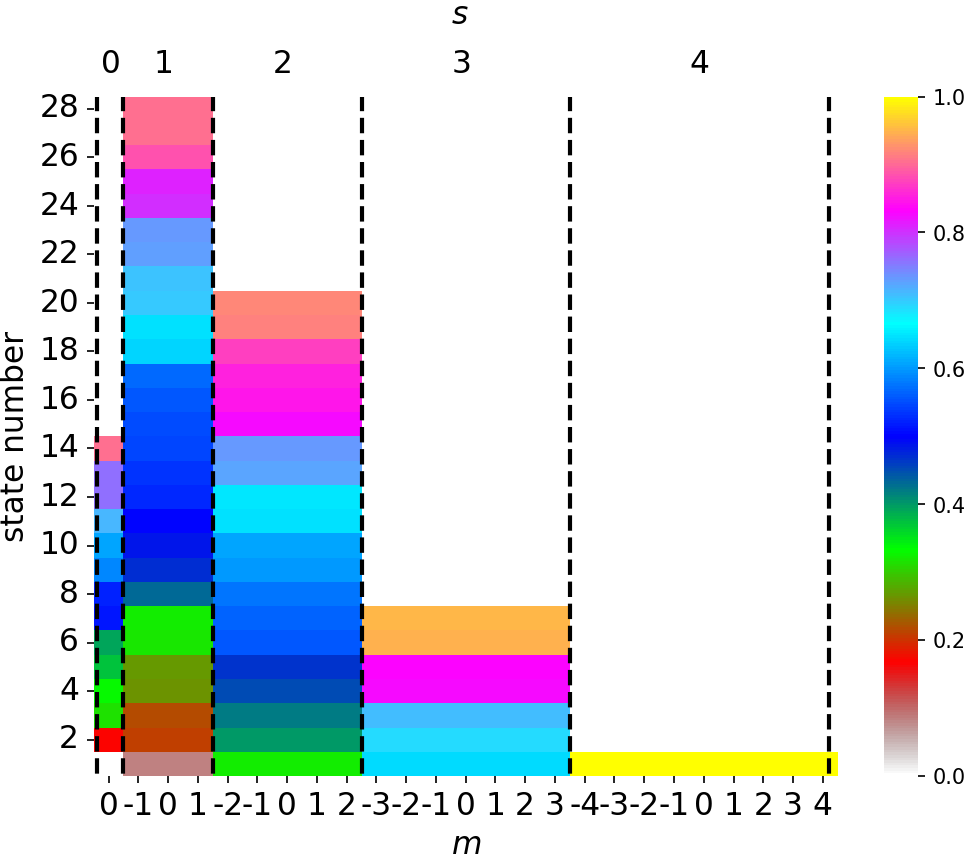} 
  (b)\includegraphics[width=0.4\textwidth]{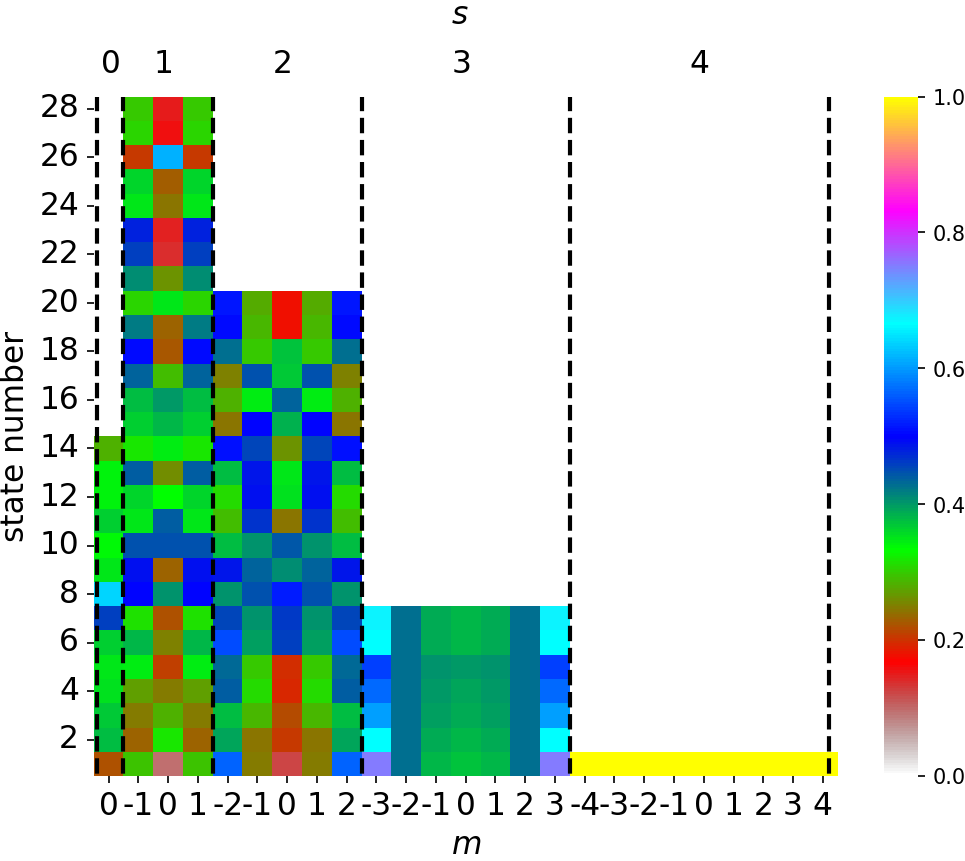}
  %(c)\includegraphics[width=0.4\textwidth]{\figdir/L_8-eigenvalues-structure-sz_0.png} 
  %(d)\includegraphics[width=0.4\textwidth]{\figdir/L_8-PR-structure-sz_0.png}
\caption{The (a) reduced energy $\epsilon$ and (b) normalized participation ratio $\mathcal{P}$ with values indicated by colors for system size $L=8$, highlighting the SU(2) symmetry structure via degeneracies in $\epsilon$ and $\mathcal{P}$. Both measures have been grouped into their different $s$ and $m$ sectors. For example, the first “0” on the horizontal axis stands for $s$=0, $m$=0, the subsequent triple “-1”, “0” and “1” corresponds to $s$=1 and so on. The vertical axis stands for the number of states\revision{, e.g.,} $14$ states have $s$=0 and $m$=0. The random exchange coupling in this case follows  $p_0(J)$ with $\Delta/J_0=0.2$. %(c) Reduced energy $\epsilon$ and (d) participation ratio $\mathcal{P}$ when $m$=0. Here, the horizontal axis stands for $s$.
} 
\label{fig-structure}
\end{figure*}
%%%%%%%%%%%%%%%%%%%%%%%%%%%%%%%%%%%%%%%%%%%%%%%%%%%%%%%%%%%%%%%%%%%%%%%%%%%%%%

In Fig.\ \ref{fig-structure}, we show the energy structure of eigenstates and their $\mathcal{P}$ values for $p_0(J)$. 
% %%%%%%%%%%%%%%%%%%%%%%%%%%%%%%%%%%%%%%%%%%%%%%%%%%%%%%%%%%%%%%%%%%%%%%%%%%%%%%
We can see that in each $s$ sector, the eigenstates are energetically degenerate while $\mathcal{P}(s,m)$ takes identical values for $\pm m$ pairs only but has overall quite different values. Hence $\mathcal{P}(s,m)= \mathcal{P}(s,-m)$.
A similar behaviour is found for $S_E$ and $\lambda$  as these are also related to the symmetry properties of the states in Fock space. 
\revision{A further peculiarity of the model lies in the pronounced asymmetry between the Fock-space localization properties of the anti-ferromagnetic ground state and the ferromagnetic spin distributions at the upper spectral edge. As shown in Fig.\ \ref{fig-structure}, this leads to, e.g., high $\mathcal{P}$ values for large $s$. As we will show later, the asymmetry in the Fock-space localization is also pronounced in $S_E$ and $\lambda$. This further distinguishes our model from the site-disordered ones \cite{Luitz2015Many-bodyChain,Naldesi2016DetectingQuenches}.} 

\begin{figure*}[tb]
 (a) \includegraphics[width=0.45\textwidth]{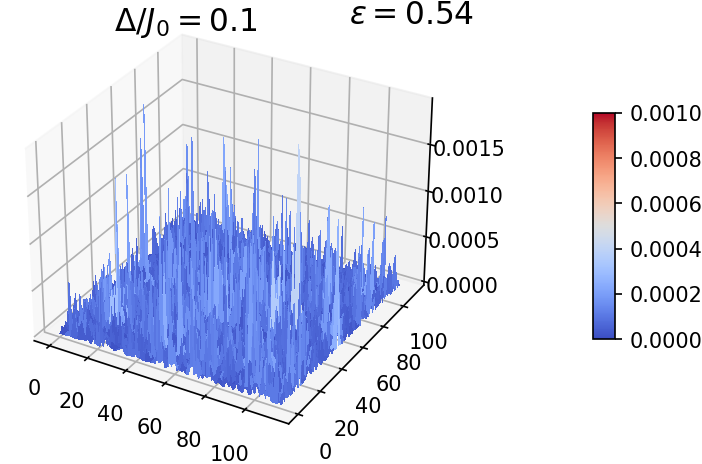}
 (b) \includegraphics[width=0.45\textwidth]{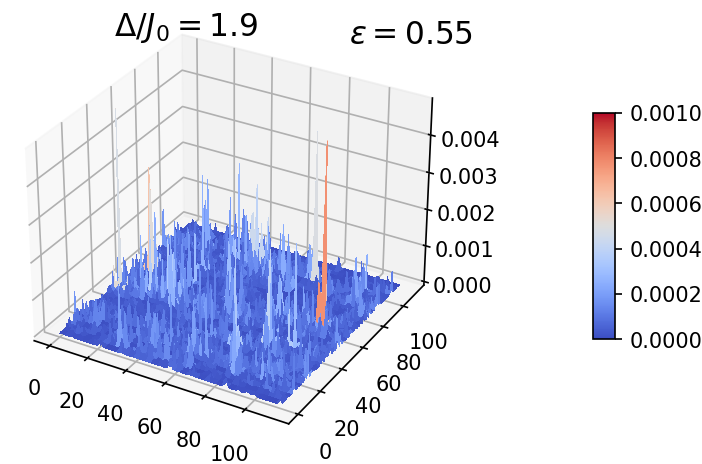}\\
 (c) \includegraphics[width=0.45\textwidth]{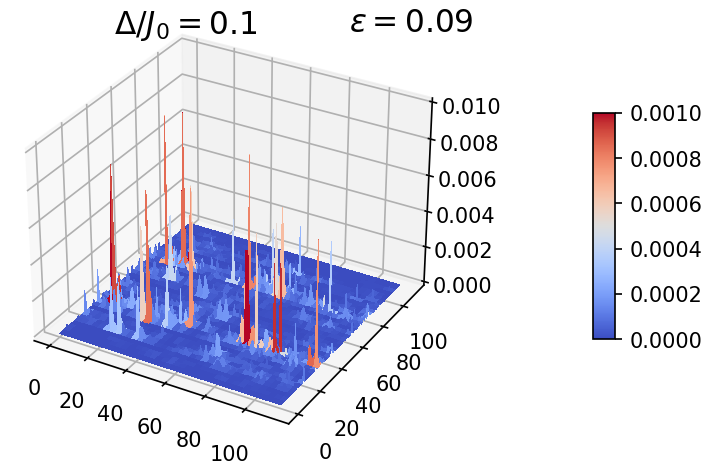}
 (d) \includegraphics[width=0.45\textwidth]{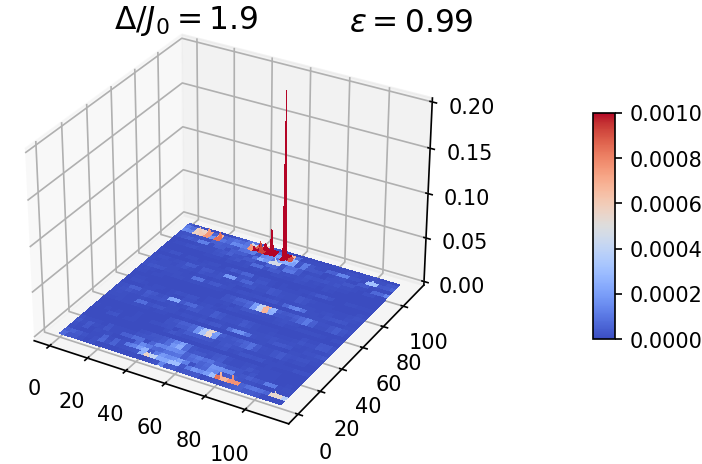}
\caption{
Wave function intensities $|\psi|^2$ for $p_0(J)$ with $L=16$ in the $m=0$ sector plotted as a two-dimensional $114^2 \sim {16 \choose 8}$ representation. 
The chosen states correspond to (a) $\Delta/J_0=0.1$ with $n=4000$, $\epsilon=0.54$ and $\mathcal{P}=0.23$, (b) $n=4000$, $\Delta/J_0=1.9$ with $\epsilon=0.55$ and $\mathcal{P}=0.13$, (c) $n=8$, $\Delta/J_0=0.1$ with $\epsilon=0.090$ and $\mathcal{P}=0.0054$, and (d) $n=12869$, $\Delta/J_0=1.9$ with $\epsilon=0.99$ and $\mathcal{P}=0.0010$ (all numbers given to two digit accuracy).
Hence the top row presents states near $\epsilon\sim 0.5$ while the bottom row gives the edge of the spectrum. The first column corresponds to weak disorder $\Delta/J_0\gtrsim 0$ that becomes much stronger in the last column.
} 
\label{fig-psi}
\end{figure*}

Before continuing to present the more quantitative results for our spectral and state statistics, let us briefly discuss intuitively how localization in Fock space can be visualized.
In Fig.\ \ref{fig-psi}, we show the wave function intensities $|\psi_{nsm}(f)|^2$ for $L=16$ computed for the $p_0(J)$ distribution at two different $\Delta/J_0$ values for the reduced energy, defined as $\epsilon={(E-E_{\text{min}})}/{(E_{\text{max}}-E_{\text{min}})}$, close to the central energy at $\epsilon=0.5$ and close to the edges of the spectrum at $\epsilon=0$ and $1$.
%%%%%%%%%%%%%%%%%%%%%%%%%%%%%%%%%%%%%%%%%%%%%%%%%%%%%%%%%%%%%%%%%%%%%%%%%%%%%%
%%%%%%%%%%%%%%%%%%%%%%%%%%%%%%%%%%%%%%%%%%%%%%%%%%%%%%%%%%%%%%%%%%%%%%%%%%%%%%
The Fock space for $L=16$ in the $m=0$ sector has $\text{dim}\mathcal{H}(0) = \binom{16}{8}= 12870$ states $\psi_{f}$. In the figures, we thus plot the states $\phi_f$, $f= 1, \ldots, \revision{12870}$, arranged as a two-dimensional grid of size $\revision{114 \times 114 = 12996}$ while setting the values $\psi_{12871}= \ldots = \psi_{12996}=0$.
We find that the more evenly extended spread of $|\psi_{nsm}(f)|^2$ values is for the state close to $\epsilon\approx 0.5$ (top row of Fig.\ \ref{fig-psi}) while close to the edge of the spectrum (bottom row), there are fewer $|\psi_{nsm}(f)|^2$ values but these are larger in value. Similarly, upon increasing the disorder, the tendency towards fewer but larger $|\psi_{nsm}(f)|^2$ values increases. 
Such behaviour is well-known from non-interacting disordered systems where it would be interpreted, when studied at much larger system sizes, as increased localization towards the edges of the spectrum and when increasing disorder \cite{Brandes2003AndersonRamifications}.
A crucial difference between both cases is of course that in the present case, spatially neighbouring states in Fock space, i.e.\ $\psi_{f}$ and $\psi_{f\pm 1}$ or $\psi_{f}$ and $\psi_{f\pm 144}$ are not necessarily connected by an off-diagonal matrix element and can correspond to a completely different structure of spin configurations in real space. Furthermore, there is no concept of boundary conditions.

%%%%%%%%%%%%%%%%%%%%%%%%%%%%%%%%%%%%%%%%%%%%%%%%%%%%%%%%%%%%%%%%%%%%%%%%%%%%%%
\subsection{\label{sec:results-statistics} Spectral statistics for the $p_j$ distributions}

%%%%%%%%%%%%%%%%%%%%%%%%%%%%%%%%%%%%%%%%%%%%%%%%%%%%%%%%%%%%%%%%%%%%%%%%%%%%%%
\subsubsection{The uniform distribution $p_0(J)$}

\begin{figure*}[tb]
    (a)\includegraphics[width=0.4\textwidth]{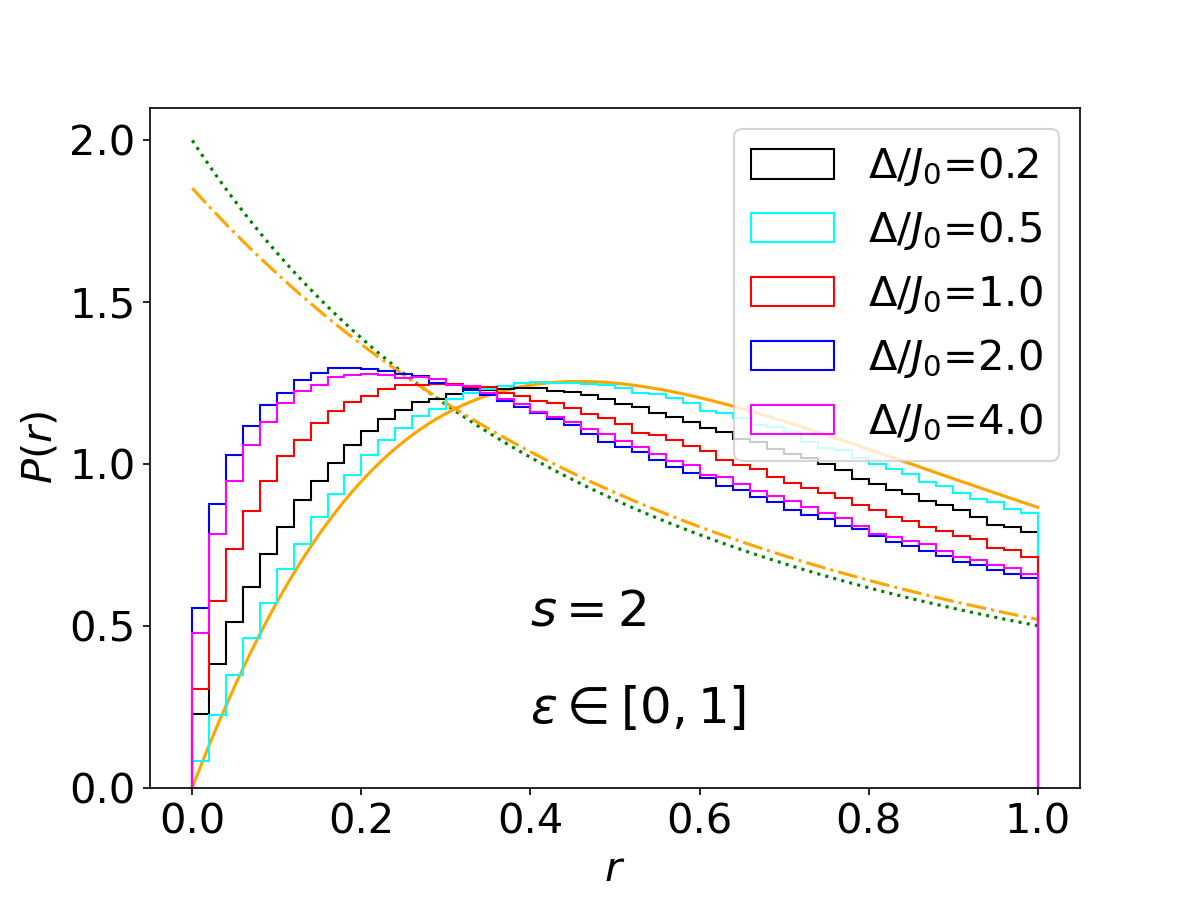}
    (b)\includegraphics[width=0.4\textwidth]{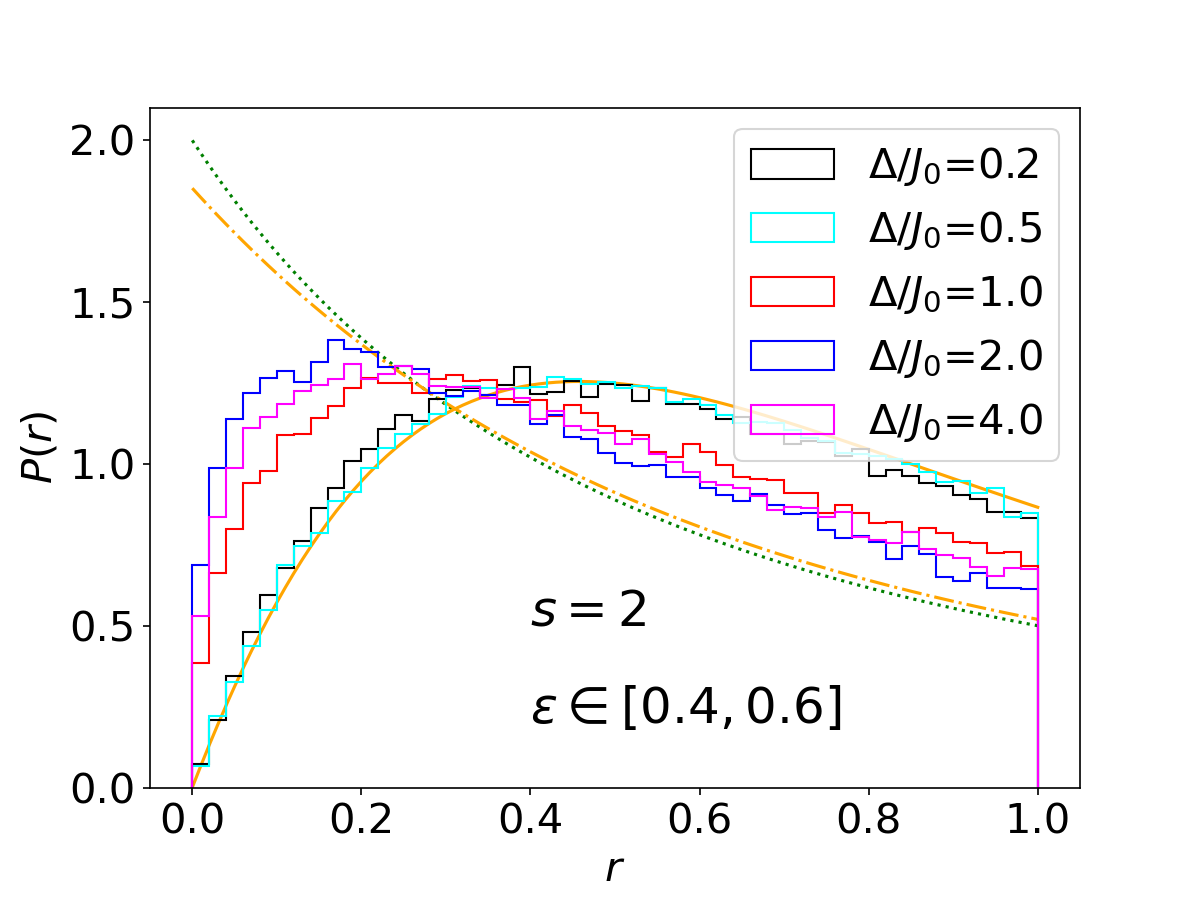}
    (c)\includegraphics[width=0.4\textwidth]{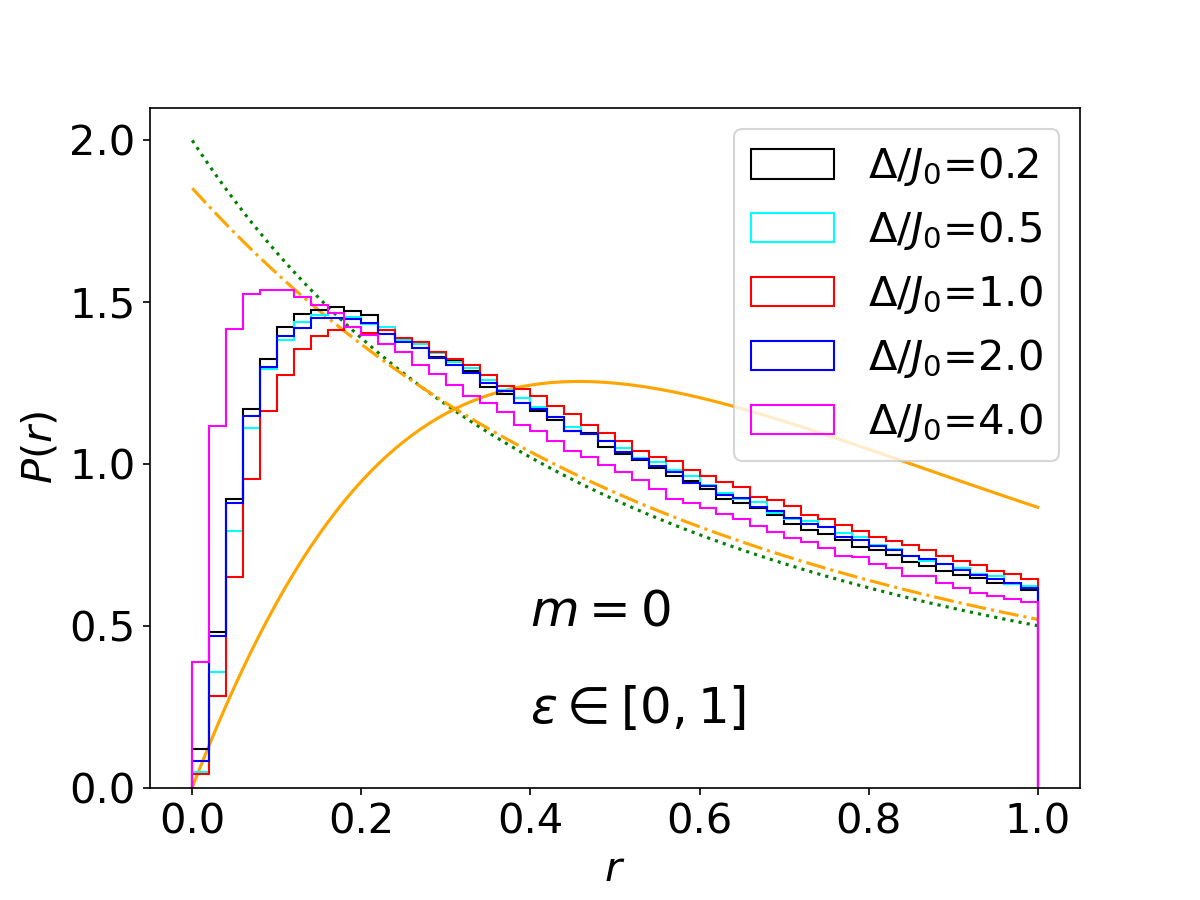}
    (d)\includegraphics[width=0.4\textwidth]{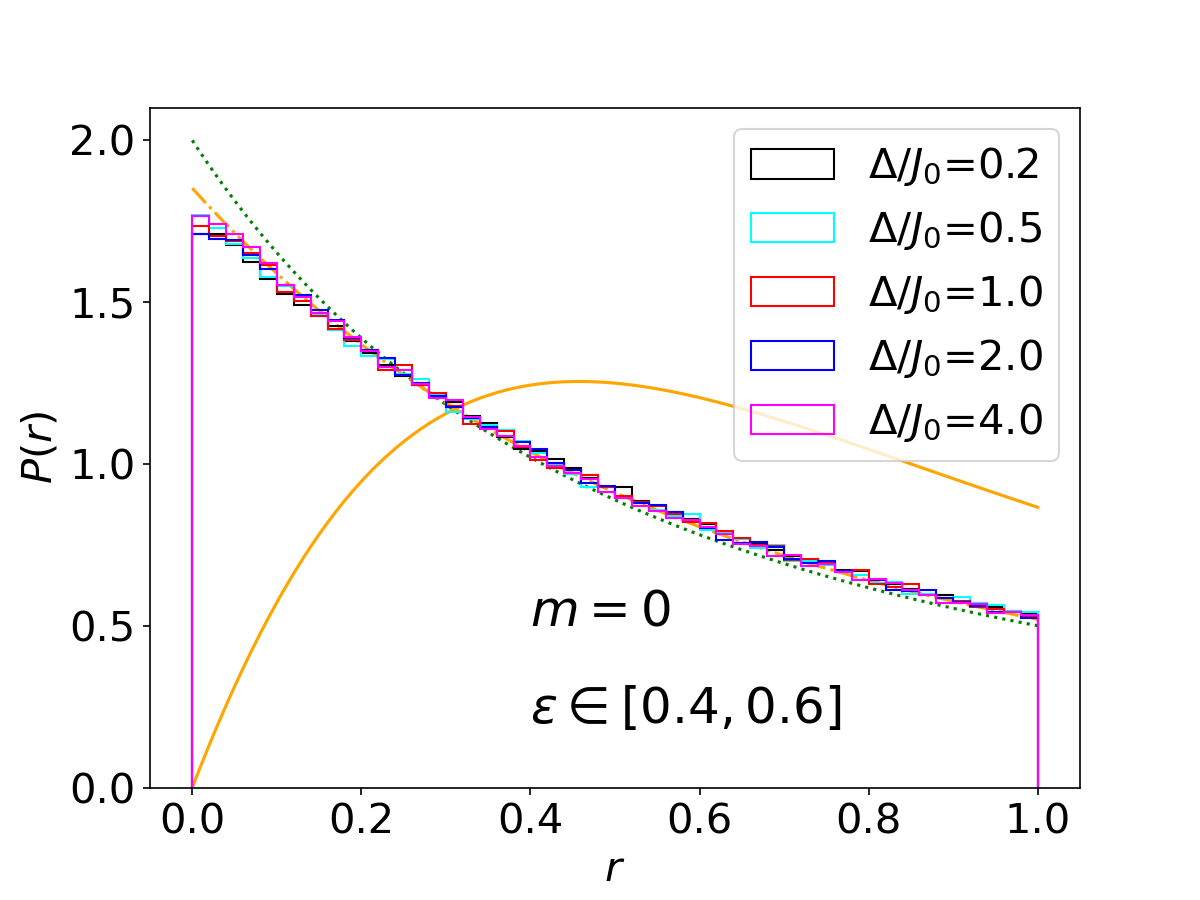}
    (e)\includegraphics[width=0.4\textwidth]{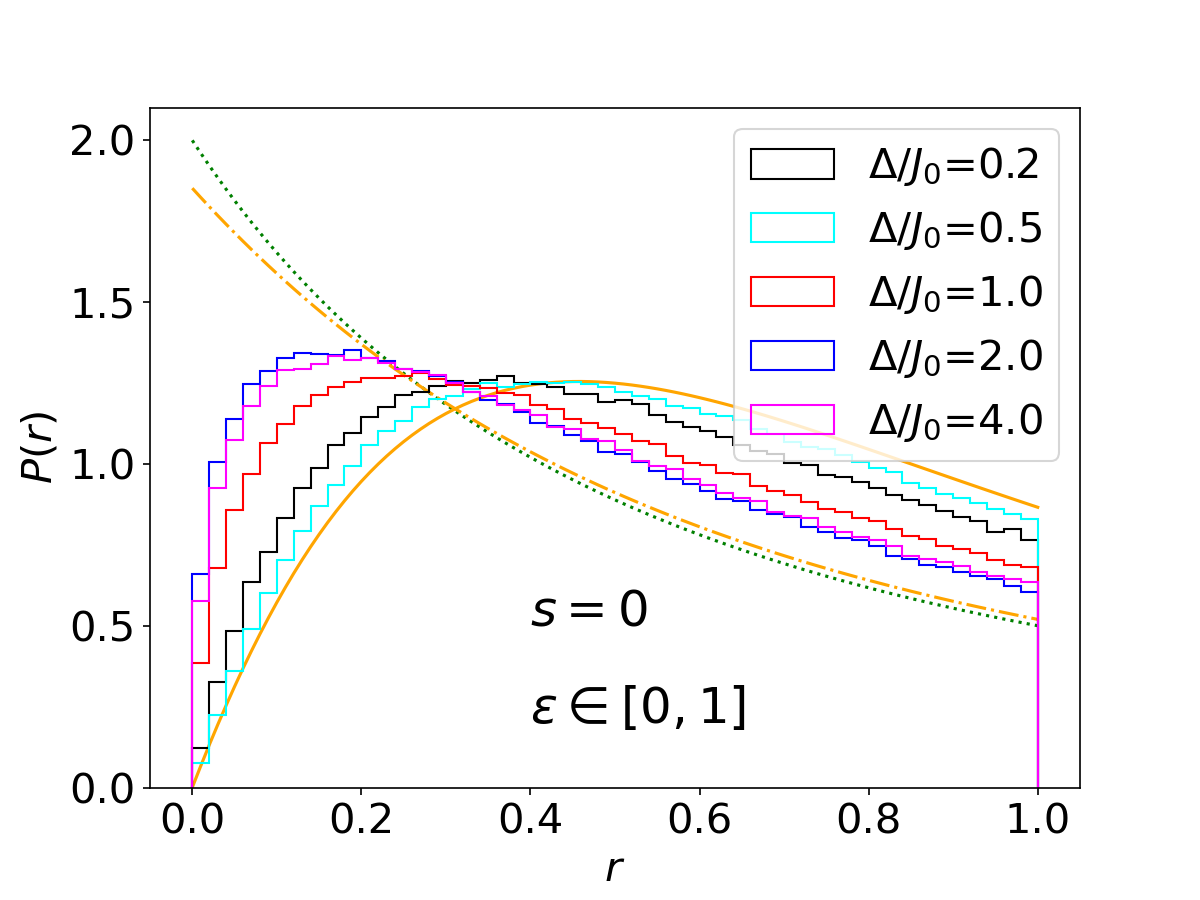}
    (f)\includegraphics[width=0.4\textwidth]{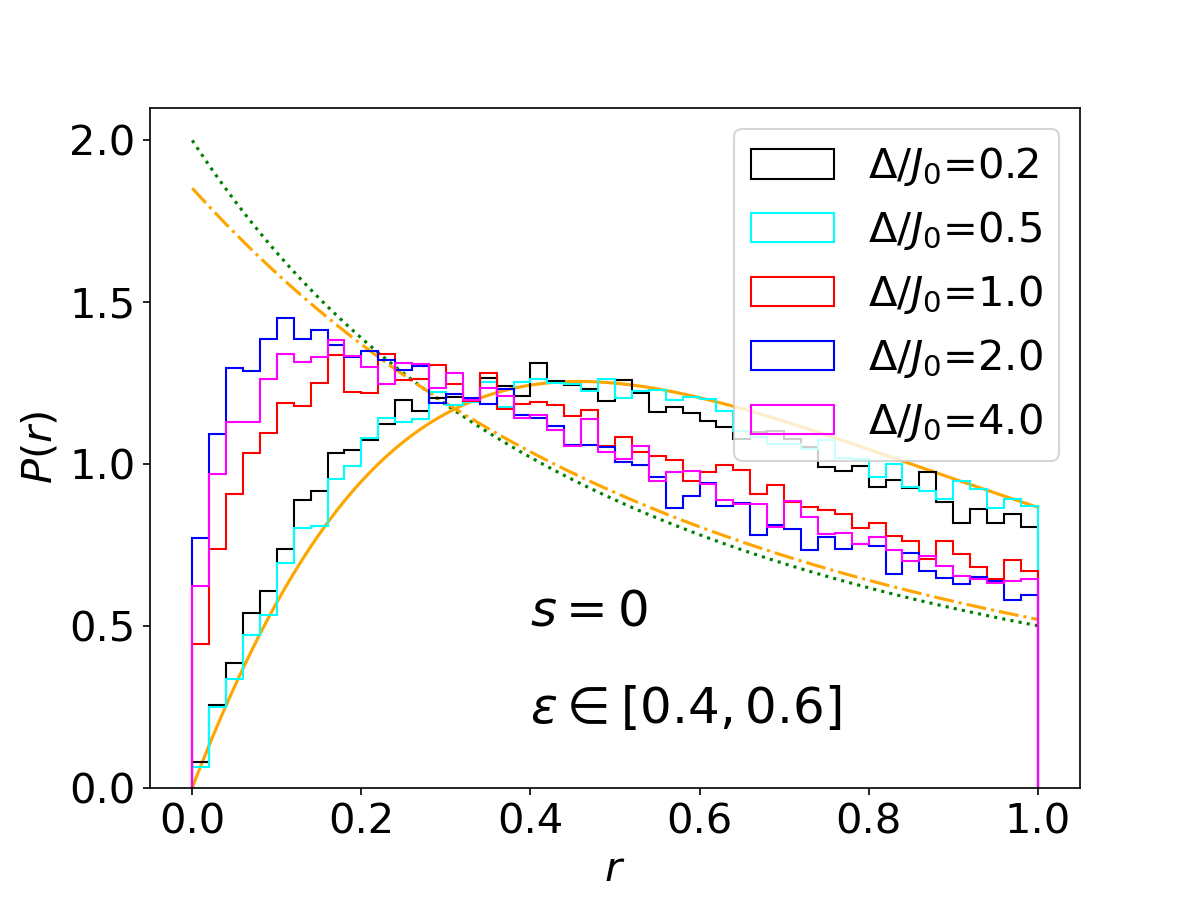}
\caption{Distribution $P(r)$ of consecutive-gap ratio when the system size is $L=16$ for disorder distribution $p_0(J)$. The top row corresponds to $s=2$ sector with (a) $\epsilon \in [0,1]$, (b) $\epsilon \in [0.4,0.6]$. The middle row (c)-(d) and the bottom row (e)-(f) follow the same order but give the $P(r)$ for $m=0$ sector and the $s=0$ sector respectively. The green dotted curve stands for $P_{\text{Poisson}}(r)$, the orange solid curve corresponds to $P_{\text{GOE}}(r)$ and the orange dash-dotted curve gives $P_{\text{GOE}}^{(9)}(r)$. } 
\label{fig-level-statistics-p0-all}
\end{figure*}

We plot the $P(r)$ for $p_0(J)$ in Fig.\ \ref{fig-level-statistics-p0-all} at various disorder strengths for both the full spectrum and in the central region $\epsilon \in [0.4,0.6]$. 
%%%%%%%%%%%%%%%%%%%%%%%%%%%%%%%%%%%%%%%%%%%%%%%%%%%%%%%%%%%%%%%%%%%%%%%%%%%%%%
%%%%%%%%%%%%%%%%%%%%%%%%%%%%%%%%%%%%%%%%%%%%%%%%%%%%%%%%%%%%%%%%%%%%%%%%%%%%%%
In doing so, we have a choice on how to combine the various sectors $(s,m)$.
We start by choosing the largest irreducible sector, which for $L=16$ contains $\text{dim}\mathcal{H}(2,0)=3640$ states as shown in Fig.\ \ref{fig-level-statistics-p0-all} (a+b). Taking all the states as done in (a), we find that the best agreement with $P_\text{GOE}$ is for $\Delta/J_0 = 0.5$ with good level repulsion. 
Moving towards the clean, and integrable limit $\Delta=0$, we see a small reduction in level repulsion for $\Delta/J_0 = 0.2$ while for strong disorders $\Delta/J_0 \geq 1$, level repulsion is progressively more decreased. 
When concentrating on the central part of the spectrum in (b) we find that these tendencies remain. While the agreement with $P_\text{GOE}$ for $\Delta/J_0 = 0.5$ is even better than before, level repulsion for $\Delta/J_0 \geq 1$ is at least equally strong. We also note that with $\sim 65\%$ fewer energies in this energy range, the statistics for $P(r)$ are somewhat deteriorated when compared to Fig.\ \ref{fig-level-statistics-p0-all}(a).
Overall, these results could be interpreted as suggesting a different behaviour for \revision{$\Delta/J_0 < 1$ and $\Delta/J_0 > 1$}.

We now stay in the $m=0$ sector, but analyse the $P(r)$ for all energies $E_n$ corresponding to the \revision{$9$ sectors $s_{\text{tot}}=0, 1, \ldots, 8$}, i.e.\ we combine $9$ irreducible blocks. Let us start our discussion with Fig.\ \ref{fig-level-statistics-p0-all}(d), i.e.\ the central spectral region. As we can see from the panel, the curves for practically all $\Delta/J_0$ values fall unto a single line. This line does not follow $P_\text{GOE}$, but seems qualitatively close to $P_\text{Poisson}$. However, as expected, the best agreement is with $P^{(9)}_\text{GOE}$.
In panel (c), we see that none of the chosen $\Delta/J_0$ values lead to go similarly good agreement with $P^{(9)}_\text{GOE}$. Clearly, the statistics close to the spectral edges must deviate from GOE behaviour more than in the central region. 
We note that instead of combining the $E_n$'s, we could have also computed the $r_n$'s in each sector and then studied their $P(r)$. This was previously done in Ref.\ \cite{Siegl2023ImperfectChains} with similar results. 

To conclude the discussion for $P(r)$, we now study the $(0,0)$ sector. This is again a single irreducible sector. The results shown in Fig.\ \ref{fig-level-statistics-p0-all} (e+f) are very similar to the sector $(2,0)$ as presented in panels (a+b) albeit with slightly more fluctuations due to the reduced number of available $r$ values ($\text{dim}\mathcal{H}=1430$). We note that the $(0,0)$ sector was studied extensively in Ref.\ \cite{Saraidaris2024Finite-sizeChains} but for a different $p(J)$.

\begin{figure*}[tb]
    (a)\includegraphics[width=0.4\textwidth]{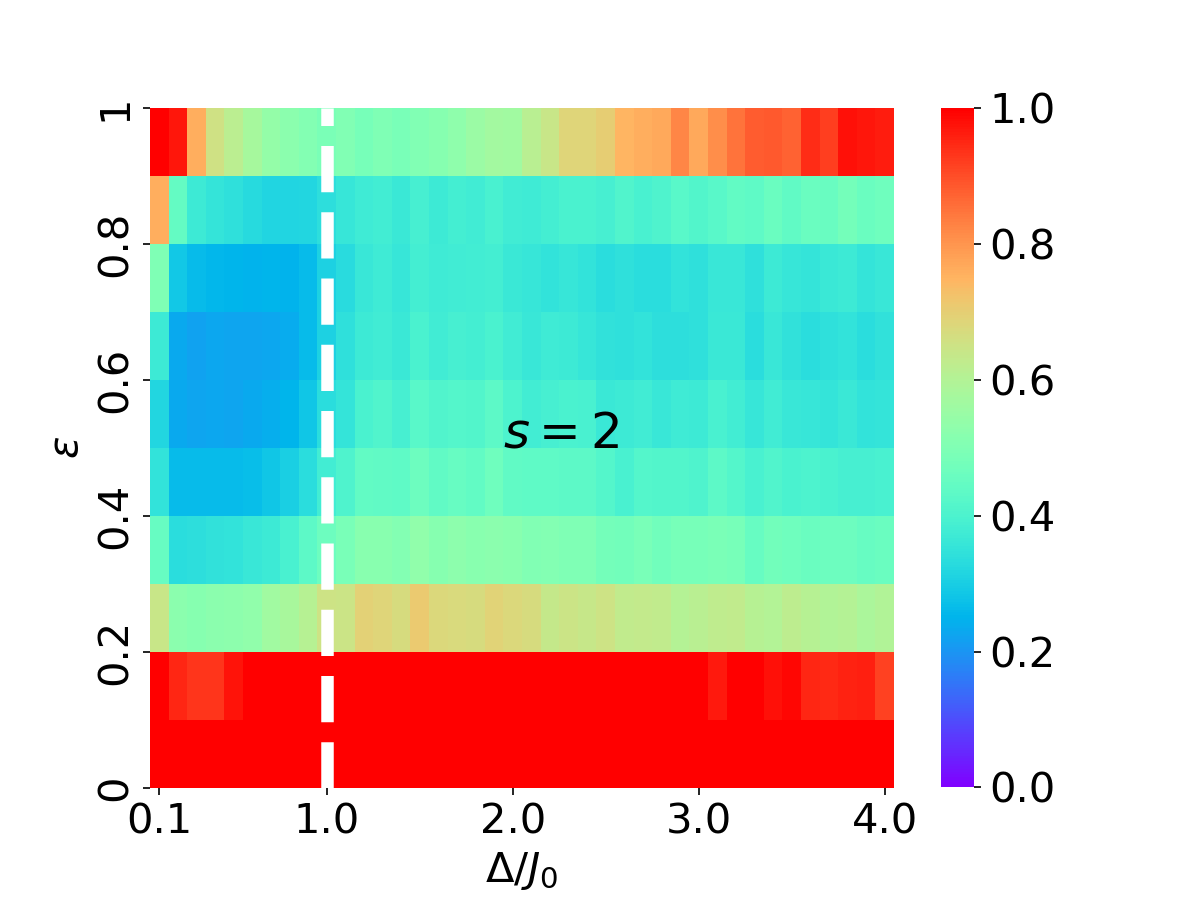}
    (b)\includegraphics[width=0.4\textwidth]{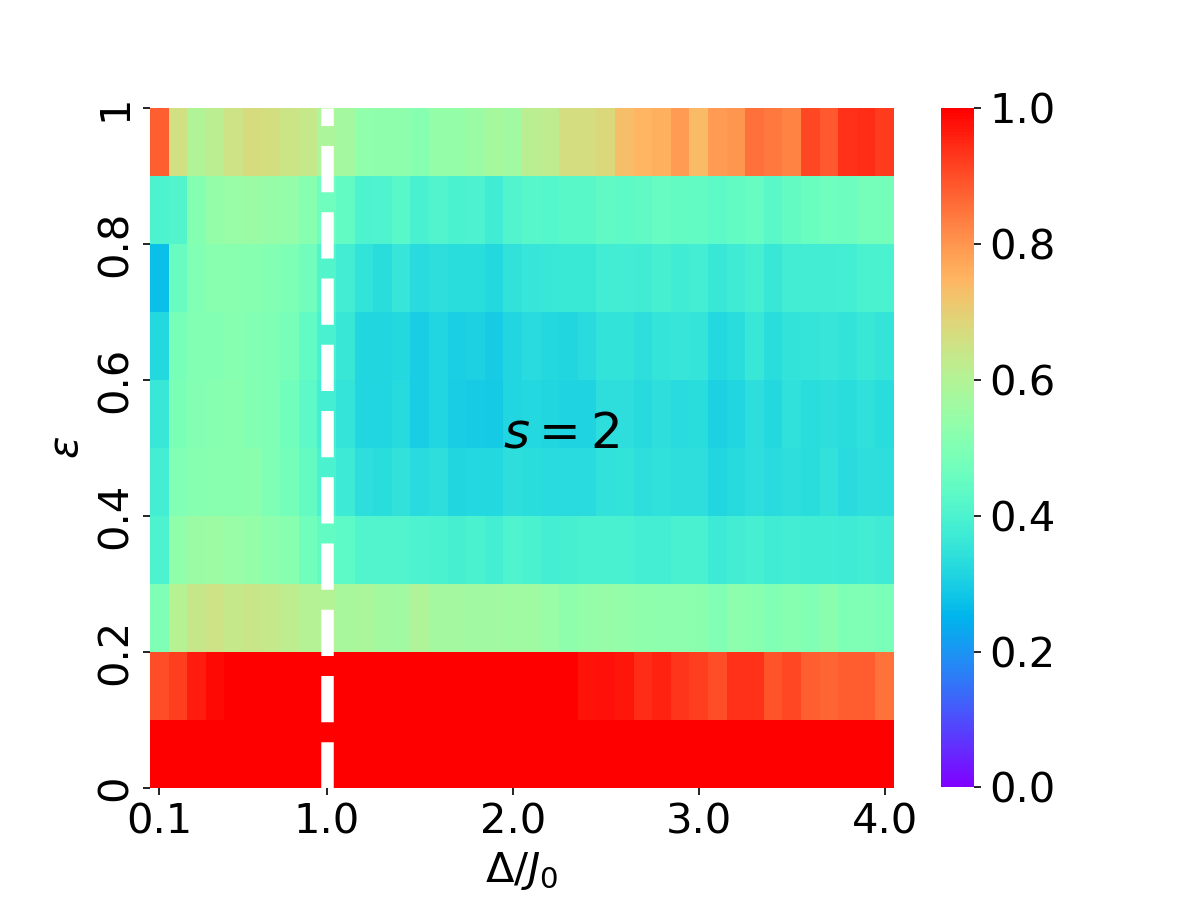}

    (c)\includegraphics[width=0.4\textwidth]{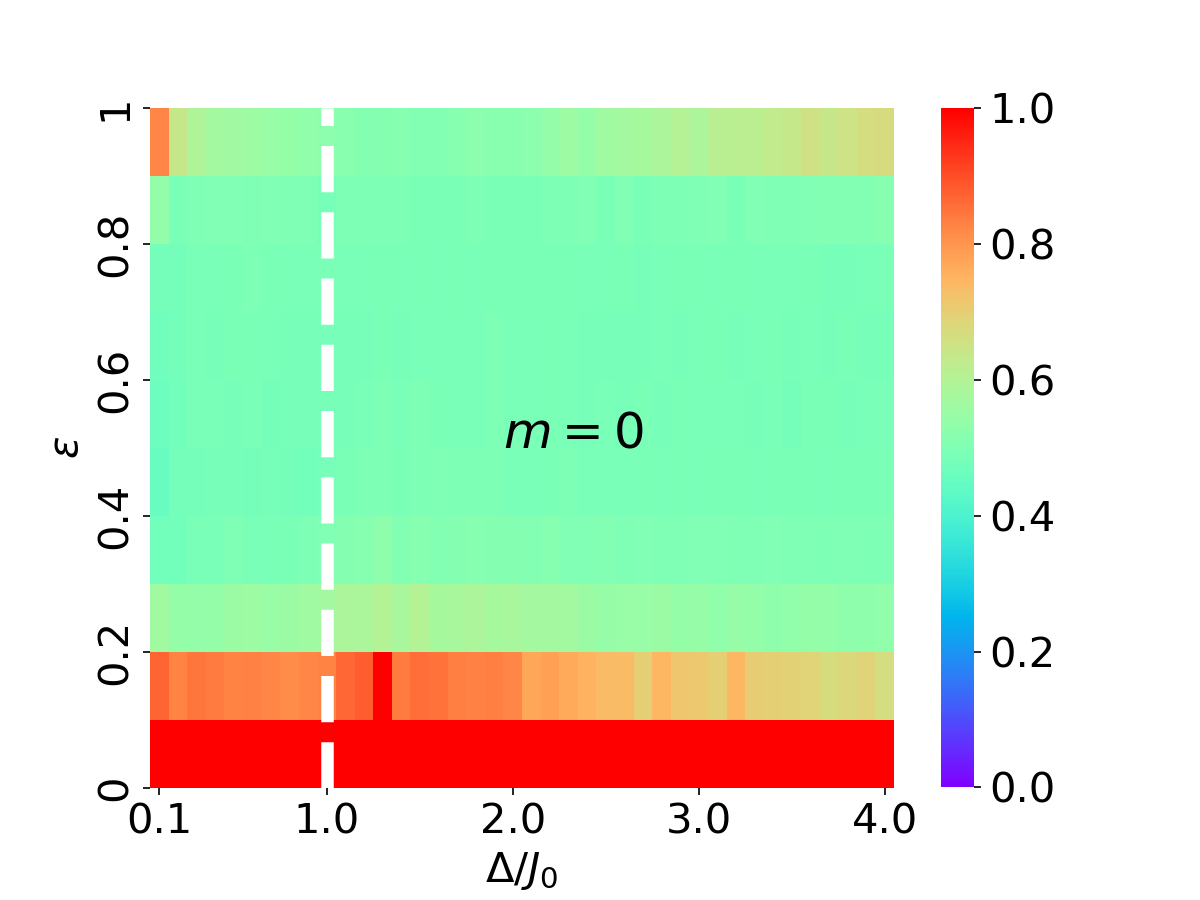}
    (d)\includegraphics[width=0.4\textwidth]{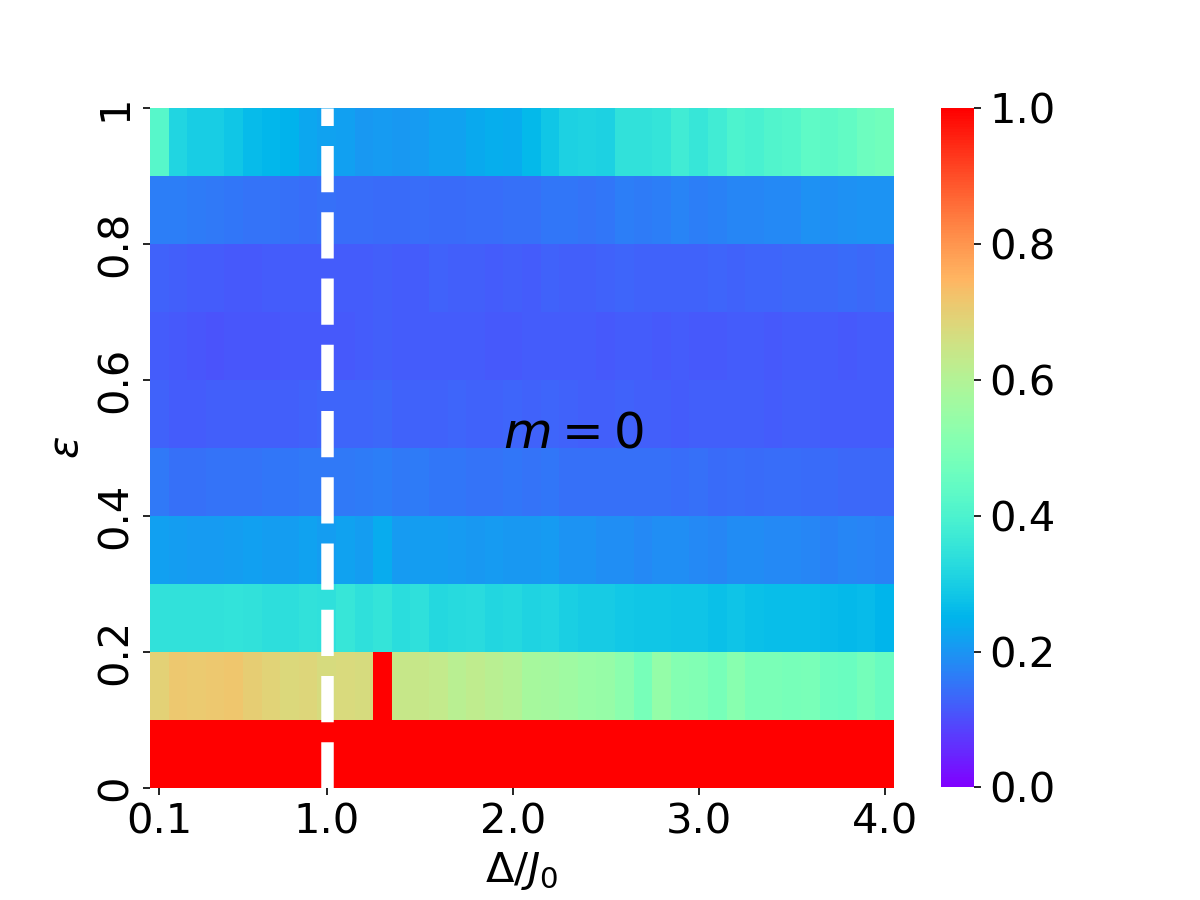}
    (e)\includegraphics[width=0.4\textwidth]{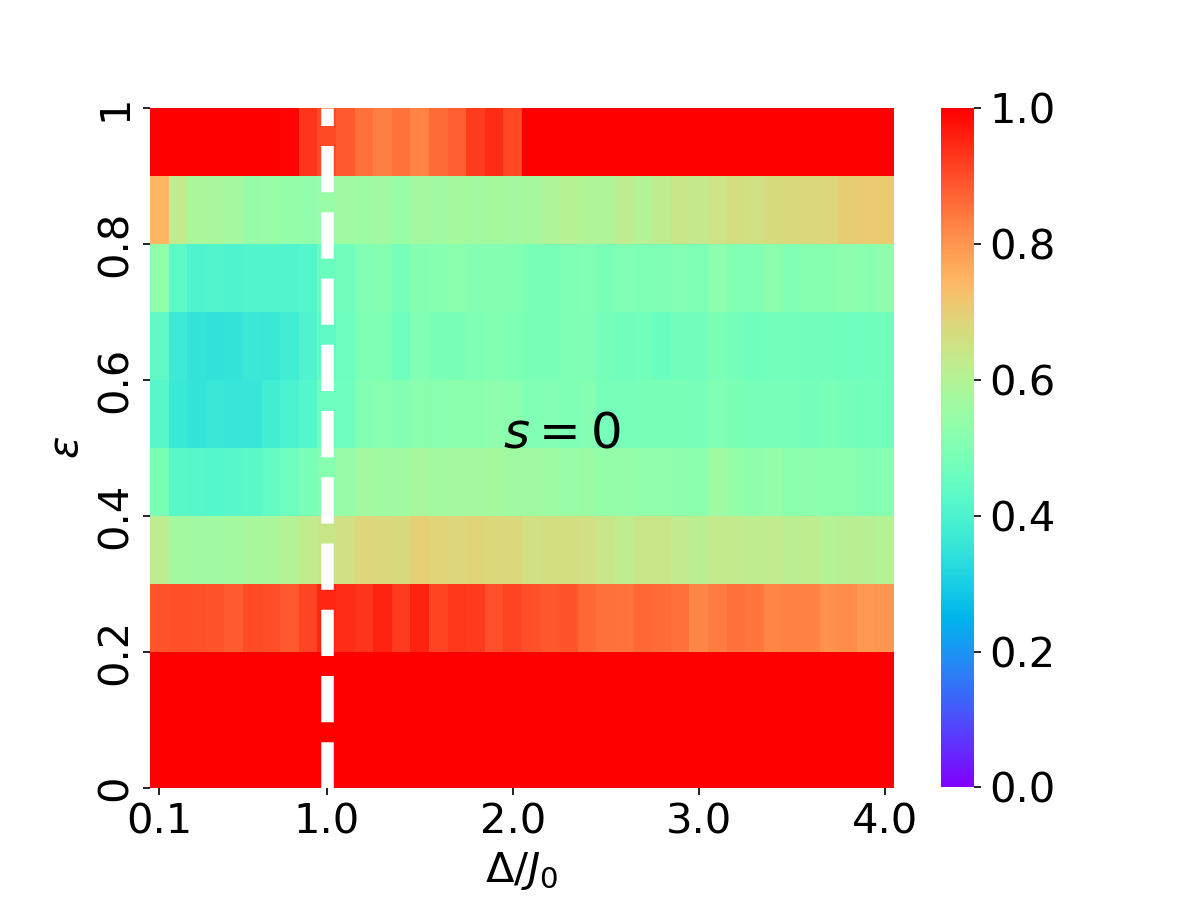}
    (f)\includegraphics[width=0.4\textwidth]{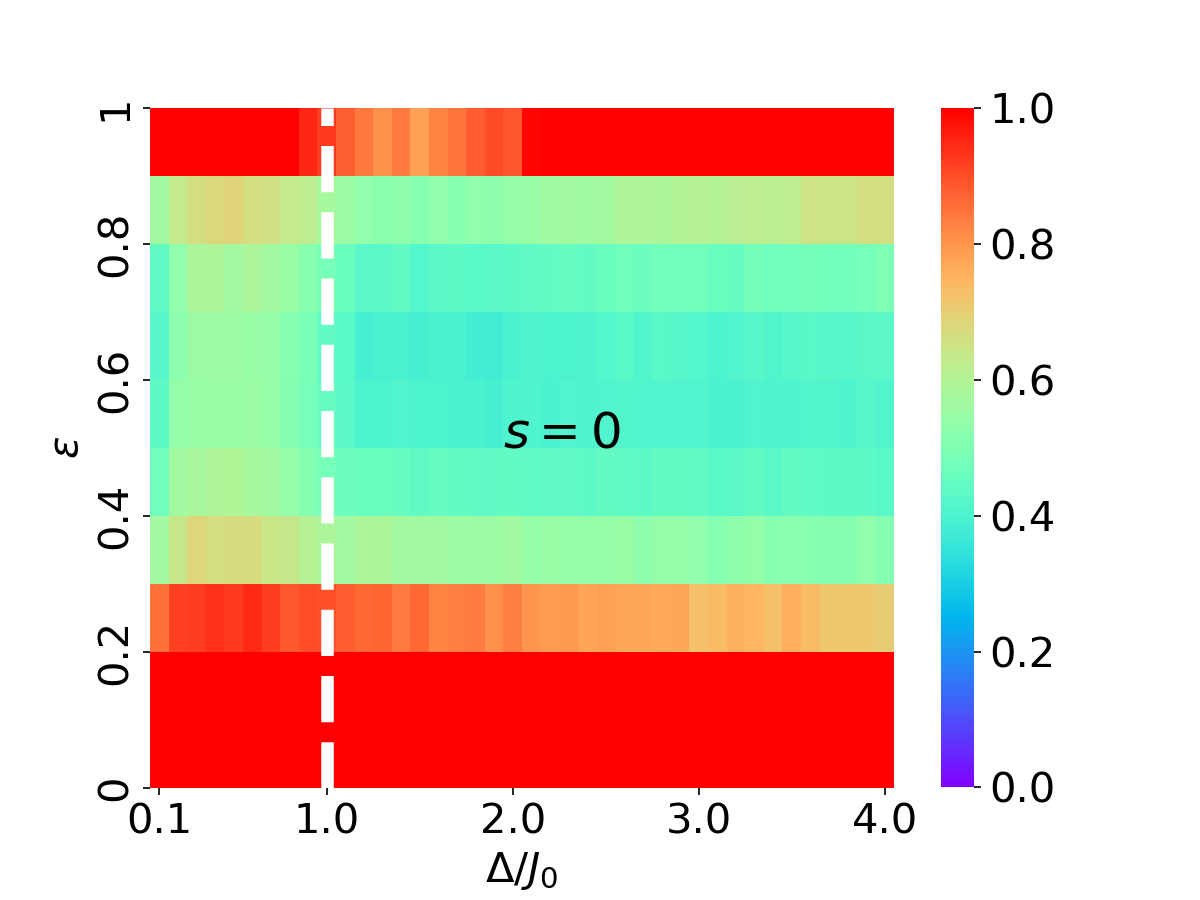}
\caption{Difference $\Delta P_{\text{GOE}}$ (left column) and $\Delta P_{\text{Poisson}}$ (right column) as a function of disorder strength $\Delta/J_0$ and reduced energy $\epsilon$ for disorder distribution $p_0(J)$. The top row gives (a) $\Delta P_{\text{GOE}}$, (b) $\Delta P_{\text{Poisson}}$ for the $s=2$ sector. The middle row (c)-(d) and bottom row (e)-(f) follow the same order but for the $m=0$ and the $s=0$ sectors, respectively. The vertical white-dashed line is at $\Delta/J_0=1.0$. } 
\label{fig-level-statistics-p0-diff}
\end{figure*}

Thus far, we have studied averages over large energy regions. Let us now instead compute the difference $\Delta P_{\text{GOE}}=\int_0^1|P(r)-P_\text{GOE}(r)|dr$ and $\Delta P_{\text{Poisson}}=\int_0^1|P(r)-P_\text{Poisson}(r)|dr$ for ten smaller subsets of the spectrum, i.e.\ $\delta\epsilon_e = [(e-1)/10, e/10]$ for $e= 1, \ldots 10$. 
In Fig.\ \ref{fig-level-statistics-p0-diff}, we show the results for $s=2$, $m=0$ and $s=0$, in the same order, i.e.\ from top to bottom, as in Fig.\ \ref{fig-level-statistics-p0-all}.
%%%%%%%%%%%%%%%%%%%%%%%%%%%%%%%%%%%%%%%%%%%%%%%%%%%%%%%%%%%%%%%%%%%%%%%%%%%%%%
%%%%%%%%%%%%%%%%%%%%%%%%%%%%%%%%%%%%%%%%%%%%%%%%%%%%%%%%%%%%%%%%%%%%%%%%%%%%%%
Clearly, in all cases, deviations from the two universal distributions are most pronounced towards the edges of the spectrum. The $m=0$ sector, comprised as discussed above of nine individual sectors, seems to follow $P_{\text{Poisson}}$ closely, but this is of course because $P_{\text{Poisson}} \sim P^{(9)}_{\text{GOE}}$ \cite{Vasseur2016Particle-holeModes, Giraud2022ProbingStatistics}.
Perhaps more interesting is to see in both the $s=2$ and the $s=0$ sectors that for $\Delta/J_0<1$ the best agreement of $P(r)$ is with $P_\text{GOE}$ while for $\Delta/J_0>1$ the agreement is better with $P_{\text{Poisson}}$. Aiming to find a possible qualitative separation between extended and localized phases at small and large disorders $\Delta/J_0$, respectively, $\Delta/J_0\approx 1$ hence appears as the most promising candidate for a possible phase boundary.

%%%%%%%%%%%%%%%%%%%%%%%%%%%%%%%%%%%%%%%%%%%%%%%%%%%%%%%%%%%%%%%%%%%%%%%%%%%%%%
\subsubsection{The distributions $p_1$ and $p_2$}

We now discuss the spectral statistics results for the disorder distributions $p_1(J)$ and $p_2(J)$. We only discuss and display those cases where the difference to $p_0(J)$ is clearly visible. A complete set of results can be found in the Supplement.

In Fig.\ \ref{fig-level-statistics-p1p2-alldiff} we show the $P(r)$'s for $s=2$ and their corresponding energy-resolved version similar to Figs.\ \ref{fig-level-statistics-p0-all} and \ref{fig-level-statistics-p0-diff}, respectively.
%%%%%%%%%%%%%%%%%%%%%%%%%%%%%%%%%%%%%%%%%%%%%%%%%%%%%%%%%%%%%%%%%%%%%%%%%%%%%%
\begin{figure*}[tb]
    (a)\includegraphics[width=0.4\textwidth]{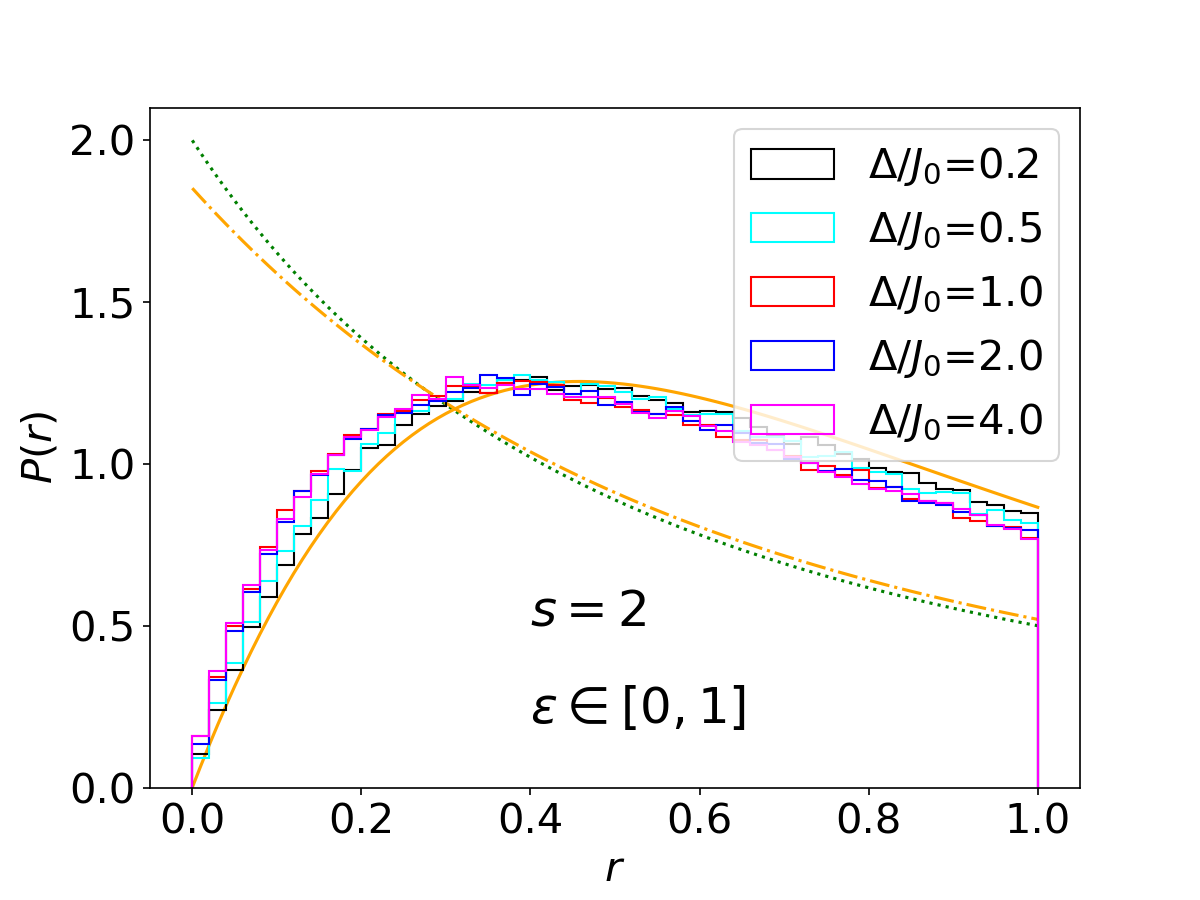}
    (b)\includegraphics[width=0.4\textwidth]{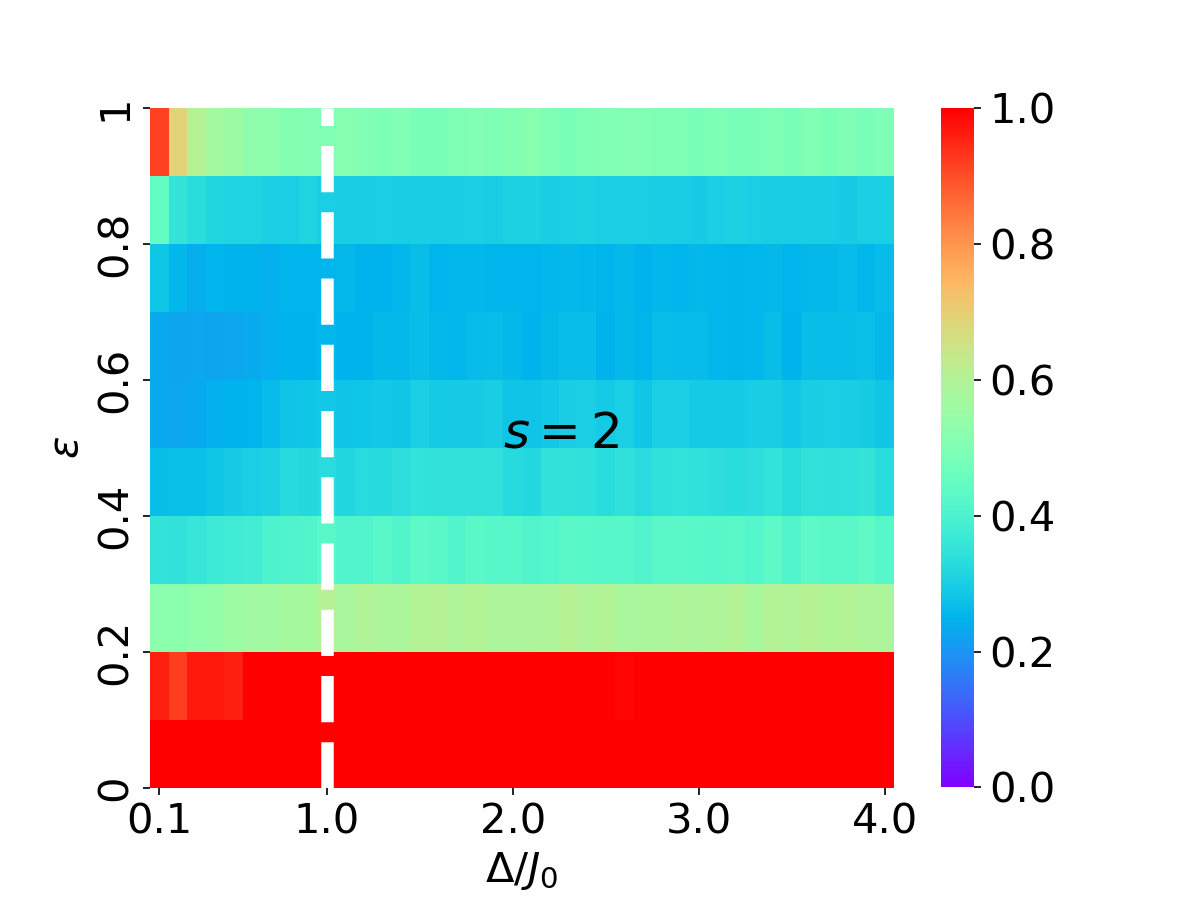}\\
    (c)\includegraphics[width=0.4\textwidth]{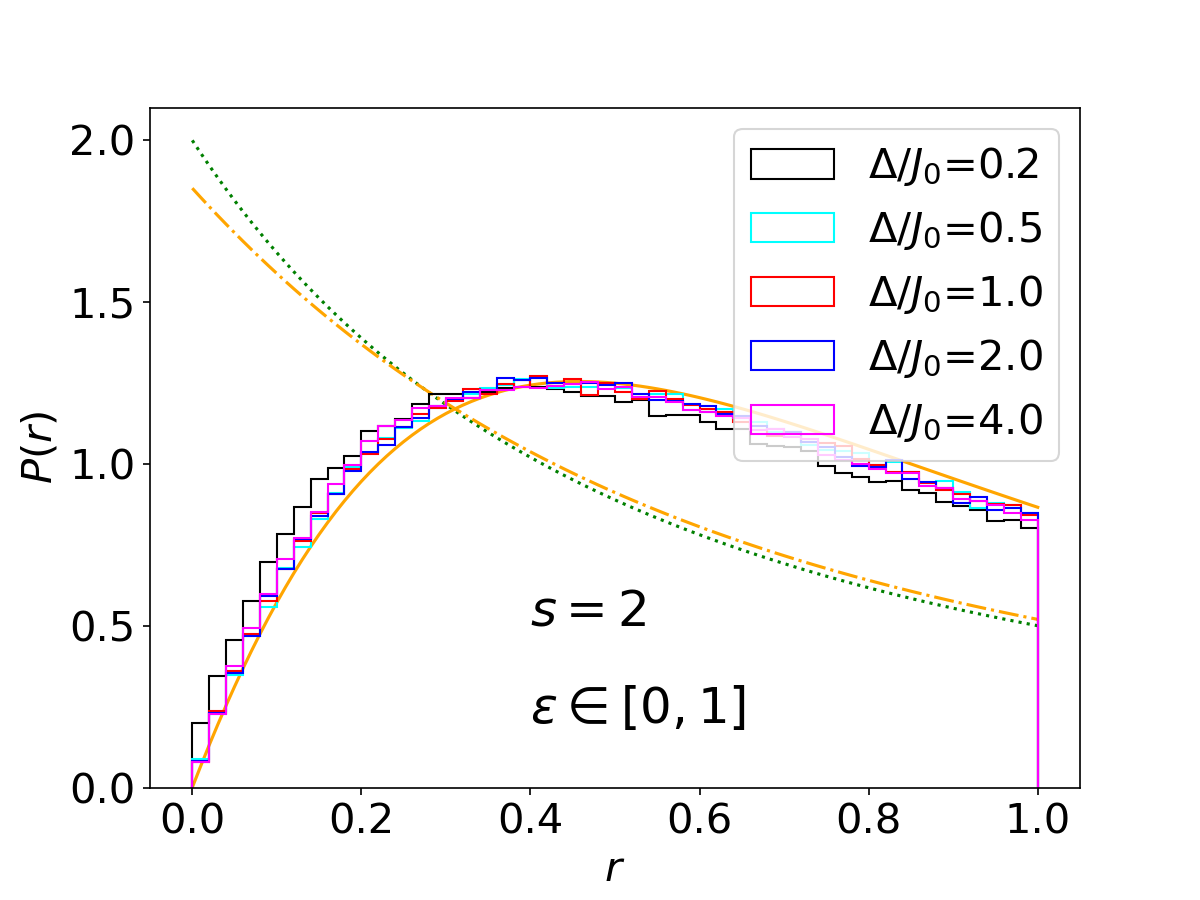}    
    (d)\includegraphics[width=0.4\textwidth]{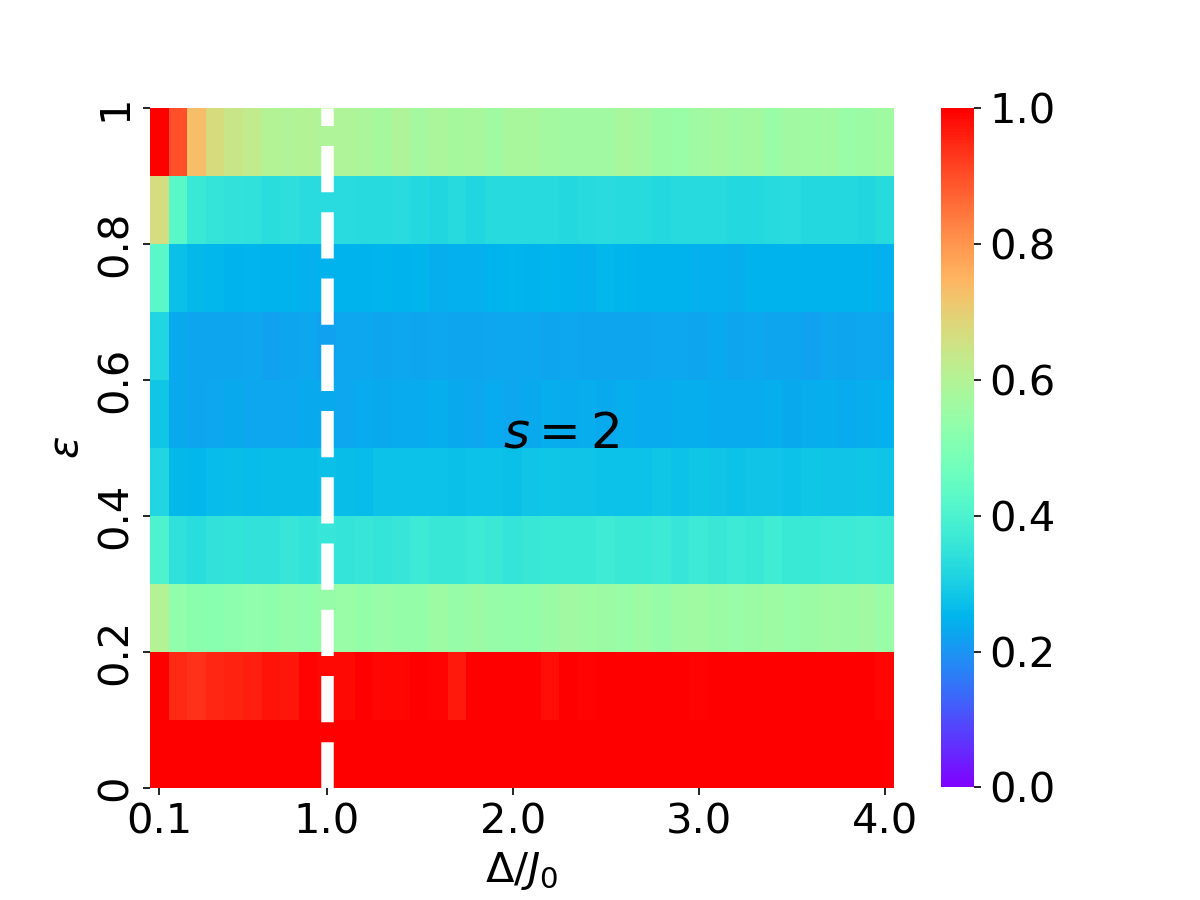}
\caption{Distribution $P(r)$ with $L=16$ and $\epsilon \in [0,1]$ in the $s=2$ sector for disorder distribution (a) $p_1(J)$ and (c) $p_2(J)$. Difference $\Delta P_{\text{GOE}}$ in terms of $\Delta/J_0$ and $\epsilon$ in the $s=2$ sector for disorder distribution (b) $p_1(J)$ and (d) $p_2(J)$. The colors of the curves for $P_{\text{Poisson}}(r)$, $P_{\text{GOE}}(r)$, $P_{\text{GOE}}^{(9)}(r)$ and $P(r)$ with different disorder strengths in (a) and (c) are the same as in Fig.\ \ref{fig-level-statistics-p0-all}. The position of the white-dashed vertical line in (b) and (d) is the same ($\Delta/J_0=1.0$) as in Fig.\ \ref{fig-level-statistics-p0-diff}.  } 
\label{fig-level-statistics-p1p2-alldiff}
\end{figure*}
%%%%%%%%%%%%%%%%%%%%%%%%%%%%%%%%%%%%%%%%%%%%%%%%%%%%%%%%%%%%%%%%%%%%%%%%%%%%%%
We find that for both $p_1(J)$ and $p_2(J)$, the overall agreement with $P_\text{GOE}$ is somewhat better than for $p_0$ for the full disorder range up to $\Delta/J_0=4$. This is true not only in the $s=2$ sector shown in Fig.\ \ref{fig-level-statistics-p1p2-alldiff}, but also for the $m=0$ and $s=2$ cases, be these energy-resolved or not.
\revision{Most importantly, the somewhat different features for $\Delta/J_0 < 1$ and $\Delta/J_0 > 1$ have now been smoothed} and no sharp change in behaviour can be seen when $\Delta/J_0 = 1$ is crossed. If there were a transition, it would have to be for $\Delta/J_0 > 4$ or develop for larger $L$.

We also note that the tendency towards more localization at the edges of the spectrum is retained for $p_1(J)$ and $p_2(J)$.

%%%%%%%%%%%%%%%%%%%%%%%%%%%%%%%%%%%%%%%%%%%%%%%%%%%%%%%%%%%%%%%%%%%%%%%%%%%%%%
\subsection{\label{sec:results-states} Measures of state spread in Fock space}

%%%%%%%%%%%%%%%%%%%%%%%%%%%%%%%%%%%%%%%%%%%%%%%%%%%%%%%%%%%%%%%%%%%%%%%%%%%%%%
\subsubsection{\label{sec:all-states} Exact diagonalization at $L=16$}

\begin{figure*}[tb]
 $\mathcal{P}$\qquad\qquad\qquad\qquad\qquad\qquad\qquad\qquad$S_E/L$ \qquad\qquad\qquad\qquad\qquad\qquad\qquad\qquad$\lambda$ \qquad\\
%\textbf{$P$ \hfill $S_E/L$ \hfill $\lambda$}
    (a)\includegraphics[width=0.3\textwidth]{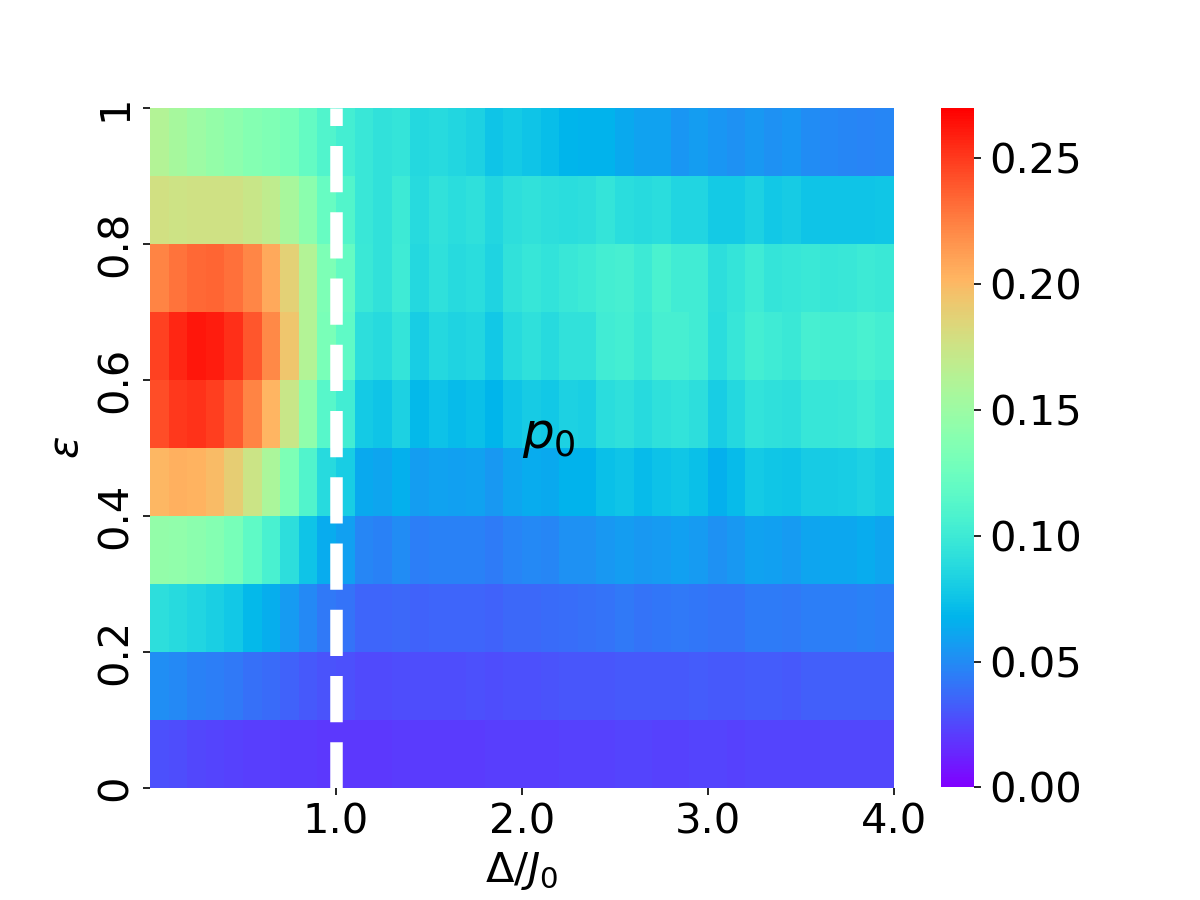}
    (b)\includegraphics[width=0.3\textwidth]{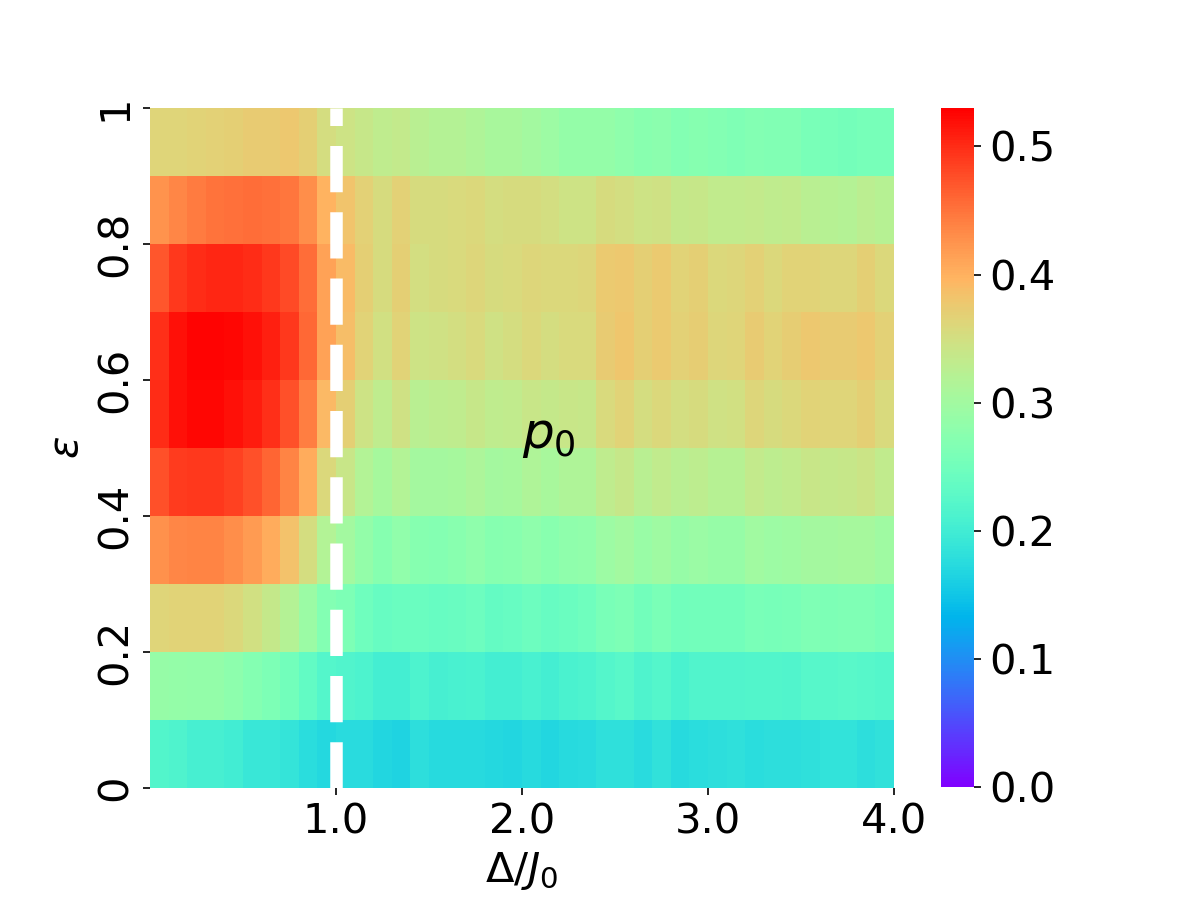}
    (c)\includegraphics[width=0.3\textwidth]{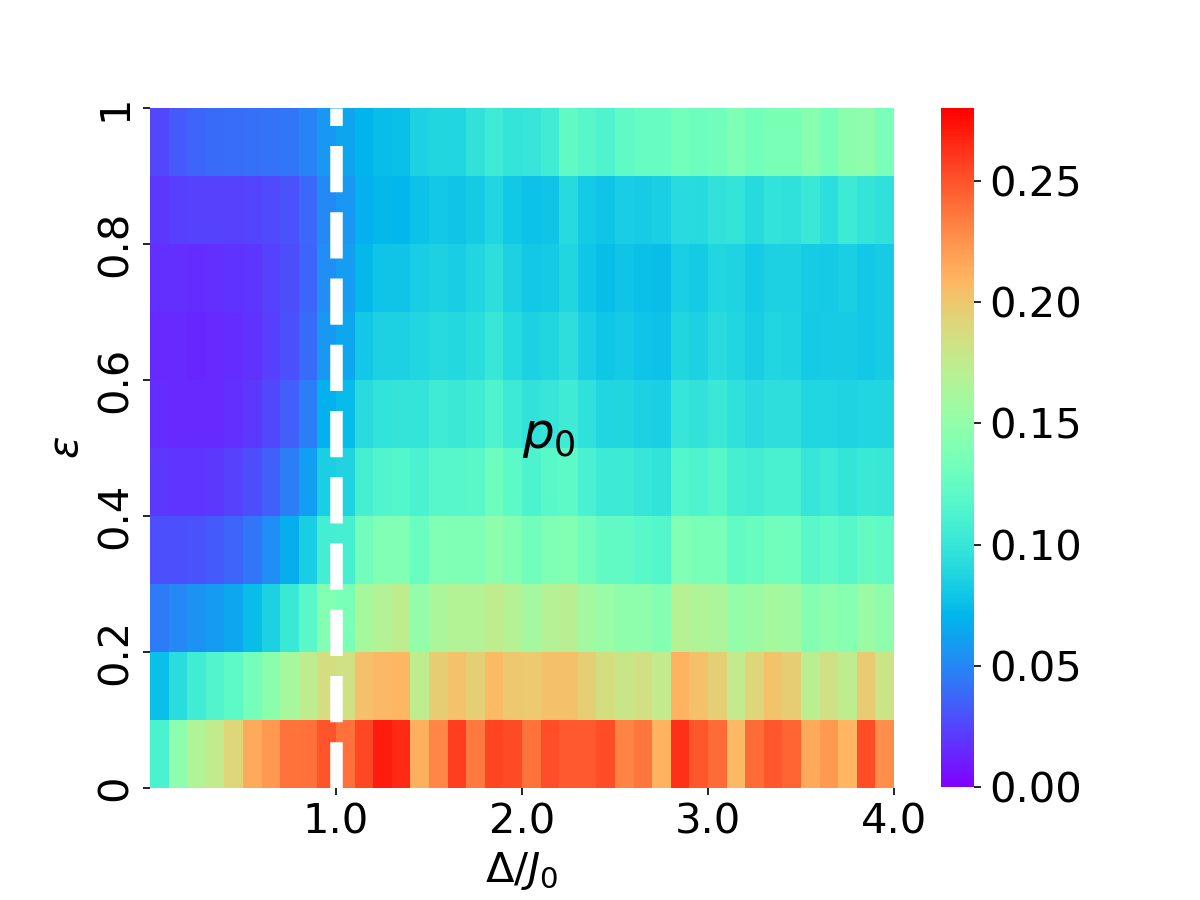}\\
    (d)\includegraphics[width=0.3\textwidth]{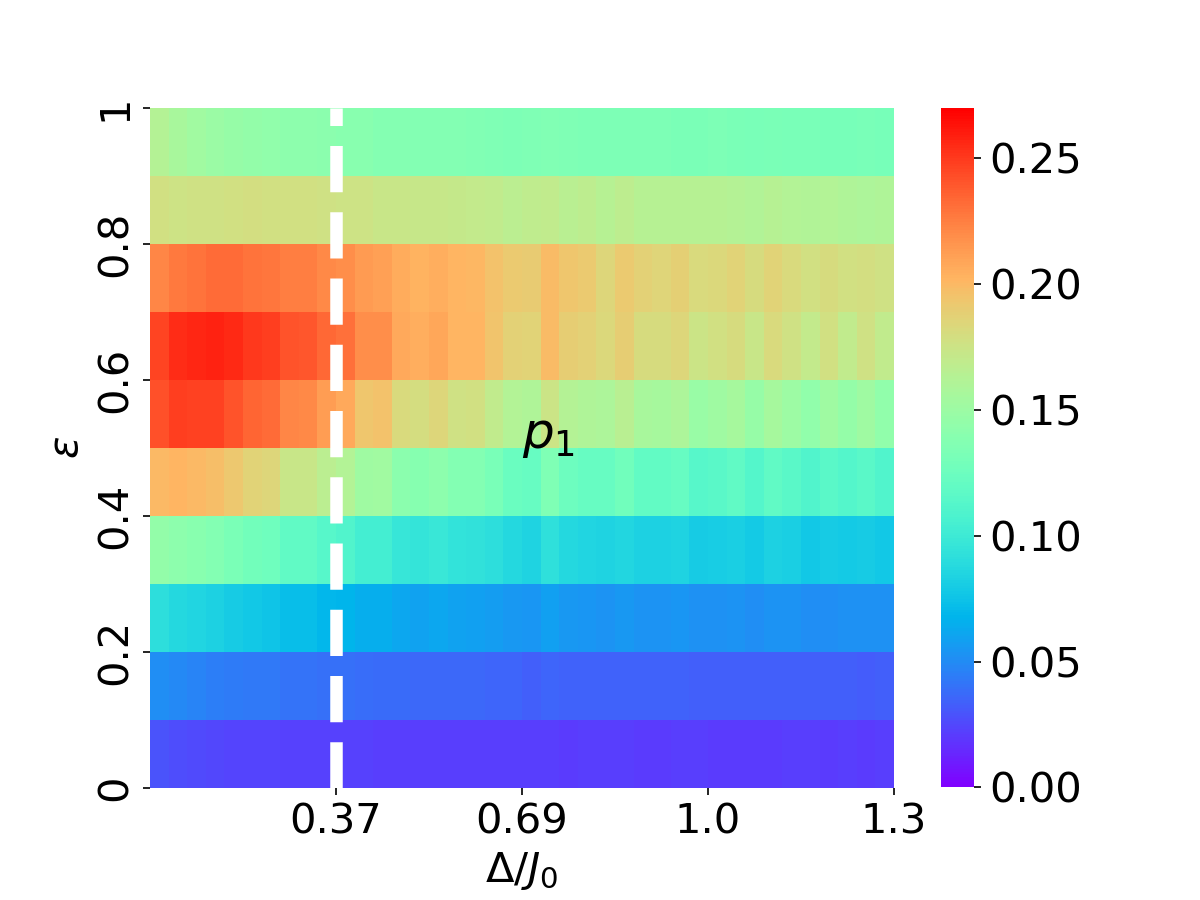}
    (e)\includegraphics[width=0.3\textwidth]{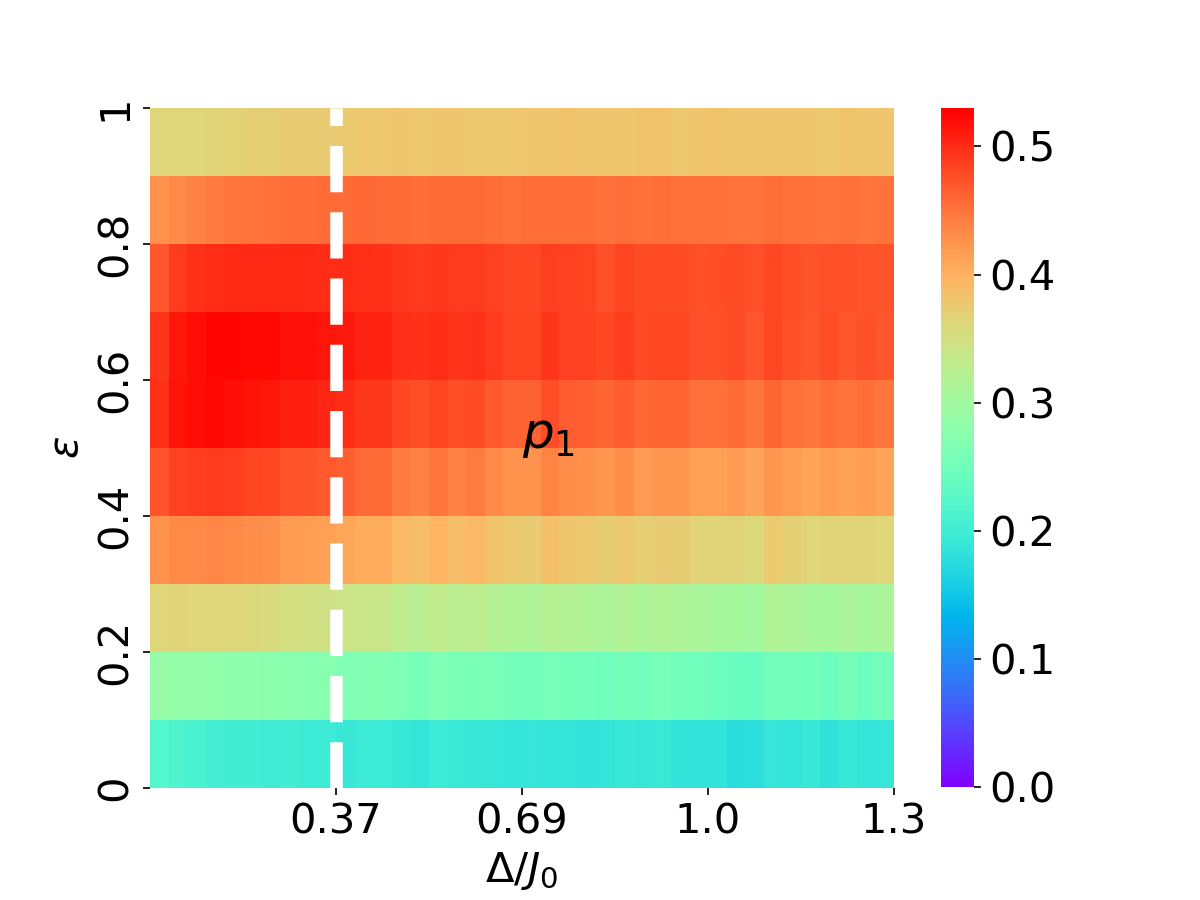}
    (f)\includegraphics[width=0.3\textwidth]{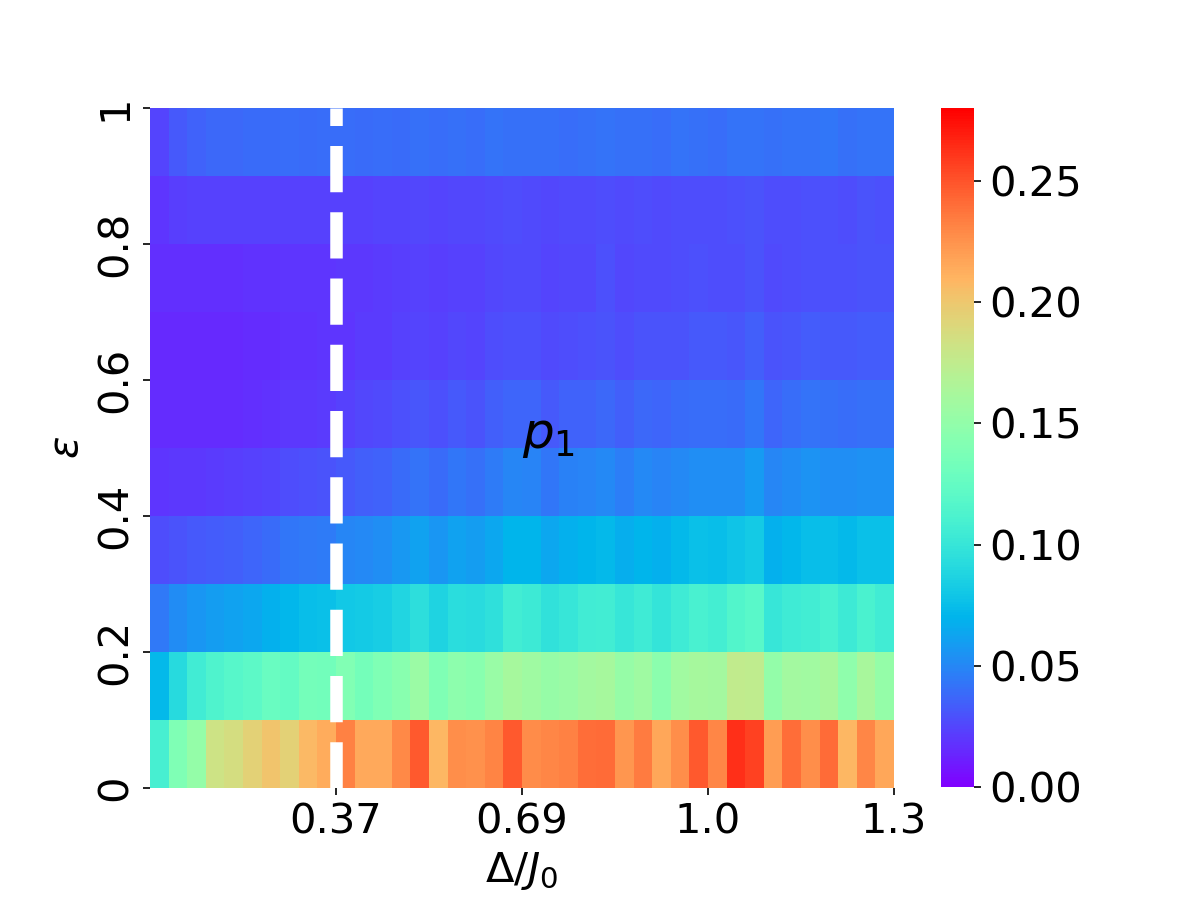}\\
    (g)\includegraphics[width=0.3\textwidth]{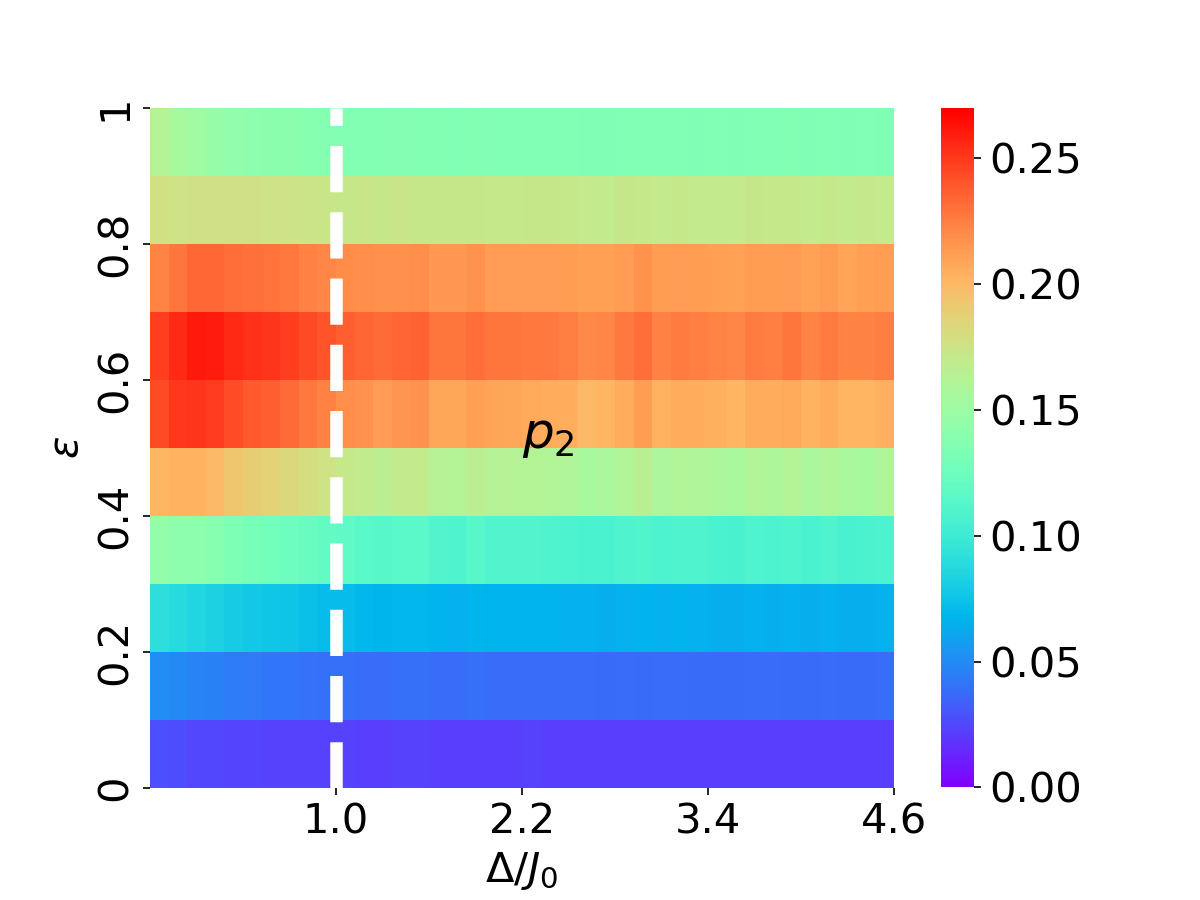}
    (h)\includegraphics[width=0.3\textwidth]{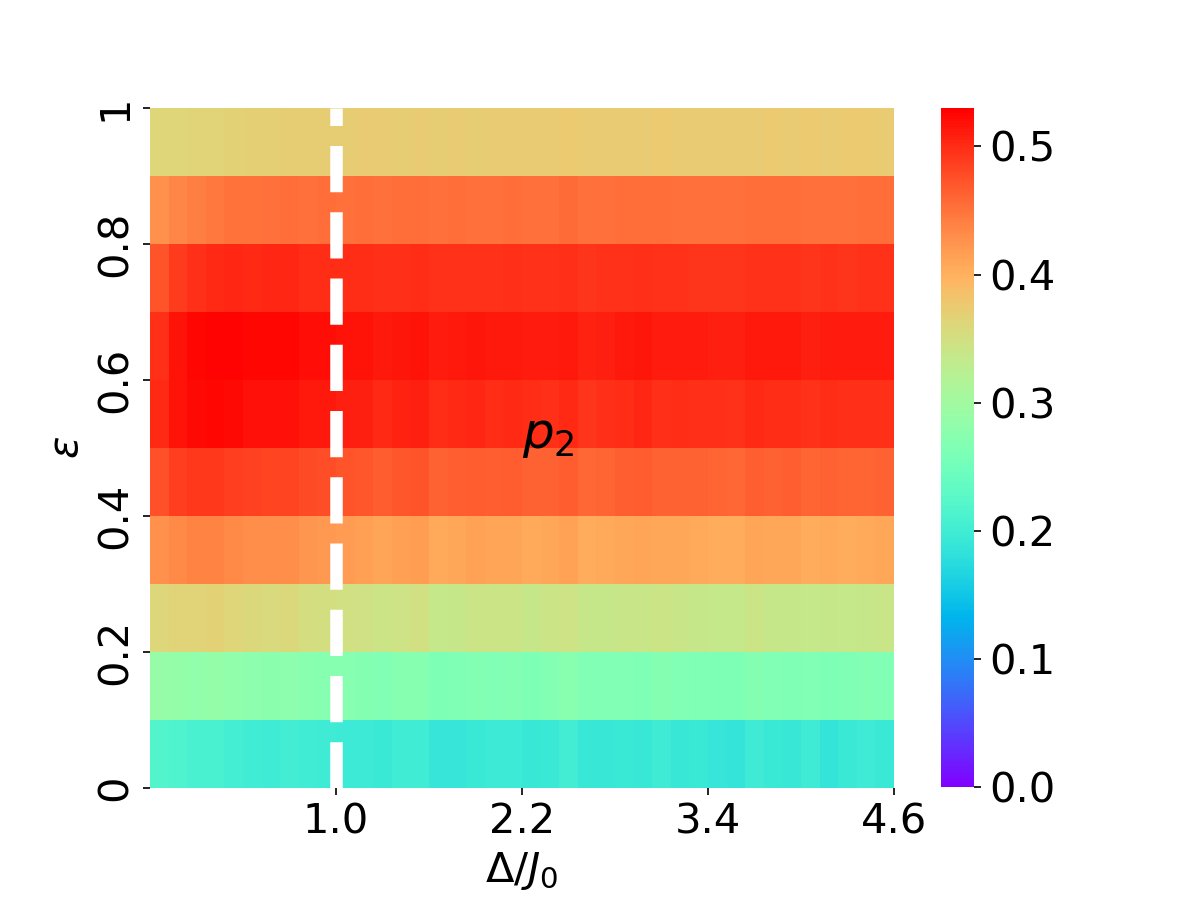}
    (i)\includegraphics[width=0.3\textwidth]{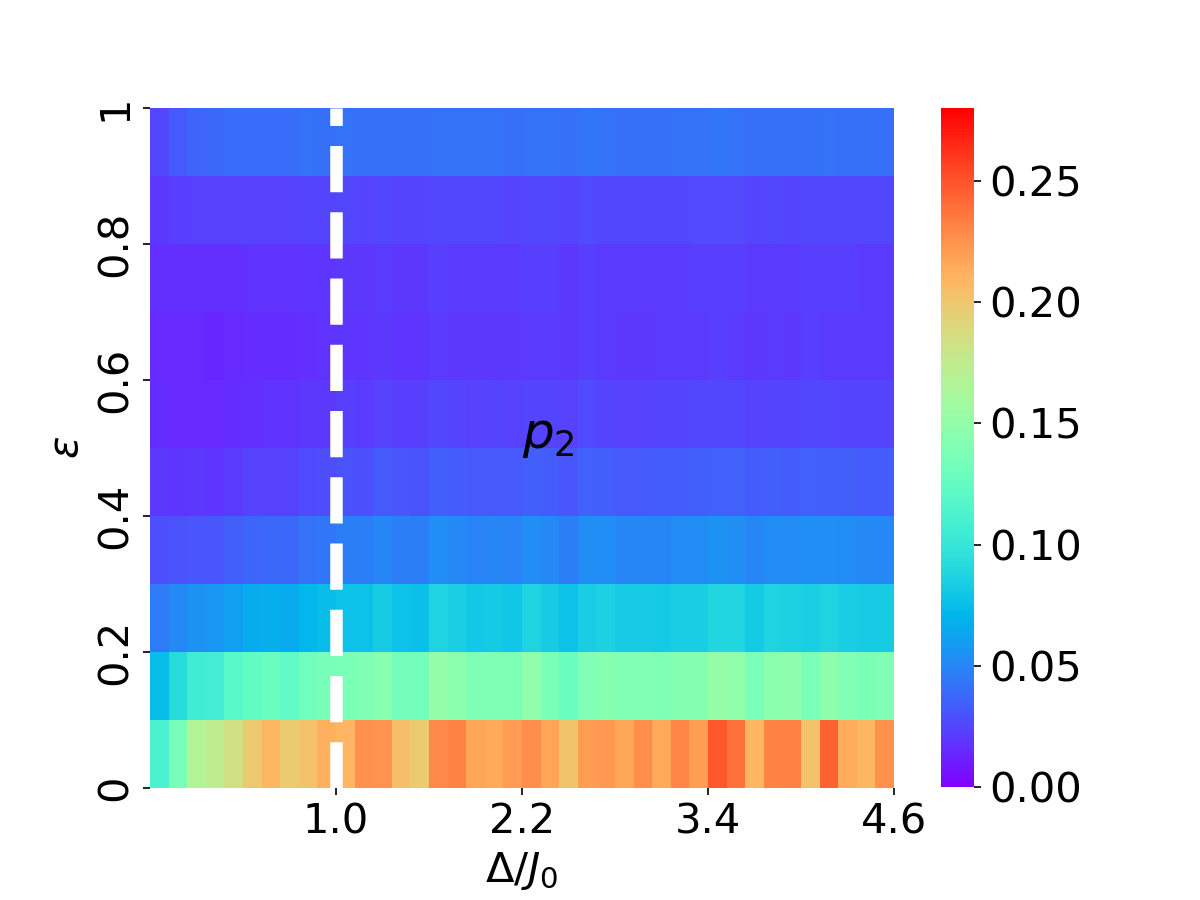}
\caption{Sample-averaged normalized participation ratio $\langle \mathcal{P} \rangle$, entanglement entropy per site $\langle S_E/L \rangle$ and entanglement spectral parameter $\langle \lambda \rangle$ as a function of disorder strength $\Delta/J_0$ and reduced energy $\epsilon$ for \emph{all non-degenerate spin sectors} at system size $L=16$. 
Top row gives (a)$\langle \mathcal{P} \rangle$, (b) $\langle S_E/L \rangle$, (c) $\langle \lambda \rangle$ for disorder distribution $p_0(J)$. Middle row (d)-(f) and bottom row (g)-(i) follow the same order but correspond to disorder distribution $p_1(J)$ and $p_2(J)$ respectively. The white-dashed vertical line is at (a)-(c) $\Delta/J_0=1.0$, (d)-(f) $\Delta/J_0=0.37$ and (g)-(i) $\Delta/J_0=0.99$, corresponding to identical variances for $p_0(J)$, $p_1(J)$ and $p_2(J)$, as shown in Fig.\ \ref{fig-disorder-distributions}.} 
\label{fig-<PR>-<SE>-<lam>}
\end{figure*}

In Fig.\ \ref{fig-<PR>-<SE>-<lam>}, we plot the $\mathcal{P}$'s, $S_E/L$'s and $\lambda$'s, again as a function of $\epsilon$ and disorder strength $\Delta/J_0$ including \emph{all non-degenerate spin sectors}, i.e.\ those for $s=0, \ldots, s_\text{tot}$ and all $m \geq 0$. This removes the degenerate values due to the spin-flip symmetry discussed in section \ref{sec:model}.
%%%%%%%%%%%%%%%%%%%%%%%%%%%%%%%%%%%%%%%%%%%%%%%%%%%%%%%%%%%%%%%%%%%%%%%%%%%%%%
%%%%%%%%%%%%%%%%%%%%%%%%%%%%%%%%%%%%%%%%%%%%%%%%%%%%%%%%%%%%%%%%%%%%%%%%%%%%%%
All three physical quantities have been averaged over all the states and all the, at least, $100$ samples. 
Large values of $\mathcal{P}$, $S_E$ and small values of $\lambda$ correspond to more extended behaviour. Looking at Fig.\ \ref{fig-<PR>-<SE>-<lam>}, we find that a regime with comparatively more extended behaviour appears at small disorder strengths when $\epsilon$ is close to the centre of the spectrum between $0.5$ to $0.7$. 
At the edges of the spectrum, we can see that the states are more localized and remain so as $\Delta/J_0$ increases.
When we compare the results of Fig.\ \ref{fig-<PR>-<SE>-<lam>} between $p_0(J)$, $p_1(J)$ and $p_2(J)$, we find that as for the spectral measures discussed in the last section, the present Fock-space-based measures again show a rapid quantitative change close to $\Delta/J_0 \approx 1$ for $p_0(J)$. However, no such rapid change is visible for $p_1(J)$ and $p_2(J)$.
We have also studied $\mathcal{P}$, $S_E$ and $\lambda$ in the $s=2$, $m=0$ and $s=0$ sectors (cp.\ the supplemental  Figs.\ S5, S6 and S7). The results for these sectors show similar features to those discussed here.
Also, when comparing the efficacy of the $\mathcal{P}$, $S_E$ and $\lambda$ measures, we find that they yield results of comparable significance. There is hence no clear reason to favour one over the others in our study.

%%%%%%%%%%%%%%%%%%%%%%%%%%%%%%%%%%%%%%%%%%%%%%%%%%%%%%%%%%%%%%%%%%%%%%%%%%%%%%
\subsubsection{\label{sec:increasedL} Results for $L=18$, $20$ and $24$}

\begin{figure*}[tb]
$\mathcal{P}$\qquad\qquad\qquad\qquad\qquad\qquad\qquad\qquad$S_E/L$ \qquad\qquad\qquad\qquad\qquad\qquad\qquad\qquad$\lambda$ \qquad\\
 (a) \includegraphics[width=0.3\textwidth]{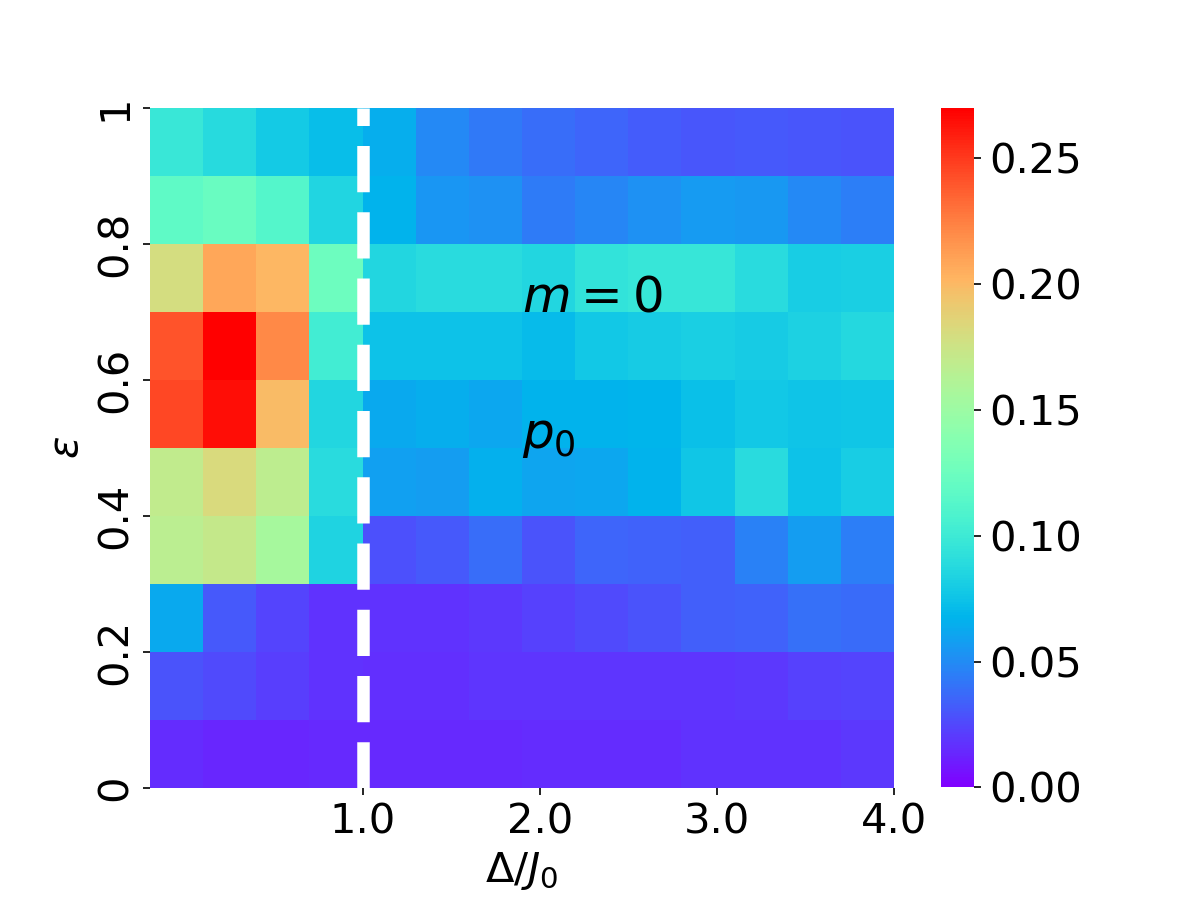}
 (b) \includegraphics[width=0.3\textwidth]{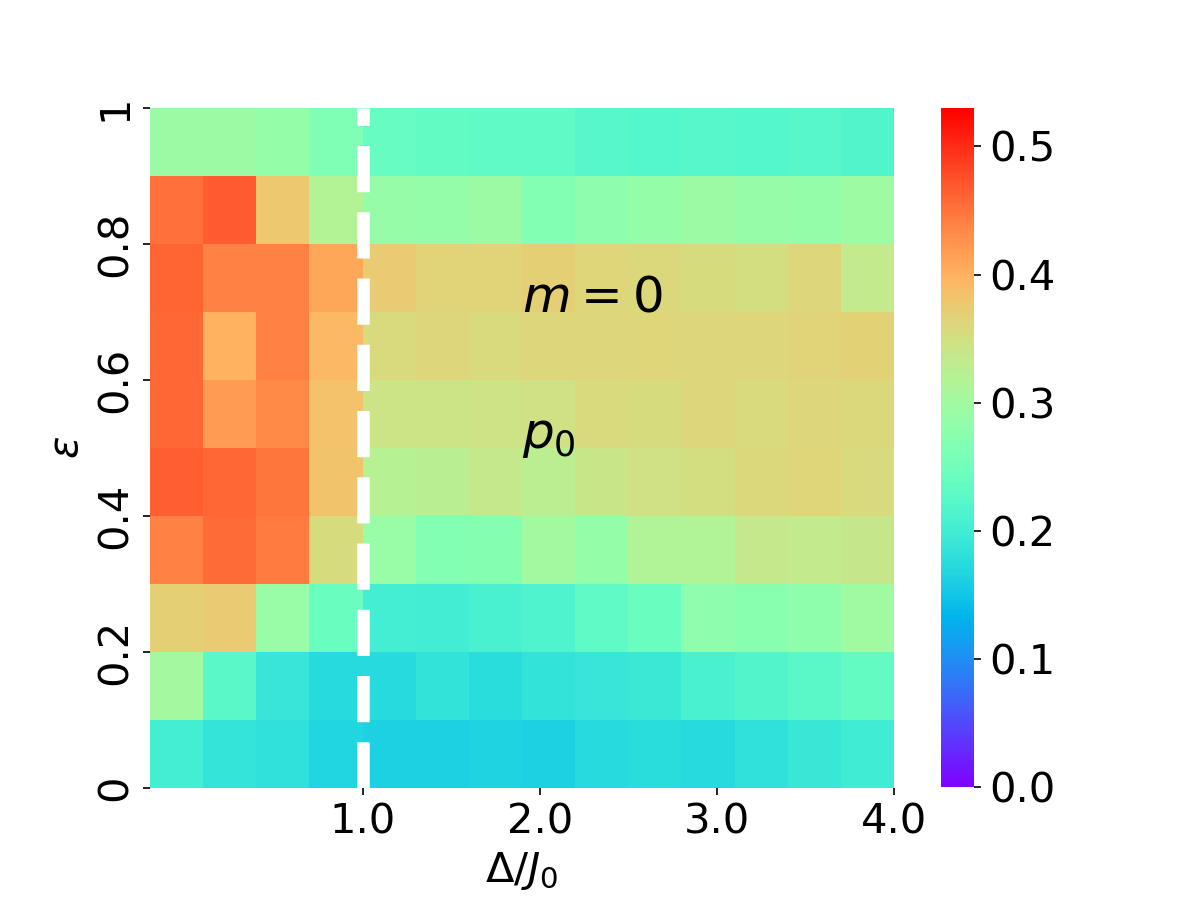}
 (c) \includegraphics[width=0.3\textwidth]{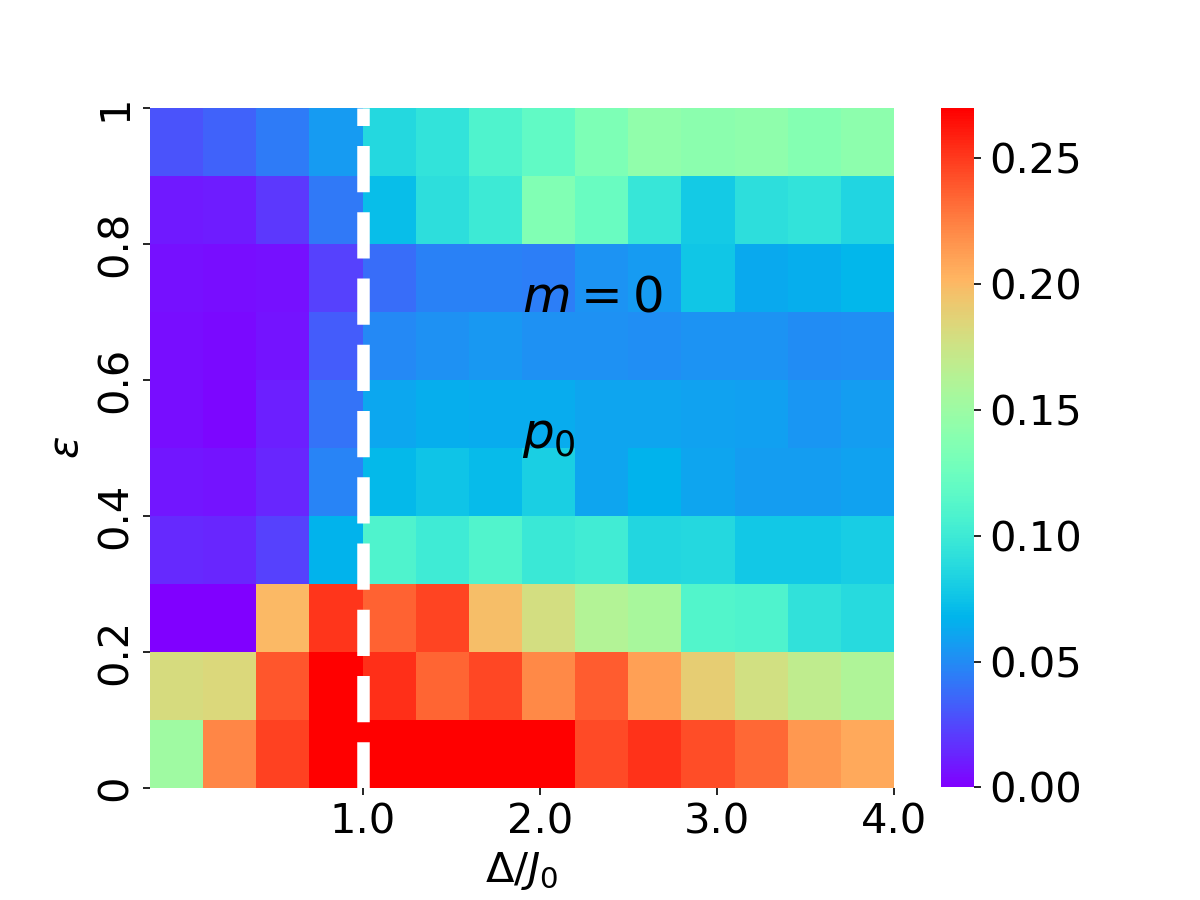}
\caption{Results from $\epsilon$-resolved sparse matrix diagonalization \textsc{JaDaMILU} at system size $L=18$ showing the sample-averaged \revision{(a) participation ratio $\langle \mathcal{P} \rangle$, (b) entanglement entropy per site $\langle S_E/L \rangle$, (c) entanglement spectral parameter $\langle \lambda \rangle$} for the $m=0$ sector and a uniform distribution $p_0(J)$. The corresponding results for exact diagonalization at $L=16$ are provided in the supplement, Fig. S6 (a),(d),(g).} 
\label{fig-JADAMILU-<PR>-<SE>-p0}
\end{figure*}

We now turn to the results for $\mathcal{P}$, $S_E$ and $\lambda$ with increased system sizes as obtained from the sparse diagonalization methods.
In Fig.\ \ref{fig-JADAMILU-<PR>-<SE>-p0} we show the sample-averaged $\langle \mathcal{P} \rangle$, $\langle S_E/L \rangle$ and $\langle \lambda \rangle$ for uniform distribution $p_0(J)$  obtained via \textsc{JaDaMILU} at $L=18$ for $p_0(J)$. 
%%%%%%%%%%%%%%%%%%%%%%%%%%%%%%%%%%%%%%%%%%%%%%%%%%%%%%%%%%%%%%%%%%%%%%%%%%%%%%
%%%%%%%%%%%%%%%%%%%%%%%%%%%%%%%%%%%%%%%%%%%%%%%%%%%%%%%%%%%%%%%%%%%%%%%%%%%%%%
The calculation has been done in the $m=0$ sector. 
We find that the most extended regime appears in the middle of the spectrum at small disorder strengths. This is in agreement with the averaged results of Fig.\ \ref{fig-<PR>-<SE>-<lam>}(a) and also similar for the $m=0$ sector from exact diagonalization at $L=16$ (cp.\ supplemental Fig. S6 (a)-(c)).
Obviously, the modest increase in system size does not seem to change the behaviour drastically.

\begin{figure*}[tb]
$\mathcal{P}$\qquad\qquad\qquad\qquad\qquad\qquad\qquad\qquad$\mathcal{P}$ \qquad\qquad\qquad\qquad\qquad\qquad\qquad\qquad$\mathcal{P}$\qquad\\
 (a) \includegraphics[width=0.3\textwidth]{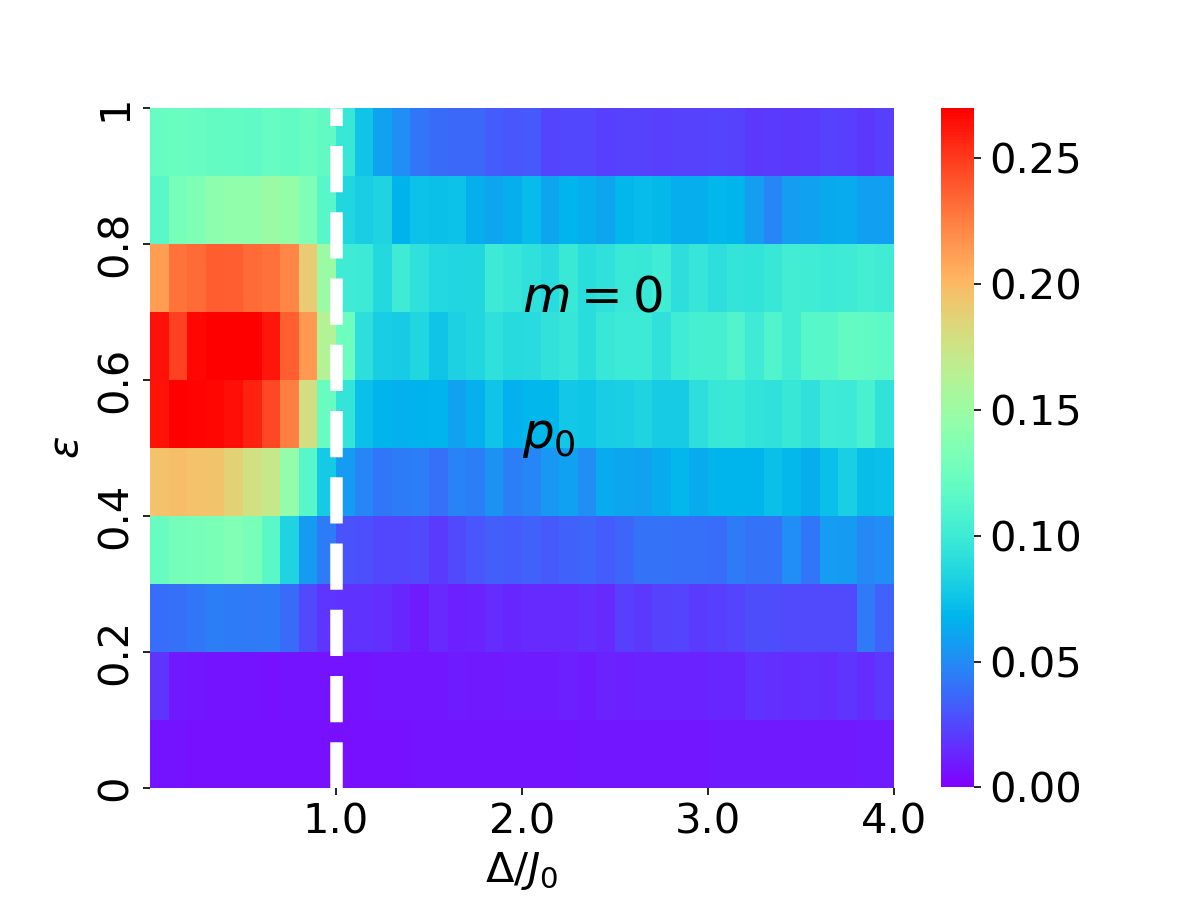}
 (b) \includegraphics[width=0.3\textwidth]{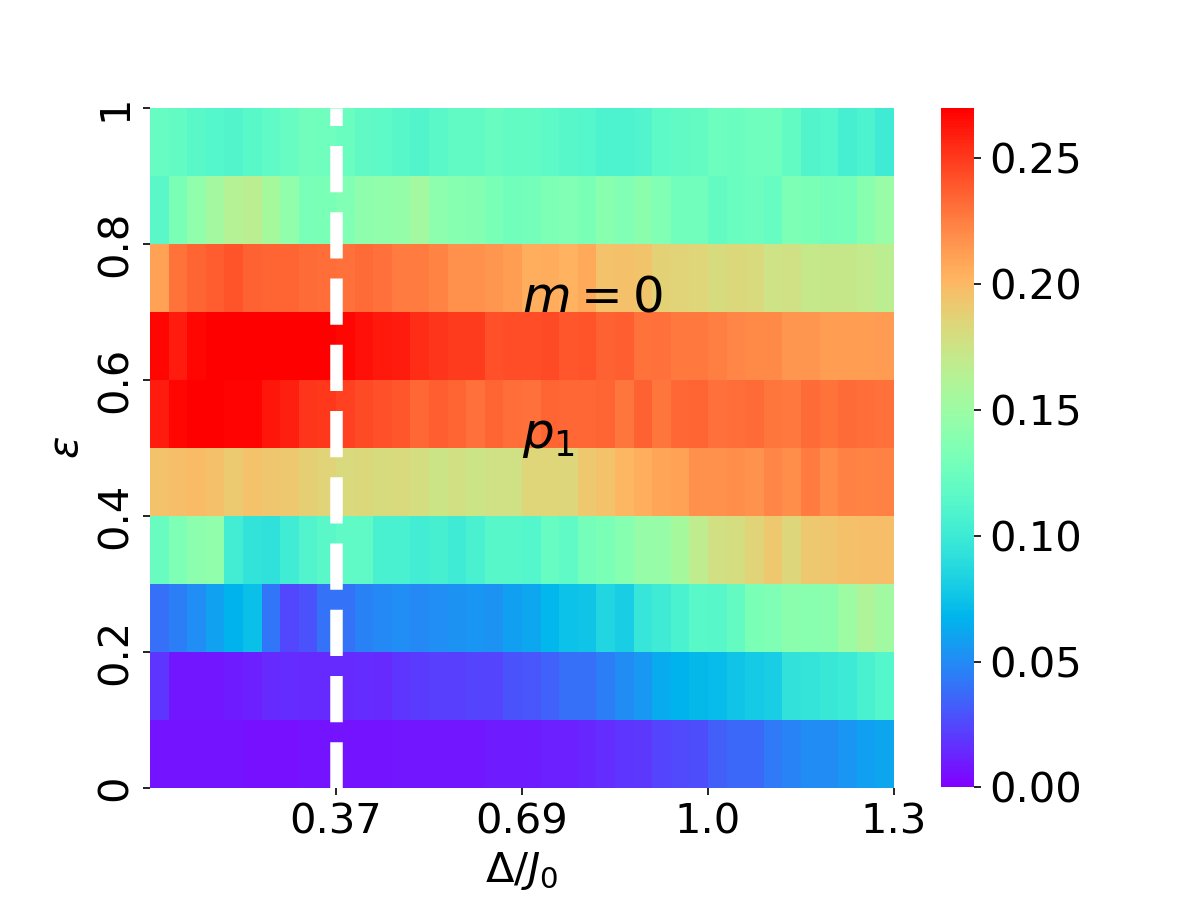}
 (c) \includegraphics[width=0.3\textwidth]{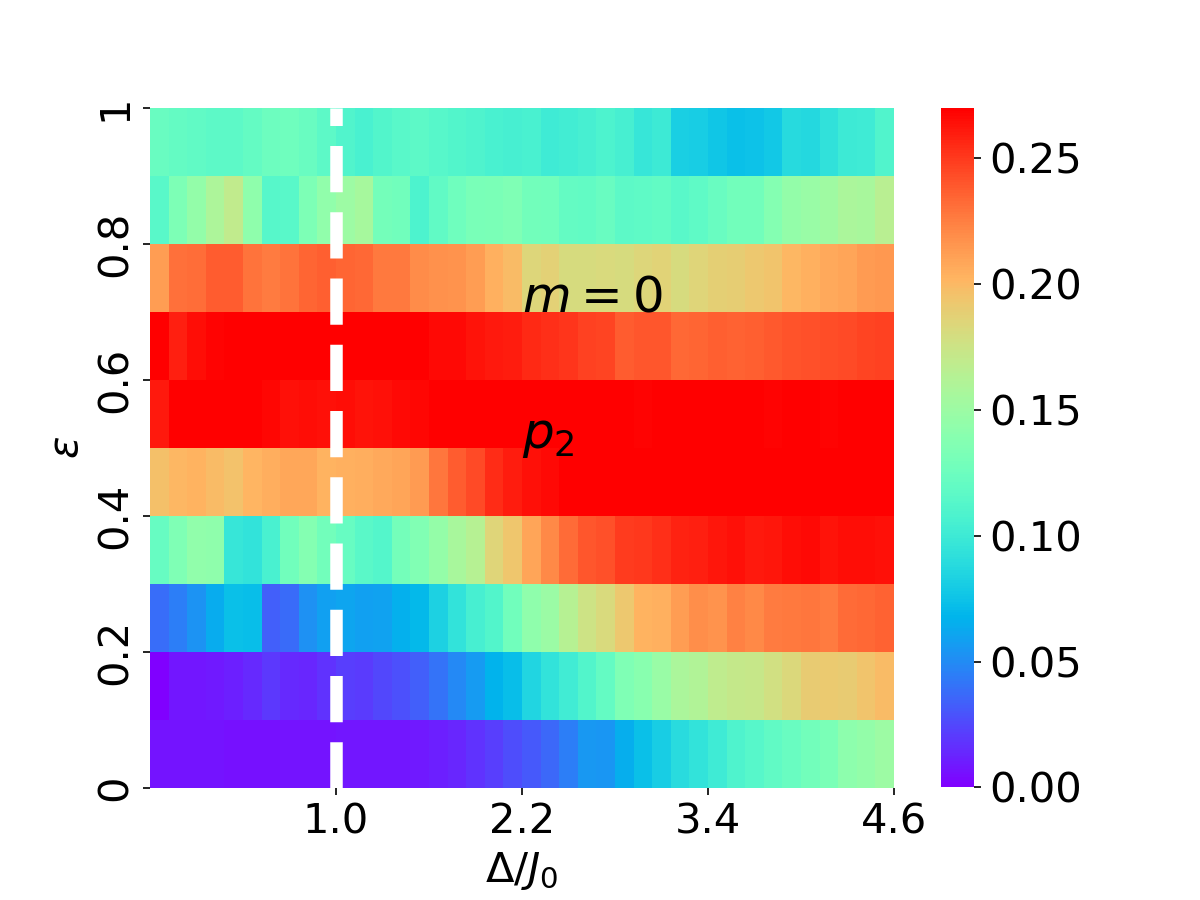}
 \\
 (d) \includegraphics[width=0.3\textwidth]{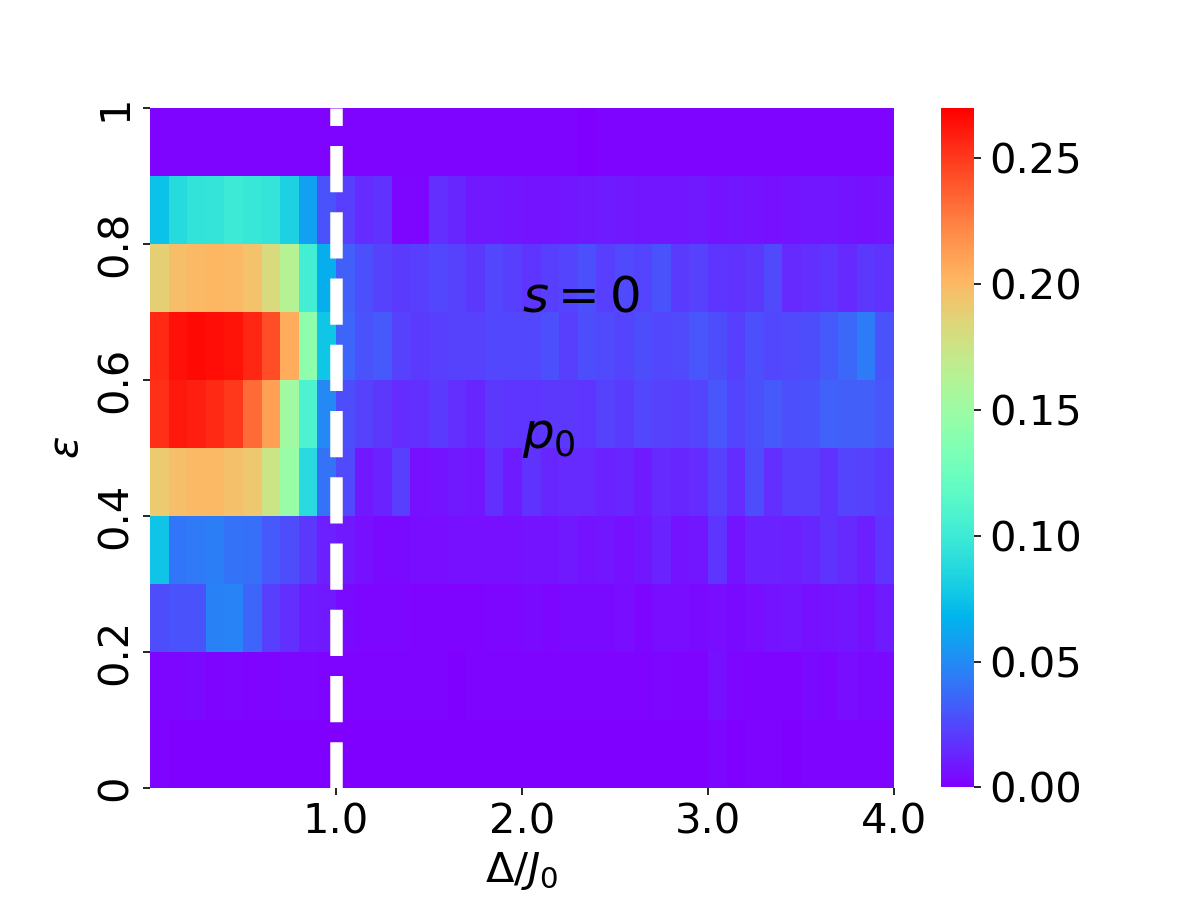}
 (e) \includegraphics[width=0.3\textwidth]{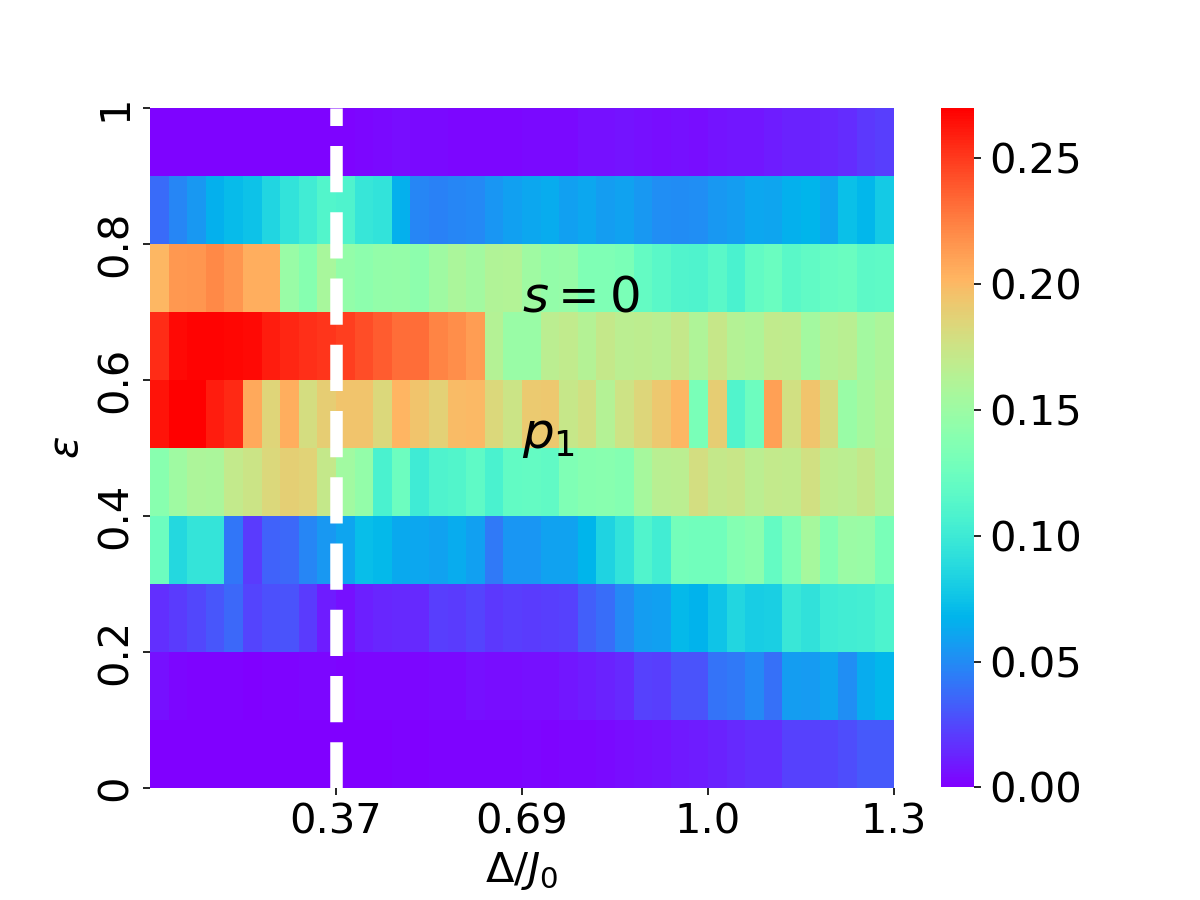}
 (f) \includegraphics[width=0.3\textwidth]{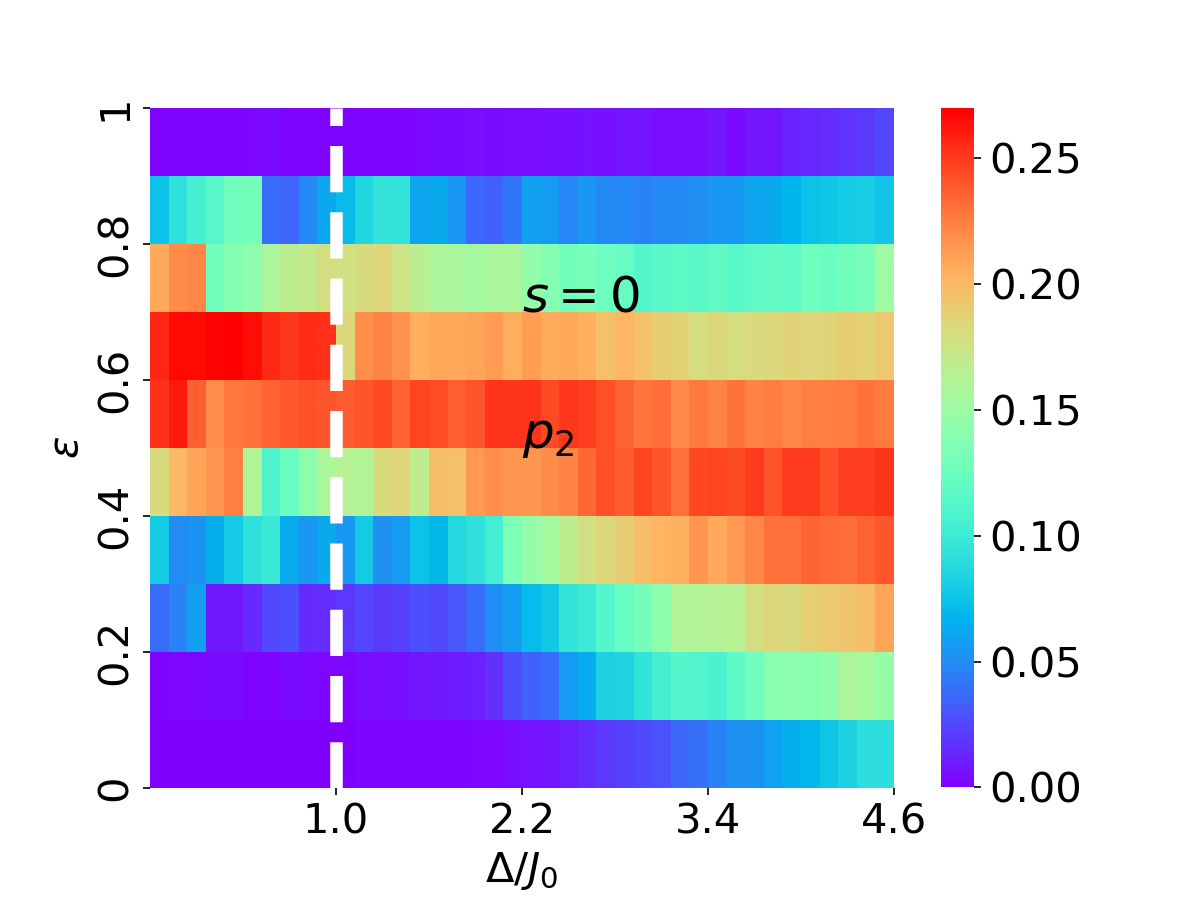}
\caption{Sample-averaged \revision{participation ratio $\langle \mathcal{P} \rangle$} in terms of $\Delta/J_0$ and $\epsilon$ for (a,d) uniform distribution $p_0(J)$, (b,e) $p_1(J)$, (c,f) $p_2(J)$ at system sizes via \textsc{SLEPc} with (a,b,c) $m=$, $L=20$ obtained in Fock basis and (d,e,f) $s=0$, $L=24$ SU(2) basis. 
We note that the SU(2)-based diagonalization uses open boundary conditions.
As in Fig.\ \ref{fig-<PR>-<SE>-<lam>}, the vertical white lines denote the variances of the $p_i(J)$'s. The corresponding results from exact diagonalization at $L=16$ are provided in the supplement, Fig.\ S6 (a)-(c).} 
\label{fig-PETSC-<PR>}
\label{fig-PETSC-<PR>-SU2-open}
\end{figure*}

We next use the sparse matrix diagonalization method of \textsc{SLEPc} in the $m=0$ sector and can increase to $L=20$. The results are shown in Fig.\ \ref{fig-PETSC-<PR>} (top row) with sample-averaged $\langle \mathcal{P} \rangle$ values. We have not computed $\langle S_E/L \rangle$ and $\langle \lambda \rangle$ although this is in principle possible \cite{Dabholkar2024PrivateCommunication}. Instead, we show $\langle \mathcal{P} \rangle$ for all three disorder distributions.
%%%%%%%%%%%%%%%%%%%%%%%%%%%%%%%%%%%%%%%%%%%%%%%%%%%%%%%%%%%%%%%%%%%%%%%%%%%%%%
%%%%%%%%%%%%%%%%%%%%%%%%%%%%%%%%%%%%%%%%%%%%%%%%%%%%%%%%%%%%%%%%%%%%%%%%%%%%%%
We can see that the participation ratios for disorder distribution $p_0(J)$ at $L=20$ are similar to the results for $L=16$ and $L=18$  of Figs.\ \ref{fig-<PR>-<SE>-<lam>}(a) and \ref{fig-JADAMILU-<PR>-<SE>-p0}(a), respectively. 
When the disorder distributions are $p_1(J)$ or $p_2(J)$, the regions at large disorders appear at least as extended for $L=20$ as at $L=16$. This is similar to the results obtained for these distributions from the energy-level statistics in section \ref{sec:results-statistics}. 
%We do not generate the data for $S_E$ and $\lambda$ since running one realization at a particular disorder strength and energy takes more than $24$ hours. Therefore, getting the data for all $40$ different disorder strengths and $10$ different energy windows with $100$ realizations would take too long.

% %%%%%%%%%%%%%%%%%%%%%%%%%%%%%%%%%%%%%%%%%%%%%%%%%%%%%%%%%%%%%%%%%%%%%%%%%%%%%%
% \begin{figure*}[tb]
% $\mathcal{P}$\qquad\qquad\qquad\qquad\qquad\qquad\qquad\qquad$\mathcal{P}$ \qquad\qquad\qquad\qquad\qquad\qquad\qquad\qquad$\mathcal{P}$\qquad\\
%  (a) \includegraphics[width=0.3\textwidth]{figure/SU2/uniform/L=24-PR-s=0-uniform-su2.png}
%  (b) \includegraphics[width=0.3\textwidth]{figure/SU2/p1/L=24-PR-s=0-p1-su2.png}
%  (c) \includegraphics[width=0.3\textwidth]{figure/SU2/p2/L=24-PR-s=0-p2-su2.png}
% \caption{Normalized participation ratio $\langle \mathcal{P} \rangle$ in the $s=0$ sector for (a) uniform distribution $p_0(J)$ (b) $p_1(J)$ (c)$p_2(J)$ at system size $L=24$. The vertical white lines are as in Fig.\ \ref{fig-<PR>-<SE>-<lam>}. Note that the Hamiltonian matrix is diagonalized in SU(2) basis in terms of \textsc{SLEPc} and open boundary condition is applied to the system.} 
% \label{fig-PETSC-<PR>-SU2-open}
% \end{figure*}
% %%%%%%%%%%%%%%%%%%%%%%%%%%%%%%%%%%%%%%%%%%%%%%%%%%%%%%%%%%%%%%%%%%%%%%%%%%%%%%

Last, we can reach $L=24$ \footnote{Further increases in $L$ are possible up to $L=28$, but require a reduction in the number of averaged samples. We have chosen not to do this here.} 
in the SU(2) basis with open boundary condition.
The resulting sample-averaged $\langle \mathcal{P} \rangle$ for all $p_i(J)$  are shown in Fig.\ \ref{fig-PETSC-<PR>-SU2-open} (bottom row). 
We find that the $\langle \mathcal{P} \rangle$ values exhibit a qualitatively similar behaviour across the whole of the $(\epsilon, \Delta/J_0$) plane when comparing $L=20$ (top row) and $24$ (bottom row), as well as $L=16$ of course. This statement is not trivial, since we are comparing different spin sectors, i.e.\ $m=0$ and $s=0$, as well as different bases, i.e.\ Fock and SU(2), respectively. 

Overall, we find that the behaviour when increasing system sizes up to $L=24$ does not change dramatically: just as for the spectral measure of section \ref{sec:results-statistics} there is no convincing evidence of a clear transition from ergodic to non-ergodic behaviour in any spin sector using $\langle \mathcal{P} \rangle$, $\langle S_E/L \rangle$, and $\langle \lambda \rangle$.

%normalized participation ratios at a particular disorder strength when $L=24$ is smaller than the result from exact diagonalization at $L=16$, with periodic boundary condition, when diagonalization is performed in each $m$ spin sector (Fig.\ \ref{fig-<PR>-<SE>-<lam>-stot=0}). We can also see that the regime with more extended behaviour appears at similar disorder strengths and $\epsilon$ at $L=24$ and $L=16$ in these two figures.

%%%%%%%%%%%%%%%%%%%%%%%%%%%%%%%%%%%%%%%%%%%%%%%%%%%%%%%%%%%%%%%%%%%%%%%%%%%%%%
\subsubsection{\label{sec:systemL} Systematic finite-size dependence}

\begin{figure*}[tb]
    (a)\includegraphics[width=0.4\textwidth]{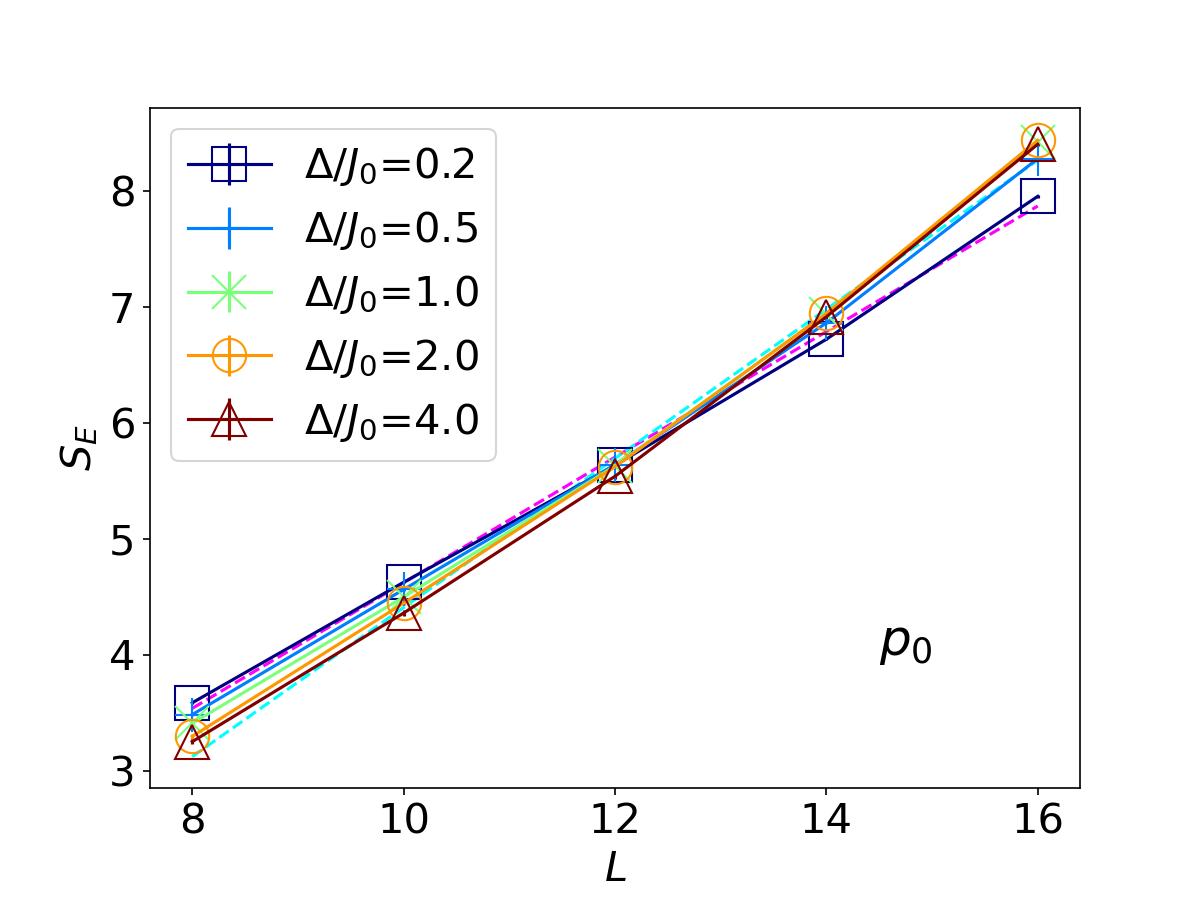}
    (b)\includegraphics[width=0.4\textwidth]{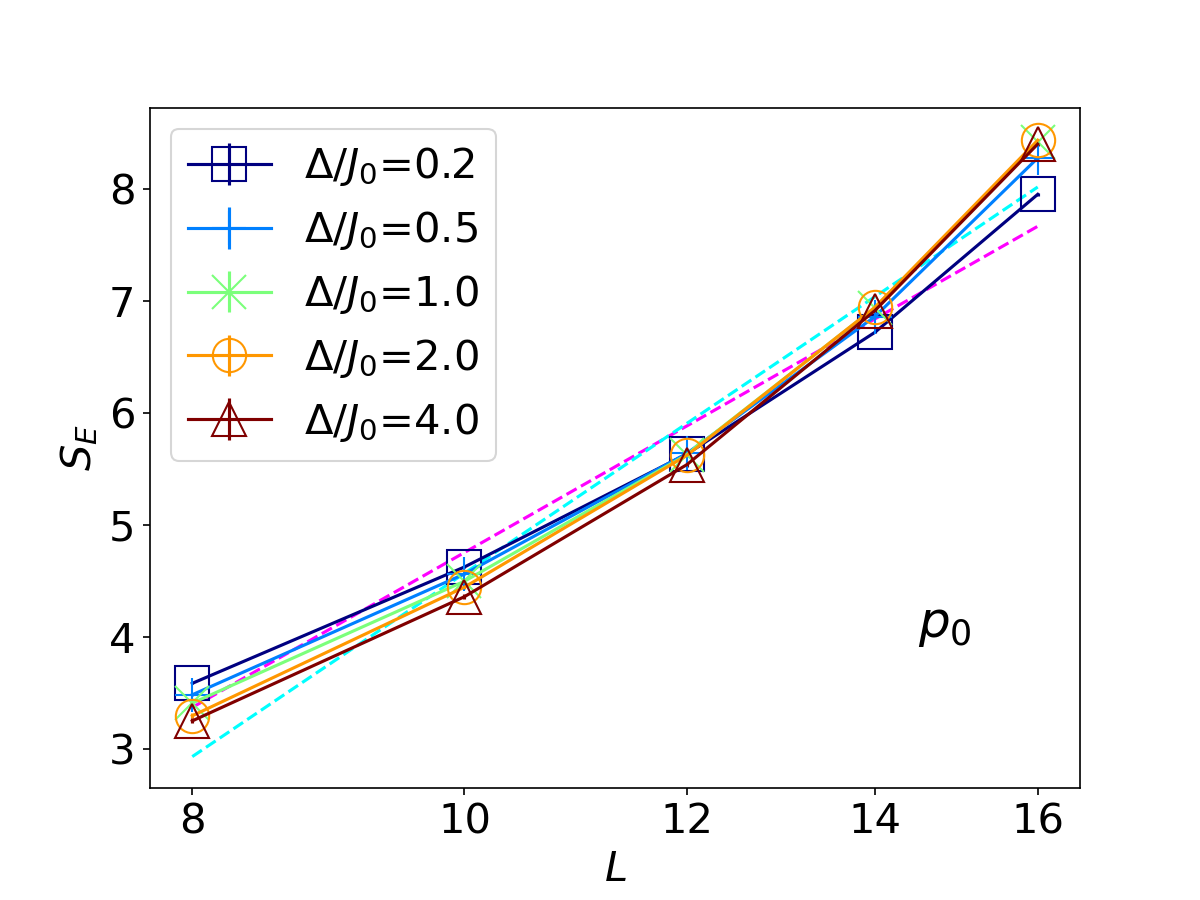}
    (c)\includegraphics[width=0.4\textwidth]{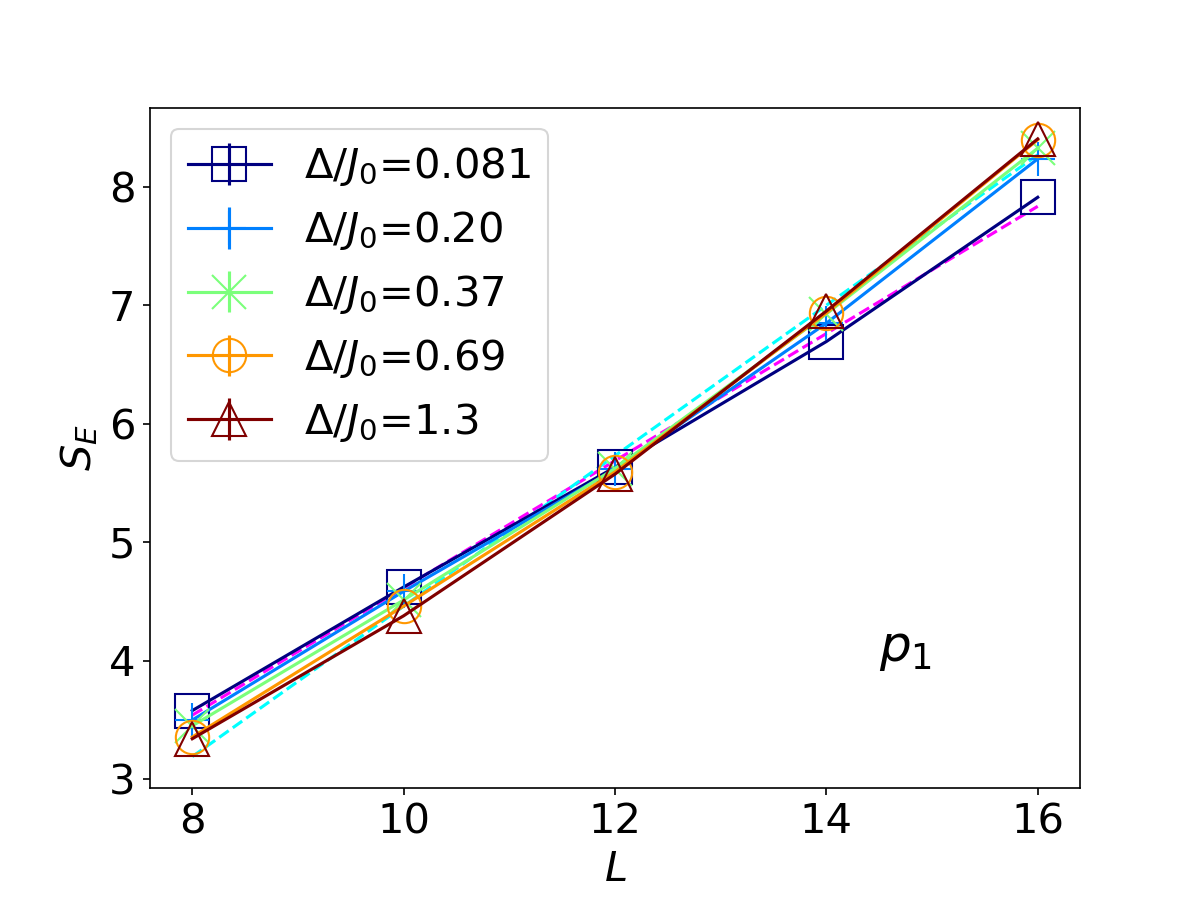}
    (d)\includegraphics[width=0.4\textwidth]{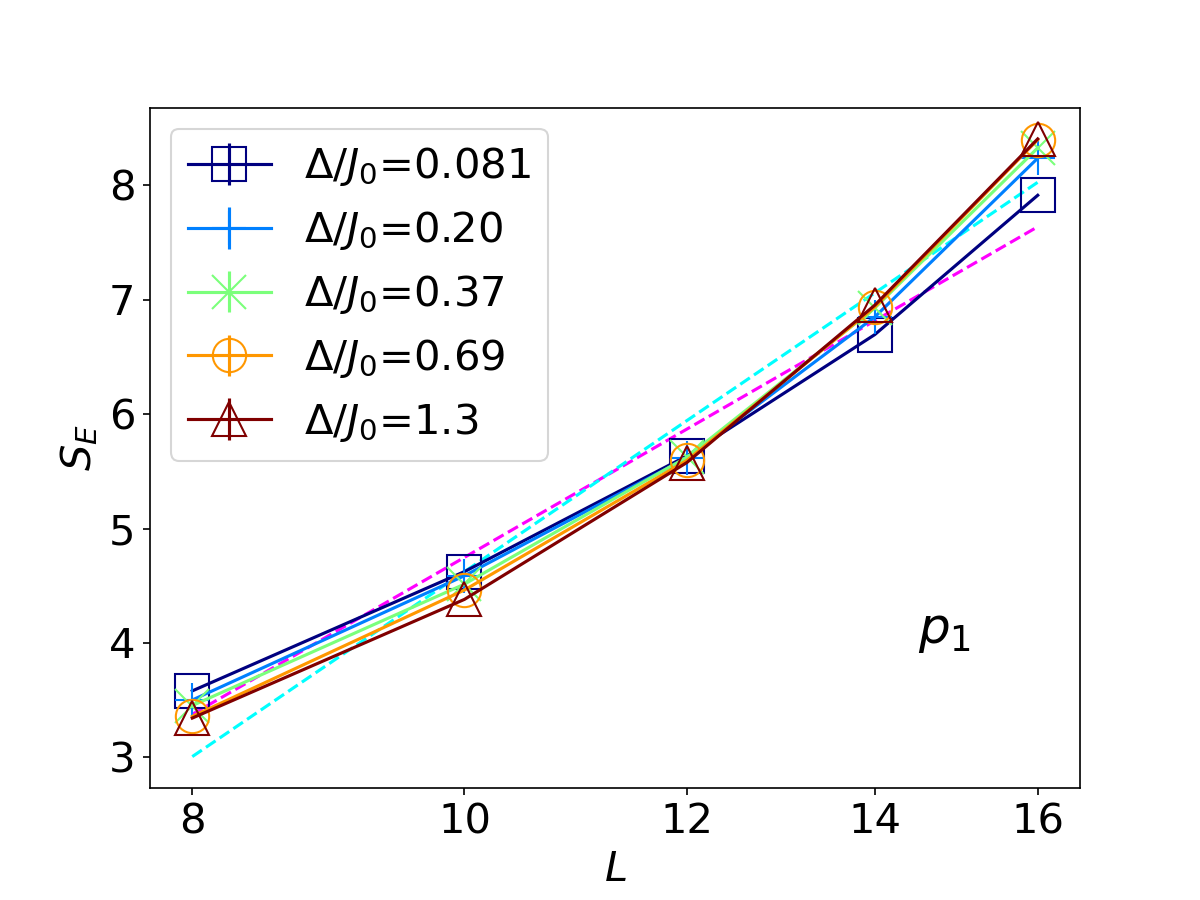}
    (e)\includegraphics[width=0.4\textwidth]{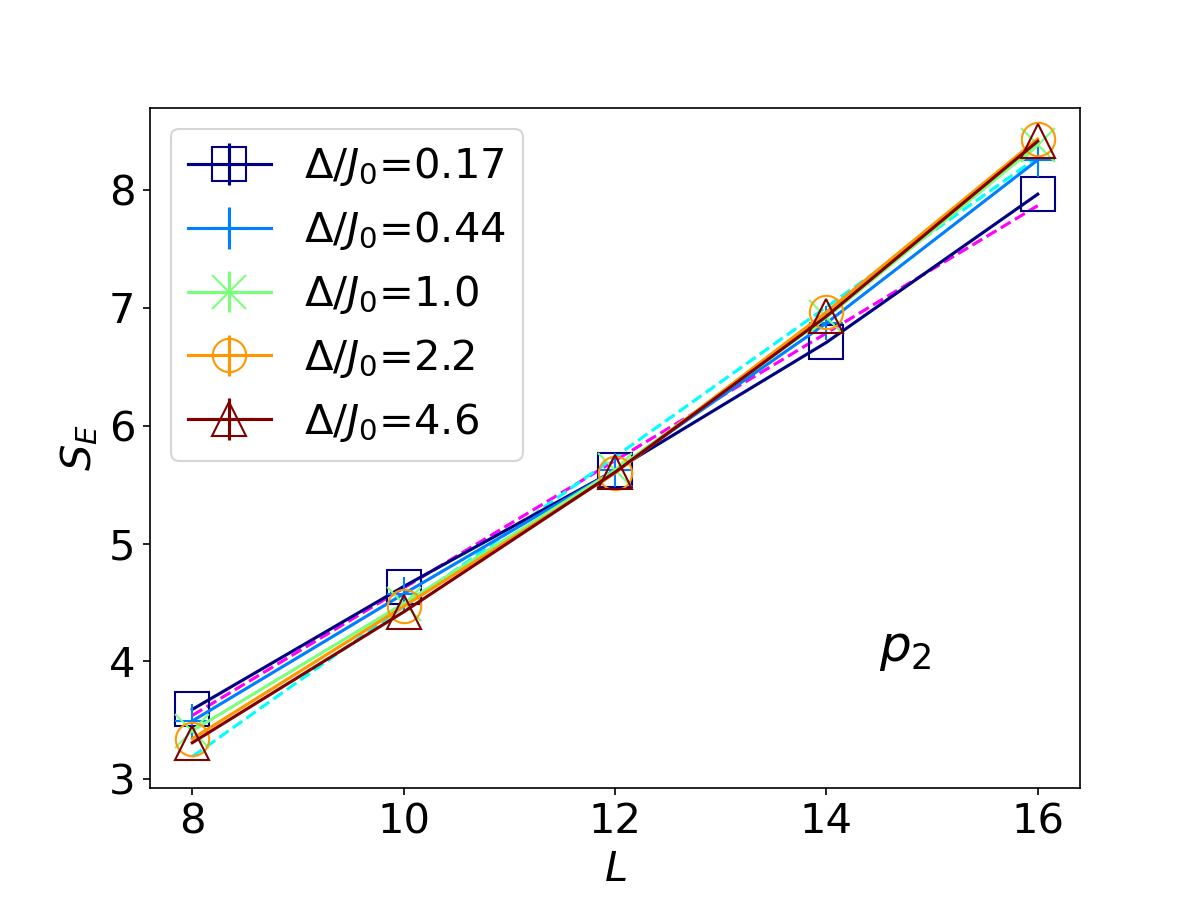}
    (f)\includegraphics[width=0.4\textwidth]{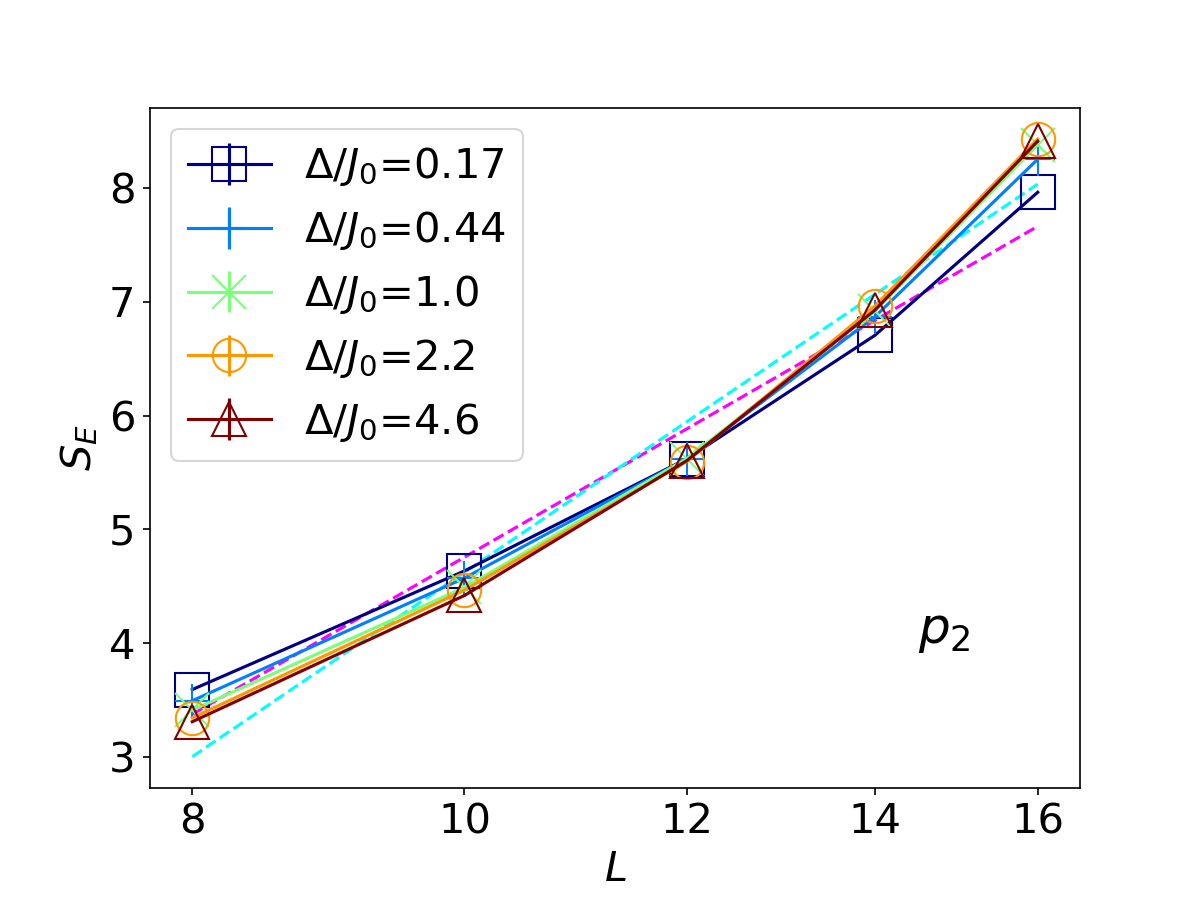}
\caption{Entanglement entropy $S_E$ as a function of system size $L$ for different disorder strengths including \emph{all non-degenerate spin sectors} for states corresponding to $\epsilon \in [0.6,0.7]$. The top row (a)-(b), the middle row (c)-(d) and the bottom row (e)-(f) corresponds to disorder distribution $p_0(J)$, $p_1(J)$ and $p_2(J)$, respectively. The magenta and cyan dashed lines in (a), (c), and (e) are the linear fit of $S_E$, i.e. $S_E \sim L$ for the smallest and largest disorder strength, respectively. The horizontal axis in (b), (d), and (f) is logarithmic. The magenta and cyan dashed lines in (b), (d), and (f) are fitted according to $S_E \sim \ln (L)$, again for the smallest and largest disorder strength, respectively.} 
\label{fig-scale-ent-allsz}
\end{figure*}

As discussed in section \ref{sec:methods}, the $L$ dependence of the entanglement entropy $S_E$ often serves as a measure of Fock space localization. 
Fig.\ \ref{fig-scale-ent-allsz} shows $S_E(L)$ when $\epsilon \in [0.6,0.7]$ and including all non-degenerate spin sectors. As can be seen from Fig.\ \ref{fig-scale-ent-allsz} (a), (c), and (e) $S_E$ scales with the volume law at weak disorders for all the disorder distributions. In contrast, the area law does not agree with $S_E$ very well even at the largest disorder strength. This result is also valid for the case of the $s=2$, $m=0$ and $s=0$ sectors, which can be found in Figs.\ S8, S9 and S10 in the supplemental material. 
%%%%%%%%%%%%%%%%%%%%%%%%%%%%%%%%%%%%%%%%%%%%%%%%%%%%%%%%%%%%%%%%%%%%%%%%%%%%%%
%%%%%%%%%%%%%%%%%%%%%%%%%%%%%%%%%%%%%%%%%%%%%%%%%%%%%%%%%%%%%%%%%%%%%%%%%%%%%%

\begin{figure*}[tb]
    (a)\includegraphics[width=0.4\textwidth]{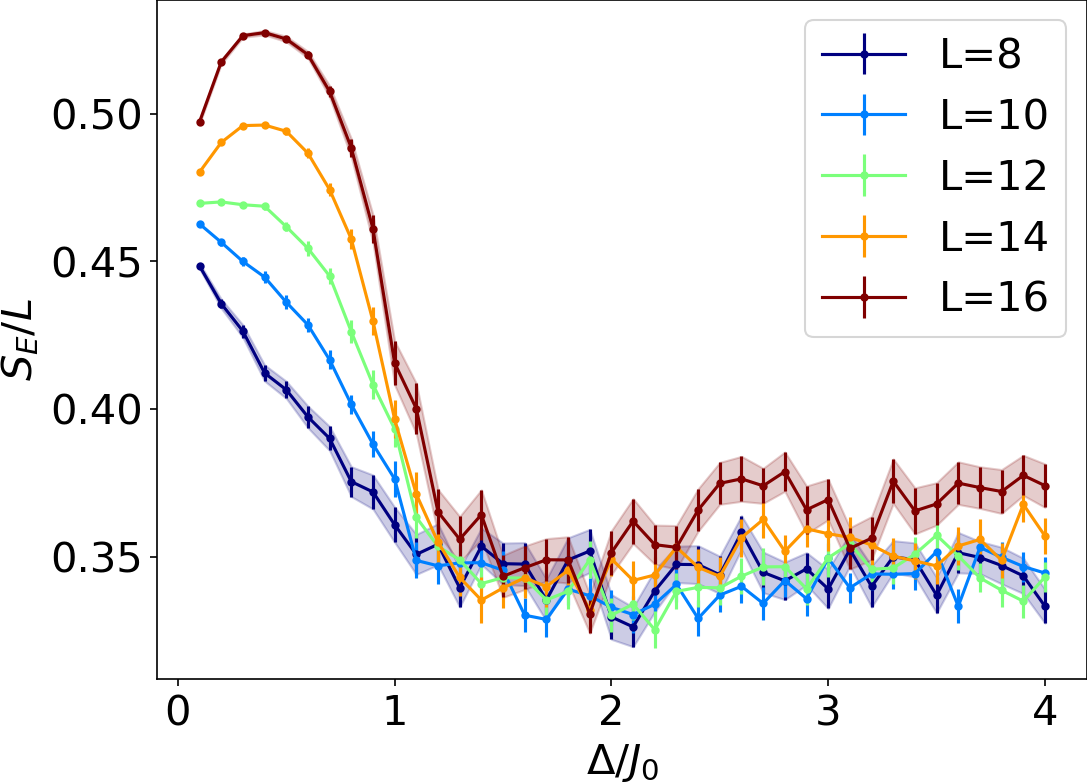}
    (b)\includegraphics[width=0.4\textwidth]{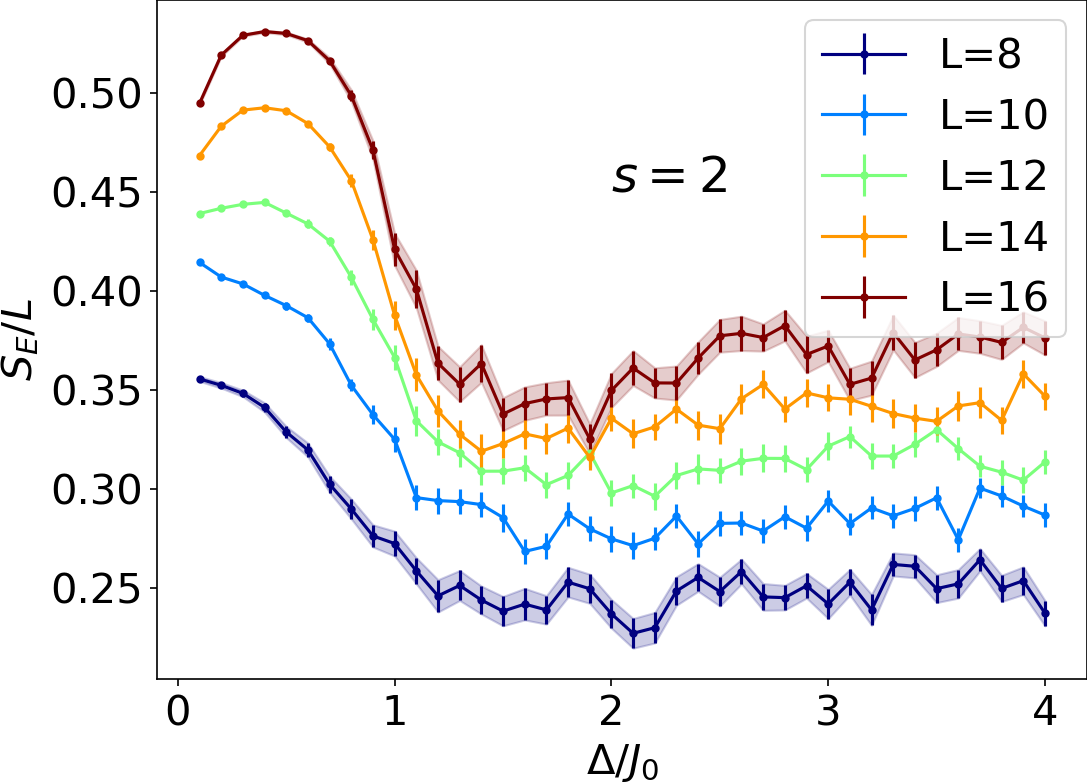}
    (c)\includegraphics[width=0.4\textwidth]{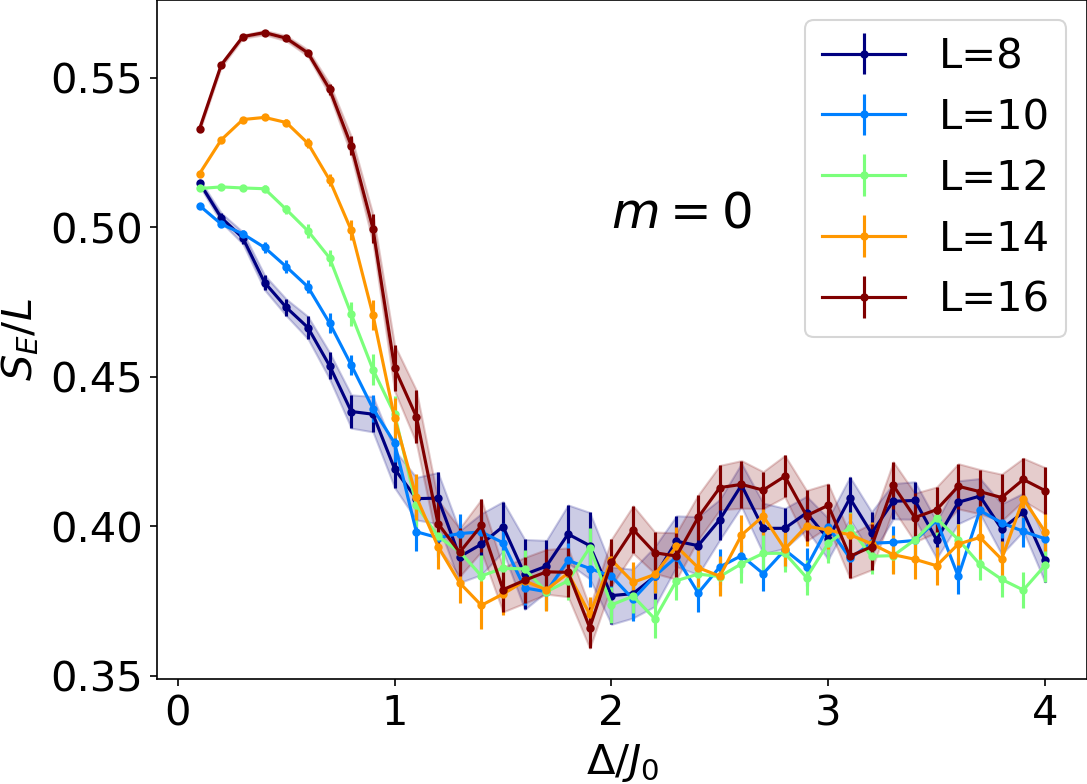}
    (d)\includegraphics[width=0.4\textwidth]{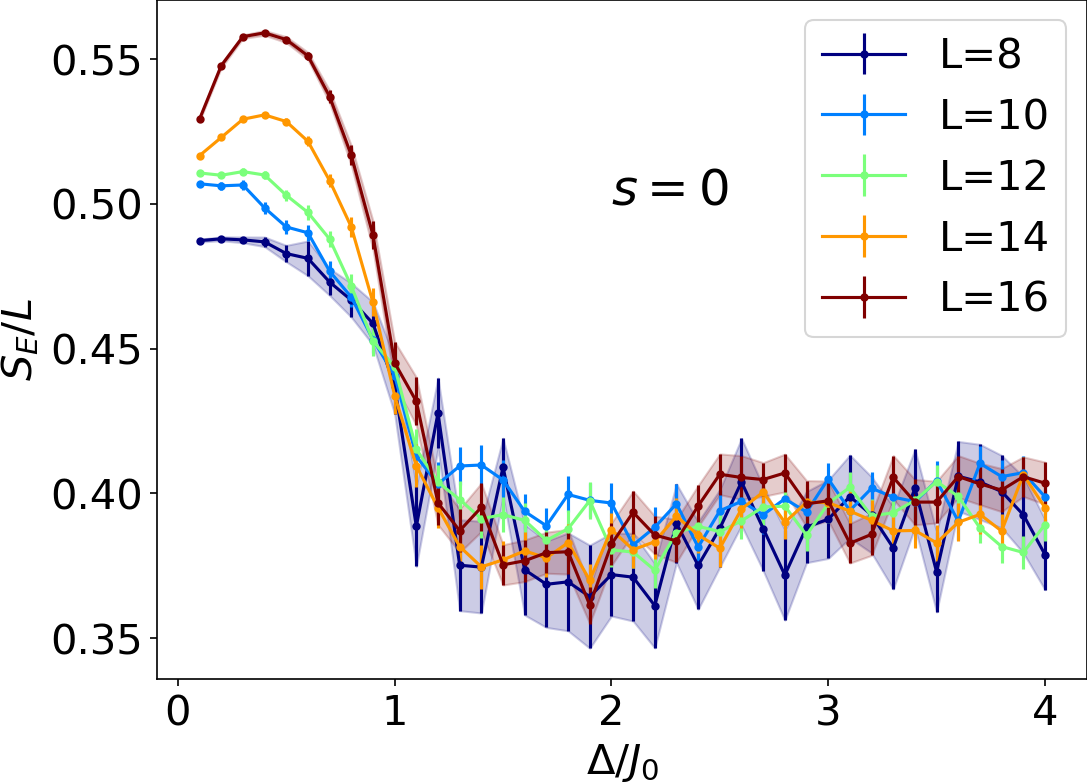}
\caption{Entanglement entropy per site $S_E/L$ as a function of disorder strength $\Delta/J_0$ for different system sizes $L$ when the disorder distribution satisfies $p_0(J)$. The corresponding spin sectors are (a) all non-degenerate spin sectors, (b) the $s=2$ sector, (c) the $m=0$ sector and (d)  the $s=0$ sector respectively. The shaded area on the \revision{data} for $L=8$ and $16$ highlights the spread of the error-of-mean region for smallest and largest system size, respectively.} 
\label{fig-scale-ent}
\end{figure*}

Fig.\ \ref{fig-scale-ent} shows the entanglement entropy per site $S_E/L$ as a function of disorder strengths $\Delta/J_0$ for $p_0(J)$. We can see that curves for $S_E/L$ corresponding to different system sizes do not have a crossover regardless of the spin sectors.
%%%%%%%%%%%%%%%%%%%%%%%%%%%%%%%%%%%%%%%%%%%%%%%%%%%%%%%%%%%%%%%%%%%%%%%%%%%%%%
%%%%%%%%%%%%%%%%%%%%%%%%%%%%%%%%%%%%%%%%%%%%%%%%%%%%%%%%%%%%%%%%%%%%%%%%%%%%%%
Hence, for the system sizes studied here, there is no support for a clear transition between small and large $\Delta/J_0$ values.

%%%%%%%%%%%%%%%%%%%%%%%%%%%%%%%%%%%%%%%%%%%%%%%%%%%%%%%%%%%%%%%%%%%%%%%%%%%%%%
\section{\label{sec:level1} Discussion and Conclusions}
%%%%%%%%%%%%%%%%%%%%%%%%%%%%%%%%%%%%%%%%%%%%%%%%%%%%%%%%%%%%%%%%%%%%%%%%%%%%%%

We have studied localization properties of the 1D random exchange model across its whole spectrum for system sizes up to chain length $L=24$. In doing so, we have taken care to separate out the localization properties in each spin sector labeled by $(s,m)$, or clearly identified when this was not done, as required by the non-Abelian ETH \cite{Murthy2023Non-AbelianHypothesis,Lasek2024NumericalHypothesis}. 
Overall, we find no firm evidence for an MBL transition for $L \rightarrow \infty$ as expected. Rather, localization measures such as $r$-value statistics, participation ratio $\mathcal{P}$, entanglement entropy $S_{E}/L$ and entanglement spectral parameter $\lambda$ suggest that for disorders of up to $\Delta/J_0=4$ much of the spectrum remains ergodic 
\footnote{For $p_0$ and $\epsilon\sim 0.5$, we have also checked that this remains true up to $\Delta/J_0=10$}.
In particular, we do not see a sharp change in behaviour for these quantities that would support the formation of a mobility edge. 
For the three disorder distributions studied here, only the uniform distribution $p_0(J)$ exhibits an apparent change in $r$-value statistics, $\mathcal{P}$, $S_{E}/L$ and $\lambda$ at $L=16$ for $\Delta/J_0 \approx 1$. However, neither of these quantities shows a clear system-size independent crossing point as would be expected if this $\Delta/J_0$ value would correspond to the mobility edge of a second-order phase transition. Furthermore, $\Delta/J_0\geq 1$ is the point when the prevalent AFM couplings become mixed with FM coupling. This changes the character of the model and we think this is the underlying reason for the quantitative change in our measure of localization: we simply enter another regime of still extended states.

For the $p_1(J)$ and $p_2(J)$ distributions, one could argue that the presence of large $J$ values, due to the tails in these distributions, effectively leads to strongly coupled neighbouring spin singlets similar to the Ma-Dasgupta RG approach \cite{Dasgupta1980Low-temperatureChain} for the ground state. Let $J_i$ have such a particularly large value, then the normalized coupling is $\tilde{J} \sim J_{i-1}J_{i+1}/(2J_i^2) \ll \langle J \rangle$ and the chain is effectively cut into two, leading to a corresponding separation into two effectively independent sectors in Fock space. But in each of the two parts, the ``spread'' of the many-body wave function can now more easily extend throughout the sector, leading to the apparent ergodic mixing of states. \revision{We believe that this argument is in agreement with the increased $\mathcal{P}$, $S_E/L$ values, while $\lambda$ decreases, when $\Delta/J_0 >0$ for the $p_1(J)$ and $p_2(J)$ distributions as shown in Fig.\ \ref{fig-<PR>-<SE>-<lam>}.}

While we do not find a clear signature of an MBL transition for our finite samples, we can of course also not rule it out for $L \rightarrow \infty$. Nevertheless, our results are still useful for experimental implementations of the random exchange model where one also has do deal with finite-size restrictions \cite{Schreiber2015ObservationLattice,Kondov2015Disorder-inducedGas,Smith2016Many-bodyDisorder}. In such realizations, the random exchange coupling can be implemented by, e.g., changing the spatial position of the spins. While this is possible for $p_0(J)$, $p_1(J)$ and $p_2(J)$, we would expect that the sharp cutoff for $p_0(J)$ is harder to implement and certainly, the change in the sign for $p_0(J)$ when $\Delta/J_0 >1$, amounts to a tricky phase shift of $\pi$. On the other hand, $p_1(J)$ and $p_2(J)$ avoid this phase shift and allow for a softer decay of large $J$ values. We expect such distributions to mimic the experimental situations more closely.
%%%%%%%%%%%%%%%%%%%%%%%%%%%%%%%%%%%%%%%%%%%%%%%%%%%%%%%%%%%%%%%%%%%%%%%%%%%%%%
\begin{acknowledgments}
We thank J.\ Schliemann and J.\ Siegl for discussions in the early stages of the project, A.\ Rodr\'iguez Gonz\'alez for suggesting to use \textsc{PETSc}/\textsc{SLEPc} and B.\ Dabholkar and F.\ Al\'et for explaining their SU($3$) diagonalization implementation to us \cite{Dabholkar2024PrivateCommunication}.
%We also appreciate B.\ Dabholkar, F.\ Alet for the advice on diagonalization of Hamiltonian matrix in SU(2) basis. 
We gratefully acknowledge the Scientific Computing Research Technology Platform (RTP) for its high-performance computing facilities and the Sulis Tier 2 HPC platform hosted by the RTP. Sulis is funded by EPSRC Grant EP/T022108/1 and the HPC Midlands+ consortium.
UK research data statement: no new data has been generated.
\end{acknowledgments}
%%%%%%%%%%%%%%%%%%%%%%%%%%%%%%%%%%%%%%%%%%%%%%%%%%%%%%%%%%%%%%%%%%%%%%%%%%%%%%

%%%%%%%%%%%%%%%%%%%%%%%%%%%%%%%%%%%%%%%%%%%%%%%%%%%%%%%%%%%%%%%%%%%%%%%%%%%%%%
\appendix

%%%%%%%%%%%%%%%%%%%%%%%%%%%%%%%%%%%%%%%%%%%%%%%%%%%%%%%%%%%%%%%%%%%%%%%%%%%%%%
\section{Size of SU(2) sectors in Fock space}
%%%%%%%%%%%%%%%%%%%%%%%%%%%%%%%%%%%%%%%%%%%%%%%%%%%%%%%%%%%%%%%%%%%%%%%%%%%%%%

To calculate the number of non-zero elements of the Hamiltonian matrix, we can separate the Hamiltonian into diagonal and off-diagonal parts \cite{Zachos1992AlteringSupersymmetry}. We choose the basis of the Hamiltonian matrix such that each spin state is represented by a number. For example, for system size $L=4$, the state $\ket{0101}(\ket{\downarrow\uparrow\downarrow\uparrow})$ stands for $f=5$. The diagonal part $\sum_{i=1}^{L-1}J_iS^z_iS^z_{i+1}$ describes the nearest spin interaction in $z$ direction and contributes overall $\binom{L}{N_{\text{up}}}$ elements to the diagonal parts of the Hamiltonian matrix. The off-diagonal term $\sum_{i=1}^{L}J_i(S^{+}_iS^{-}_{i+1}+S^{-}_iS^{+}_{i+1})$ describes the spin-exchange process between nearest spins. In this case, if we pick any two of the neighboring spins, for example, $i$ and $i+1$, they need to have opposite spins to be able to exchange. Therefore, the rest of spins would have $2\binom{L-2}{N_{\text{up}}-1}$ choice of alignments. In total, there are $2L\binom{L-2}{N_{\text{up}}-1}$ as there are $L$ pairs of neighboring spins for periodic boundary condition. 

%%%%%%%%%%%%%%%%%%%%%%%%%%%%%%%%%%%%%%%%%%%%%%%%%%%%%%%%%%%%%%%%%%%%%%%%%%%%%%
\section{$r$-statistics for multiple irreducible blocks}
\label{sec:rosenzweigporter}
%%%%%%%%%%%%%%%%%%%%%%%%%%%%%%%%%%%%%%%%%%%%%%%%%%%%%%%%%%%%%%%%%%%%%%%%%%%%%%

When $k$ irreducible sectors mix, we have to apply the Rosenzweig-Porter approach to determine the corresponding $P^{(k)}(r)$ and $\left<r\right>^{(k)}$ \cite{Giraud2022ProbingStatistics}. We briefly present the method in the case when $L=16$ and $m=0$. The $m=0$ sector can be split into $9$ irreducible $s$ sectors with $s$ going from $0$ to $8$ in integer steps. The formulas to calculate $P^{(9)}(r)$ and $\left<r\right>^{(9)}$ are given by 
\begin{eqnarray}
P^{(9)}(r) & = & 2\int_{0}^{\infty}xP(x,rx)dx, \\
\left<r\right>^{(9)} & = & \int_{0}^{1}rP(r)dr,
\end{eqnarray}
where $x$ stands for a nearest neighbour level spacing, and $P(x,y)=\partial_x\partial_yH(x,y)$ is the joint distribution of consecutive nearest neighbour spacings with \revision{$H(x,y)=\sum_{s=0}^{8}\mu_s h(\mu_k x,\mu_k y)\Pi_{j\neq k}^8 g\left[ \mu_j (x+y) \right]$}. %In this case, $m$ stands for the number of irreducible sectors which in our $L=16$ case is equal to $9$. 
The weighting factors $\mu_s$, $s = 0, \ldots, 8$ can be calculated to be
$(s,\mu_s)= (0,1/9)$, $(1, 4/15)$, $(2, 28/99)$, $(3, 98/495)$, $(4, 14/143)$, $(5, 4/117)$, $(6, 4/495)$, $(7, 1/858)$ and $(8, 1/12870)$.
The functions $g(x)$ and $h(x,y)$ in the GOE case are given by
\begin{flalign}
& g(x)= && \nonumber \\ 
%\MoveEqLeft %\quad \mbox{ }
%\lefteqn{g(x)=} \nonumber \\ \mbox{ }
 & \quad
 e^{-\frac{9x^2}{4\pi}}-\frac{x}{2}\text{Erfc}\left(\frac{3x}{2\sqrt{\pi}}\right)-\frac{x}{2}e^{-\frac{27x^2}{16\pi}}\text{Erfc}\left(\frac{3x}{4\sqrt{\pi}}\right) &&
\end{flalign}
and
\begin{flalign}
%\lefteqn{h(x,y)=} \nonumber \\ \mbox{ }
& h(x)= && \nonumber \\
& \quad
\frac{9(x+y)}{4\pi}e^{-\frac{9(x^2+xy+y^2)}{4\pi}}\nonumber && \\
&+\frac{8\pi-27x^2}{16\pi}e^{-\frac{-27x^2}{16\pi}}\text{Erfc}\left[\frac{3(x+2y)}{4\sqrt{\pi}}\right]\nonumber && \\
&+\frac{8\pi-27y^2}{16\pi}e^{-\frac{-27y^2}{16\pi}}\text{Erfc}\left[\frac{3(2x+y)}{4\sqrt{\pi}}\right] && .
\end{flalign}
We find $\left<r\right>^{(9)}=0.395053$ and $P^{(9)}_\text{GOE}(r)$ has been used in Figs.\ \ref{fig-level-statistics-p0-all}, \ref{fig-level-statistics-p1p2-alldiff} (a) and \ref{fig-level-statistics-p1p2-alldiff} (c) (as well as in supplementary Figs.\ S1 and S2).

%%%%%%%%%%%%%%%%%%%%%%%%%%%%%%%%%%%%%%%%%%%%%%%%%%%%%%%%%%%%%%%%%%%%%%%%%%%%%%
\section{Normalizing the $p_1$ and $p_2$ disorder distributions}
\label{sec:disorderdistribution}
%%%%%%%%%%%%%%%%%%%%%%%%%%%%%%%%%%%%%%%%%%%%%%%%%%%%%%%%%%%%%%%%%%%%%%%%%%%%%%

The normalization factor of $p_1(J)$ is given by $\mathcal{N}_1(\Delta)= \Delta^3(4-2e^{-\frac{J_0}{\Delta}})+2\Delta J_0^2$. 
For $p_2(J)$, we have $\mathcal{N}_2(\Delta) = \Delta {\left\{2\Delta J_0e^{-\frac{J_0^2}{\Delta^2}}+(\Delta^2+2J_0^2)\sqrt{\pi}\left[1+\text{Erfc}(\frac{J_0}{\Delta})\right]\right\}}/{4}$. 
For these normalized distributions we can then compute the means $\langle J \rangle_{i}= \int_{0}^{\infty} J p_i(J) dJ$ giving
\begin{eqnarray}
\langle J \rangle_{1}(\Delta)
&= & \mathcal{N}_1(\Delta) \int_{0}^{\infty} J^3e^{-|J-J_0|/\Delta}dJ \nonumber \\
&= & \frac{6\Delta^4e^{-\frac{J_0}{\Delta}}+12\Delta^3J_0+2\Delta J_0^3}{\Delta^3(4-2e^{-\frac{J_0}{\Delta}})+2\Delta J_0^2}
\end{eqnarray}
and similarly 
\begin{widetext}
\begin{eqnarray}
\langle J \rangle_{2}(\Delta)
&= & \frac{2\Delta(\Delta^2+J_0^2)e^{-\frac{J_0^2}{\Delta^2}}+J_0(3\Delta^2+2J_0^2)\sqrt{\pi}\left[1+\text{Erfc}(\frac{J_0}{\Delta})\right]}{2\Delta J_0e^{-\frac{J_0^2}{\Delta^2}}+(\Delta^2+2J_0^2)\sqrt{\pi}\left[1+\text{Erfc}(\frac{J_0}{\Delta})\right]} \ .
\end{eqnarray}
The respective variances are
\begin{eqnarray}
\sigma^2_{p_1}(\Delta)
&= &\frac{\Delta^2[3\Delta^4+2e^{\frac{2J_0}{\Delta}}(24\Delta^4+6\Delta^2J_0^2+J_0^4)-e^{\frac{J_0}{\Delta}}(48\Delta^4+36\Delta^3J_0+24\Delta^2J_0^2+6\Delta J_0^3+J_0^4)]}{[\Delta^2-e^{\frac{J_0}{\Delta}}(2\Delta^2+J_0^2)]^2}
\end{eqnarray}
and
\begin{eqnarray}
\sigma^2_{p_2}(\Delta)
&= &\frac{\Delta^2\left\{-8\Delta^4+4\Delta^2J_0^2-8\Delta e^{\frac{J_0^2}{\Delta^2}}(\Delta-J_0)J_0(\Delta+J_0)\sqrt{\pi}\left[1+\text{Erfc}(\frac{J_0}{\Delta})\right]+e^{\frac{2J_0^2}{\Delta^2}}(3\Delta^4+4J_0^4)\pi\left[1+\text{Erfc}(\frac{J_0}{\Delta})\right]^2\right\} }{\left\{2\Delta J_0e^{-\frac{J_0^2}{\Delta^2}}+(\Delta^2+2J_0^2)\sqrt{\pi}\left[1+\text{Erfc}(\frac{J_0}{\Delta})\right]\right\}^2}
\end{eqnarray}
\end{widetext}
Clearly, these expressions become increasingly cumbersome but are readily implemented in our code. Since we set $J_0=1$ in the main text, we have only highlighted the $\Delta$ dependence in $\mathcal{N}_i(\Delta)$, $\langle J \rangle_{p_i}(\Delta)$ and $\sigma_{p_i}^2(\Delta)$. 
%%%%%%%%%%%%%%%%%%%%%%%%%%%%%%%%%%%%%%%%%%%%%%%%%%%%%%%%%%%%%%%%%%%%%%%%%%%%%%

\section{Diagonalization in each $s$ sector}
\label{sec:SU(2)basis}
%%%%%%%%%%%%%%%%%%%%%%%%%%%%%%%%%%%%%%%%%%%%%%%%%%%%%%%%%%%%%%%%%%%%%%%%%%%%%%
The Hamiltonian can also be diagonalized in each $s$ sector based on the Young tableaux. Here, we give an example of constructing the basis and matrix elements in terms of $s=0$ sector for system size $L=6$. In this case, the Young tableaux have two rows and three columns filled with numbers from 1 to 6 in increasing order. The Hamiltonian can be rewritten as \cite{Saito1989PossibleModel}
\begin{equation}
H=\sum_{i=1}^{L-1}\frac{1}{2}J_i(P_{i,i+1}-\frac{1}{2}),
\end{equation}
where $P_{i,i+1}$ is the permutation operator that permutes the states at site $i$ and $i+1$ and open boundary condition is applied. 
We start with a Young tableau and continuously apply permutation operators $P_{i,i+1}$ ($i \in [1,4]$) until we get all the basis states. The dimension of the $s$ basis is given by \cite{Nataf2014ExactModels}
\begin{equation}
\text{dim}\mathcal{H}(s)=\frac{L!}{\Pi_{i=1}^{L}l_i},
\end{equation}
where $l_i$ is the hook length on the site $i$ defined as the number of boxes on the right plus the number of boxes below plus $1$. 
To construct a matrix element of the Hamiltonian matrix, we apply the permutation operator to each state in the $s$ basis. When the indices $i$ and $i+1$ of the permutation operator $P_{i,i+1}$ belong to the same row or column of the Young tableau state, applying $P_{i,i+1}$ to that state returns the same state \cite{Nataf2014ExactModels,Dabholkar2024ErgodicChains,Dabholkar2024PrivateCommunication}. For example,
\begin{equation}
P_{2,3} \ 
\ytableausetup{centertableaux}
\begin{ytableau}
      1 & 2 & 3 \\
      4 & 5 & 6
\end{ytableau}
=
\ytableausetup{centertableaux}
\begin{ytableau}
      1 & 2 & 3 \\
      4 & 5 & 6
\end{ytableau}
\end{equation}
When the indices $i$ and $i+1$ of the permutation operator $P_{i,i+1}$ are neither on the same row or same column, applying $P_{i,i+1}$ to that state returns a superposition of two states. To be specific, $P_{i,i+1}\ket{\psi}=-\frac{1}{d}\ket{\psi}+\sqrt{1-\frac{1}{d^2}}\ket{\bar{\psi}}$, where $\ket{\bar{\psi}}$ stands for the Young tableau that exchanges $i$ and $i+1$ in $\ket{\psi}$. For example, applying $P_{3,4}$ to \ytableausetup{centertableaux}
\begin{ytableau}
      1 & 2 & 3 \\
      4 & 5 & 6
\end{ytableau}
we get
\begin{equation}
P_{3,4} \ 
\ytableausetup{centertableaux}
\begin{ytableau}
      1 & 2 & 3 \\
      4 & 5 & 6
\end{ytableau}
=
\ytableausetup{centertableaux}
-\frac{1}{3}\ \begin{ytableau}
      1 & 2 & 3 \\
      4 & 5 & 6
\end{ytableau}
+\frac{2\sqrt{2}}{3}\ \begin{ytableau}
      1 & 2 & 4 \\
      3 & 5 & 6.
\end{ytableau}
\end{equation}
By applying all the permutation operators to each state in the basis using the above method, a Hamiltonian matrix can be constructed. We emphasize that the SU$(2)$ basis is different from the Fock basis customarily used in MBL studies. Furthermore, open boundary conditions have to be used \cite{Dabholkar2024PrivateCommunication}.

% %%%%%%%%%%%%%%%%%%%%%%%%%%%%%%%%%%%%%%%%%%%%%%%%%%%%%%%%%%%%%%%%%%%%%%%%%%%%%%%%%%%%%%%%%%%%%%%%%%%%%%%%%%%%%%%%%%%%%%%%%%%%%%%%%%%%%%%%%%%%%%%%%%%%%%%%%%%%
%\bibliographystyle{apsrev4-2}
%\bibliography{referencesRAR.bib}
%\bibliography{references.bib}


\ifNOSUP\end{document}\else%

%%%%%%%%%%%%%%%%%%%%%%%%%%%%%%%%%%%%%%%%%%%%%%%%%%%%%%%%%%%%
%%
%% Supporting Information
%%
%% to be moved to its own file later
%%
%%%%%%%%%%%%%%%%%%%%%%%%%%%%%%%%%%%%%%%%%%%%%%%%%%%%%%%%%%%%
\clearpage\newpage
\setcounter{figure}{0}
\setcounter{table}{0}
\def\thefigure{S\arabic{figure}}
\def\thetable{S\arabic{table}}
\setcounter{page}{1}
\pagestyle{plain}
%%%%%%%%%%%%%%%%%%%%%%%%%%%%%%%%%%%%%%%%%%%%%%%%%%%%%%%%%%%%

\section*{Supplemental Material}

{\center
\textbf{Spectral and Entanglement Properties of the Random Exchange Heisenberg Chain}\\[2ex]

\noindent%
Yilun Gao and Rudolf A R\"{o}mer\\[2ex]

{Department of Physics, University of Warwick, Coventry, CV4 7AL, UK, R.Roemer@warwick.ac.uk}\\
}
\hspace*{2ex}

\begin{figure*}[tb]
    (a)\includegraphics[width=0.4\textwidth]{\figdir/p1-P-L_16-Stot_2-allb-Rosenzweig-Porter.png}
    (b)\includegraphics[width=0.4\textwidth]{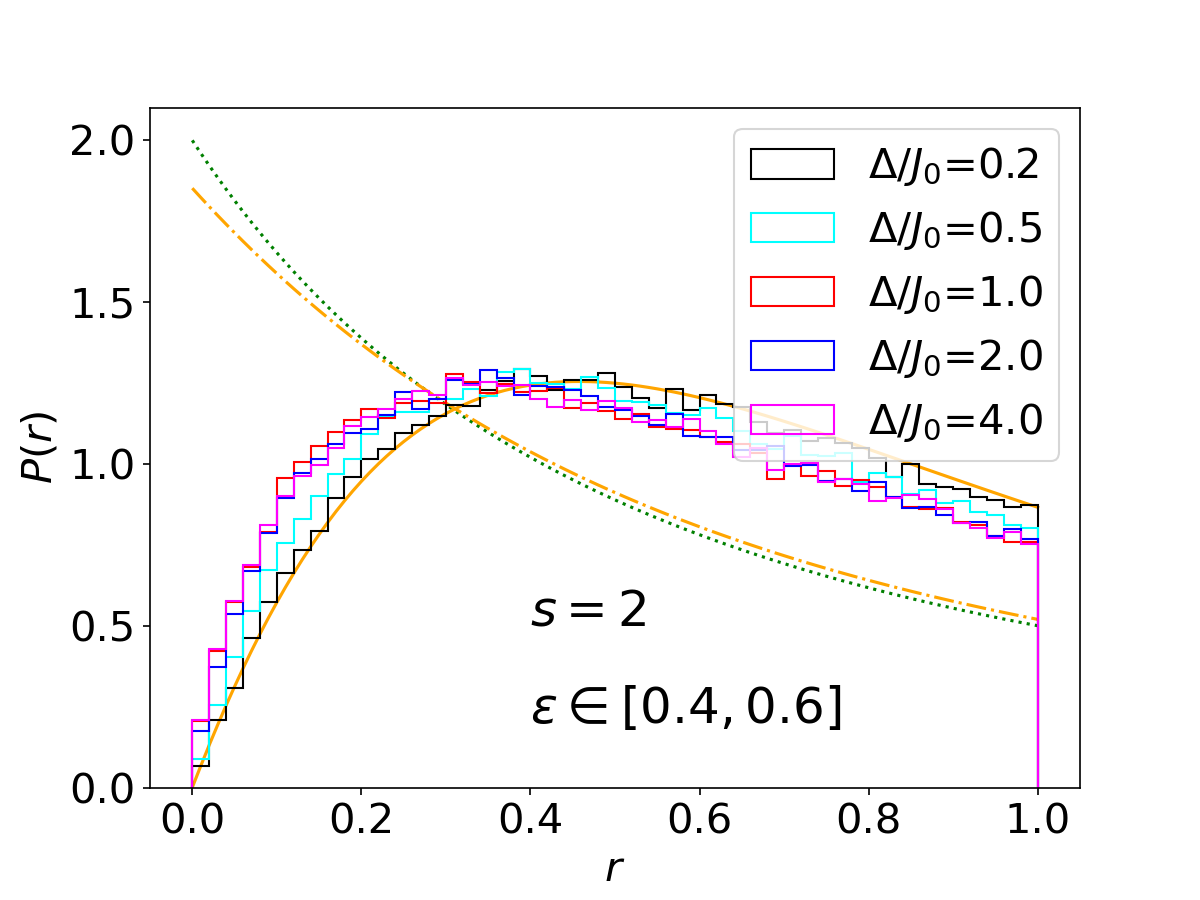}
    (c)\includegraphics[width=0.4\textwidth]{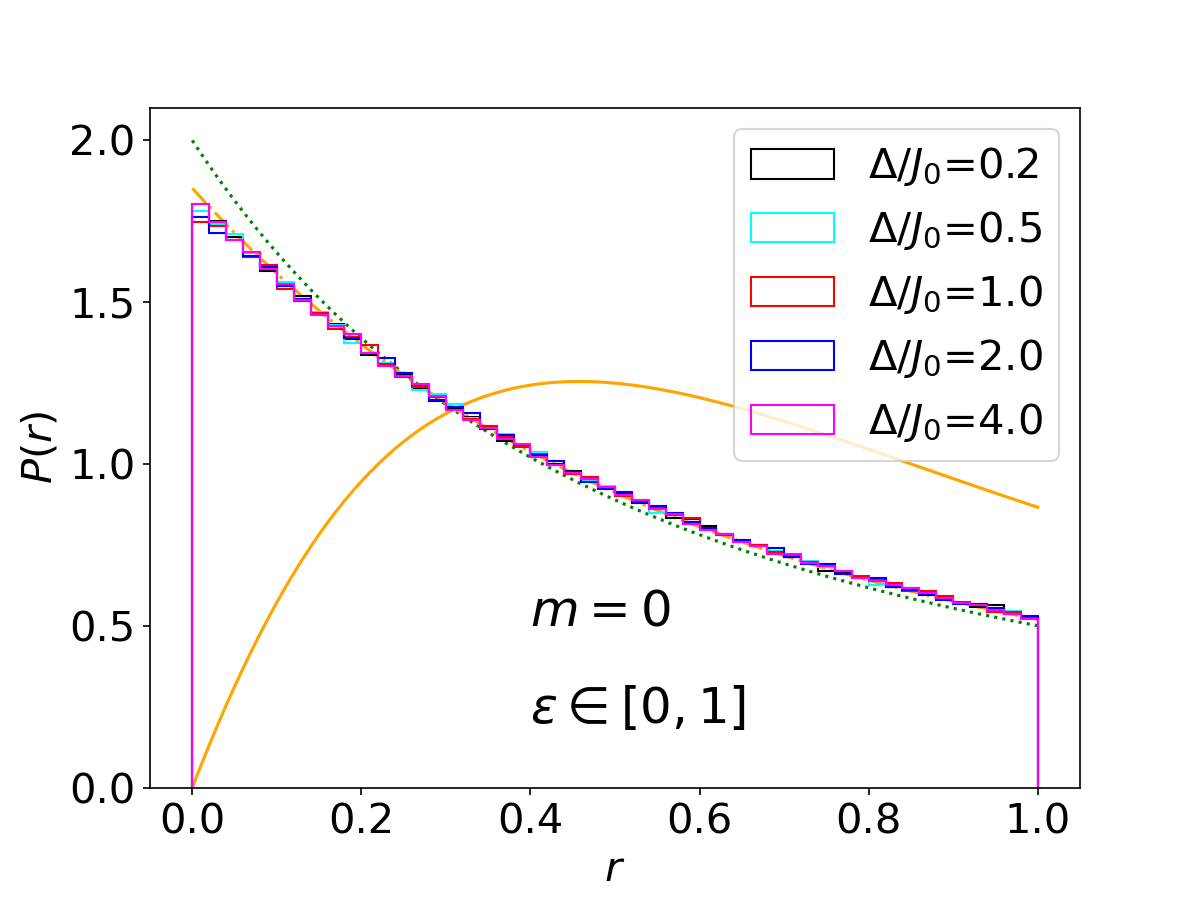}
    (d)\includegraphics[width=0.4\textwidth]{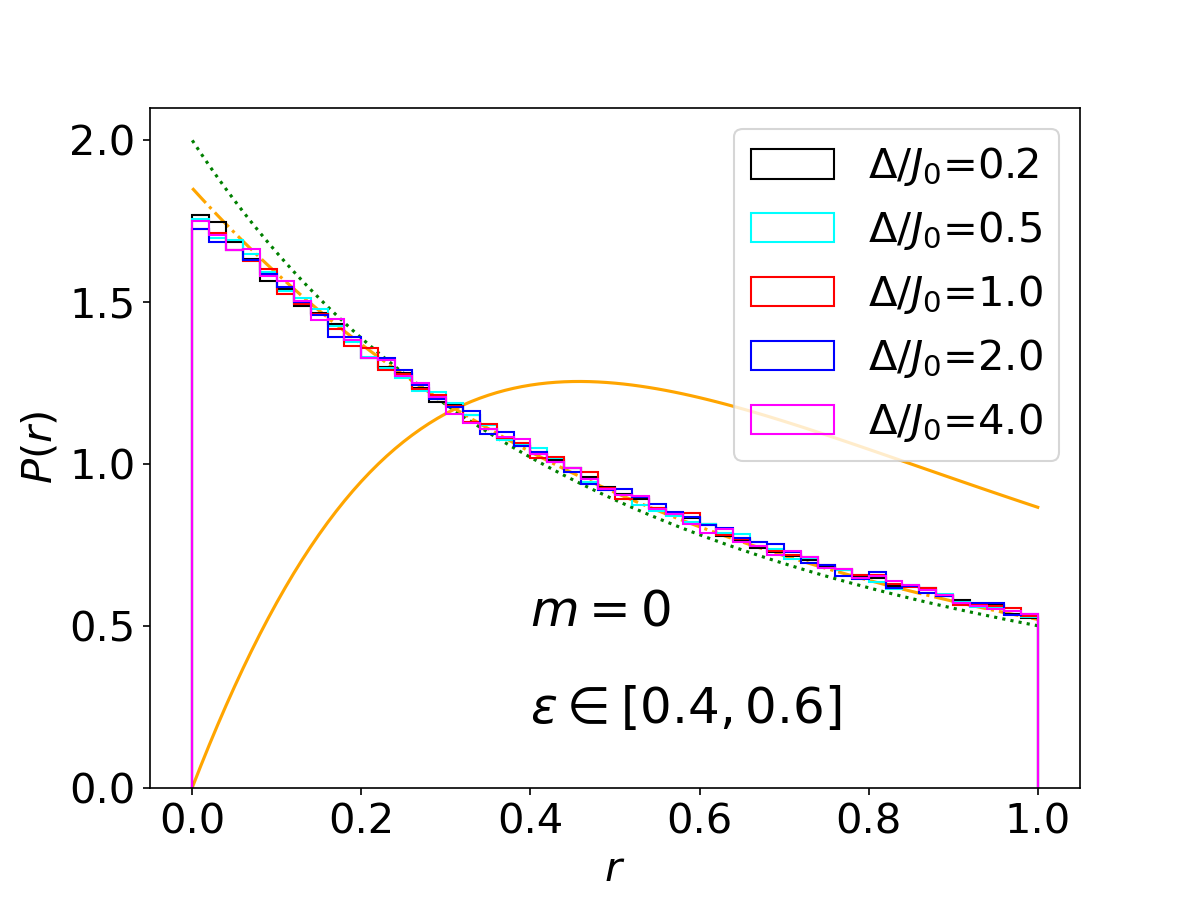}
    (e)\includegraphics[width=0.4\textwidth]{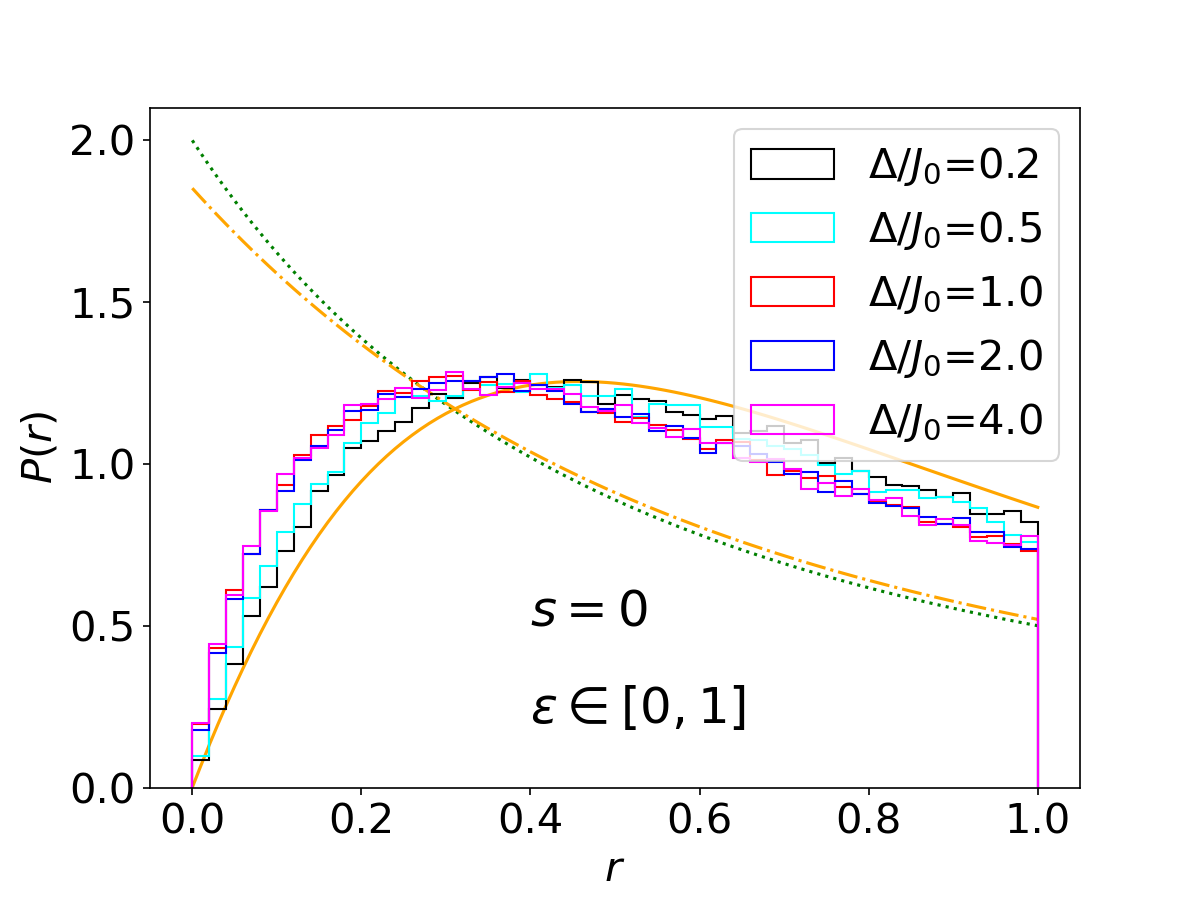}
    (f)\includegraphics[width=0.4\textwidth]{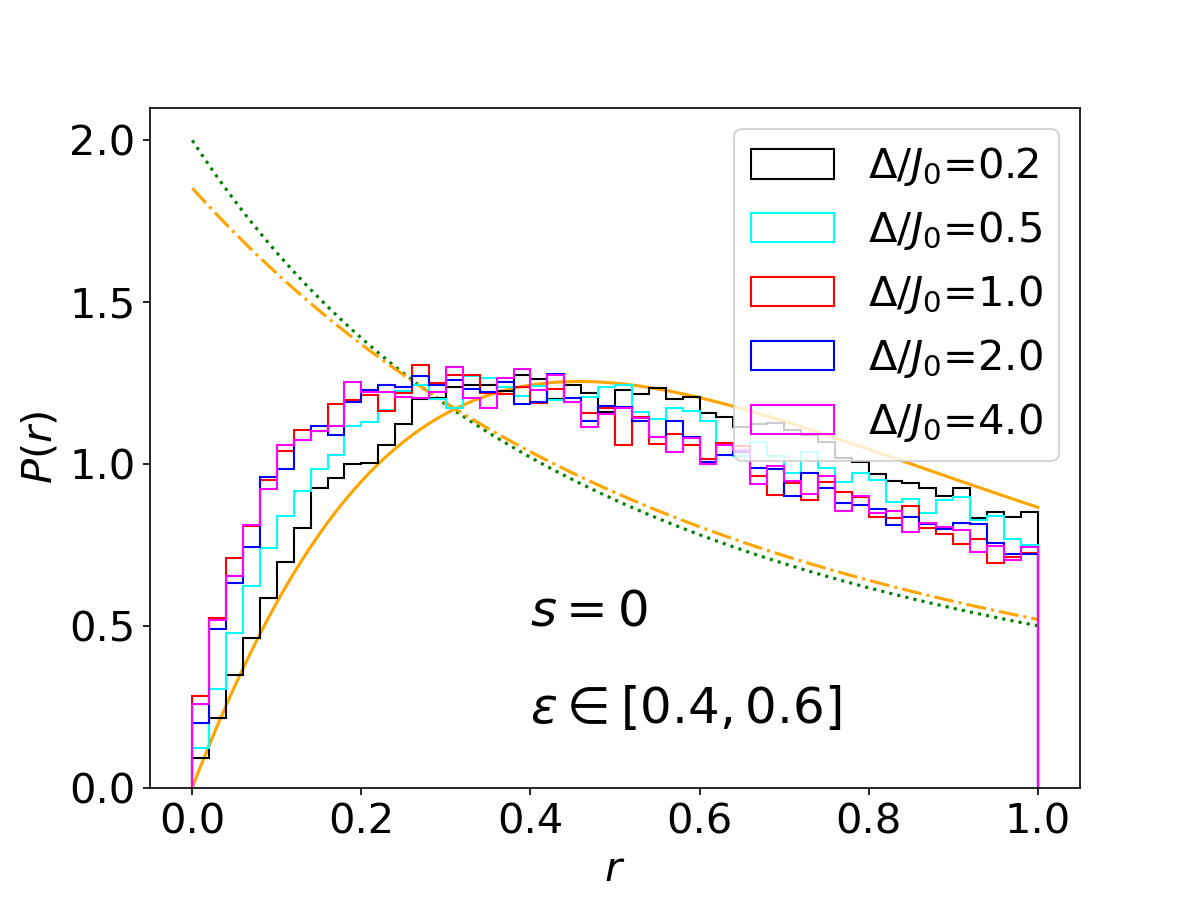}
\caption{Distribution $P(r)$ of consecutive-gap ratio for disorder distribution \emph{$p_1(J)$}. The order of the plots for spin sectors, the colors of the curves for different disorder strengths $\Delta/J_0$, $P_{\text{GOE}}$, $P_{\text{Poisson}}$ as well as $P_{\text{GOE}}^{(9)}$, the system size $L$ are the same as in Fig.\ \ref{fig-level-statistics-p0-all}. Panel (a) is the same as Fig.\ \ref{fig-level-statistics-p0-diff} (a).} 
\label{fig-level-statistics-p1-all}
\end{figure*}

\begin{figure*}[tb]
    (a)\includegraphics[width=0.4\textwidth]{\figdir/p2-P-L_16-Stot_2-allb-Rosenzweig-Porter.png}
    (b)\includegraphics[width=0.4\textwidth]{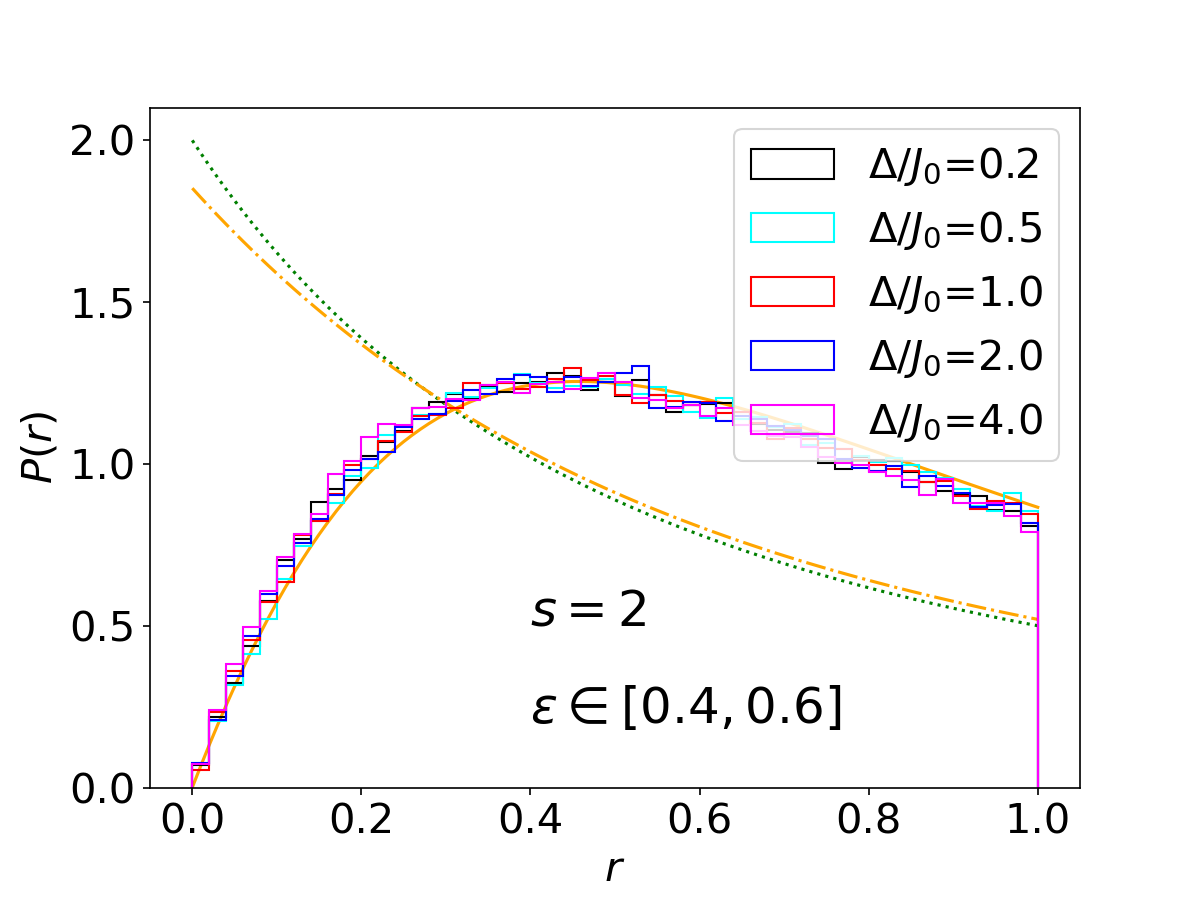}
    (c)\includegraphics[width=0.4\textwidth]{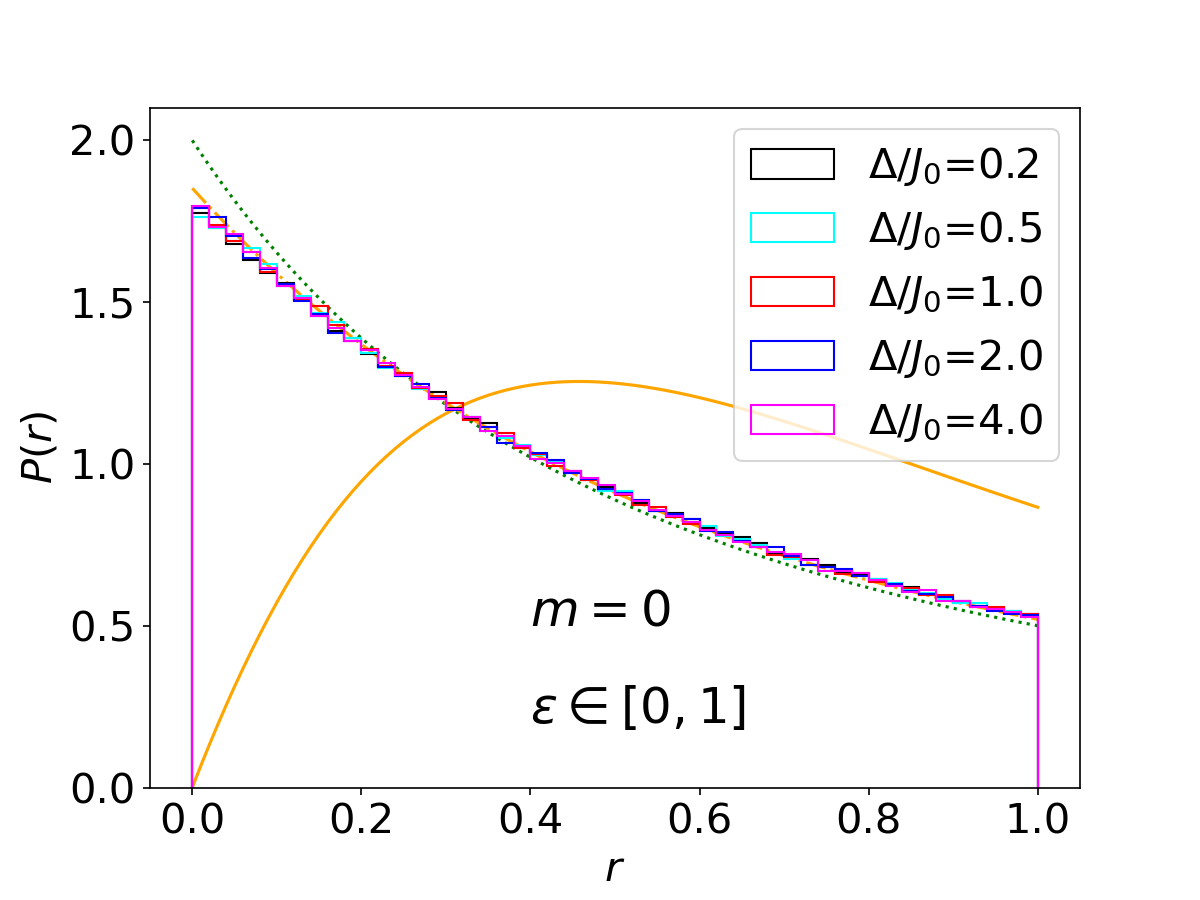}
    (d)\includegraphics[width=0.4\textwidth]{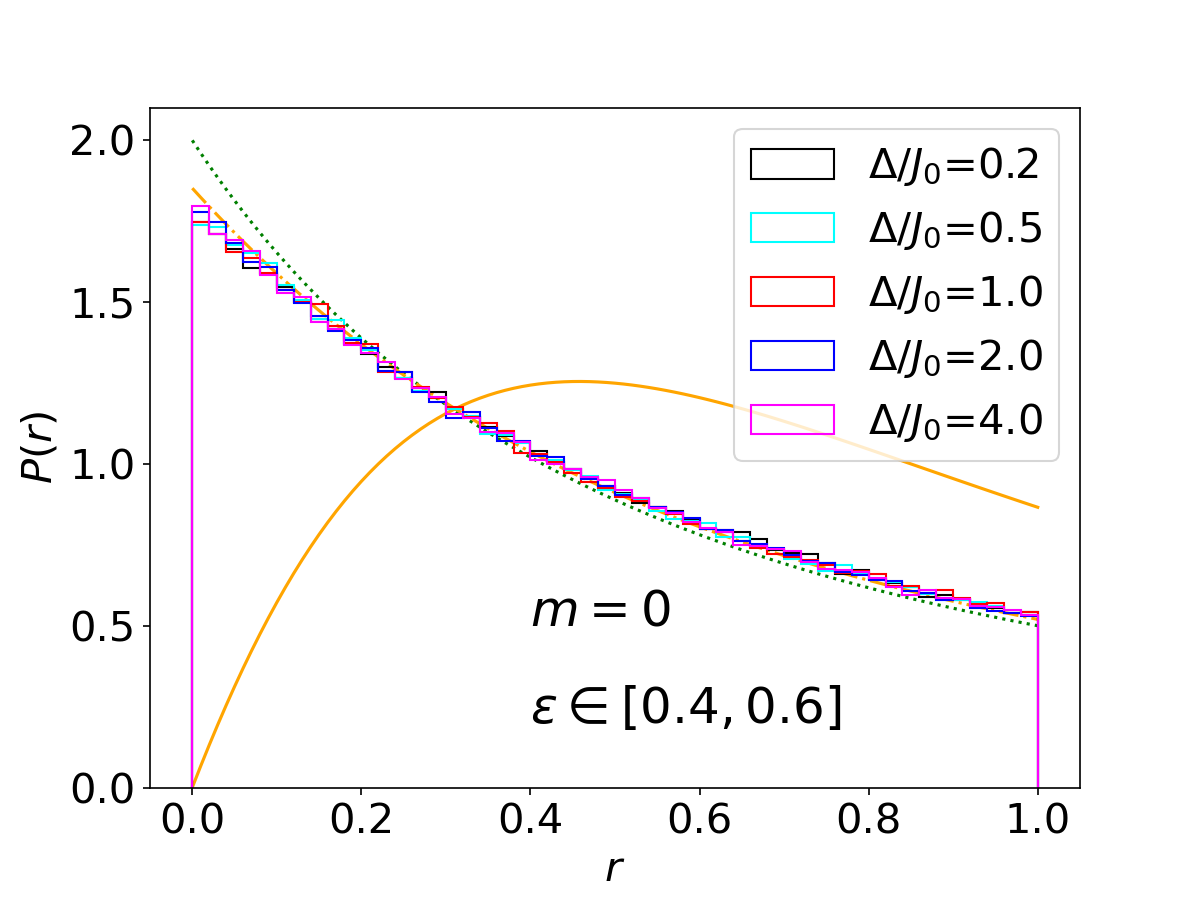}
    (e)\includegraphics[width=0.4\textwidth]{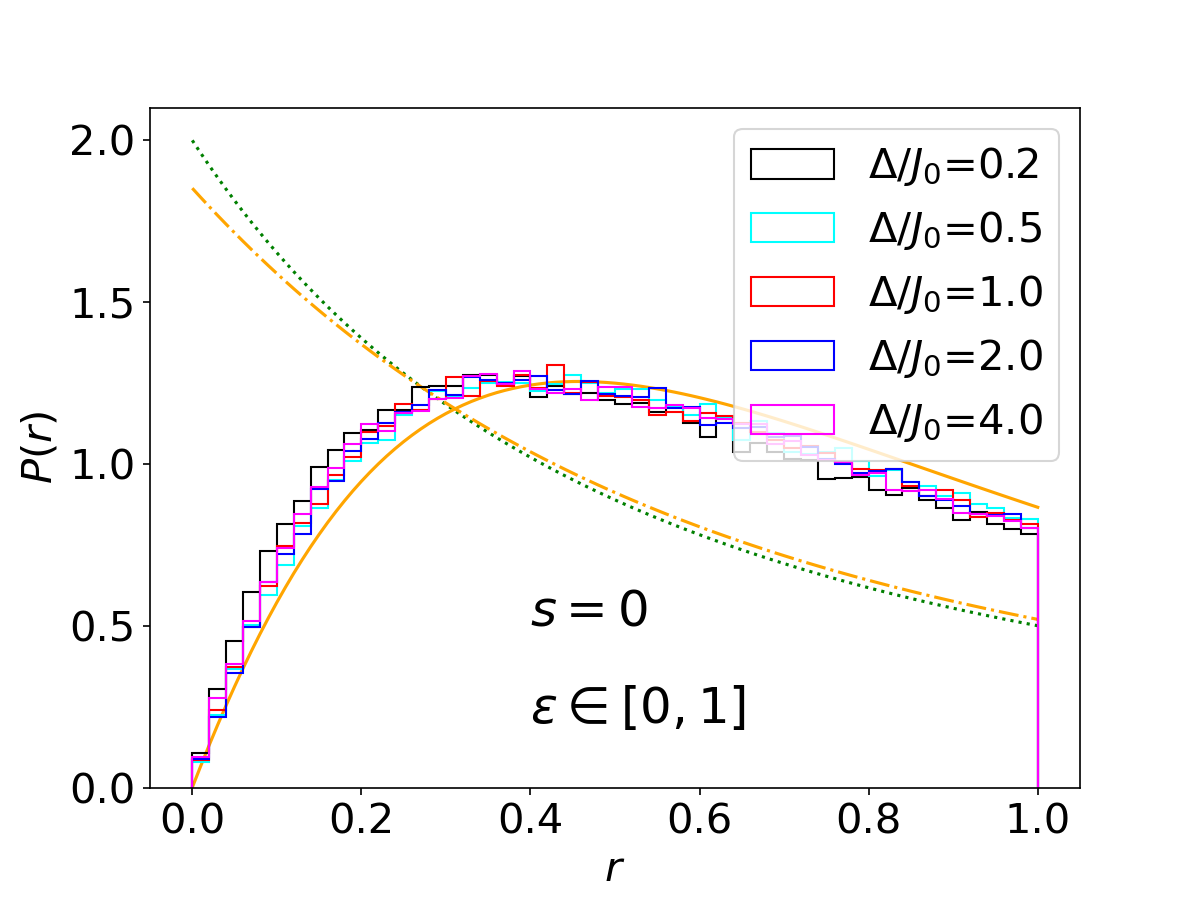}
    (f)\includegraphics[width=0.4\textwidth]{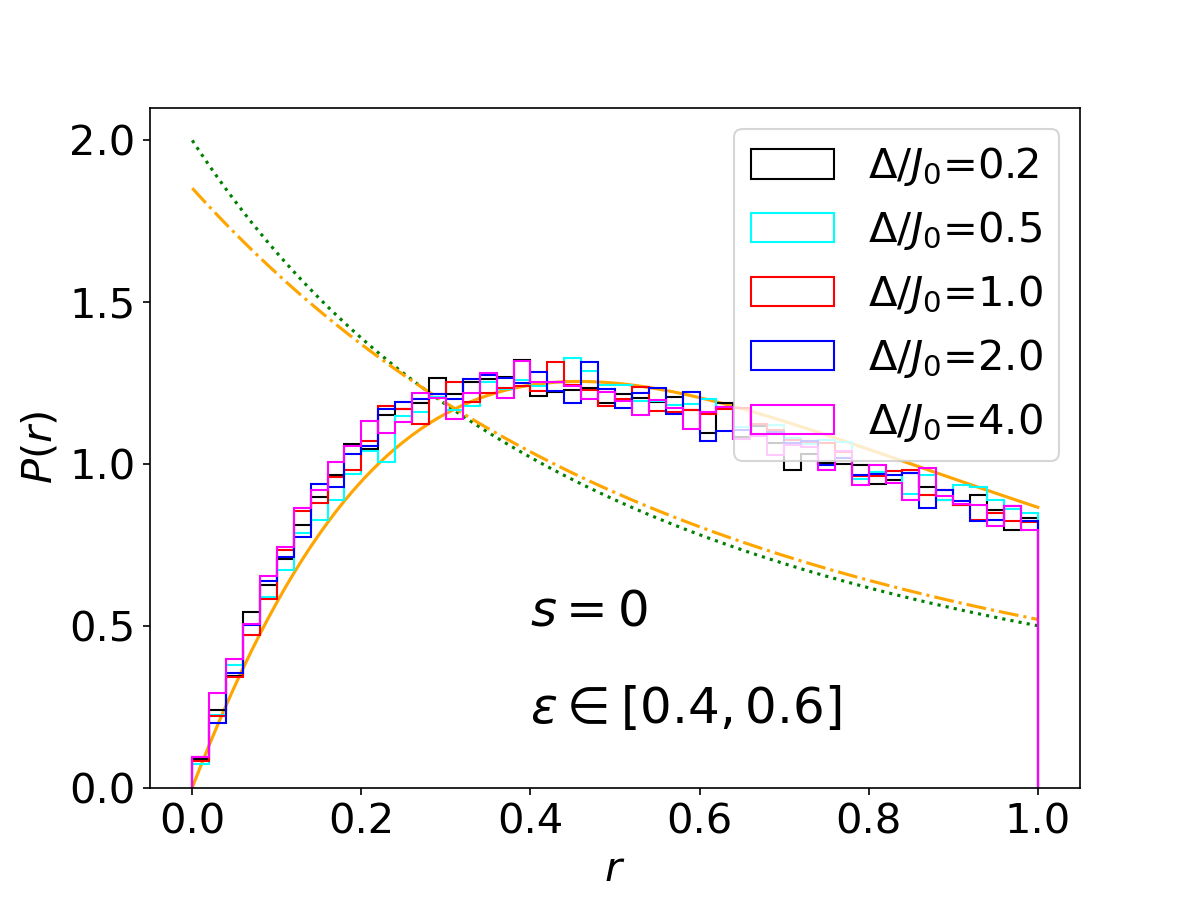}
\caption{Distribution $P(r)$ of consecutive-gap ratio for disorder distribution \emph{$p_2(J)$}. The order of the plots for spin sectors, the colors of the curves for different disorder strengths $\Delta/J_0$, $P_{\text{GOE}}$, $P_{\text{Poisson}}$ as well as $P_{\text{GOE}}^{(9)}$, and the system size $L$ are the same as in Figs.\ \ref{fig-level-statistics-p0-all} and \ref{fig-level-statistics-p1-all}. Panel (a) is the same as Fig.\ \ref{fig-level-statistics-p0-diff} (c).} 
\label{fig-level-statistics-p2-all}
\end{figure*}

In Fig.\ \ref{fig-level-statistics-p1-all} and Fig.\ \ref{fig-level-statistics-p2-all}, we show $P(r)$ for $p_1(J)$ and $p_2(J)$. Similar to Fig.\ \ref{fig-level-statistics-p0-all}, both the full spectrum and the central region are plotted for $s=2$, $m=0$, and $s=0$ sector. In $m=0$ sector, $P(r)$ is close to $P^{(9)}_{\text{GOE}}$ for both $p_1(J)$ and $p_2(J)$ as can be seen in Fig.\ \ref{fig-level-statistics-p1-all} (c+d) and Fig.\ \ref{fig-level-statistics-p2-all} (c+d). We also find that $P(r)$ in $s=2$ sector and $s=0$ sector agrees better with $P_{\text{GOE}}$ for both $p_1(J)$ and $p_2(J)$ than $p_0(J)$.

\begin{figure*}[tb]
    (a)\includegraphics[width=0.4\textwidth]{\figdir/p1-L_16-P-GOE-stot_2.png}
    (b)\includegraphics[width=0.4\textwidth]{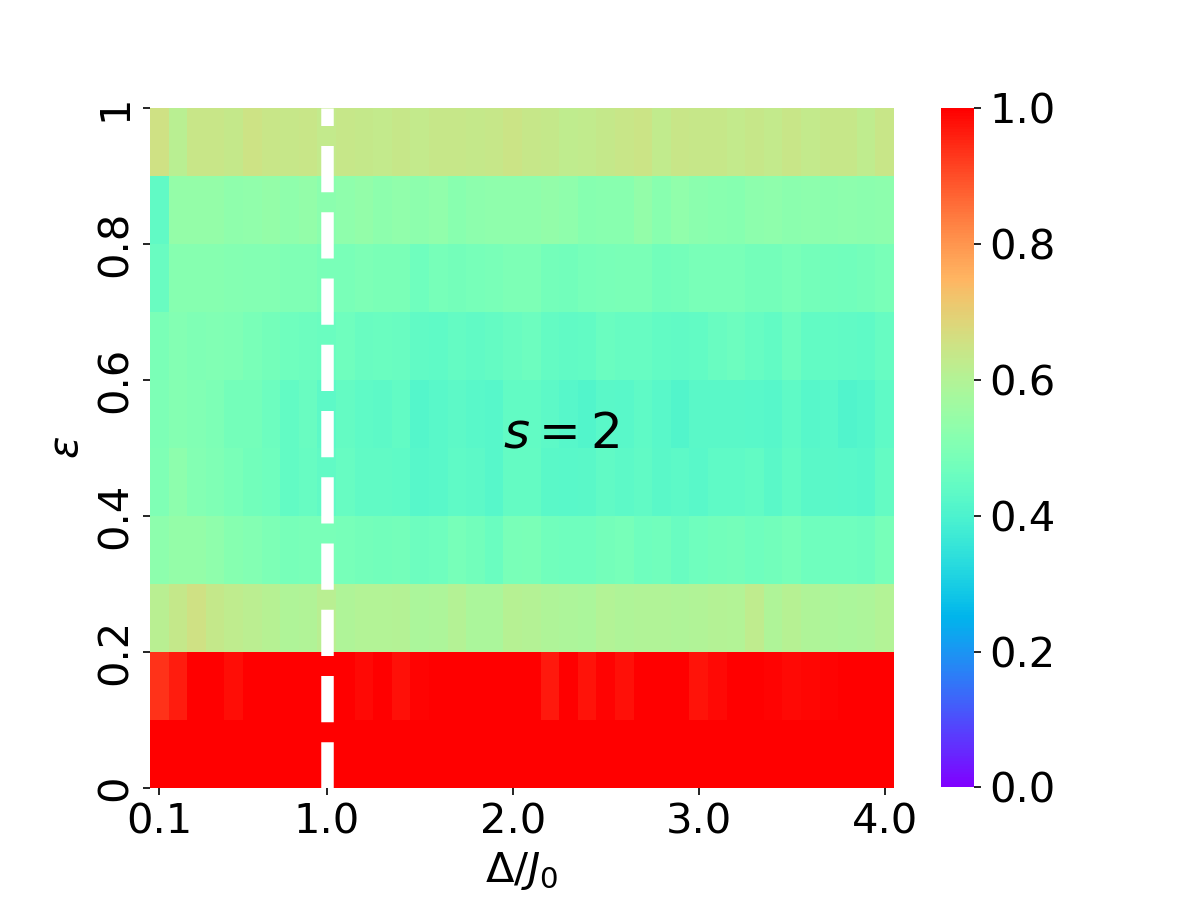}
    (c)\includegraphics[width=0.4\textwidth]{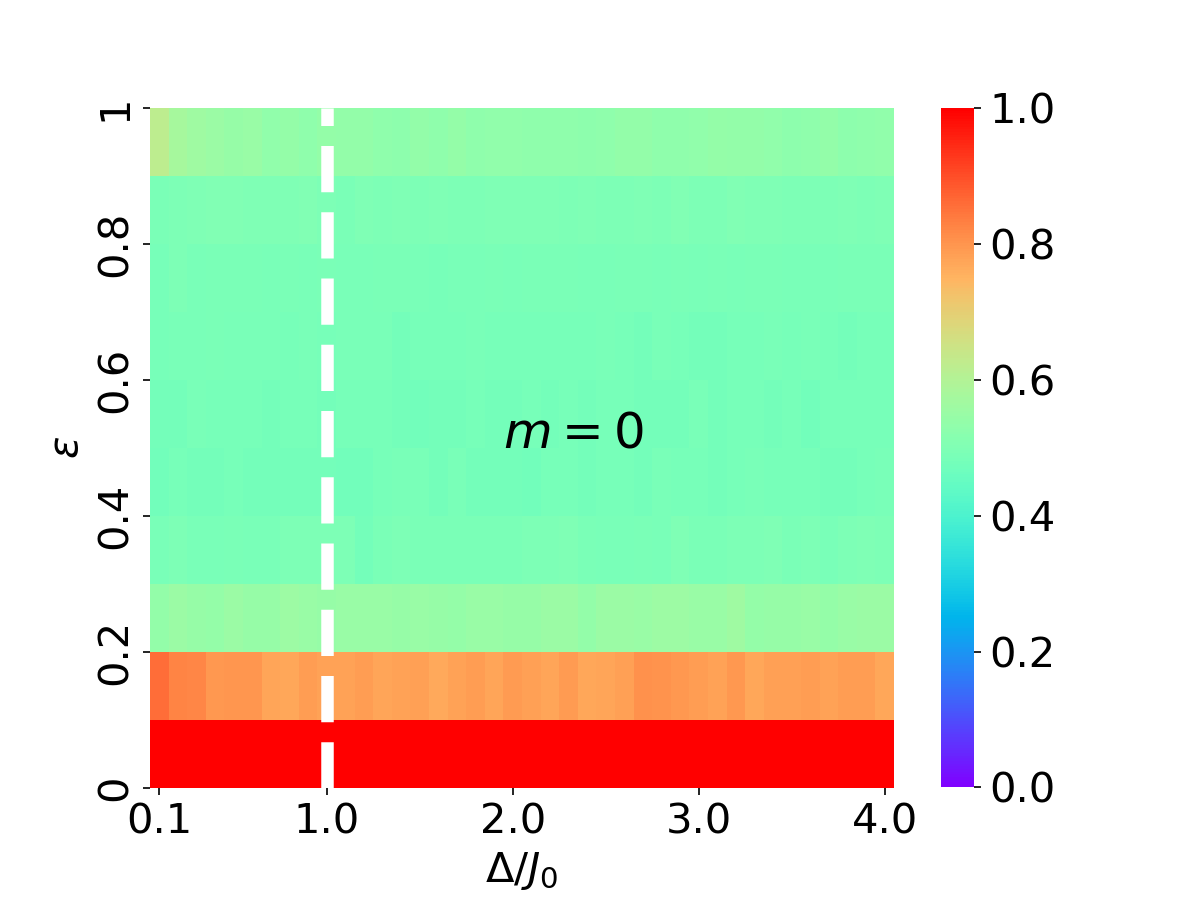}
    (d)\includegraphics[width=0.4\textwidth]{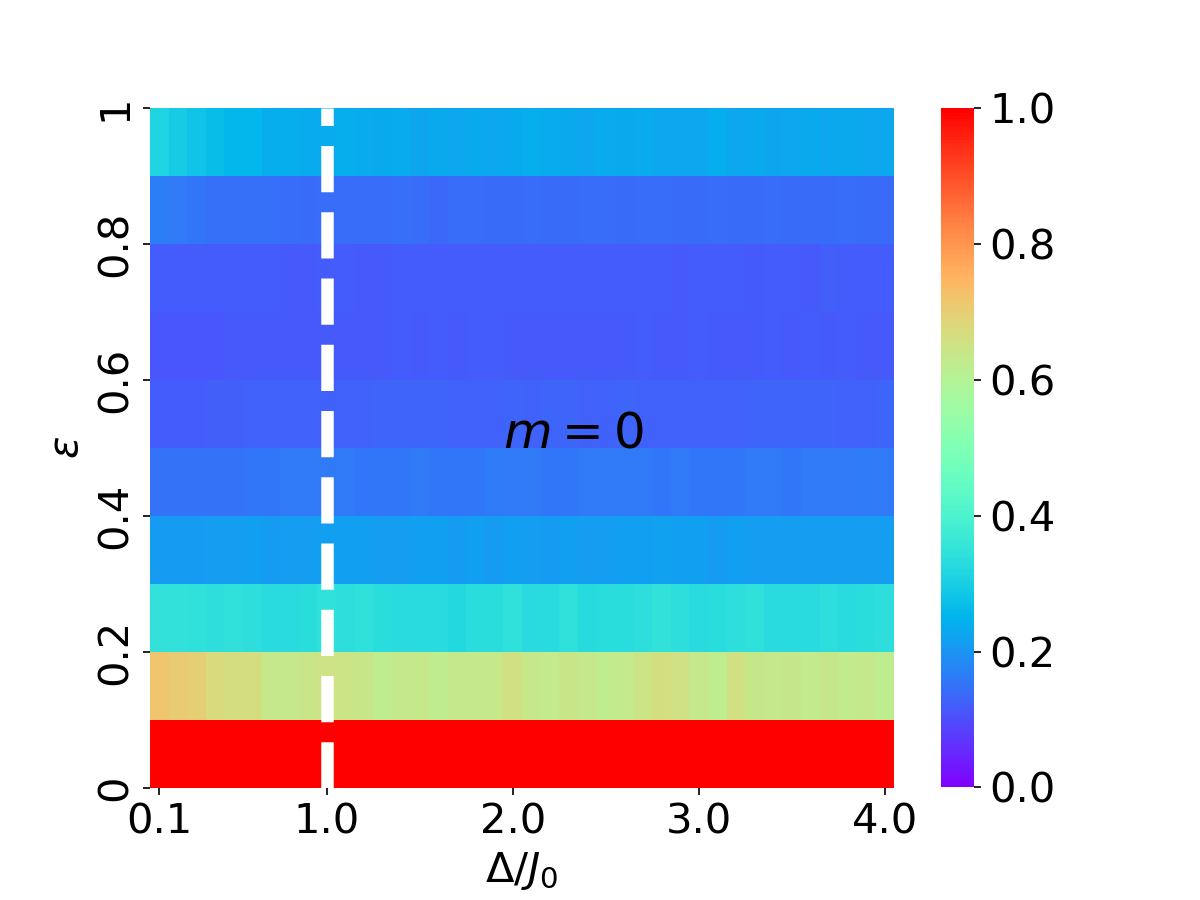}
    (e)\includegraphics[width=0.4\textwidth]{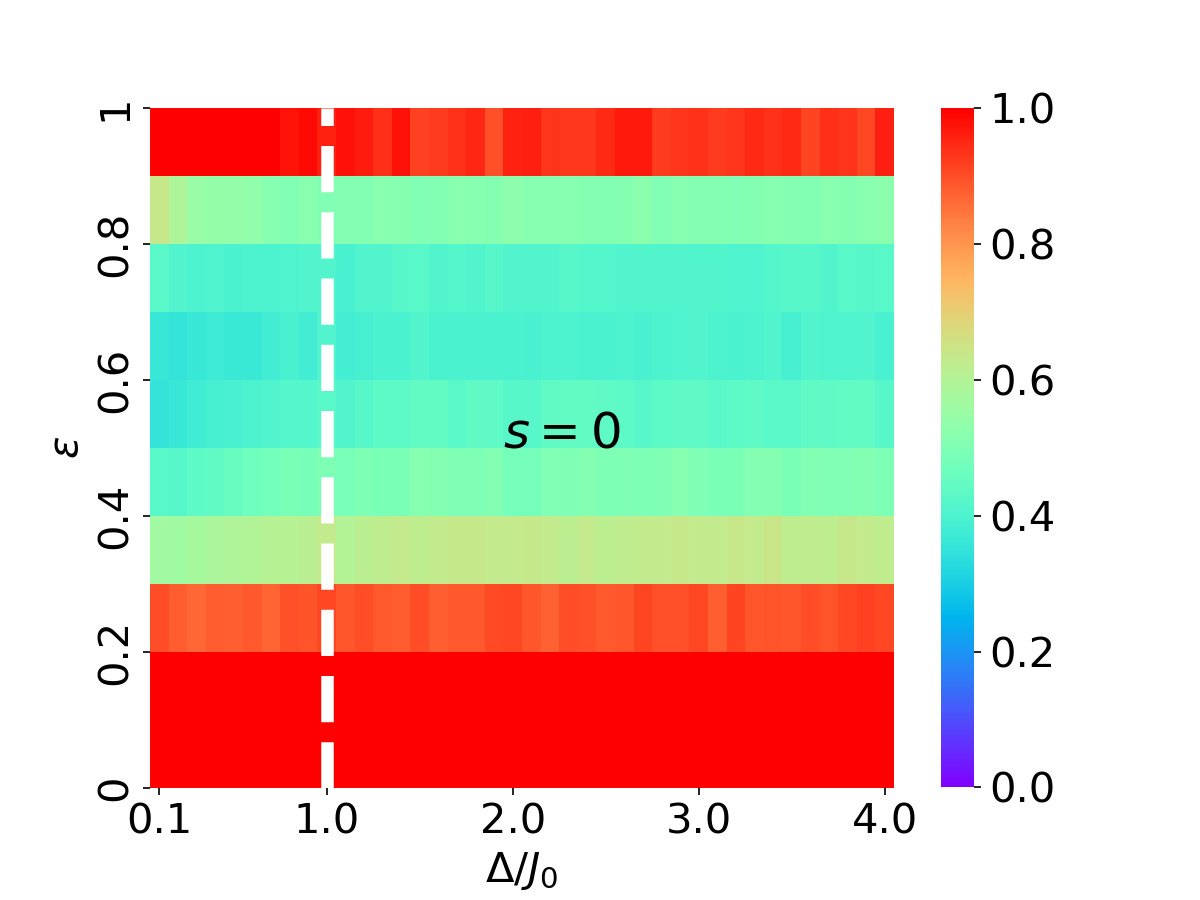}
    (f)\includegraphics[width=0.4\textwidth]{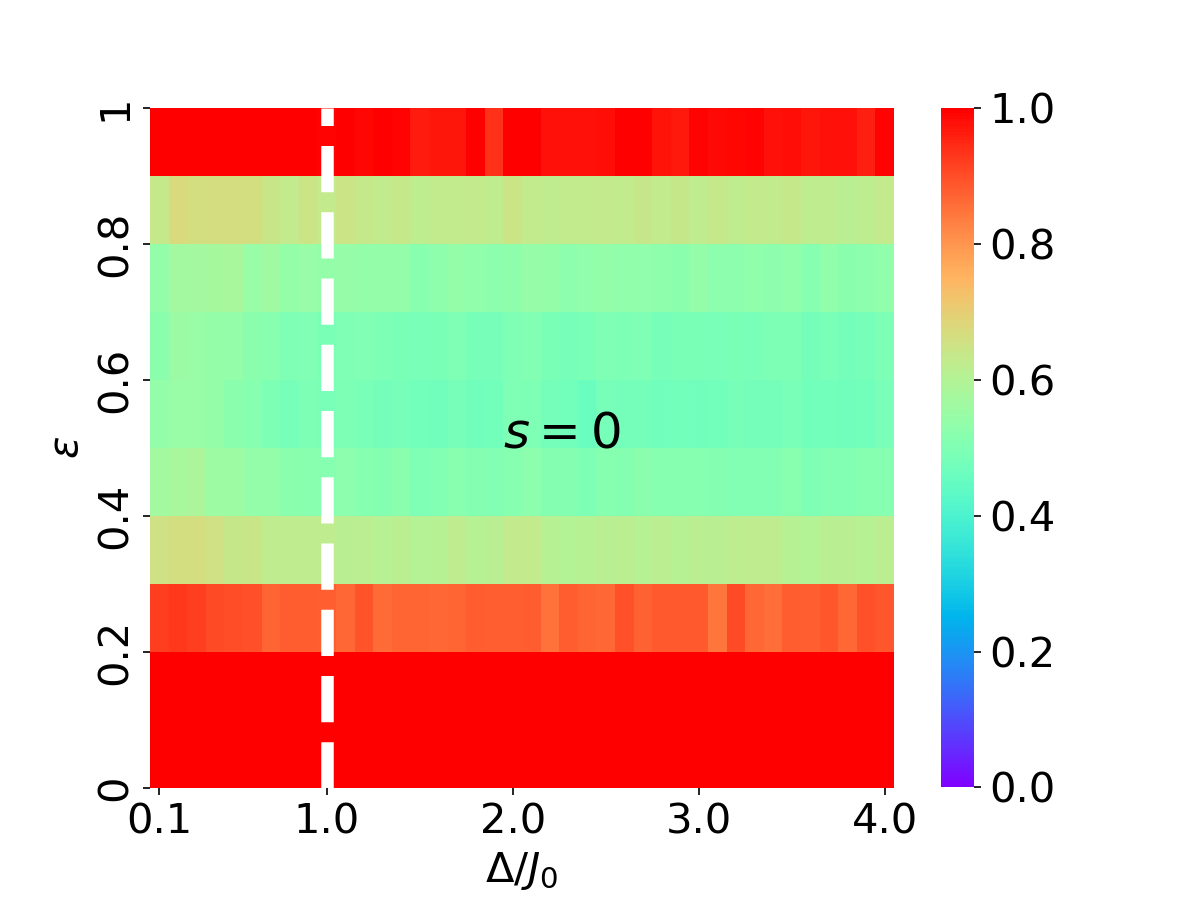}
\caption{Difference $\Delta P_{\text{GOE}}$ and $\Delta P_{\text{Poisson}}$ as a function of disorder strength $\Delta/J_0$ and reduced energy $\epsilon$ for disorder distribution $p_1(J)$. The order of the plots for different spin sectors and the position of the white-dashed vertical line is the same as in Fig.\ \ref{fig-level-statistics-p0-diff}. Panel (a) is the same as Fig.\ \ref{fig-level-statistics-p0-diff} (b).} 
\label{fig-level-statistics-diff-p1}
\end{figure*}

\begin{figure*}[tb]
    (a)\includegraphics[width=0.4\textwidth]{\figdir/p2-L_16-P-GOE-stot_2.png}
    (b)\includegraphics[width=0.4\textwidth]{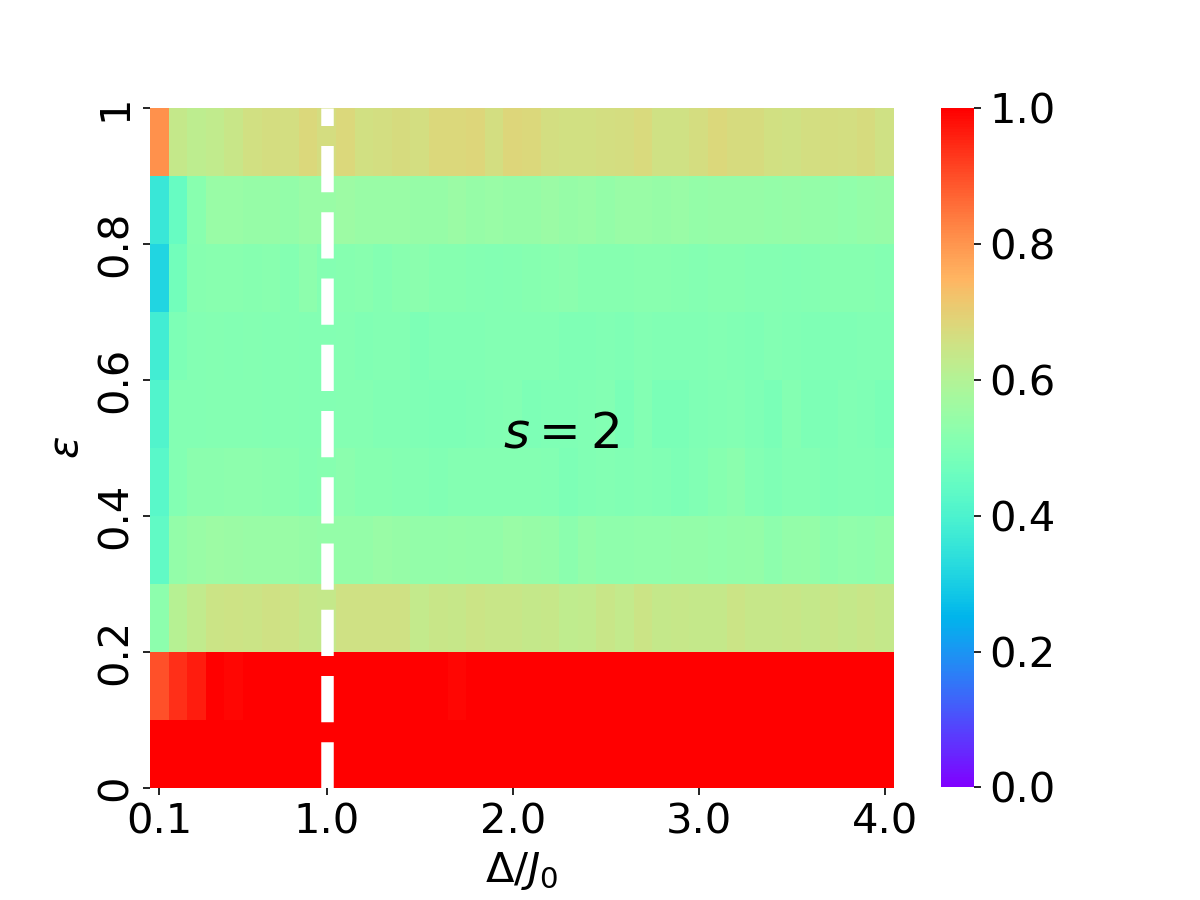}
    (c)\includegraphics[width=0.4\textwidth]{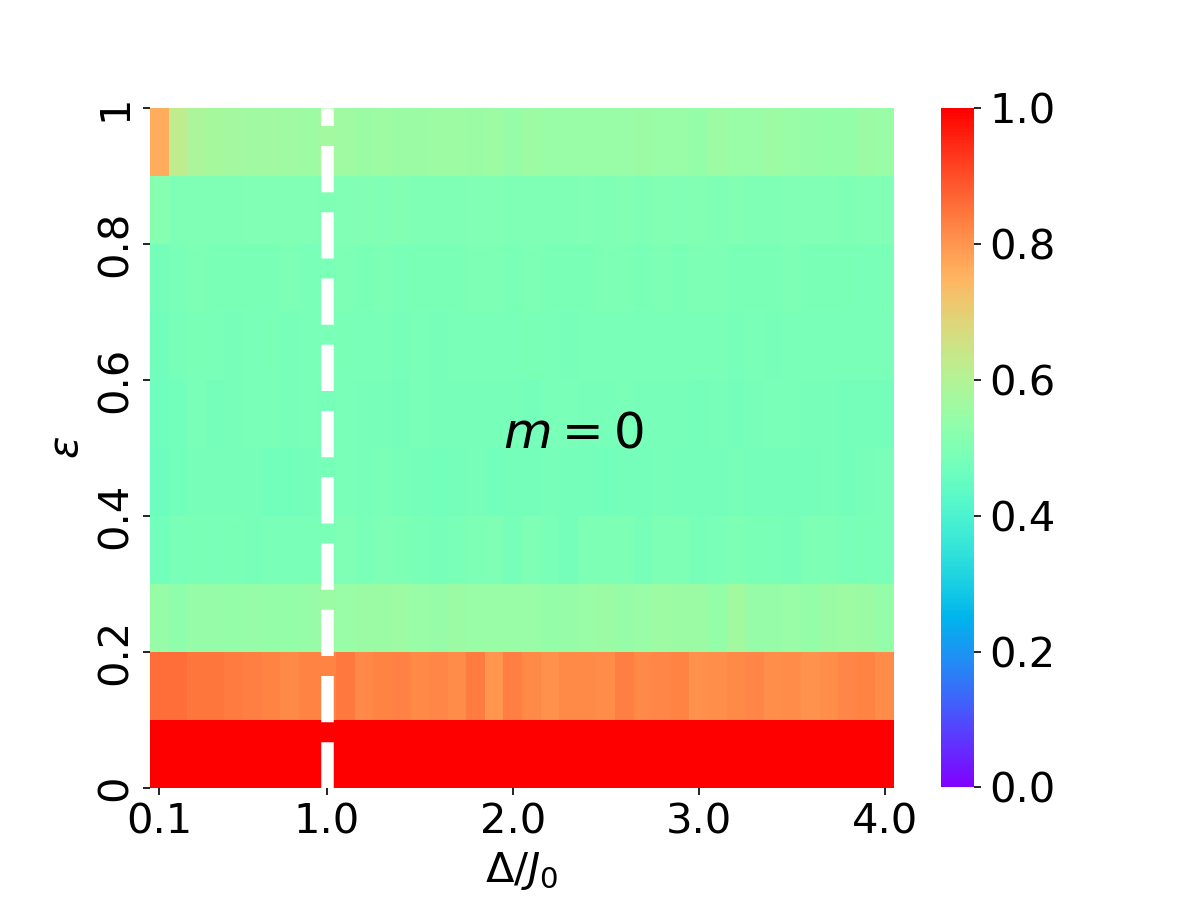}
    (d)\includegraphics[width=0.4\textwidth]{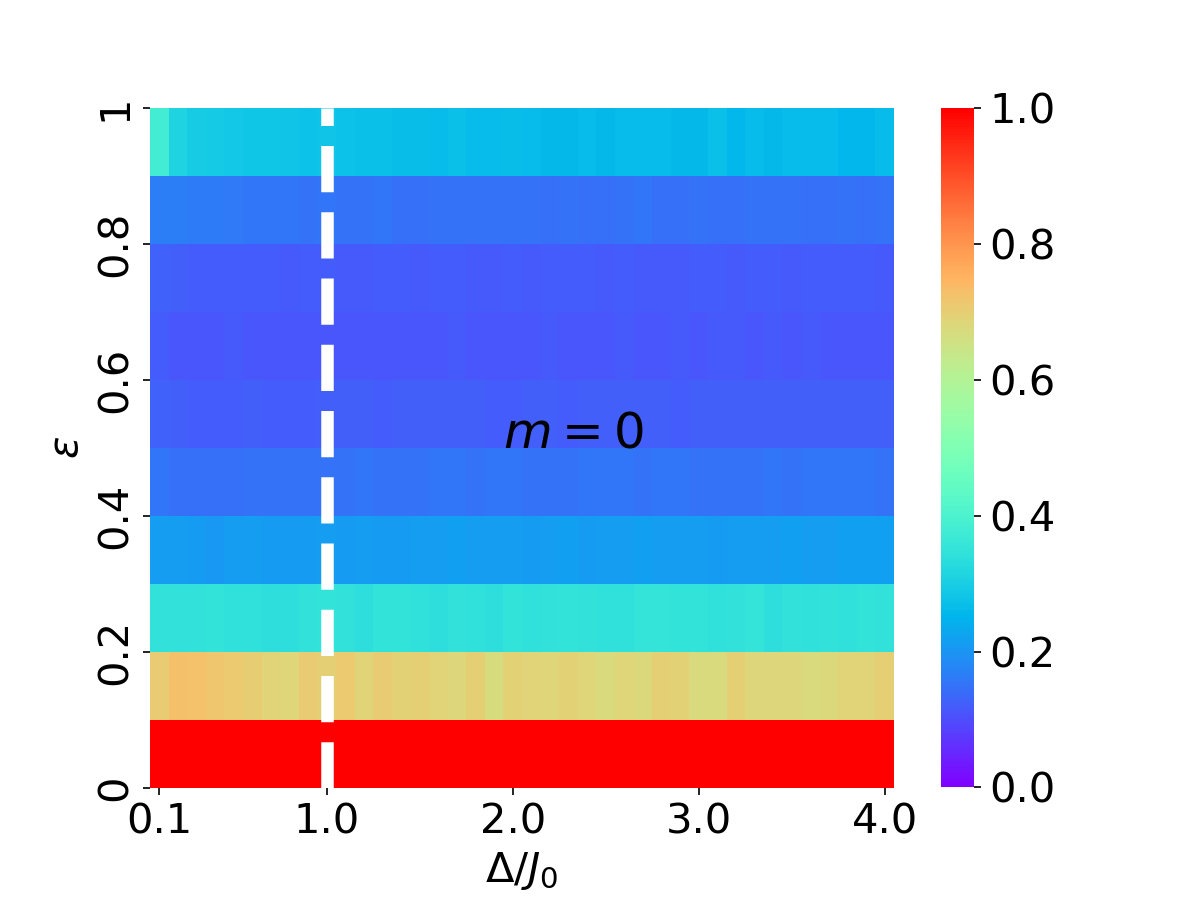}
    (e)\includegraphics[width=0.4\textwidth]{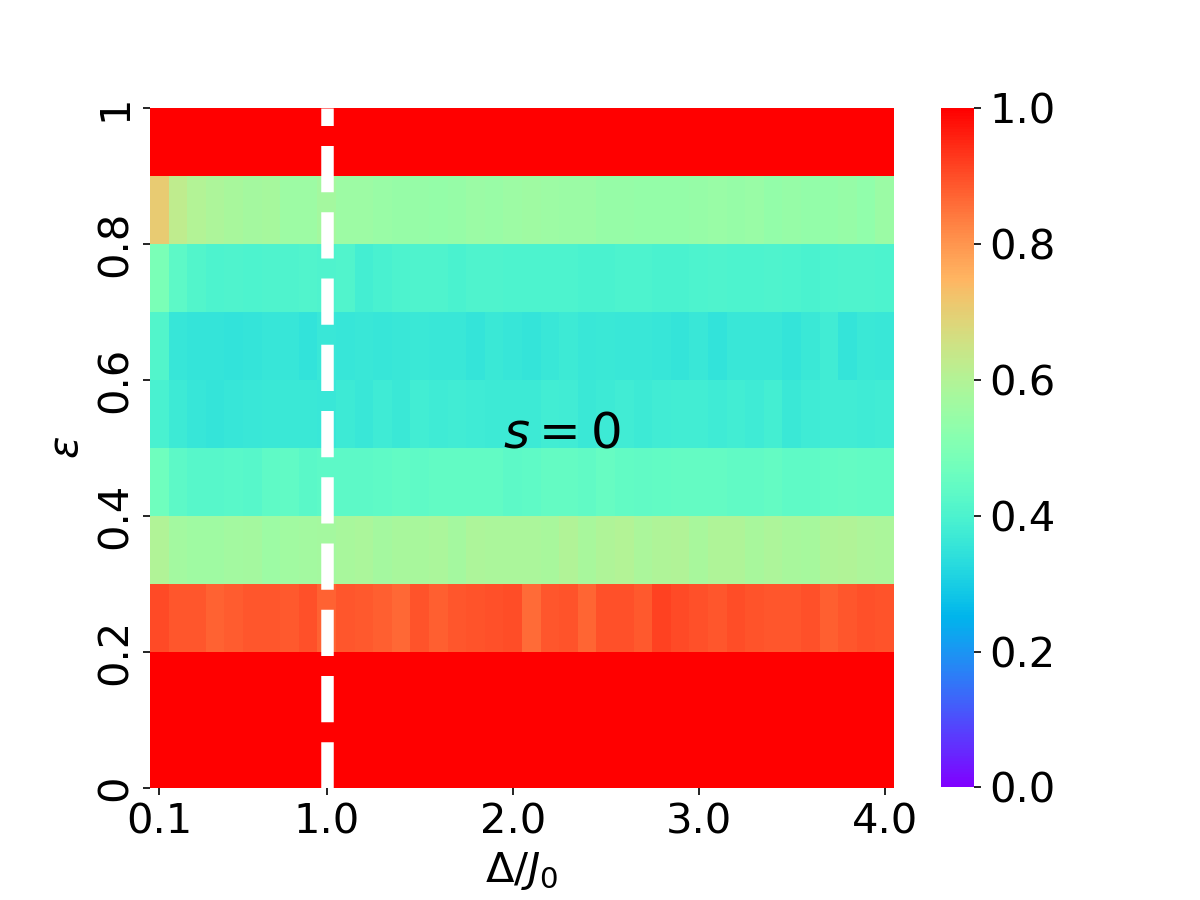}
    (f)\includegraphics[width=0.4\textwidth]{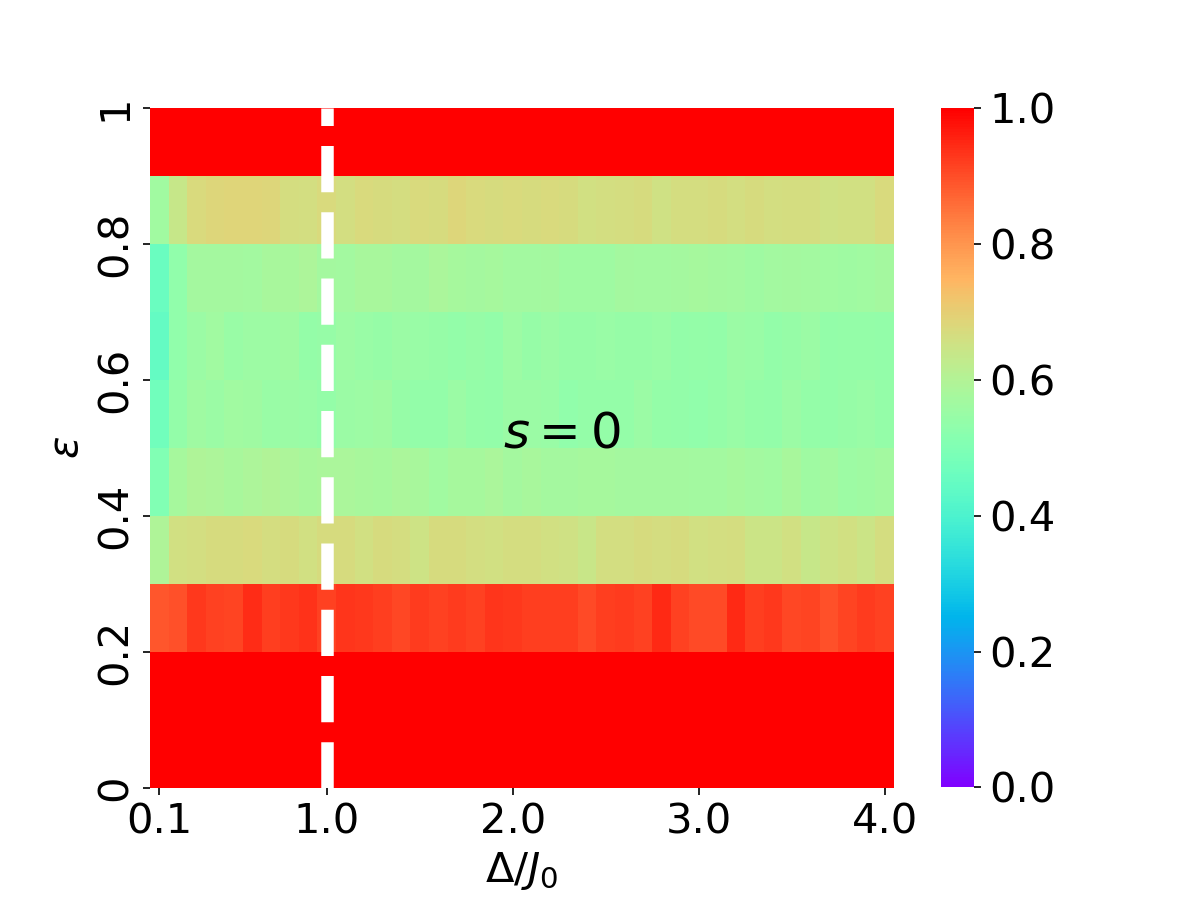}
\caption{Difference $\Delta P_{\text{GOE}}$ and $\Delta P_{\text{Poisson}}$ as a function of disorder strength $\Delta/J_0$ and reduced energy $\epsilon$ for disorder distribution $p_2(J)$. The order of the plots for different spin sectors and the position of the white-dashed vertical line is the same as in Figs.\ \ref{fig-level-statistics-p0-diff} and \ref{fig-level-statistics-diff-p1}. Panel (a) is the same as Fig.\ \ref{fig-level-statistics-p0-diff} (d).} 
\label{fig-level-statistics-diff-p2}
\end{figure*}

The difference $\Delta P_{\text{GOE}}$ and $\Delta P_{\text{Poisson}}$ in different energy regimes as a function of $\Delta$ for both $p_1(J)$ and $p_2(J)$ are shown in Fig.\ \ref{fig-level-statistics-diff-p1} and Fig.\ \ref{fig-level-statistics-diff-p2}. As mentioned in the main text, the change in differences as we cross $\Delta/J_0=1$ is much less obvious for $p_1(J)$ and $p_2(J)$ compared with the case in $p_0(J)$.

\begin{figure*}[tb]
$\mathcal{P}$\qquad\qquad\qquad\qquad\qquad\qquad\qquad\qquad$S_E/L$ \qquad\qquad\qquad\qquad\qquad\qquad\qquad\qquad$\lambda$ \qquad\\
    (a)\includegraphics[width=0.3\textwidth]{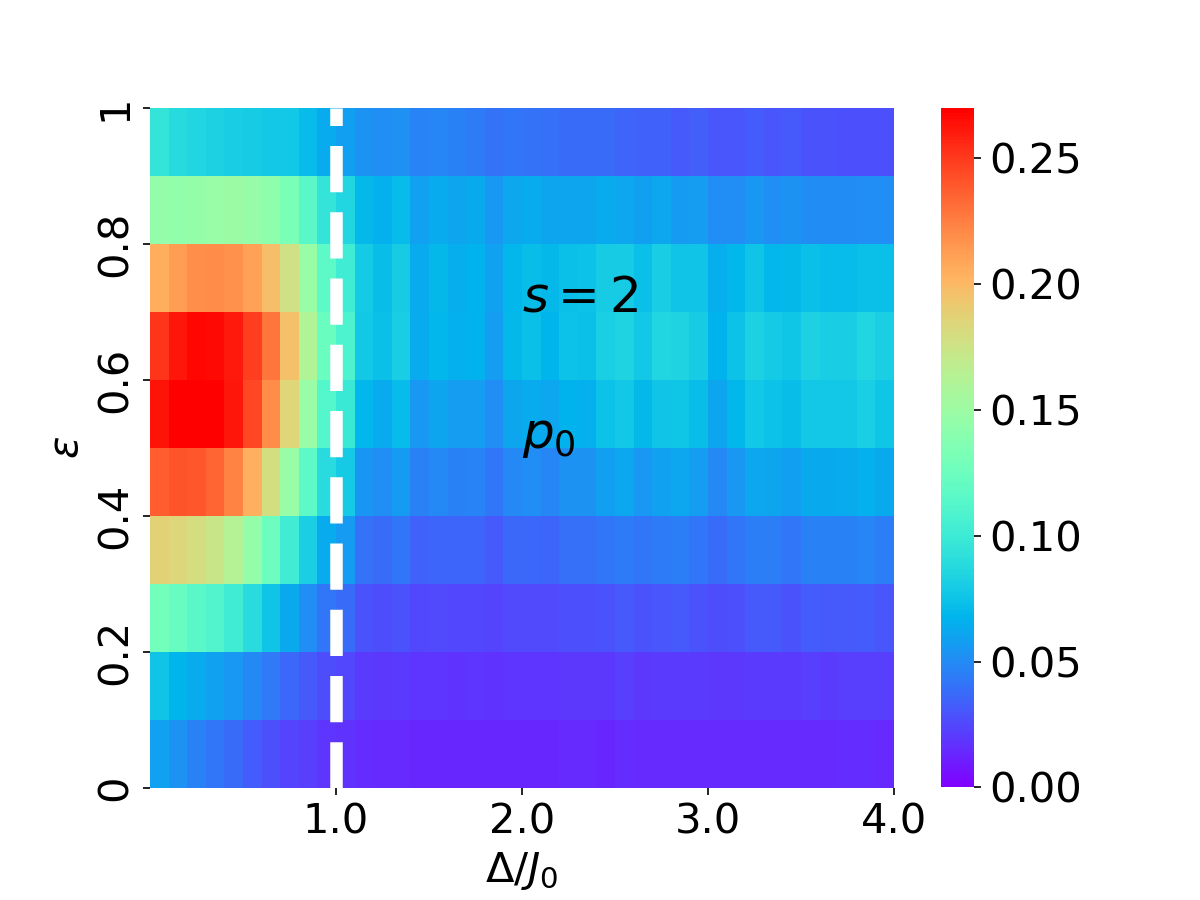}
    (b)\includegraphics[width=0.3\textwidth]{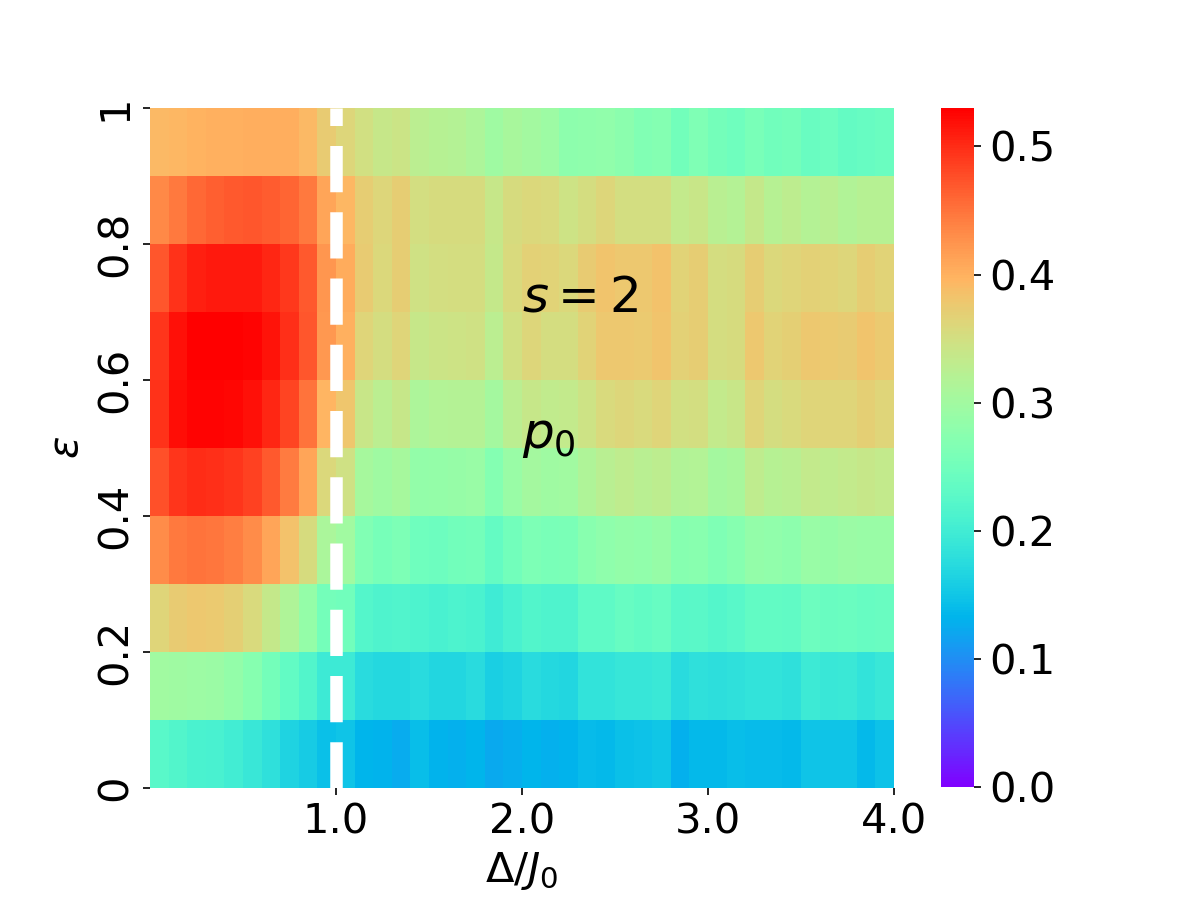}
    (c)\includegraphics[width=0.3\textwidth]{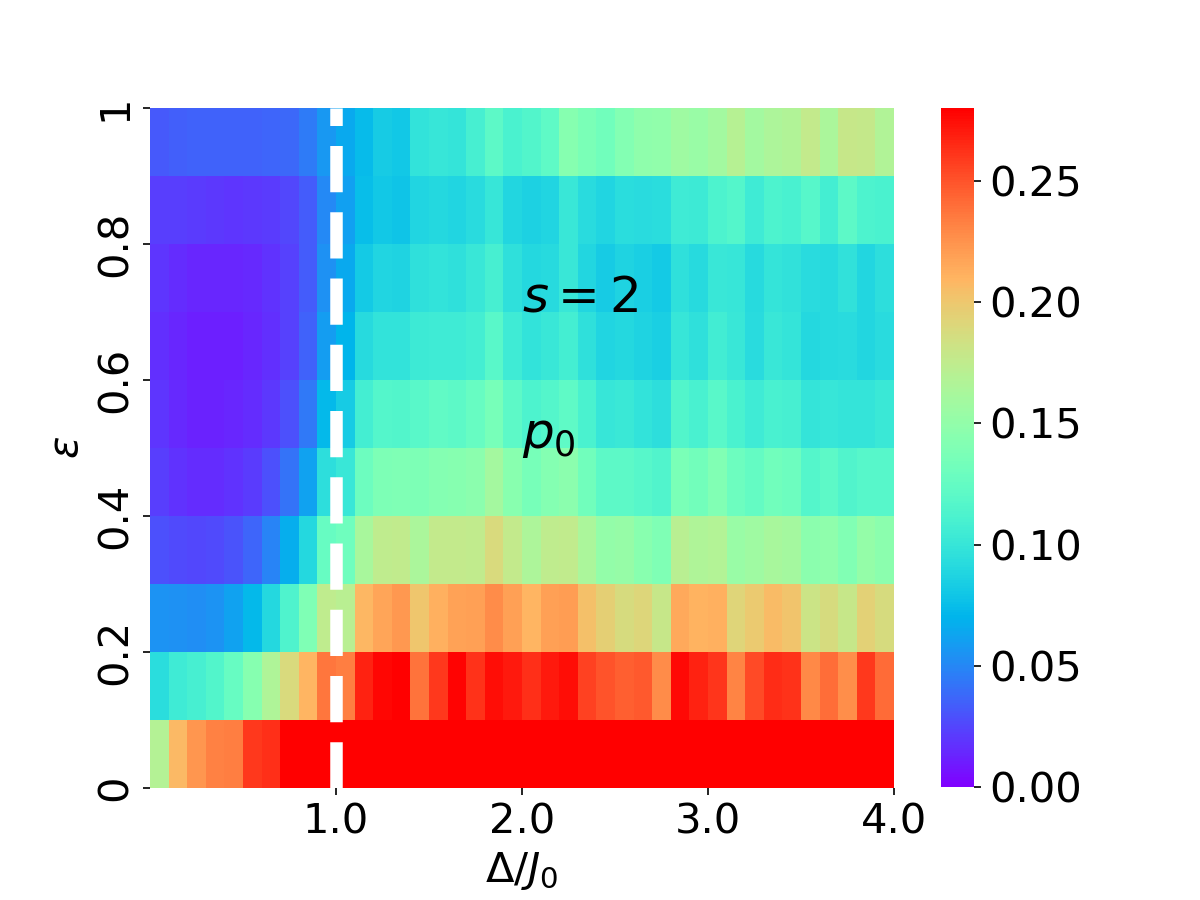}\\
    (d)\includegraphics[width=0.3\textwidth]{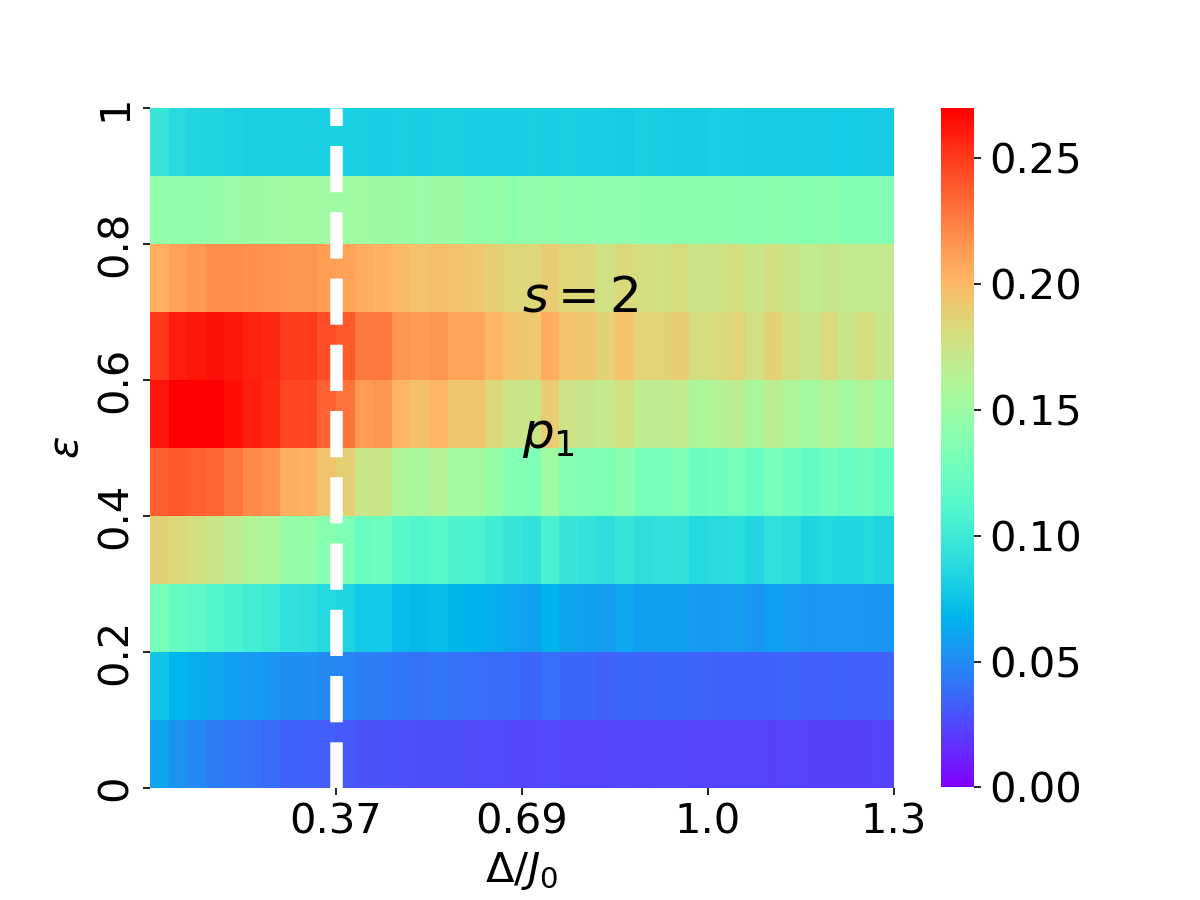}
    (e)\includegraphics[width=0.3\textwidth]{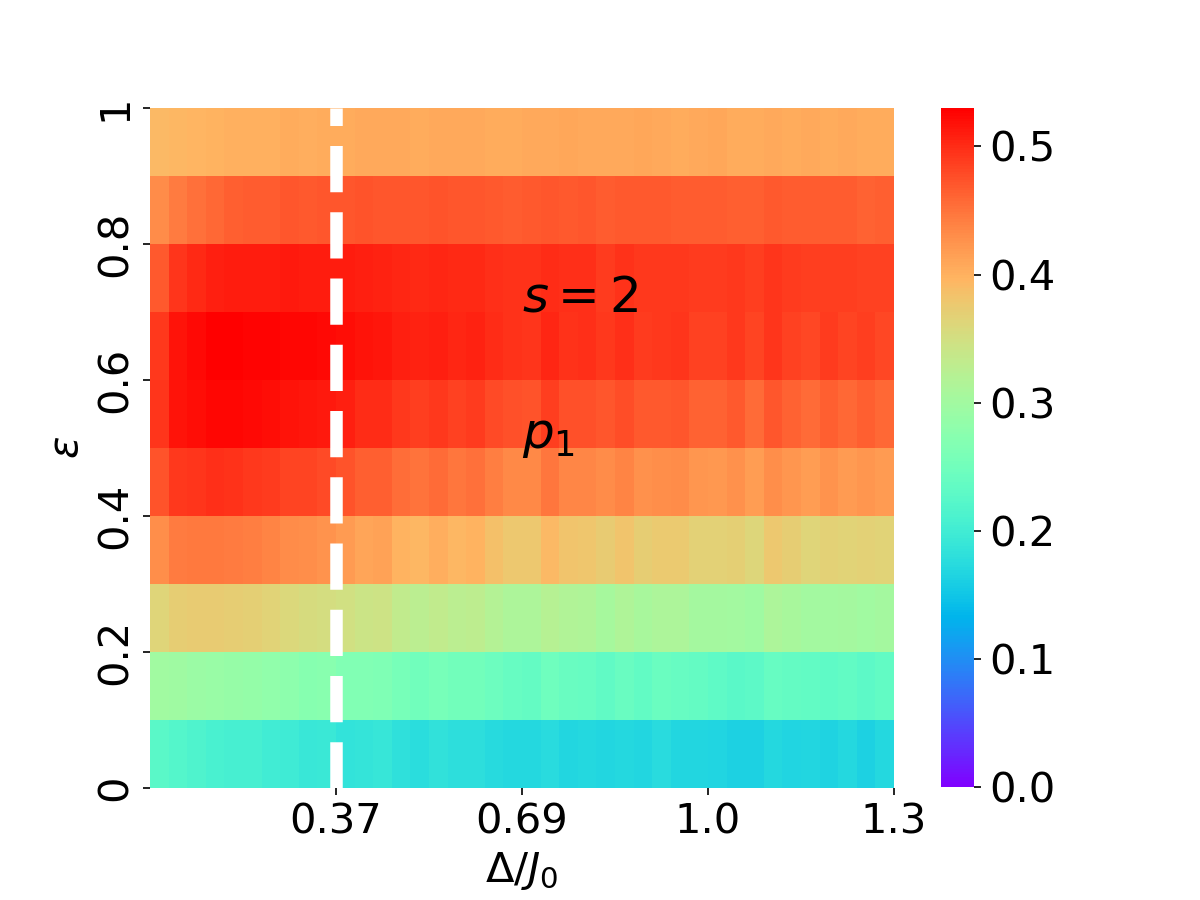}
    (f)\includegraphics[width=0.3\textwidth]{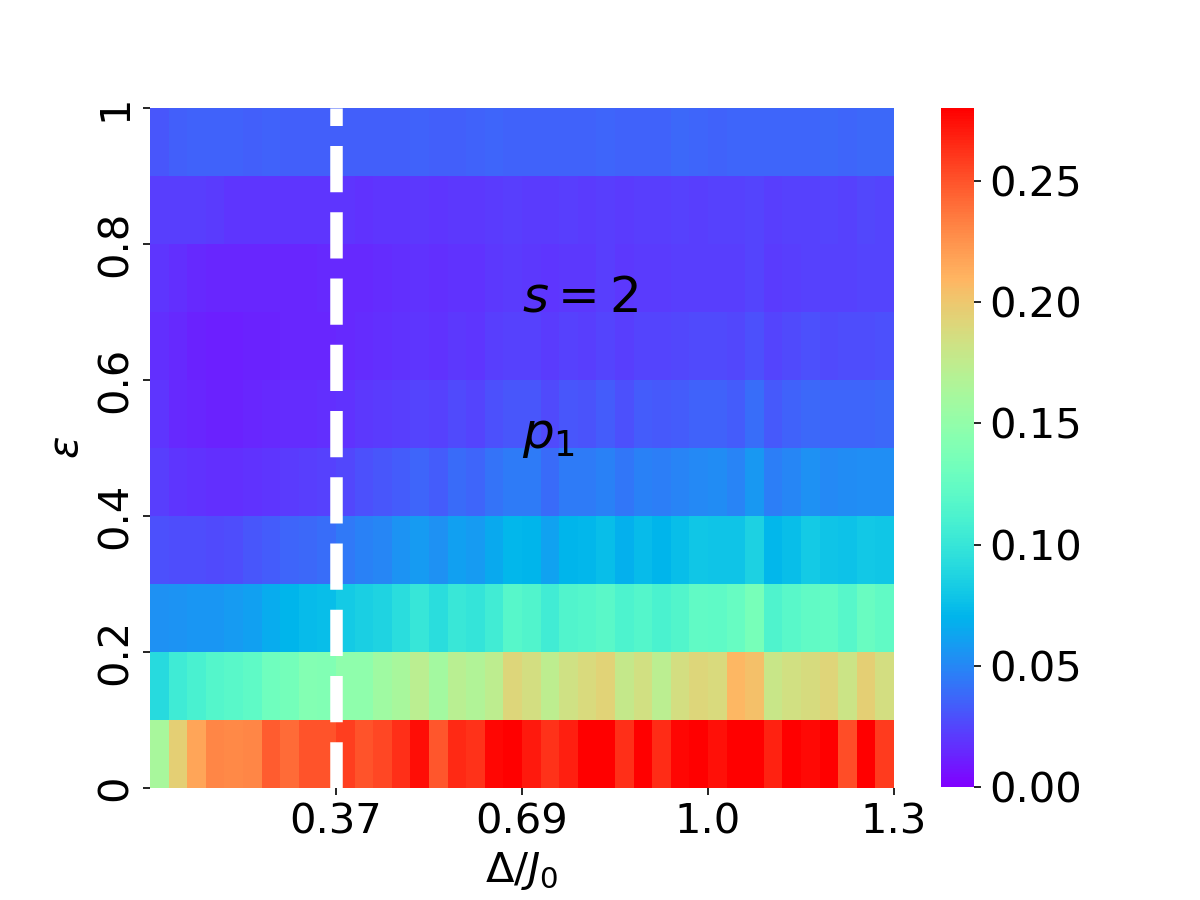}\\
    (g)\includegraphics[width=0.3\textwidth]{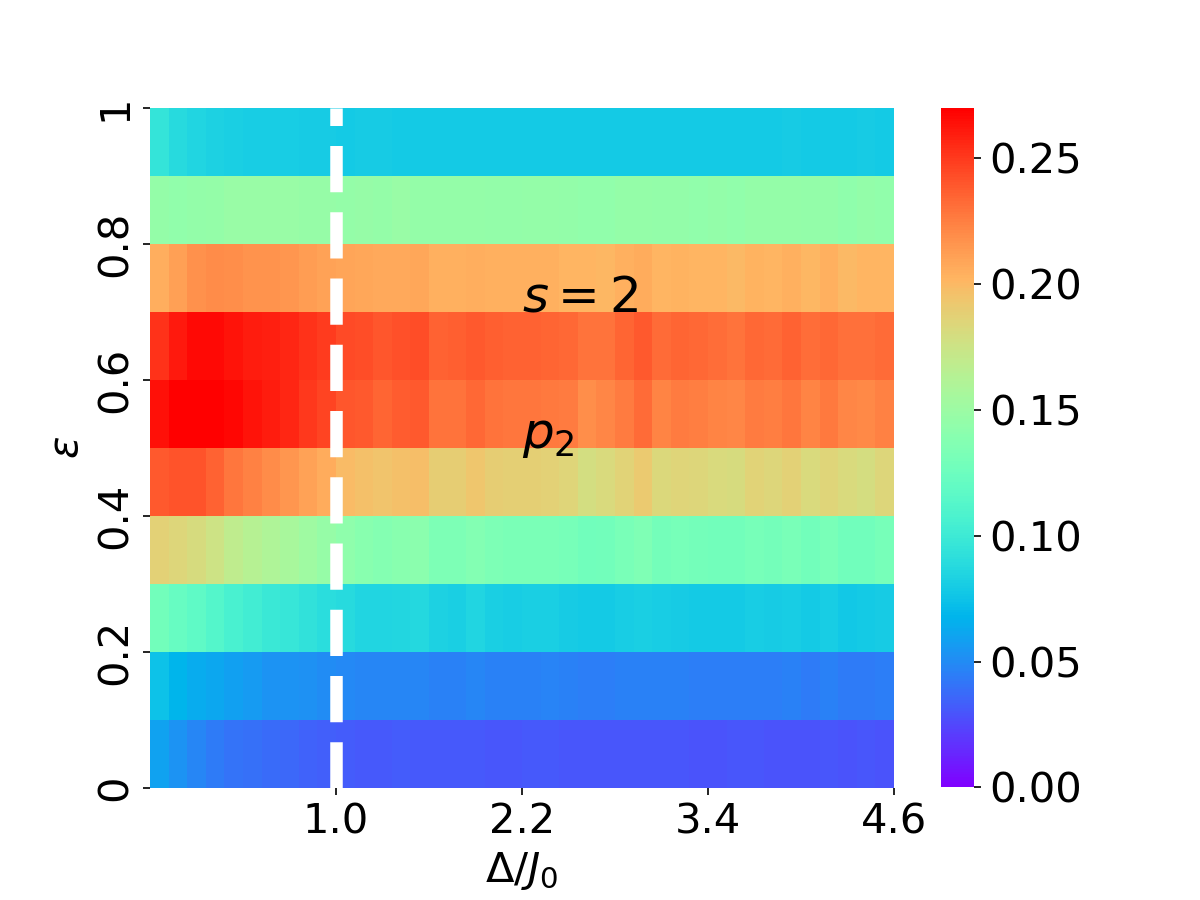}
    (h)\includegraphics[width=0.3\textwidth]{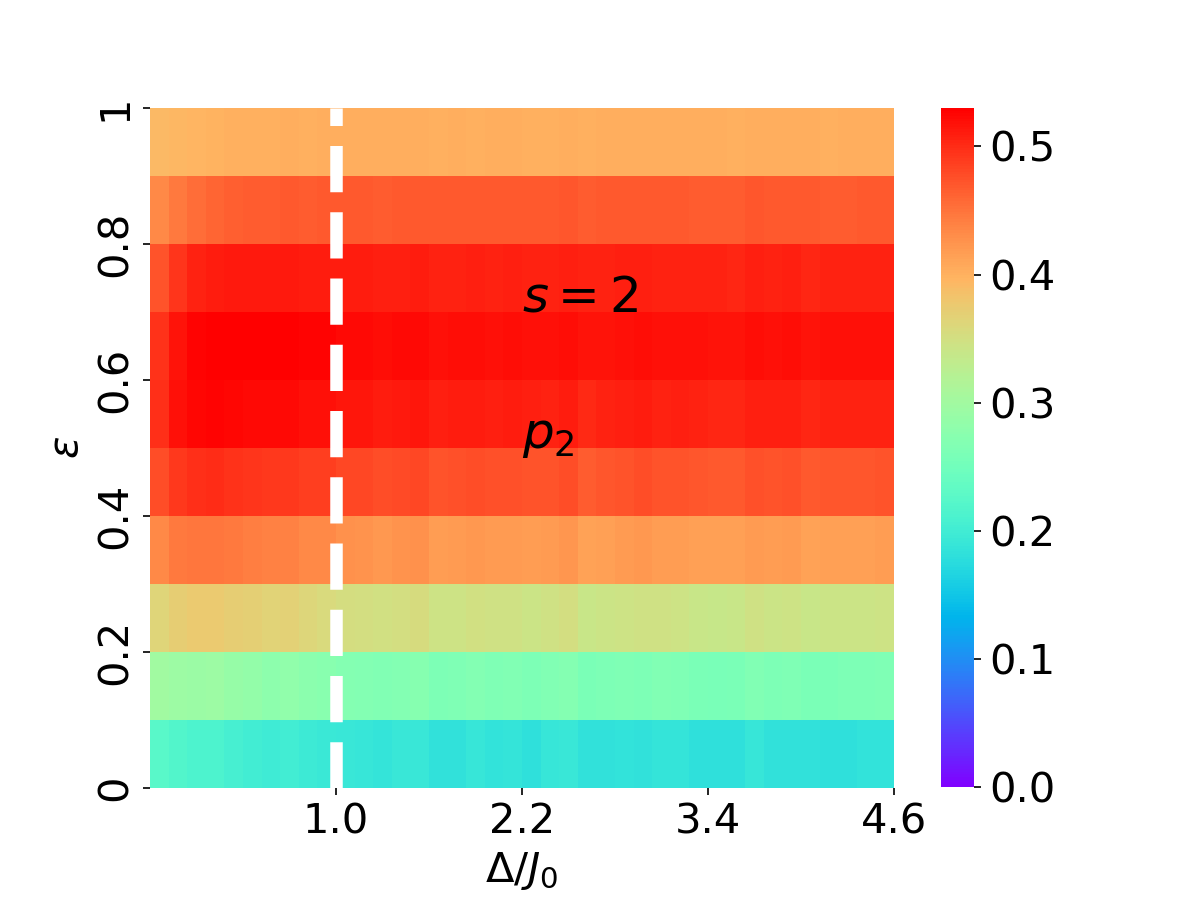}
    (i)\includegraphics[width=0.3\textwidth]{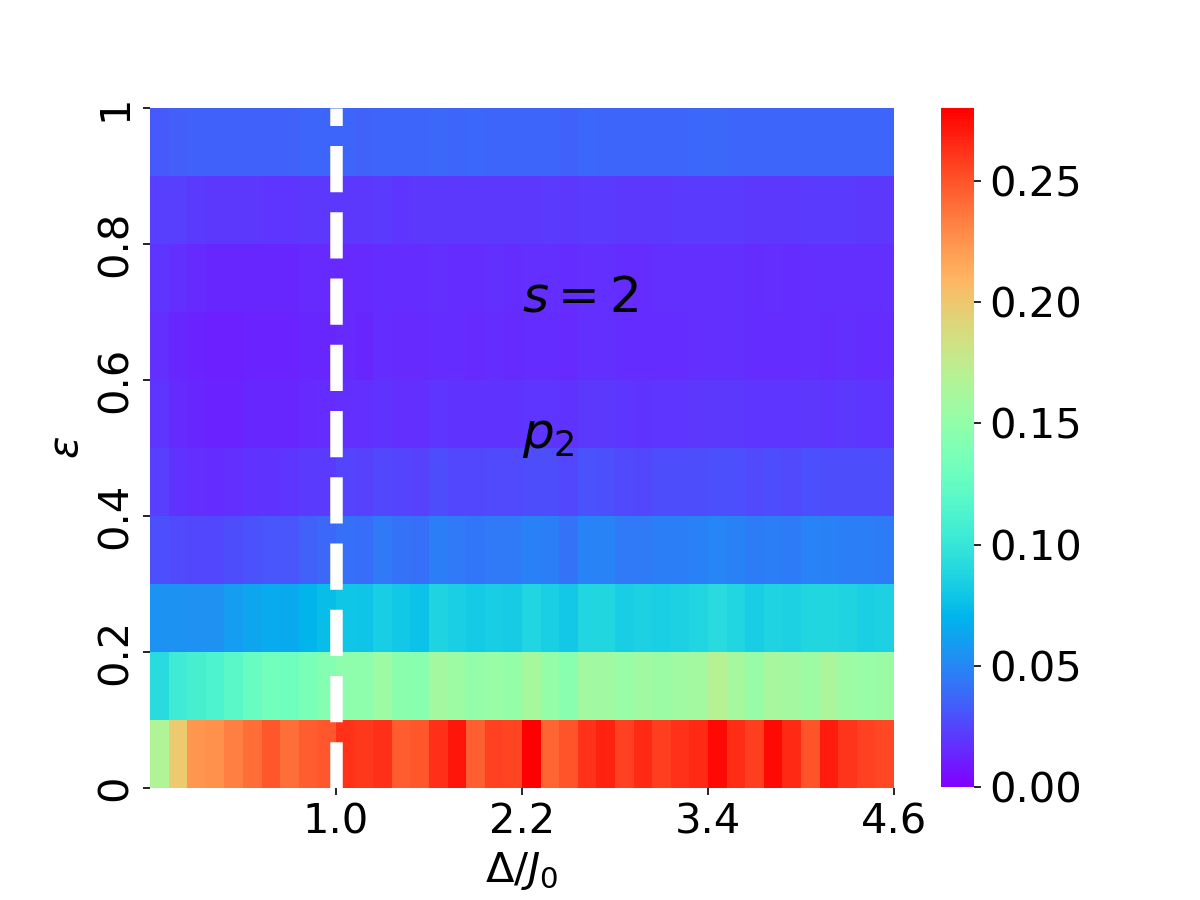}
\caption{Sample-averaged normalized participation ratio $\langle \mathcal{P} \rangle$, entanglement entropy per site $\langle S_E/L \rangle$ and entanglement spectral parameter $\langle \lambda \rangle$ as a function of disorder strength $\Delta/J_0$ and reduced energy $\epsilon$ for $s=2$ sector at system size $L=16$. The order of the plots for different disorder distributions, the position of the white-dashed line, and the scale of the color bar are the same as in Fig.\ \ref{fig-<PR>-<SE>-<lam>}.} 
\label{fig-<PR>-<SE>-<lam>-stot=2}
\end{figure*}

\begin{figure*}[tb]
$\mathcal{P}$\qquad\qquad\qquad\qquad\qquad\qquad\qquad\qquad$S_E/L$ \qquad\qquad\qquad\qquad\qquad\qquad\qquad\qquad$\lambda$ \qquad\\
    (a)\includegraphics[width=0.3\textwidth]{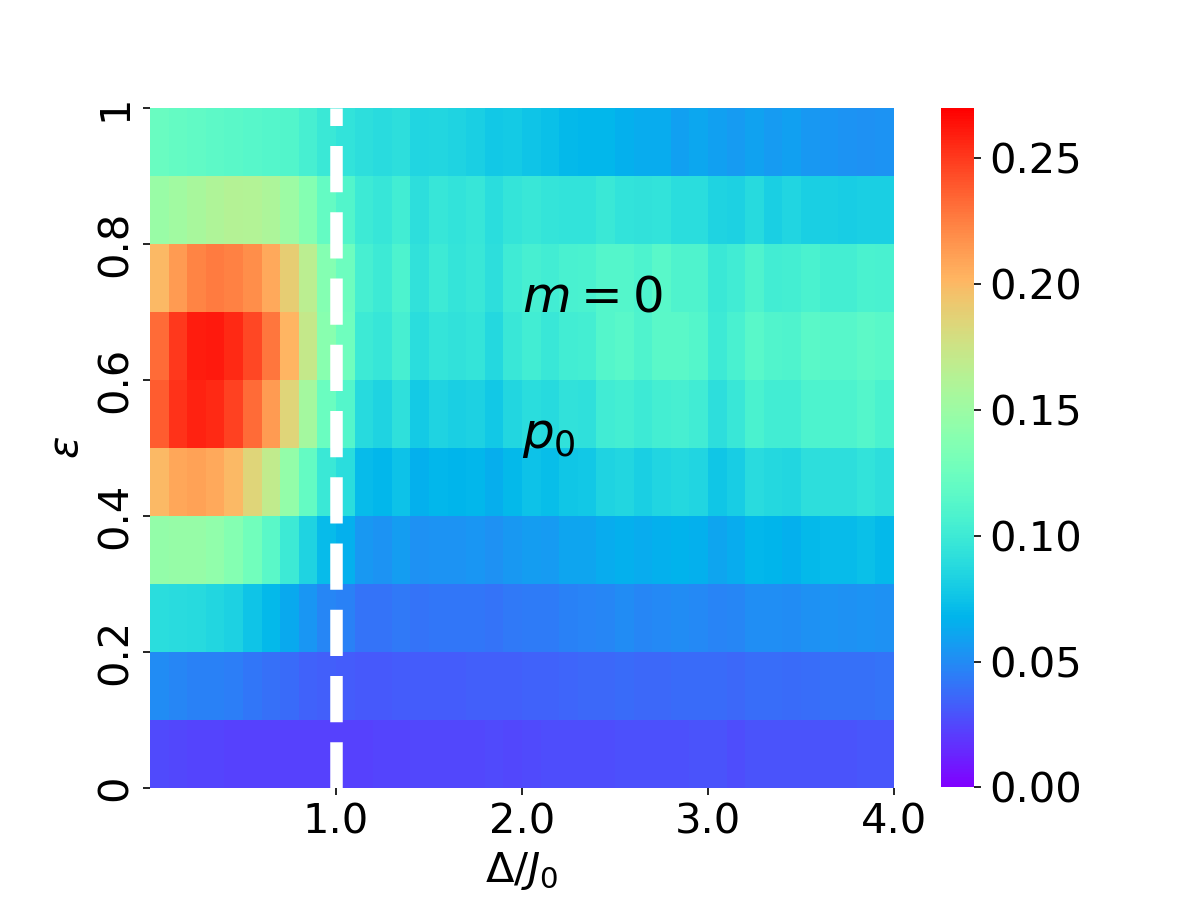}
    (b)\includegraphics[width=0.3\textwidth]{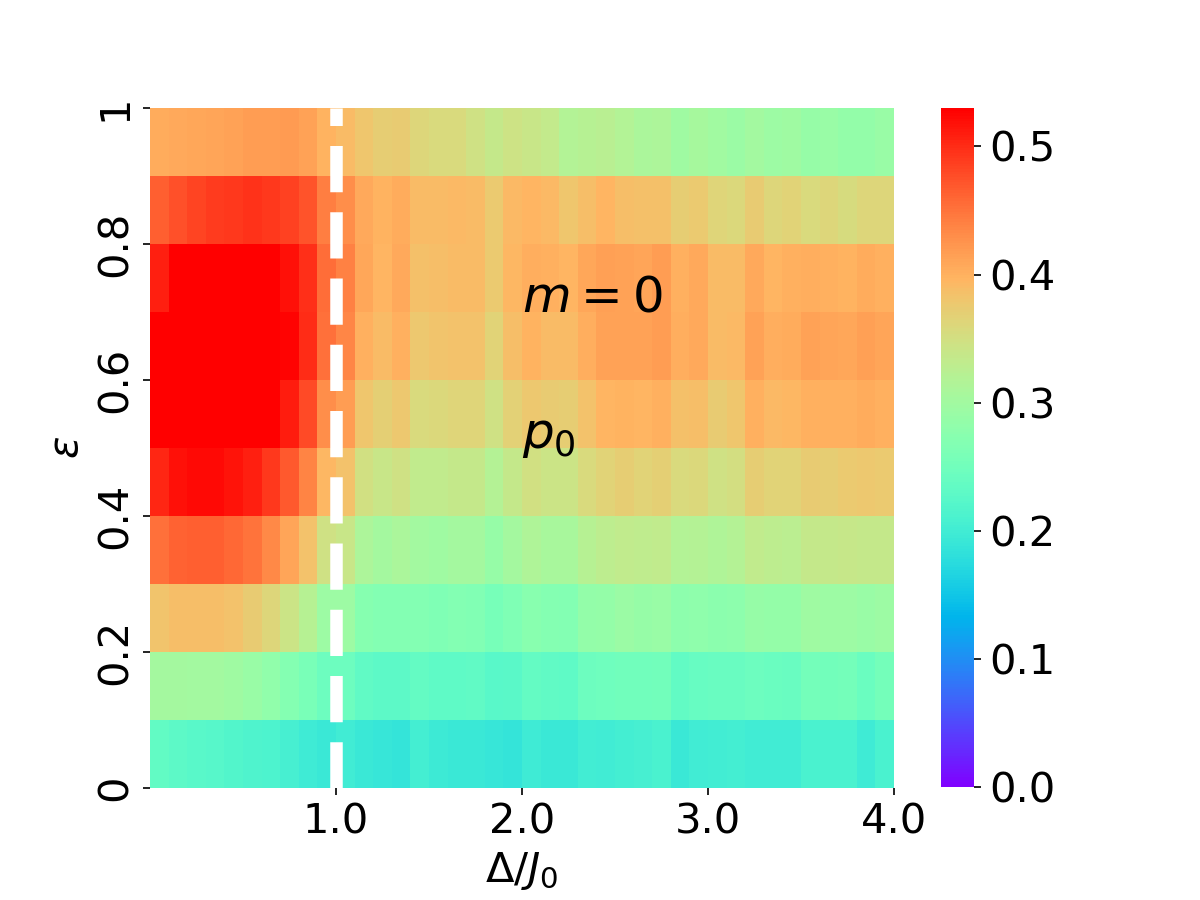}
    (c)\includegraphics[width=0.3\textwidth]{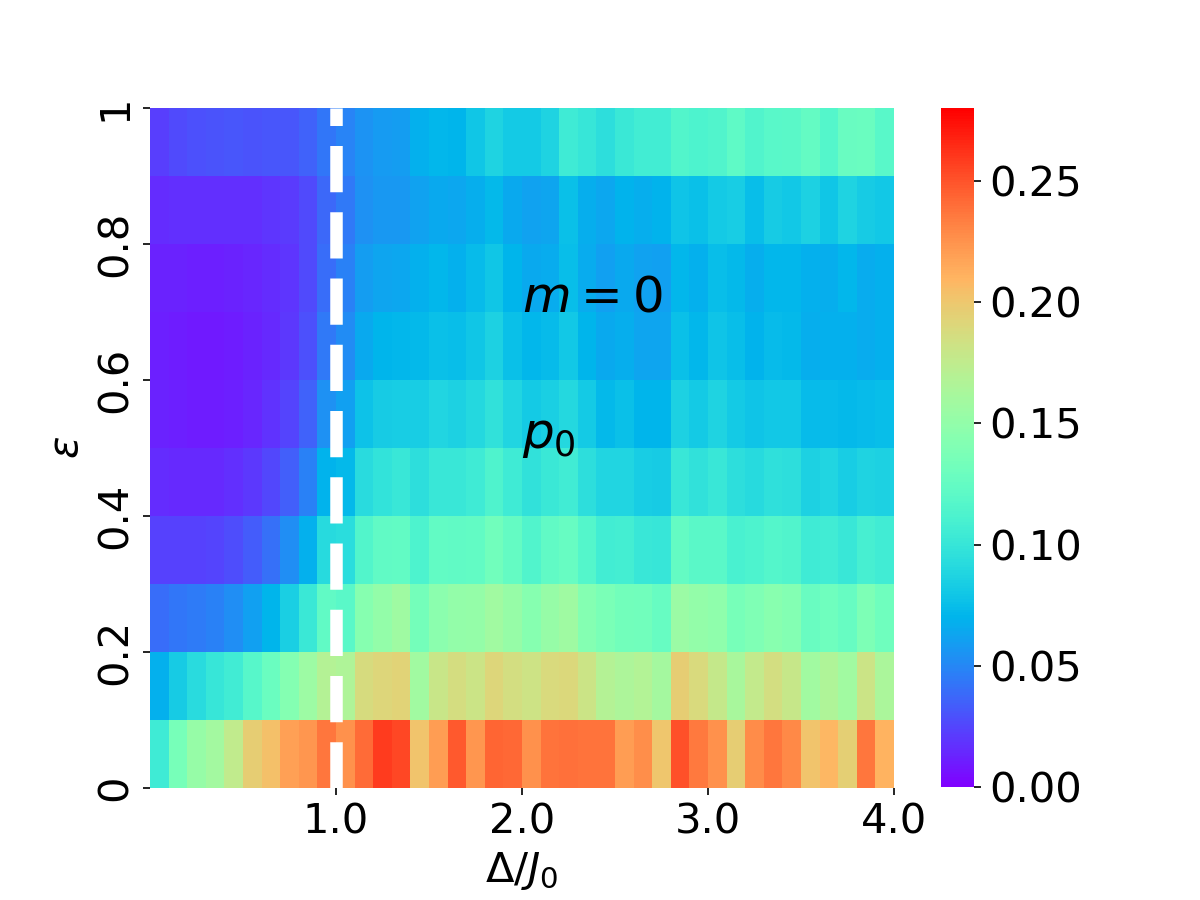}\\
    (d)\includegraphics[width=0.3\textwidth]{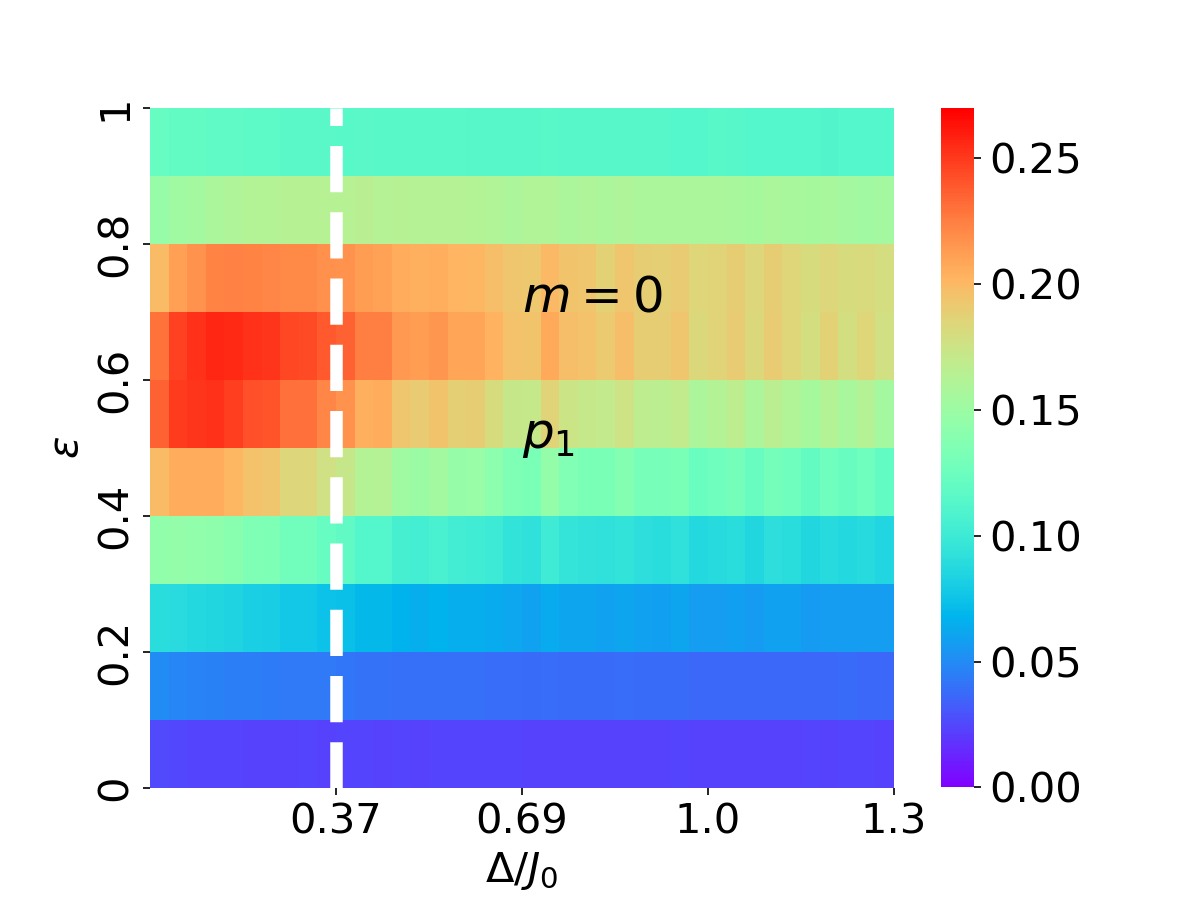}
    (e)\includegraphics[width=0.3\textwidth]{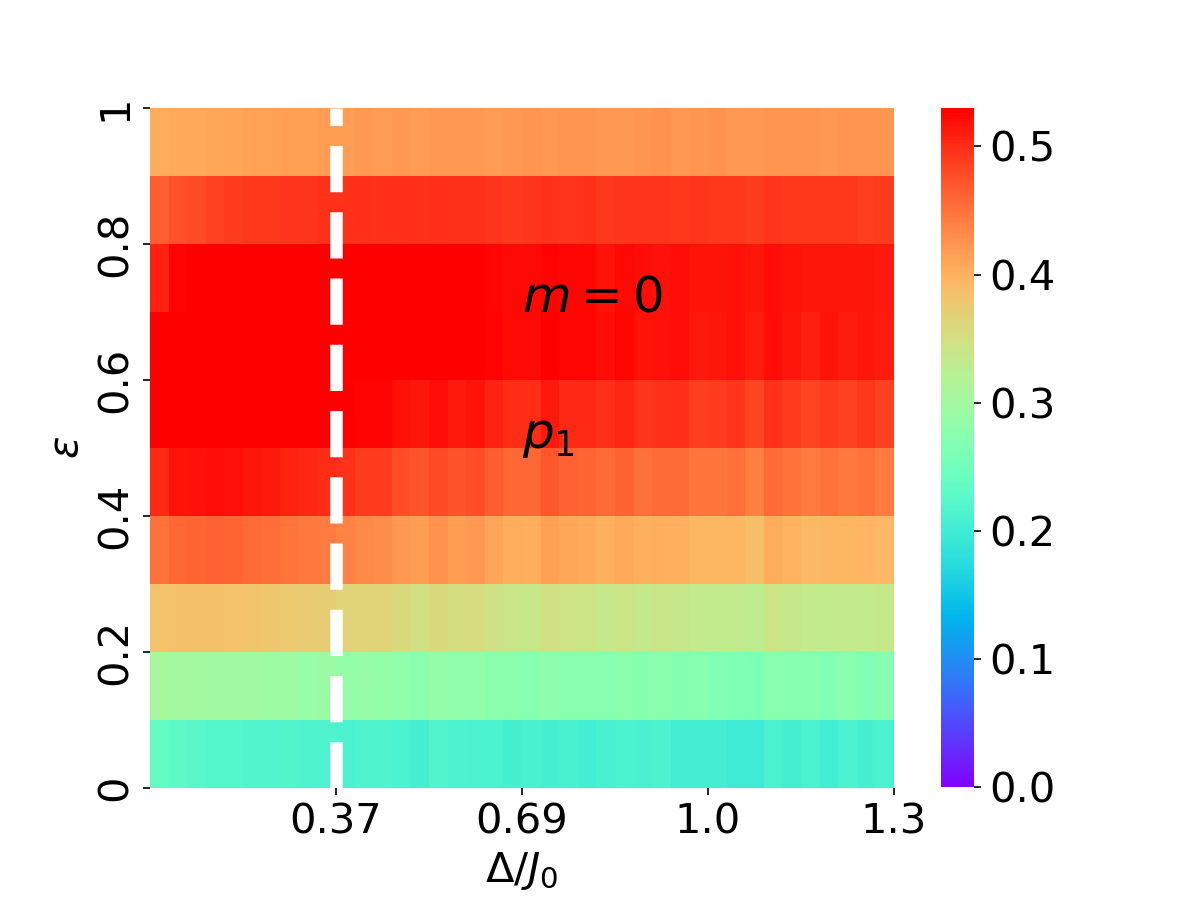}
    (f)\includegraphics[width=0.3\textwidth]{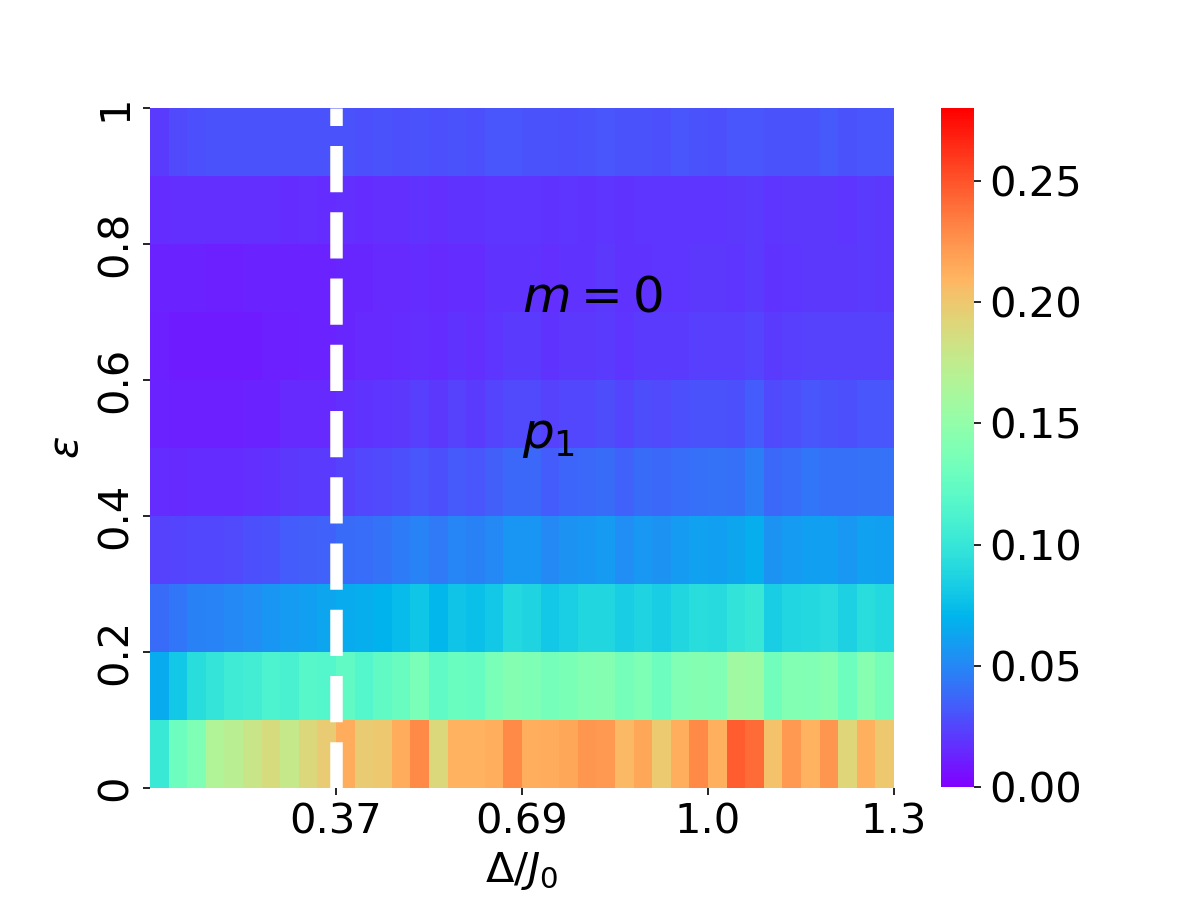}\\
    (g)\includegraphics[width=0.3\textwidth]{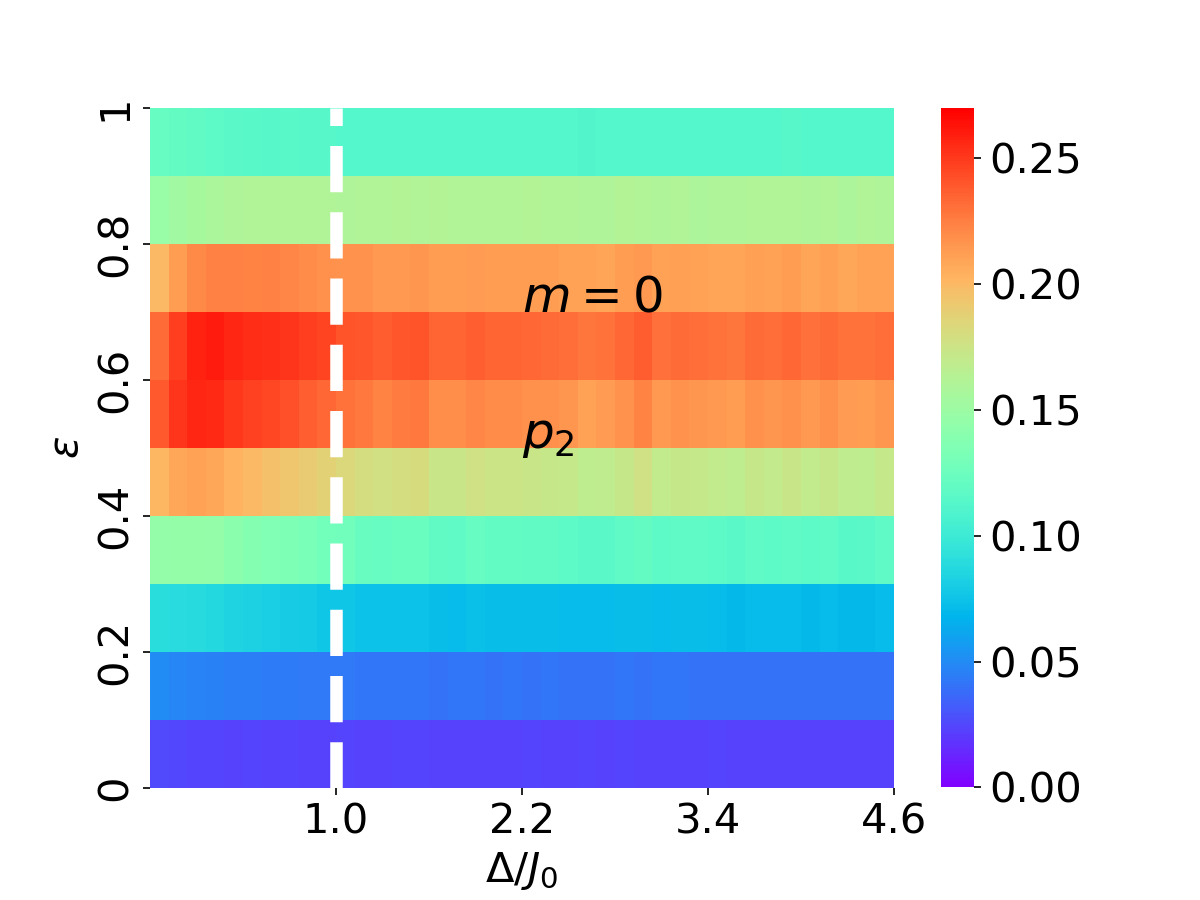}
    (h)\includegraphics[width=0.3\textwidth]{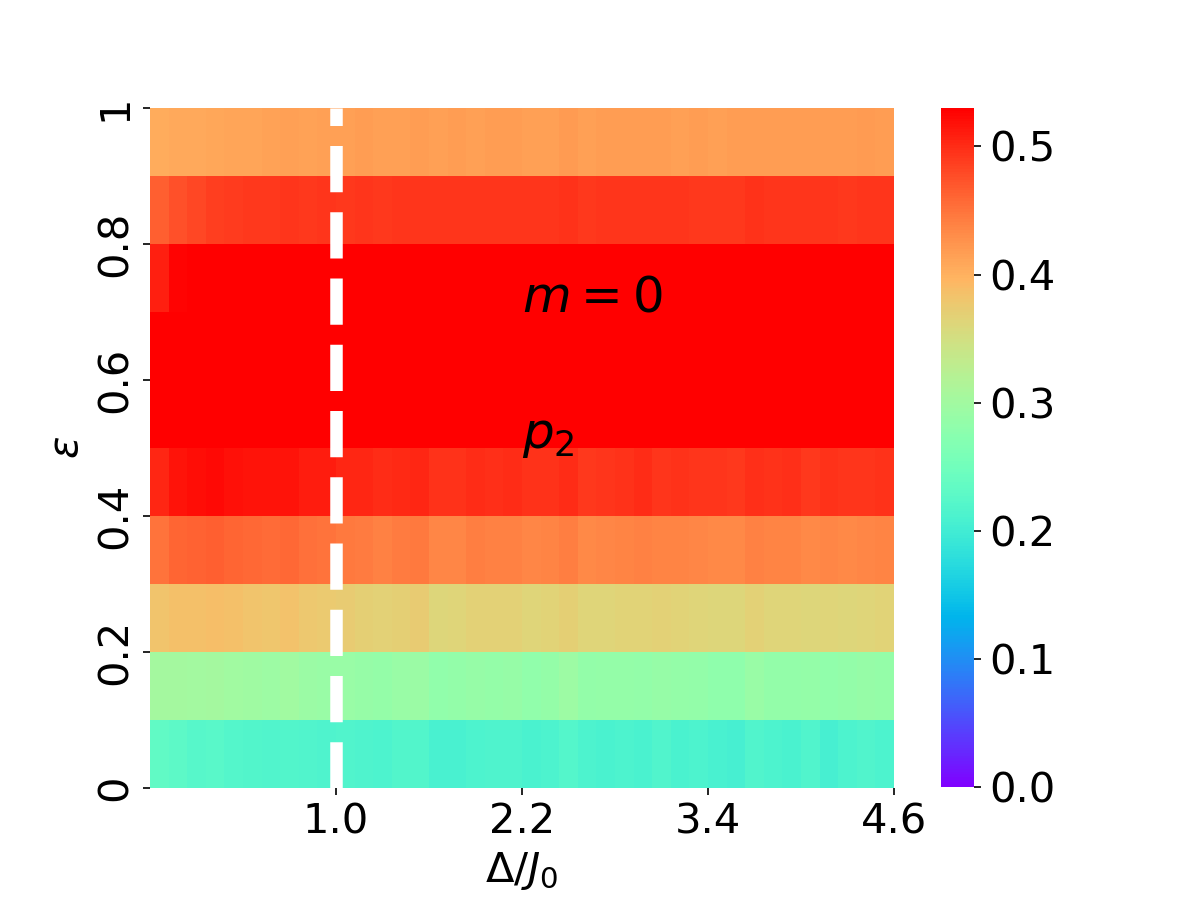}
    (i)\includegraphics[width=0.3\textwidth]{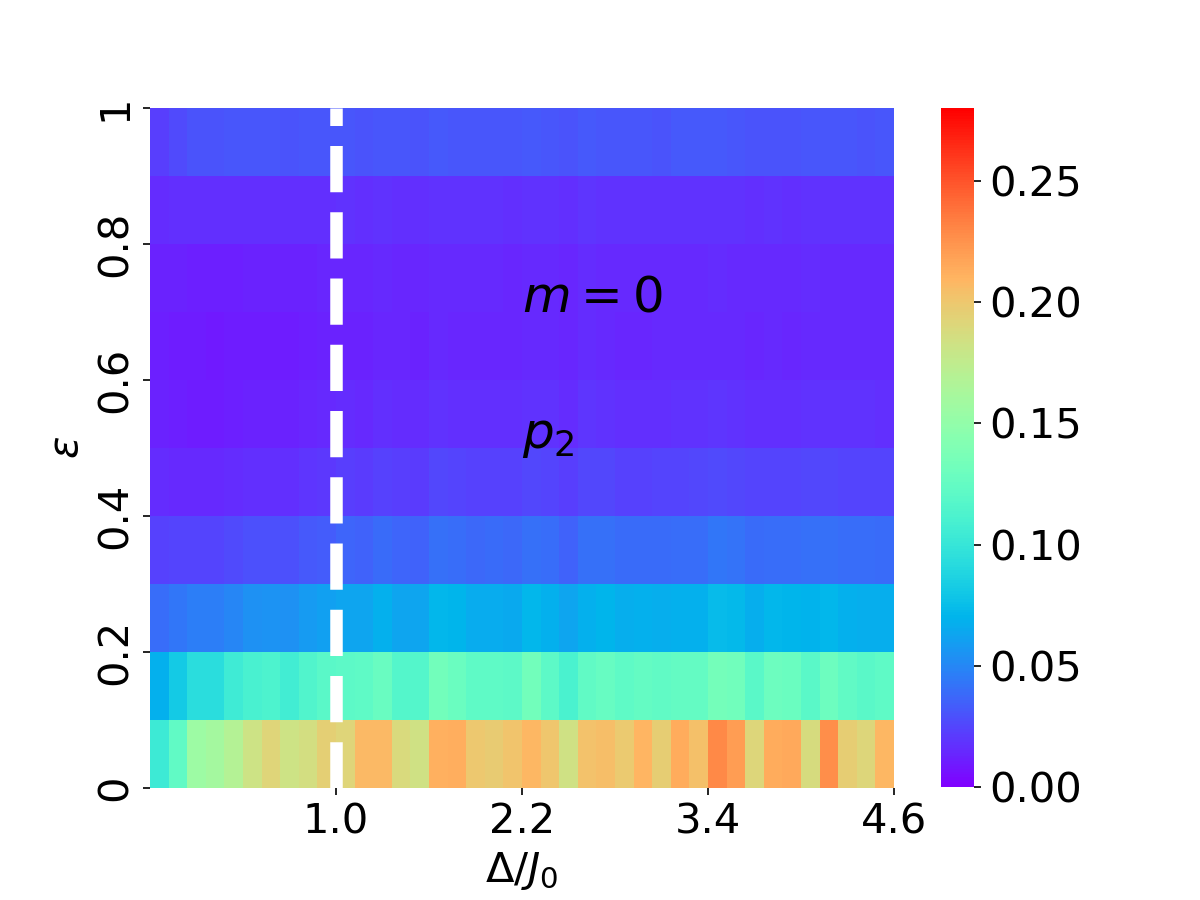}
\caption{Sample-averaged normalized participation ratio $\langle \mathcal{P} \rangle$, entanglement entropy per site $\langle S_E/L \rangle$ and entanglement spectral parameter $\langle \lambda \rangle$ as a function of disorder strength $\Delta/J_0$ and reduced energy $\epsilon$ for $m=0$ sector at system size $L=16$. The order of the plots for different disorder distributions, the position of the white-dashed line, and the scale of the color bar are the same as in Figs.\ \ref{fig-<PR>-<SE>-<lam>} and \ref{fig-<PR>-<SE>-<lam>-stot=2}.}
\label{fig-<PR>-<SE>-<lam>-sz=0}
\end{figure*}

\begin{figure*}[tb]
$\mathcal{P}$\qquad\qquad\qquad\qquad\qquad\qquad\qquad\qquad$S_E/L$ \qquad\qquad\qquad\qquad\qquad\qquad\qquad\qquad$\lambda$ \qquad\\
    (a)\includegraphics[width=0.3\textwidth]{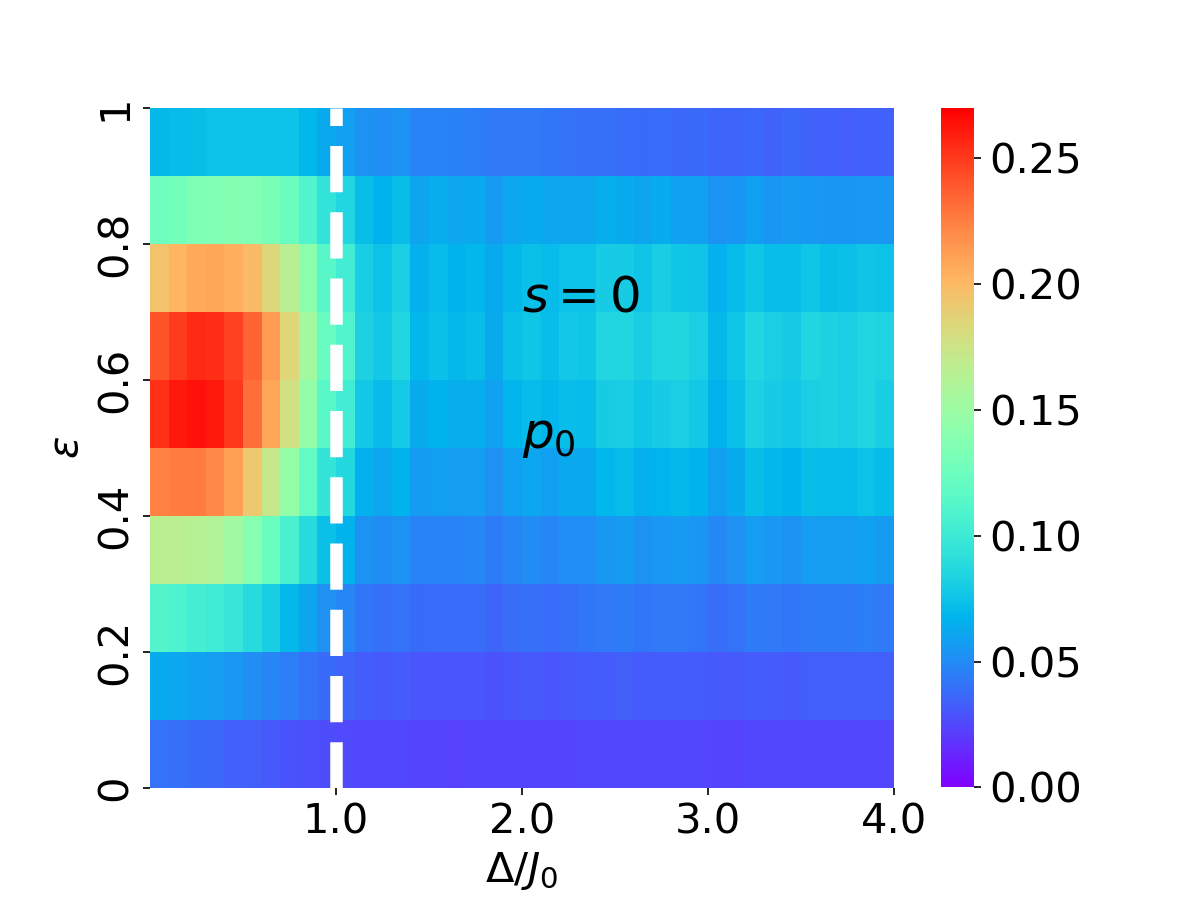}
    (b)\includegraphics[width=0.3\textwidth]{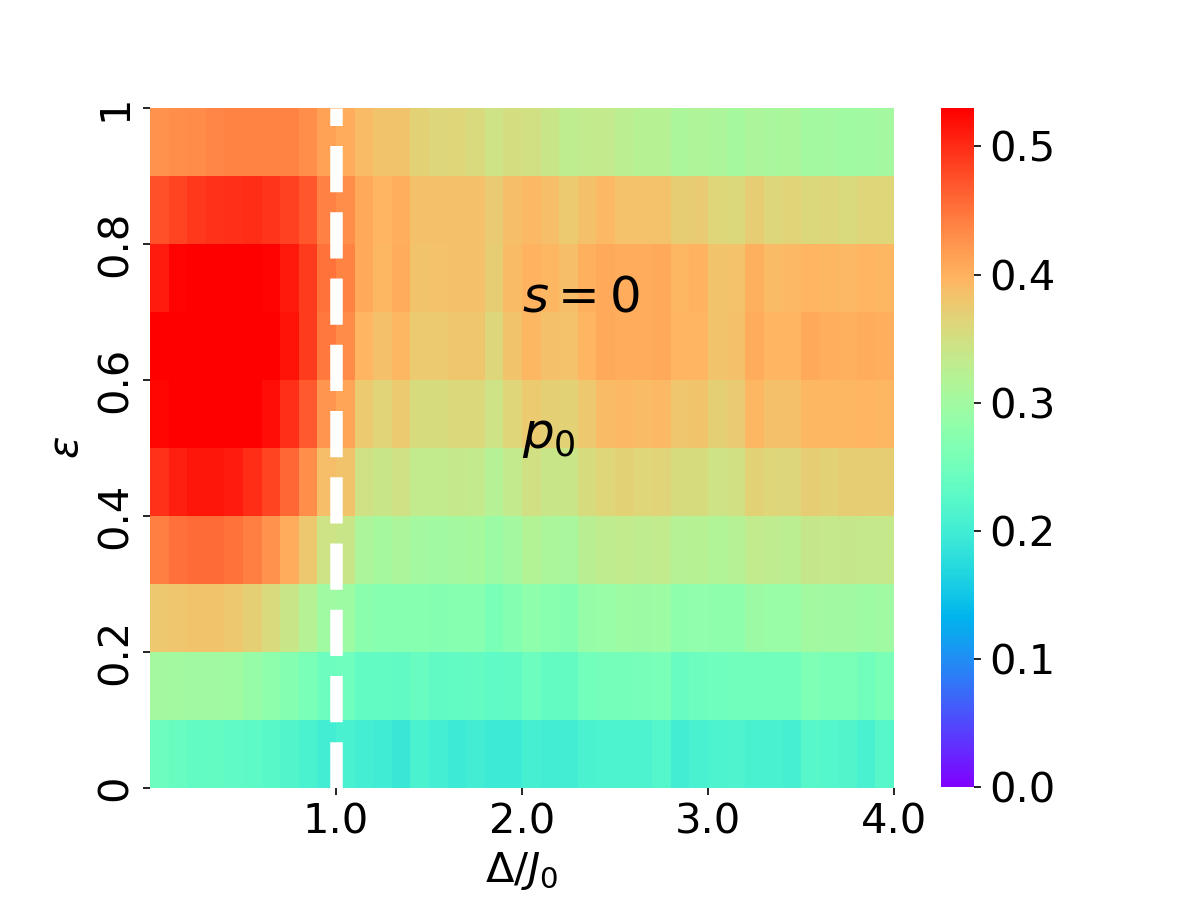}
    (c)\includegraphics[width=0.3\textwidth]{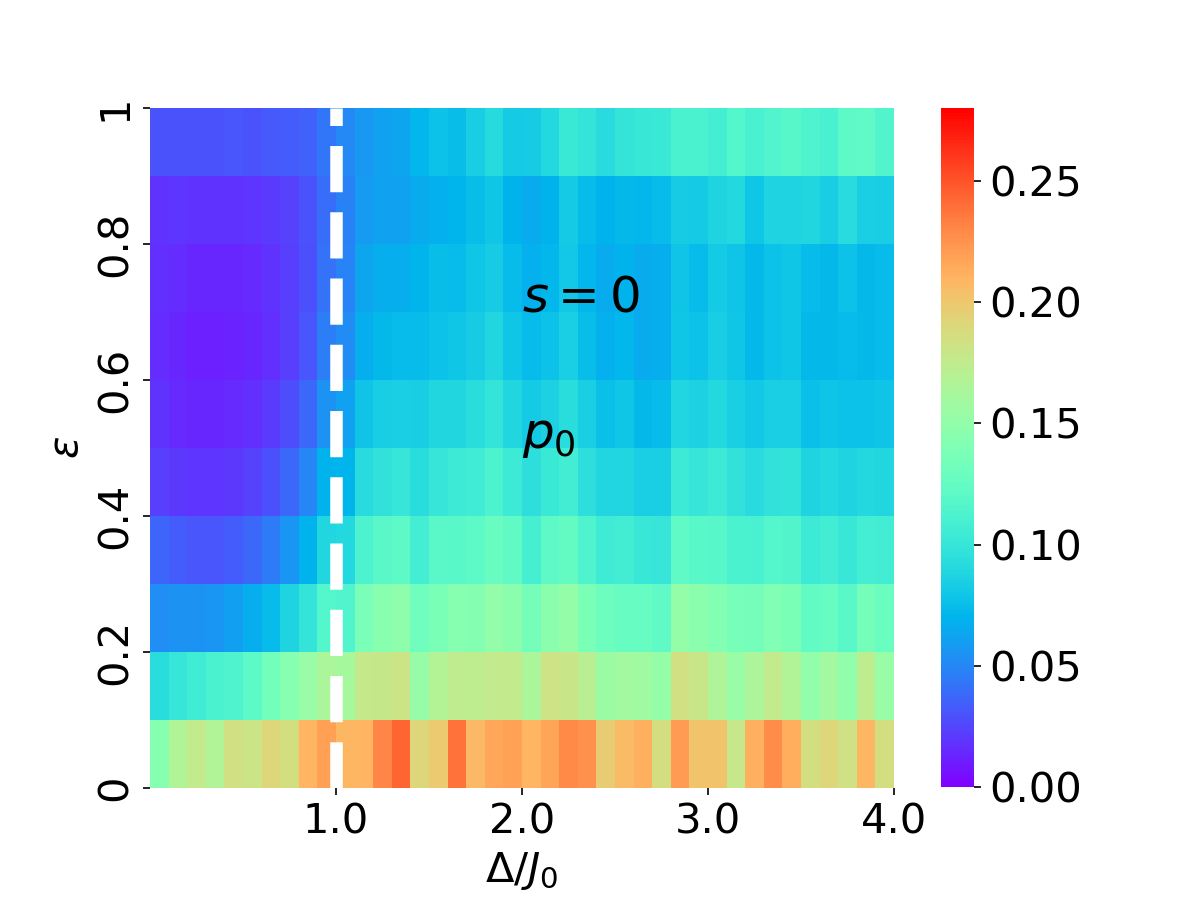}\\
    (d)\includegraphics[width=0.3\textwidth]{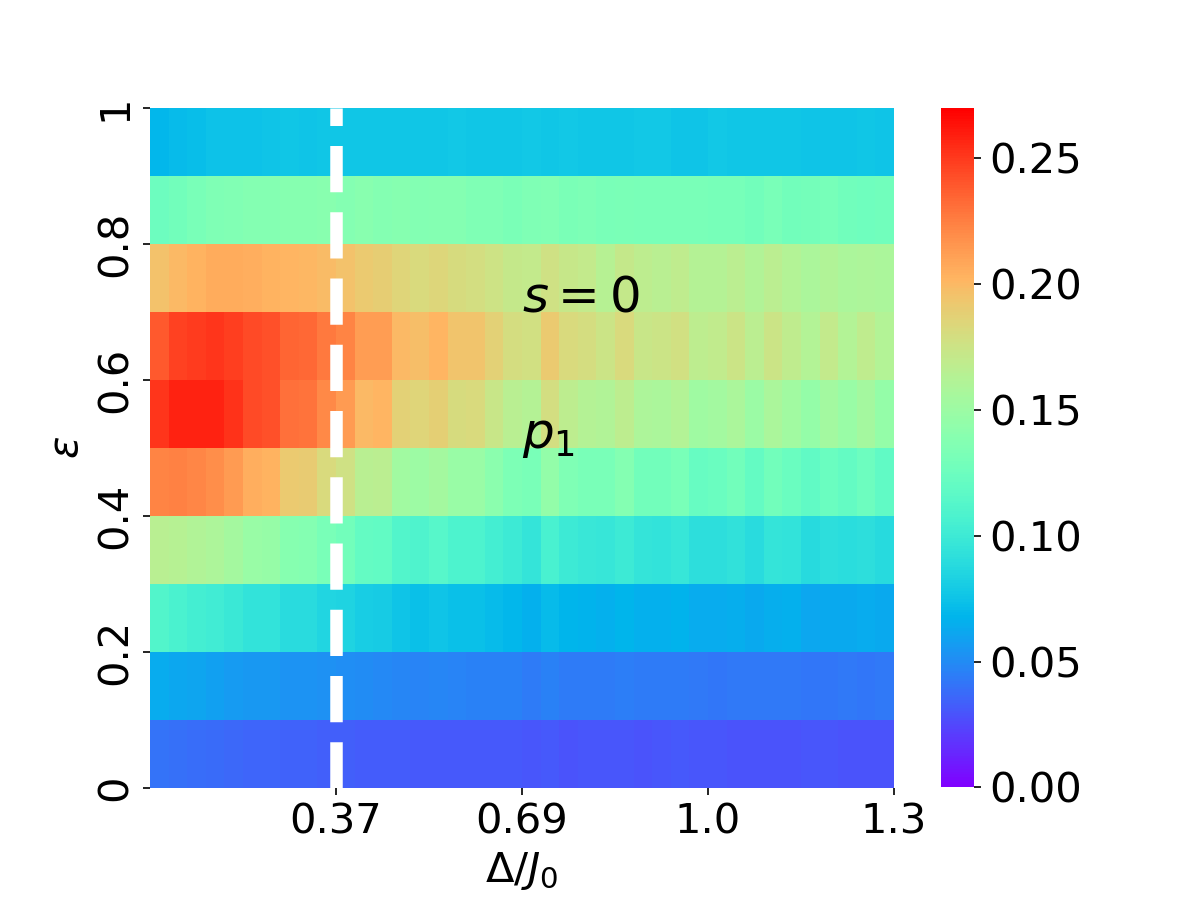}
    (e)\includegraphics[width=0.3\textwidth]{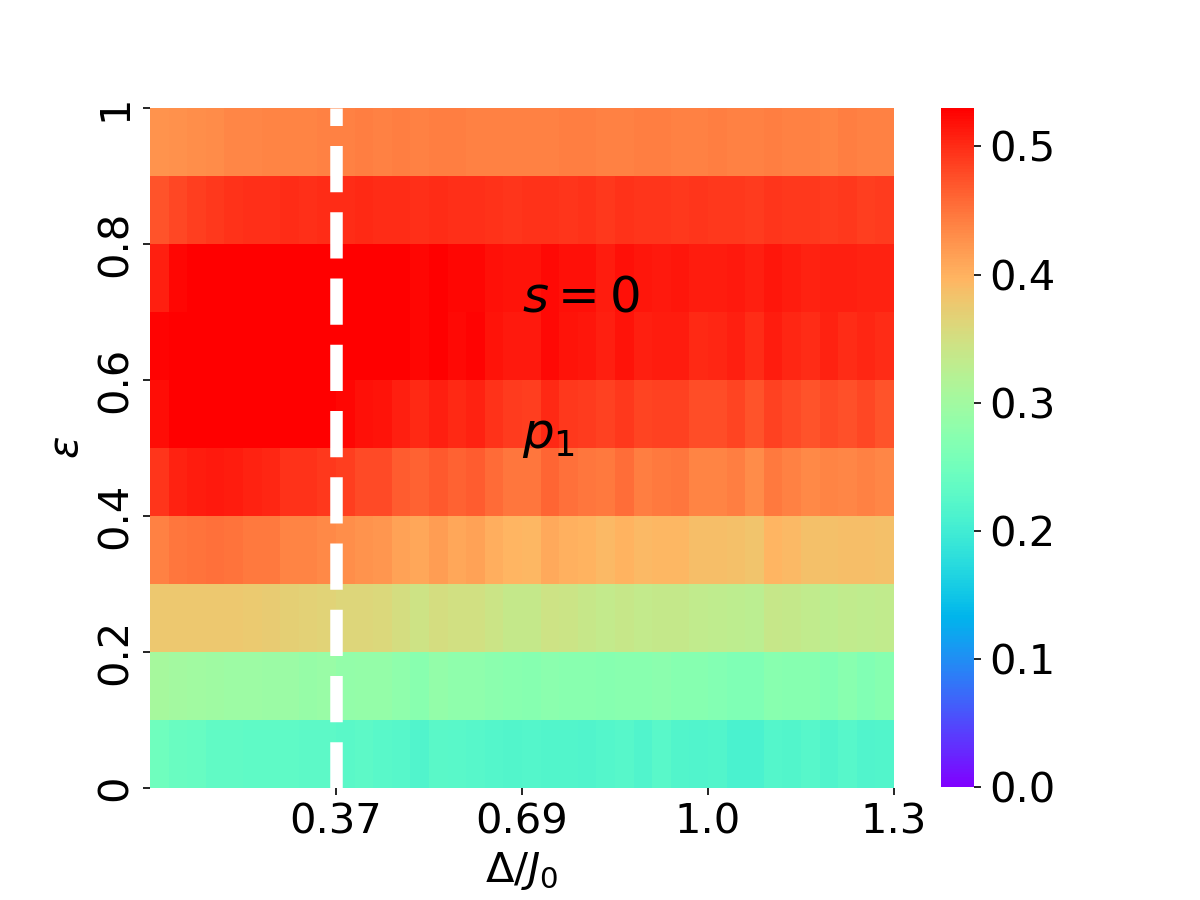}
    (f)\includegraphics[width=0.3\textwidth]{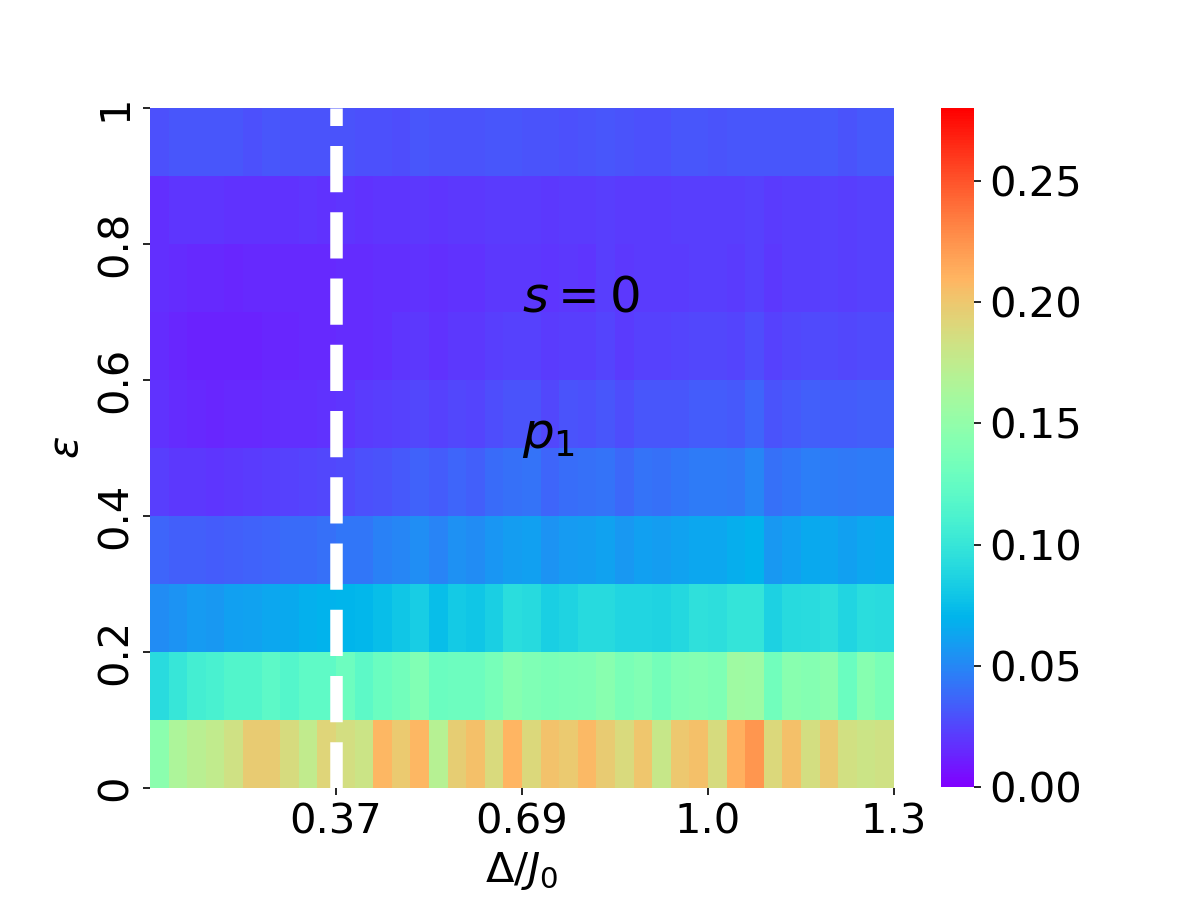}\\
    (g)\includegraphics[width=0.3\textwidth]{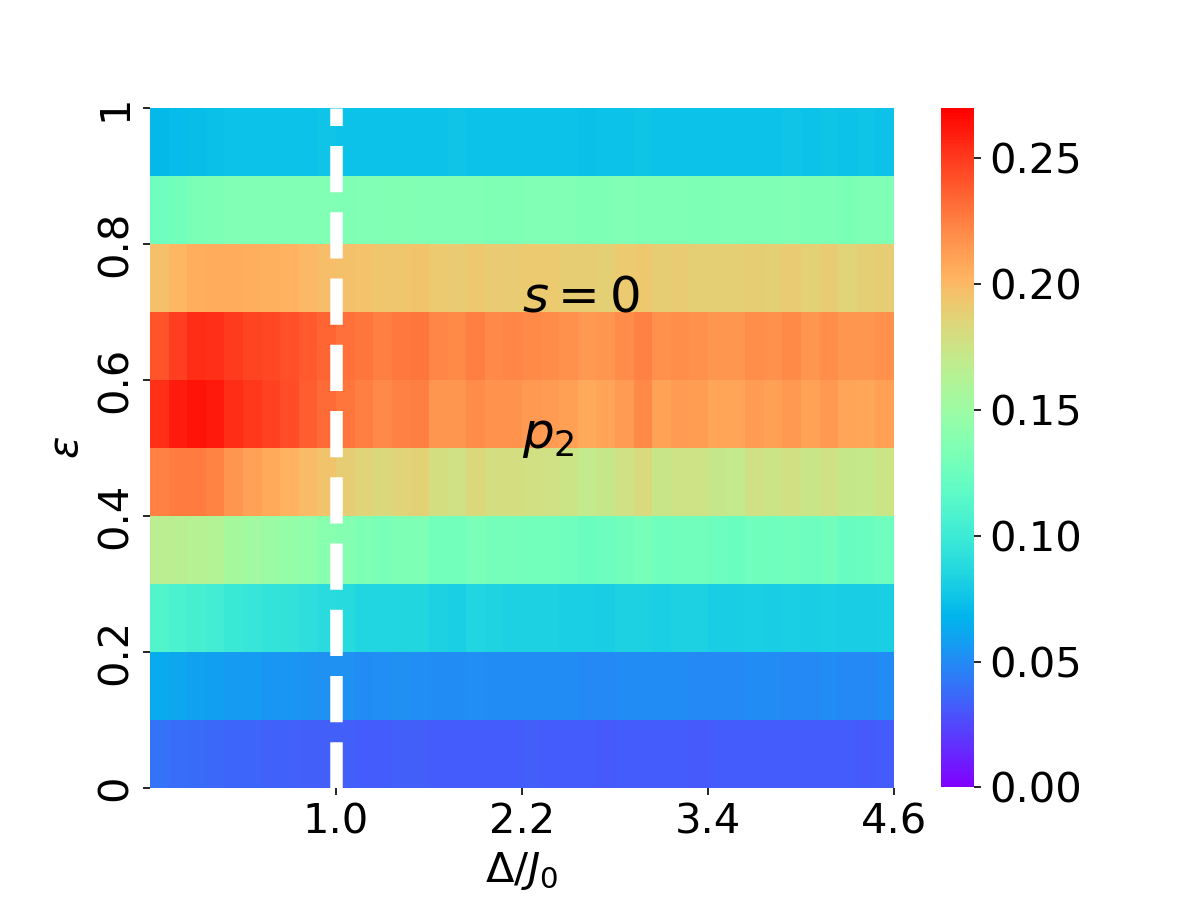}
    (h)\includegraphics[width=0.3\textwidth]{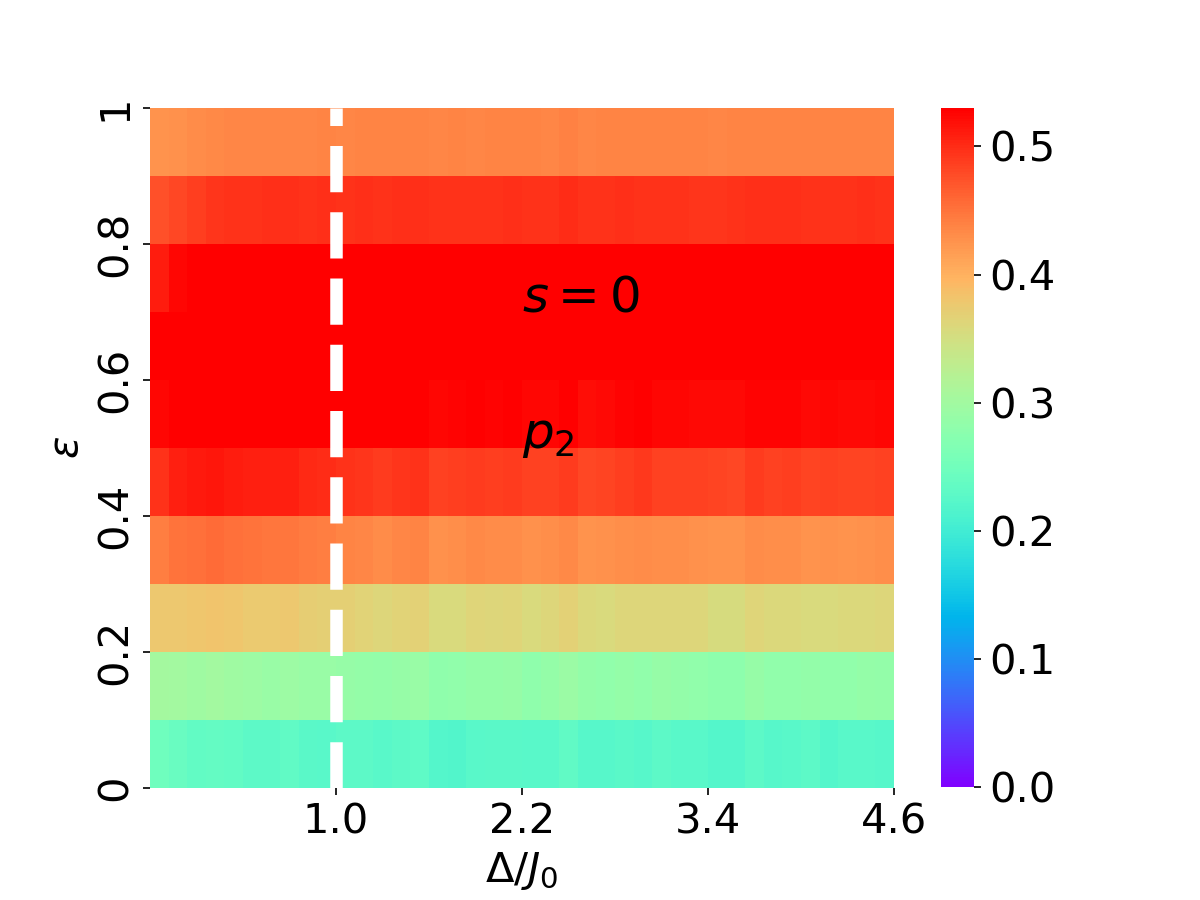}
    (i)\includegraphics[width=0.3\textwidth]{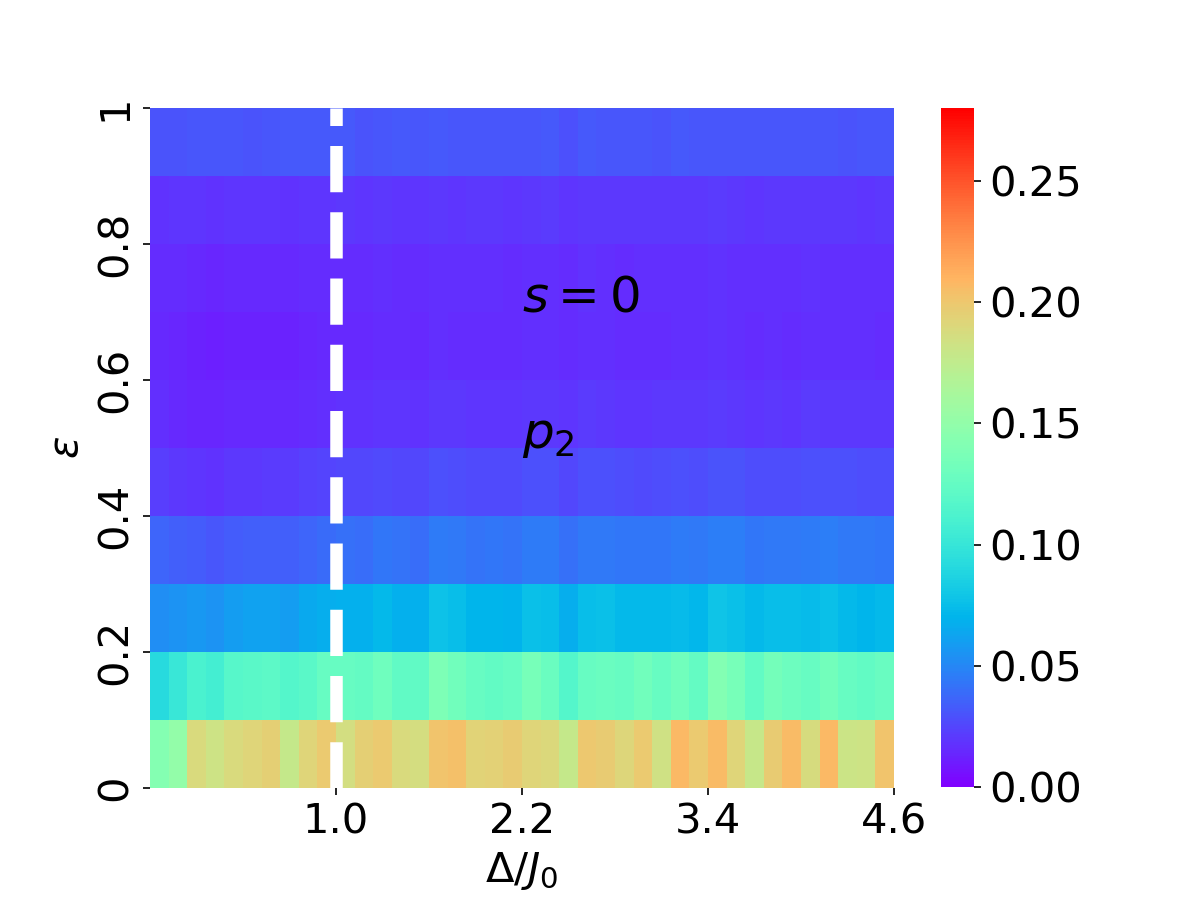}
\caption{Sample-averaged normalized participation ratio $\langle \mathcal{P} \rangle$, entanglement entropy per site $\langle S_E/L \rangle$ and entanglement spectral parameter $\langle \lambda \rangle$ as a function of disorder strength $\Delta/J_0$ and reduced energy $\epsilon$ for $s=0$ sector at system size $L=16$. The order of the plots for different disorder distributions, the position of the white-dashed line, and the scale of the color bar are the same as in Figs.\ \ref{fig-<PR>-<SE>-<lam>}, \ref{fig-<PR>-<SE>-<lam>-stot=2} and \ref{fig-<PR>-<SE>-<lam>-sz=0}.} 
\label{fig-<PR>-<SE>-<lam>-stot=0}
\end{figure*}

The plots for $\mathcal{P}$, $S_E$ and $\lambda$ measures in the $s=2$, $m=0$ and $s=0$ sectors are shown in Fig.\ \ref{fig-<PR>-<SE>-<lam>-stot=2}, Fig.\ \ref{fig-<PR>-<SE>-<lam>-sz=0} and Fig.\ \ref{fig-<PR>-<SE>-<lam>-stot=0}. The results are similar to what has been discussed in \ \ref{fig-<PR>-<SE>-<lam>} when $\emph{all non-degenerate spin sectors}$ are included.

\begin{figure*}[tb]
    (a)\includegraphics[width=0.4\textwidth]{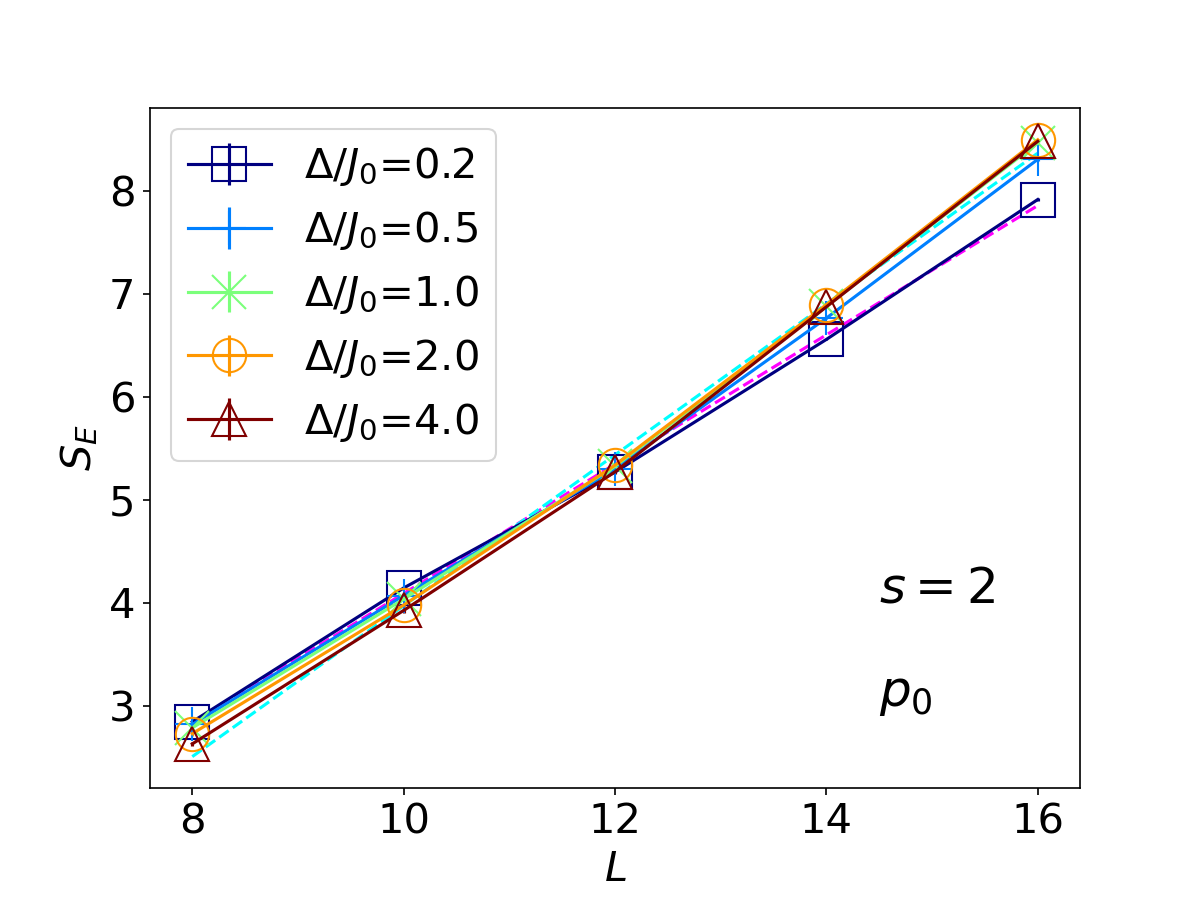}
    (b)\includegraphics[width=0.4\textwidth]{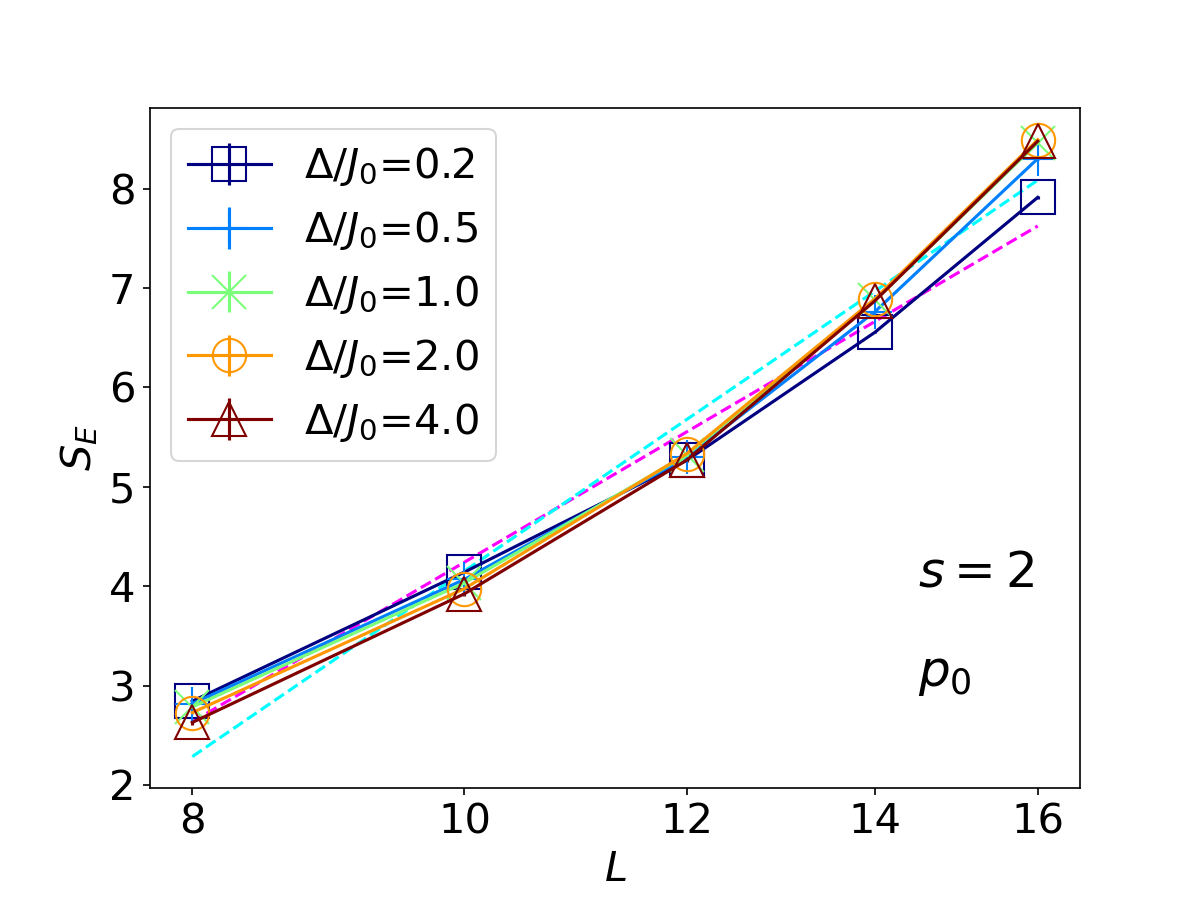}
    (c)\includegraphics[width=0.4\textwidth]{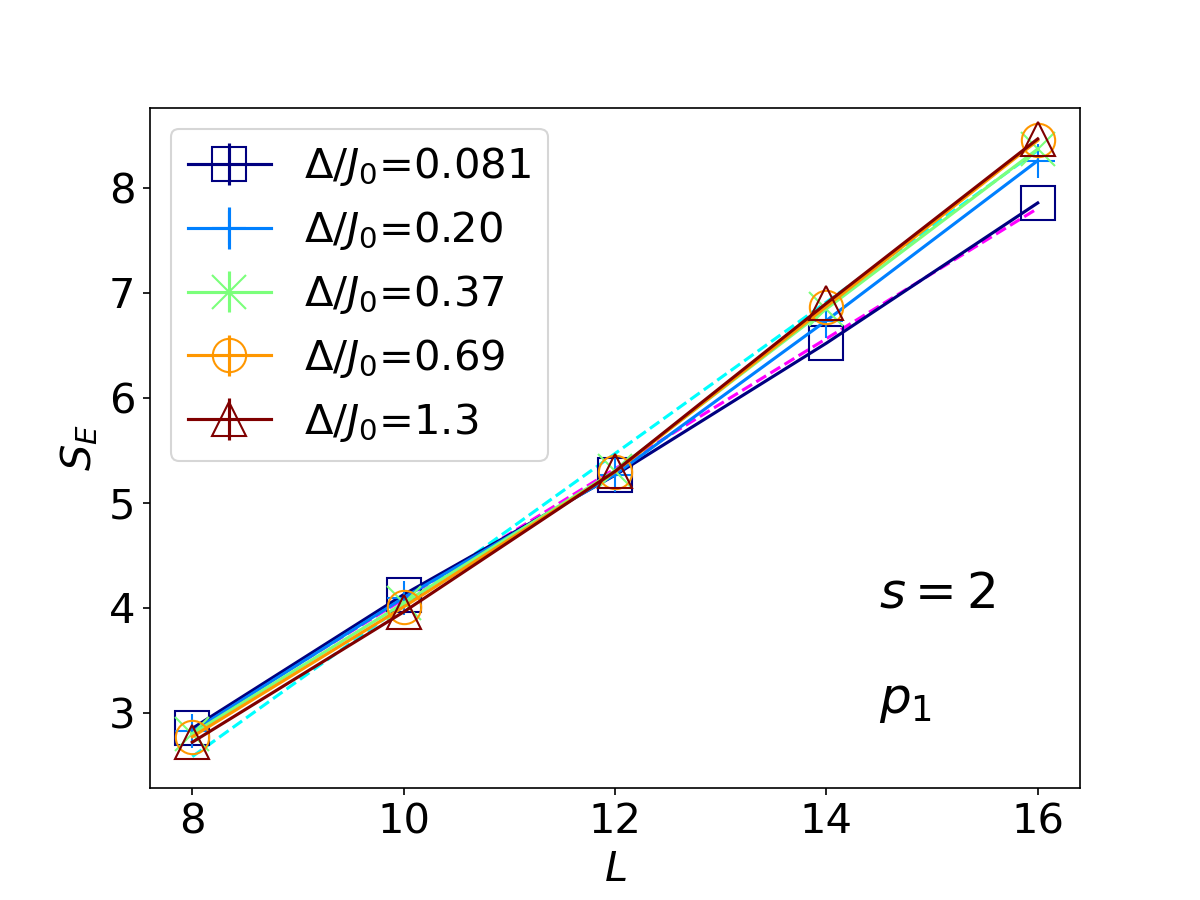}
    (d)\includegraphics[width=0.4\textwidth]{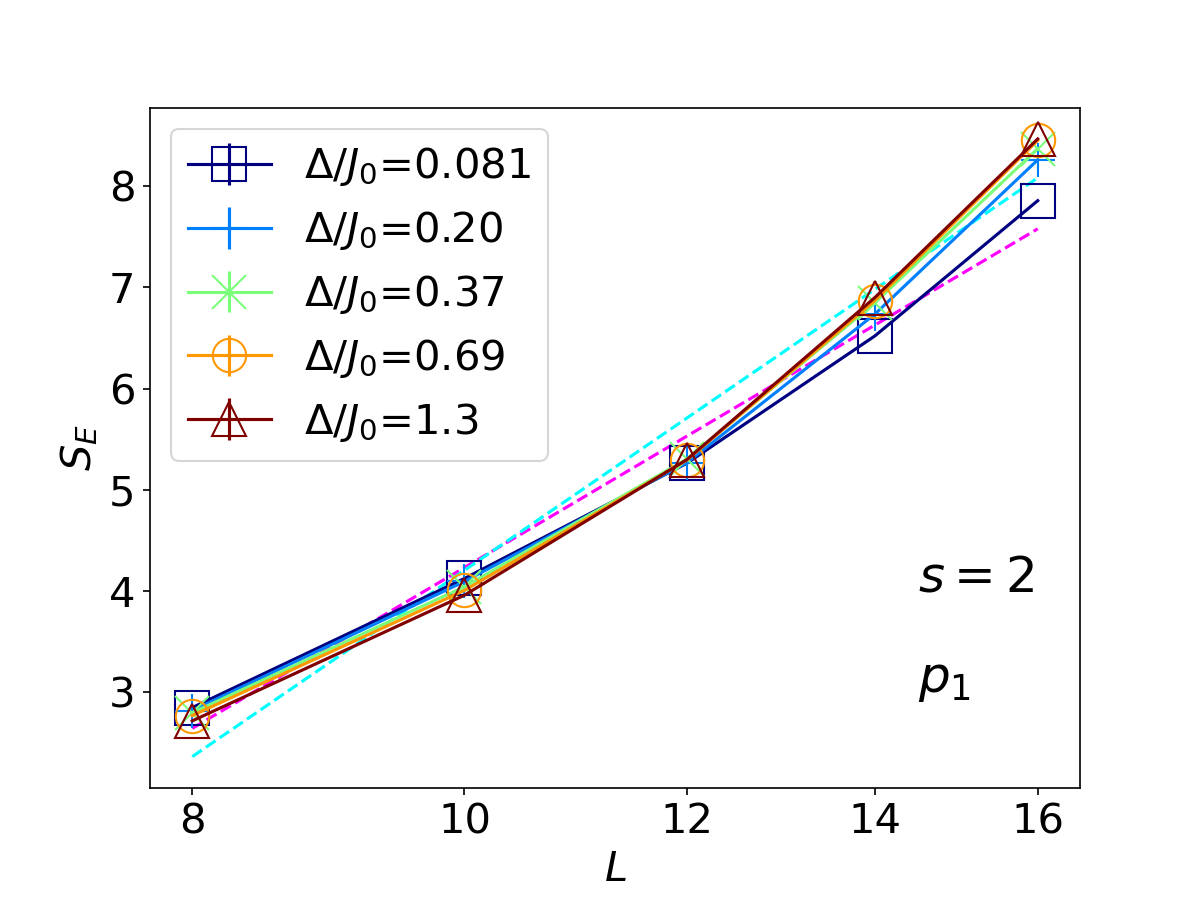}
    (e)\includegraphics[width=0.4\textwidth]{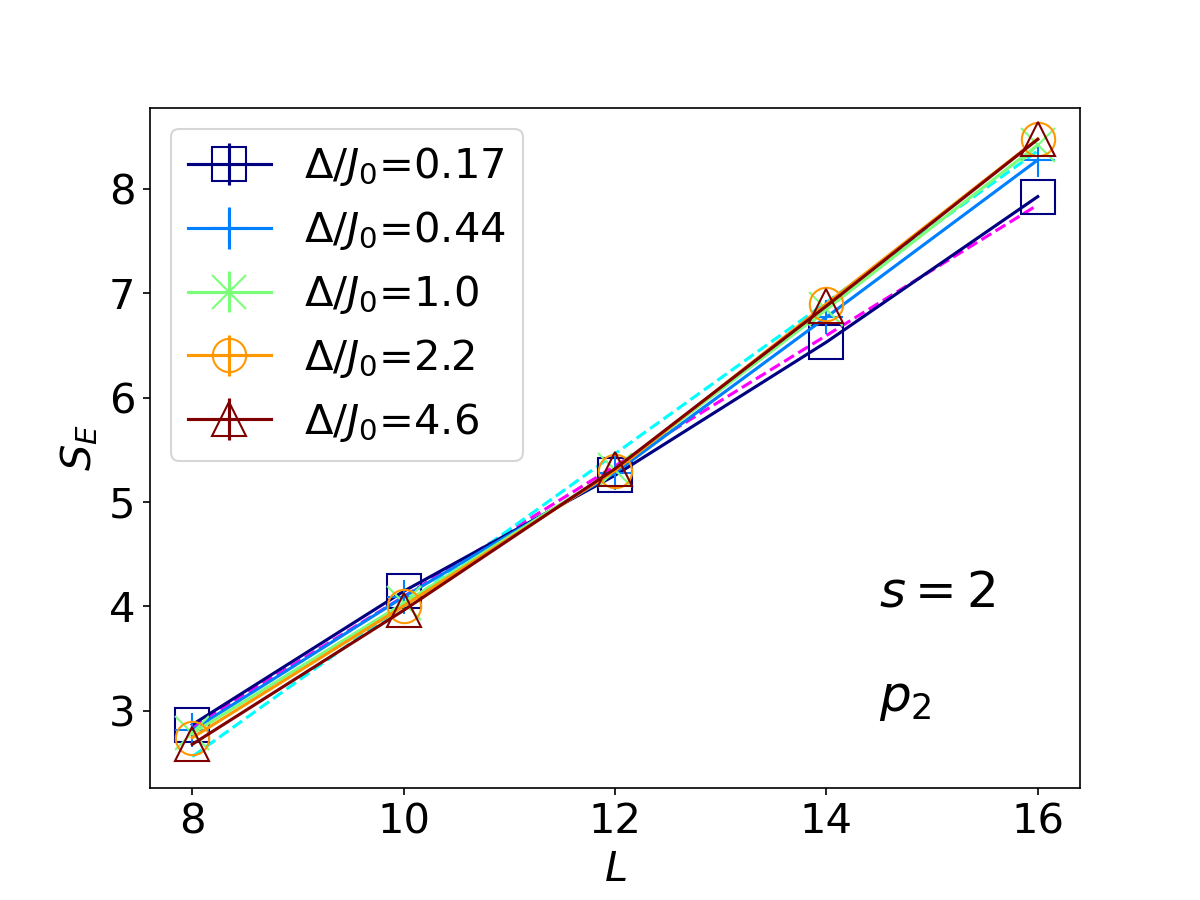}
    (f)\includegraphics[width=0.4\textwidth]{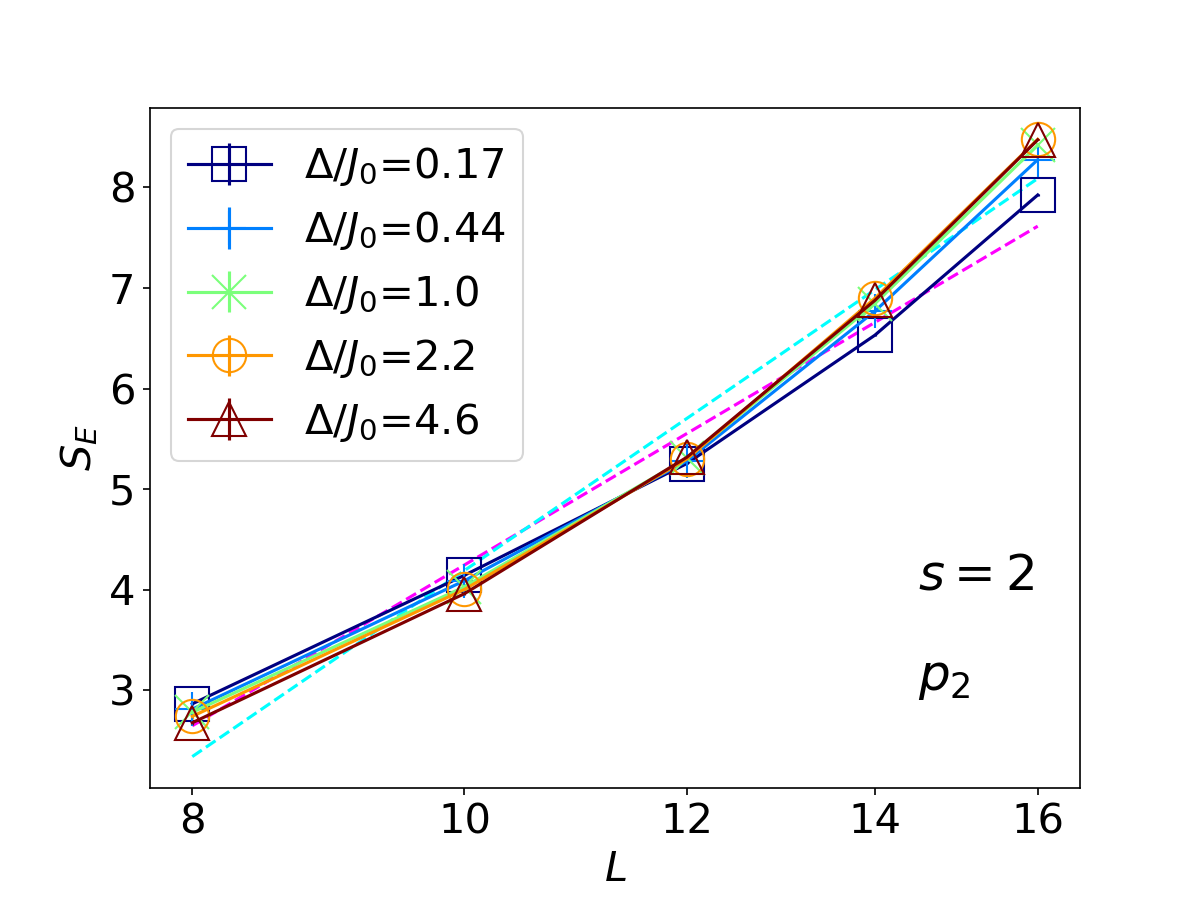}
\caption{Entanglement entropy $S_E$ as a function of system size $L$ for different disorder strengths including $s=2$ sector for states corresponding to $\epsilon \in [0.6,0.7]$. Plots (a) and (b) are for $p_0(J)$, (c) and (d) are for $p_1(J)$, (e) and (f) are for $p_2(J)$. The meaning of the magenta dashed line and the color of the curves corresponding to different disorder strengths are the same as in Fig.\ \ref{fig-scale-ent-allsz}.} 
\label{fig-scale-ent-s=2}
\end{figure*}

\begin{figure*}[tb]
    (a)\includegraphics[width=0.4\textwidth]{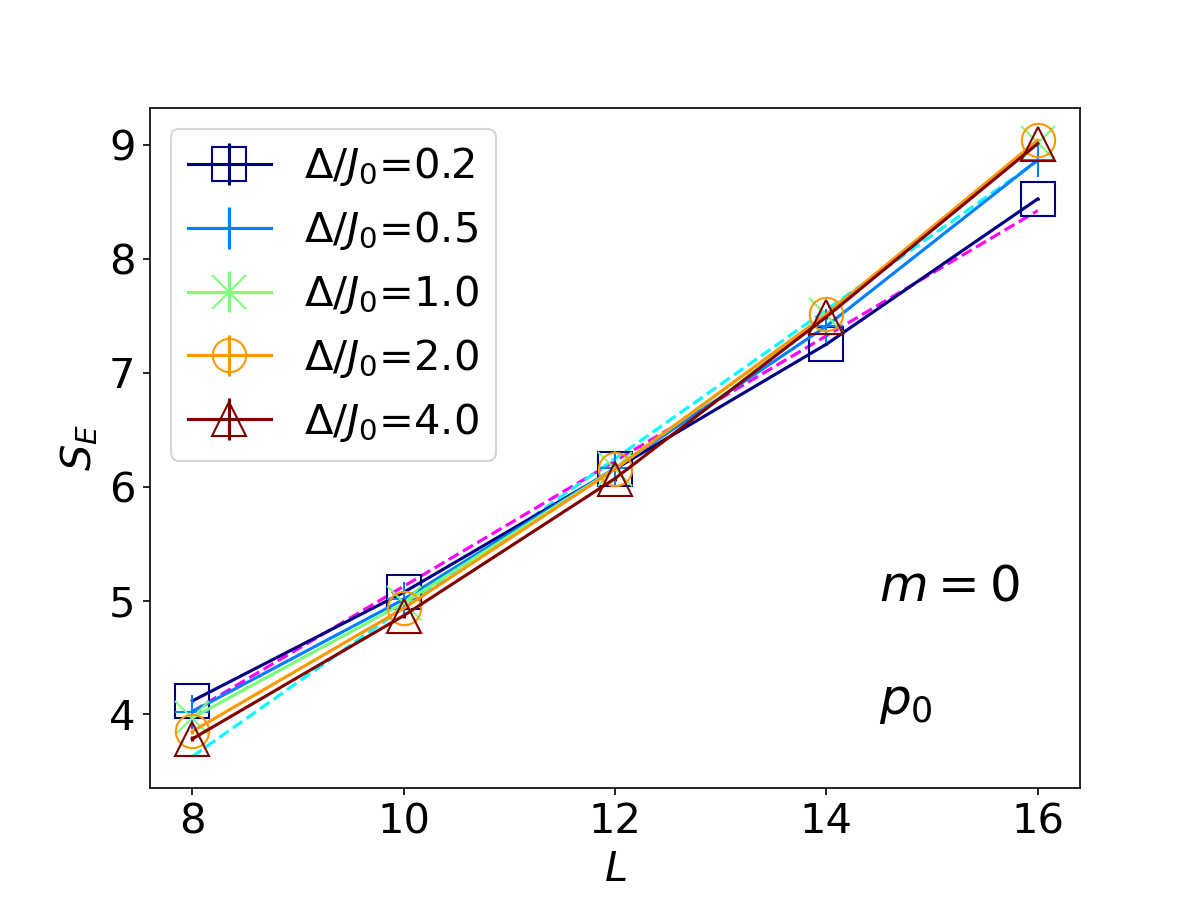}
    (b)\includegraphics[width=0.4\textwidth]{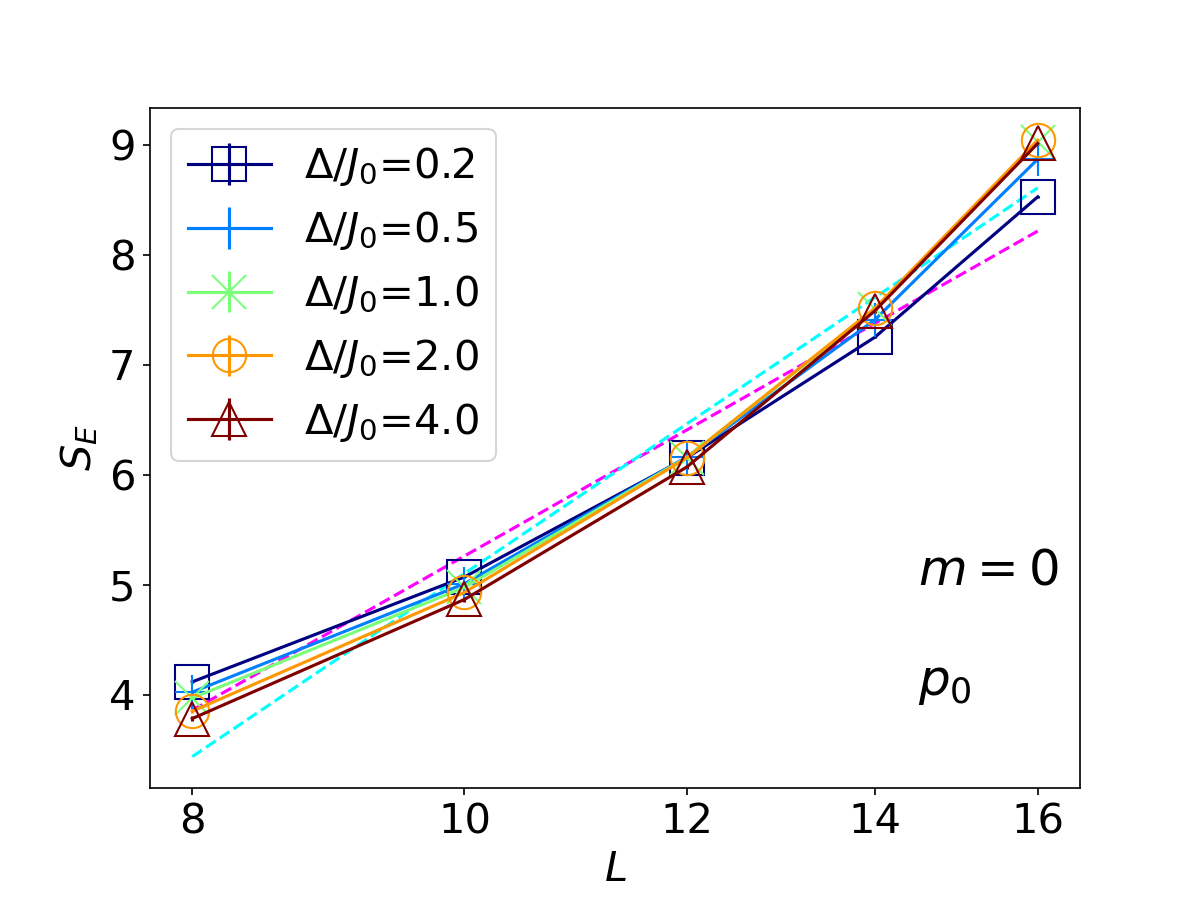}
    (c)\includegraphics[width=0.4\textwidth]{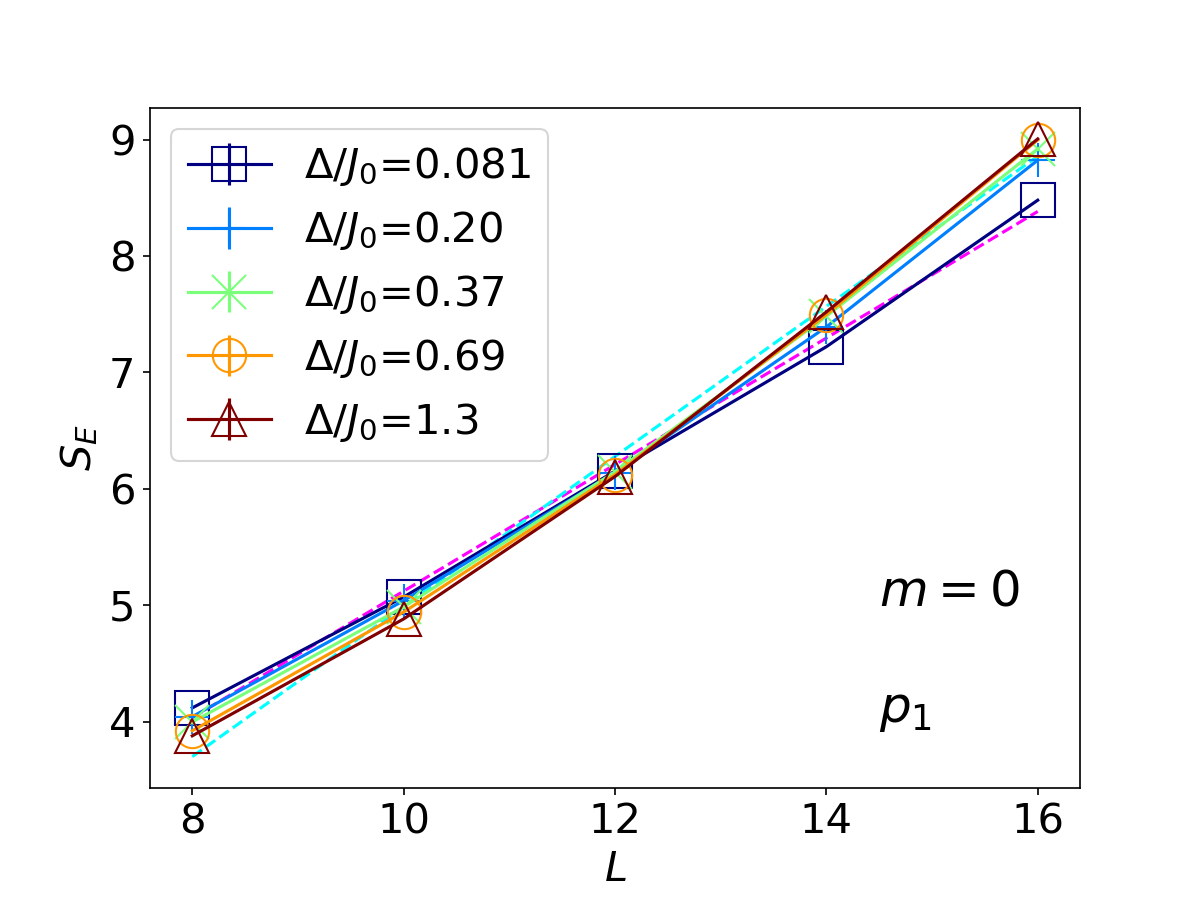}
    (d)\includegraphics[width=0.4\textwidth]{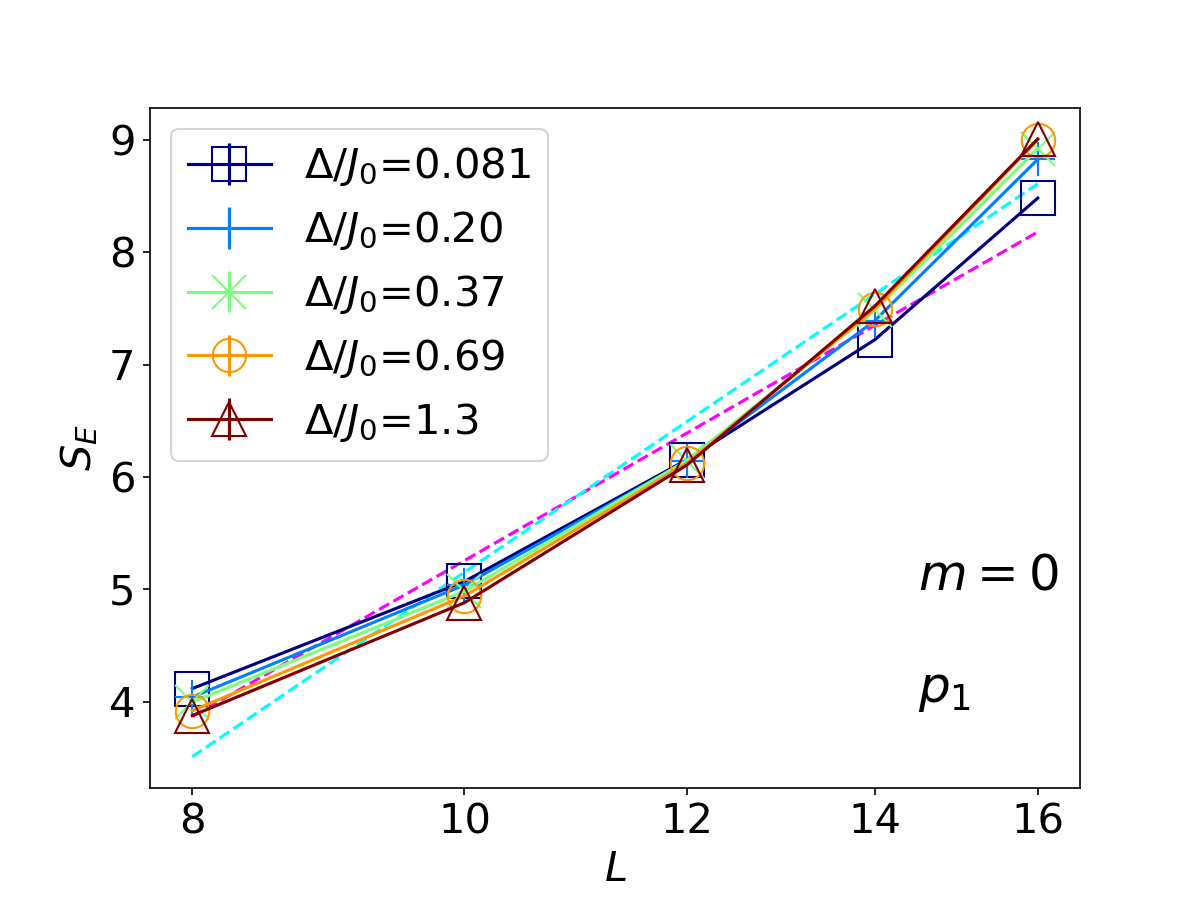}
    (e)\includegraphics[width=0.4\textwidth]{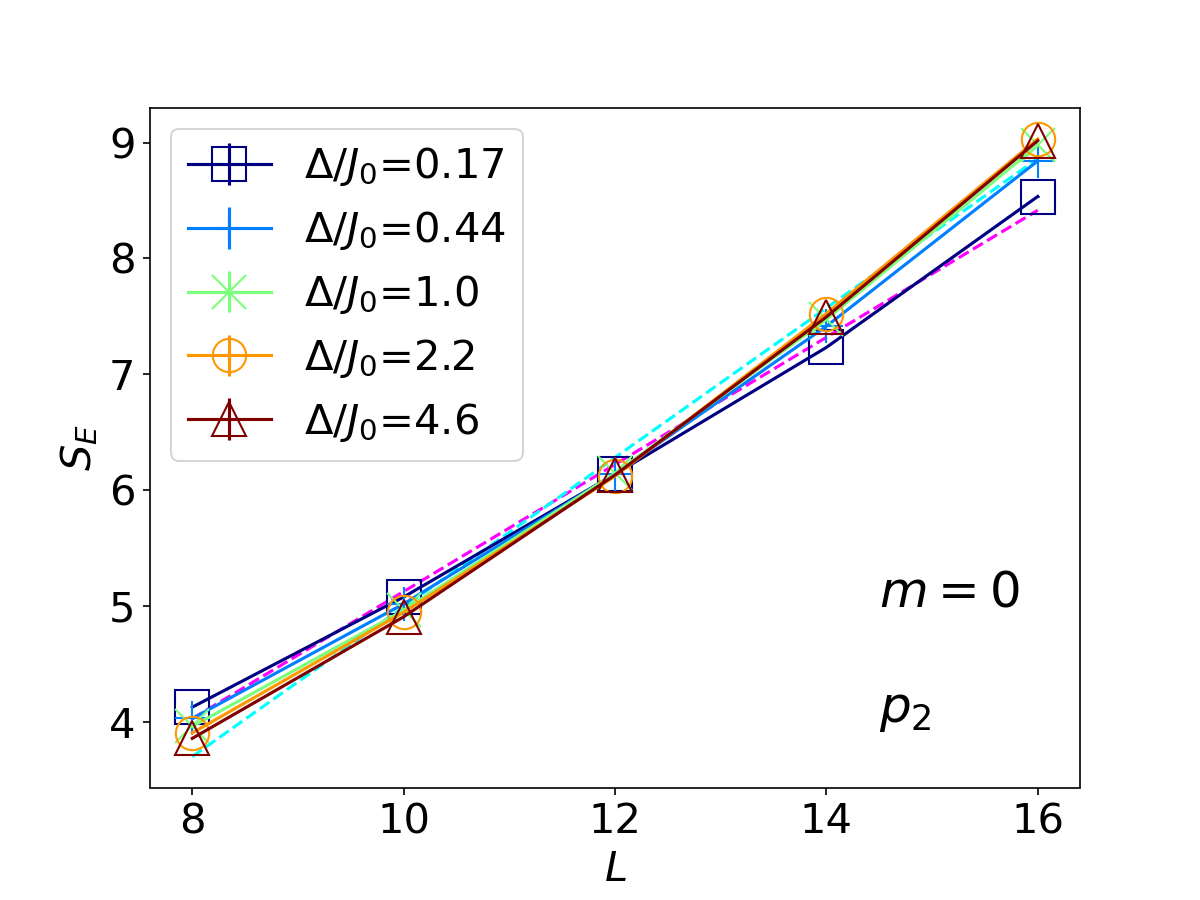}
    (f)\includegraphics[width=0.4\textwidth]{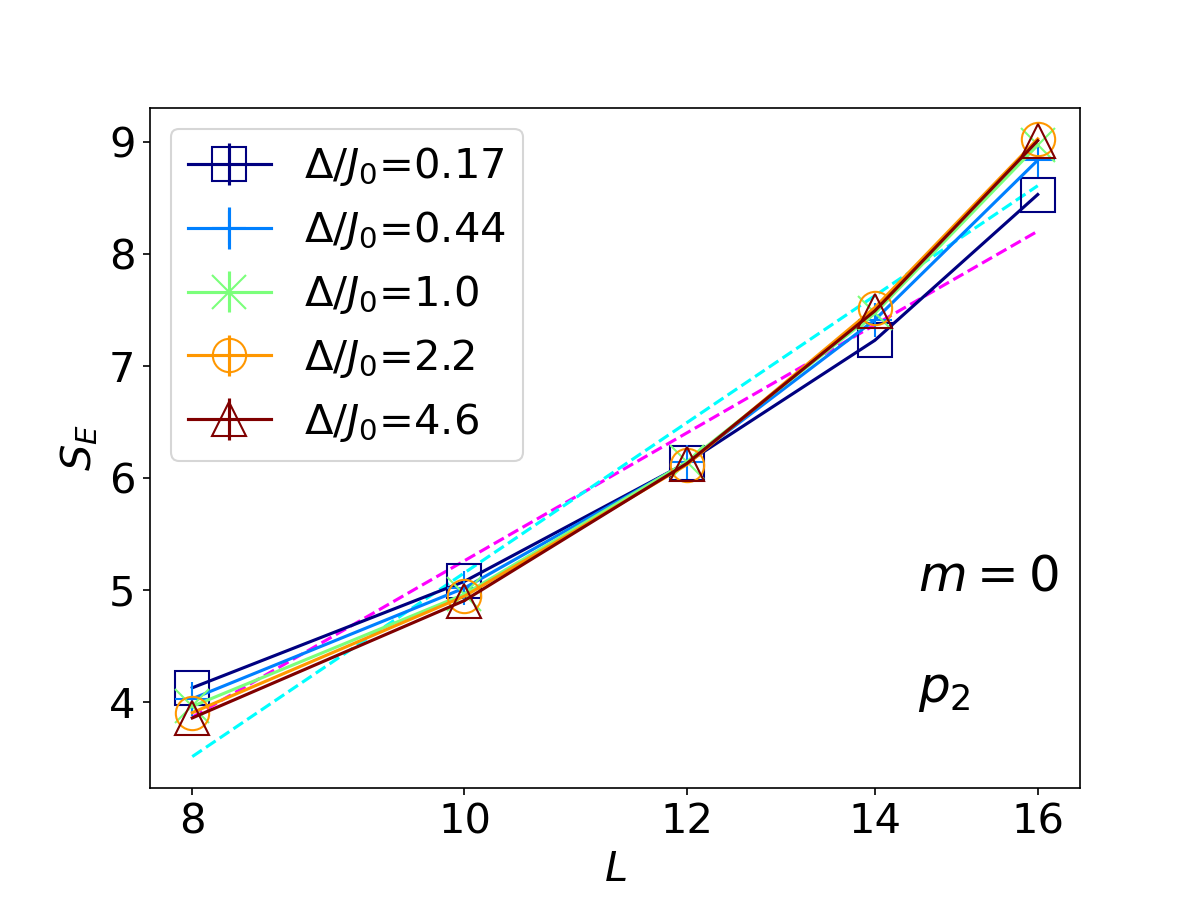}
\caption{Entanglement entropy $S_E$ as a function of system size $L$ for different disorder strengths including $m=0$ sector for states corresponding to $\epsilon \in [0.6,0.7]$. The order of the plots, the meaning of the magenta dashed line and the color of the curves corresponding to different disorder strengths are the same as in Fig.\ \ref{fig-scale-ent-allsz} and Fig.\ \ref{fig-scale-ent-s=2}.} 
\label{fig-scale-ent-m=0}
\end{figure*}

\begin{figure*}[tb]
    (a)\includegraphics[width=0.4\textwidth]{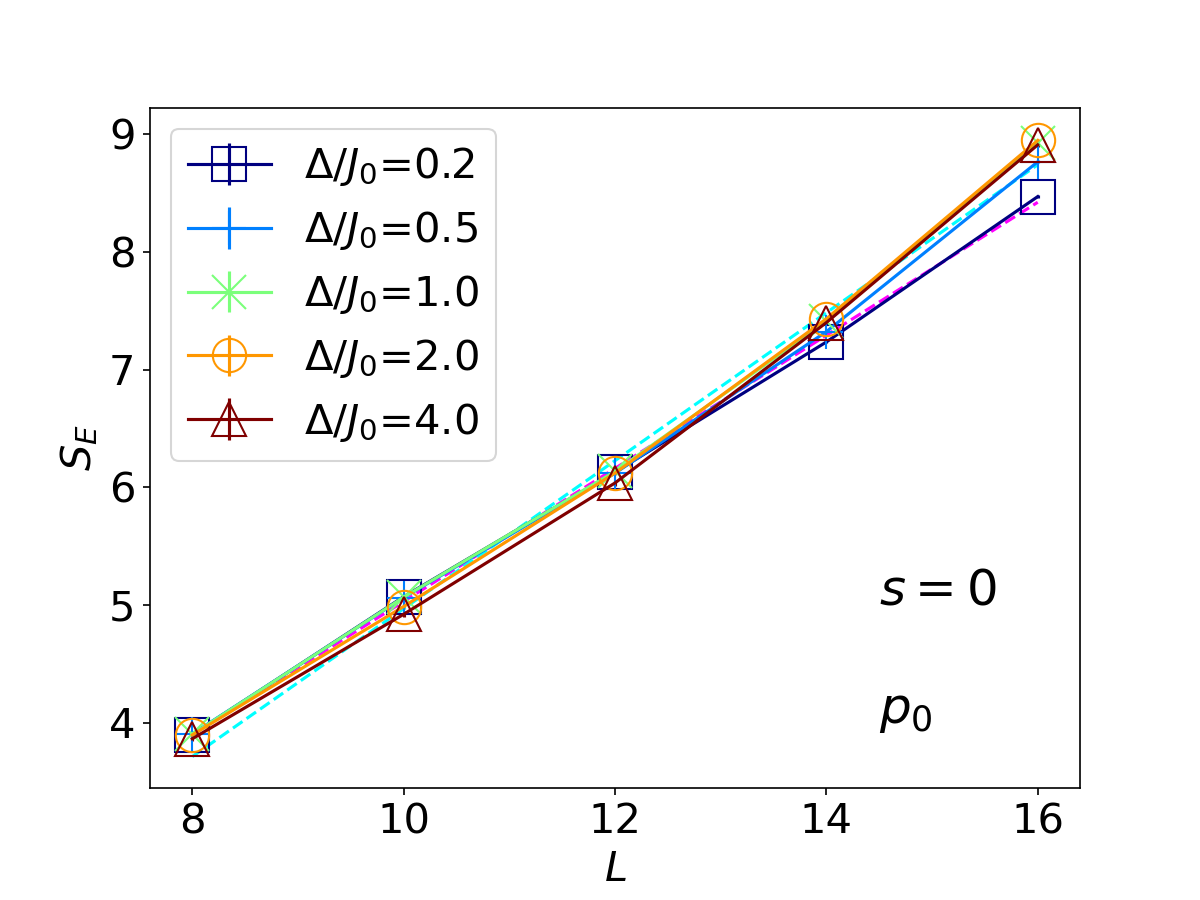}
    (b)\includegraphics[width=0.4\textwidth]{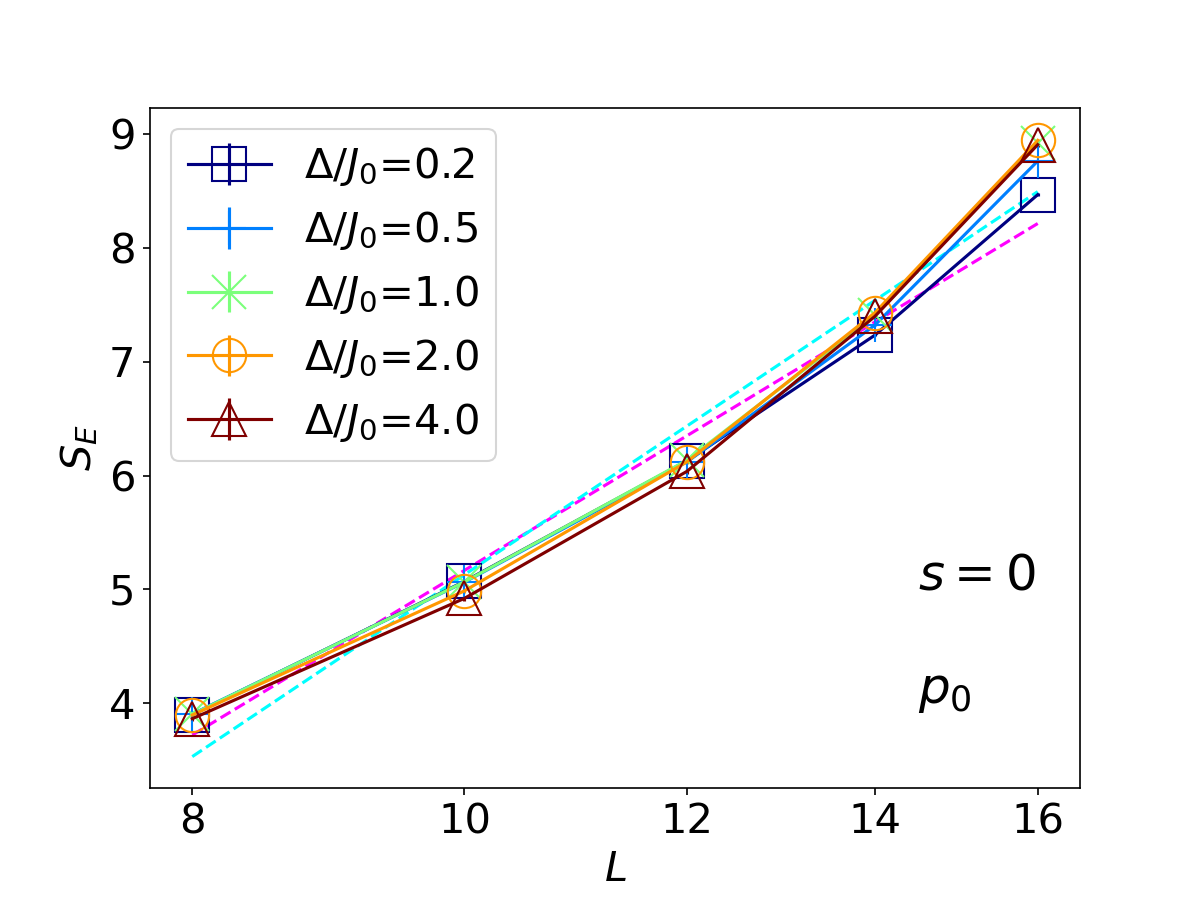}
    (c)\includegraphics[width=0.4\textwidth]{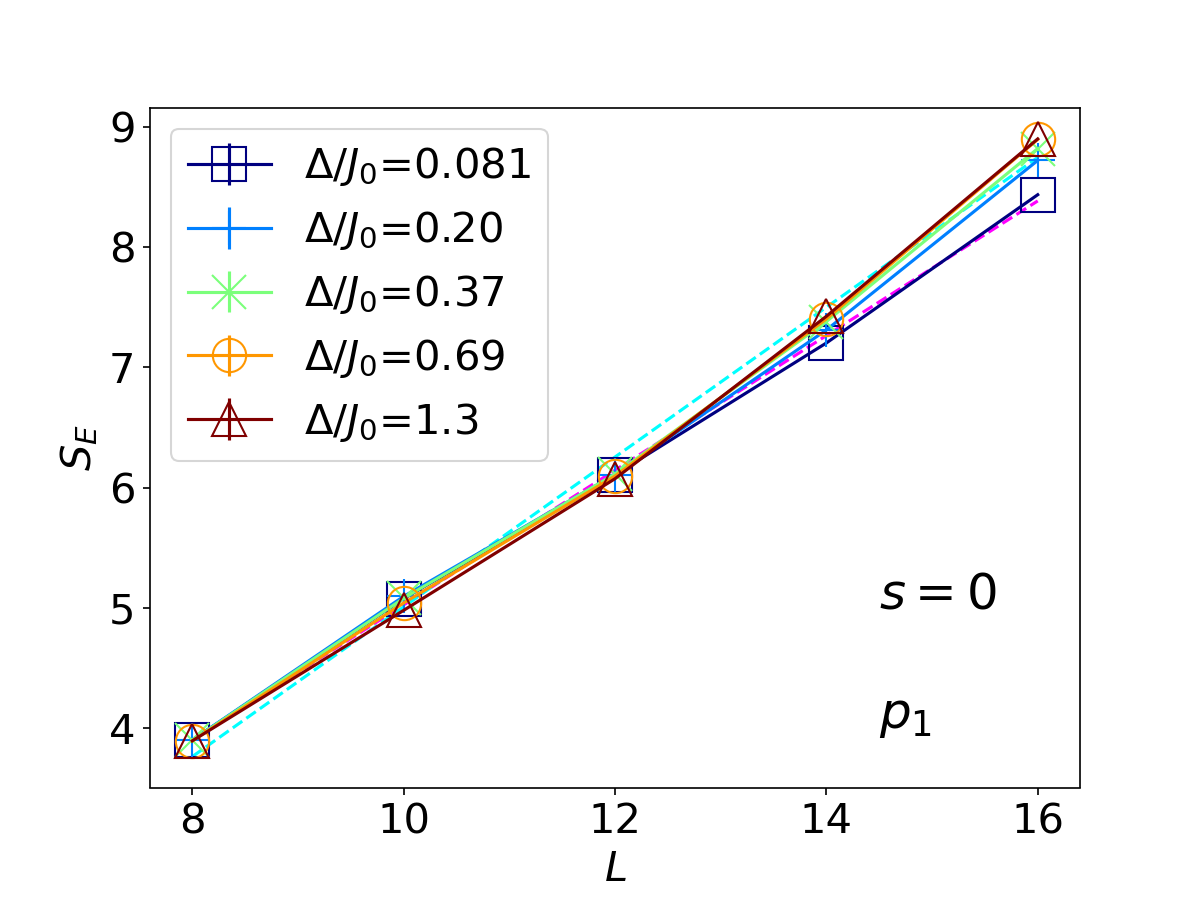}
    (d)\includegraphics[width=0.4\textwidth]{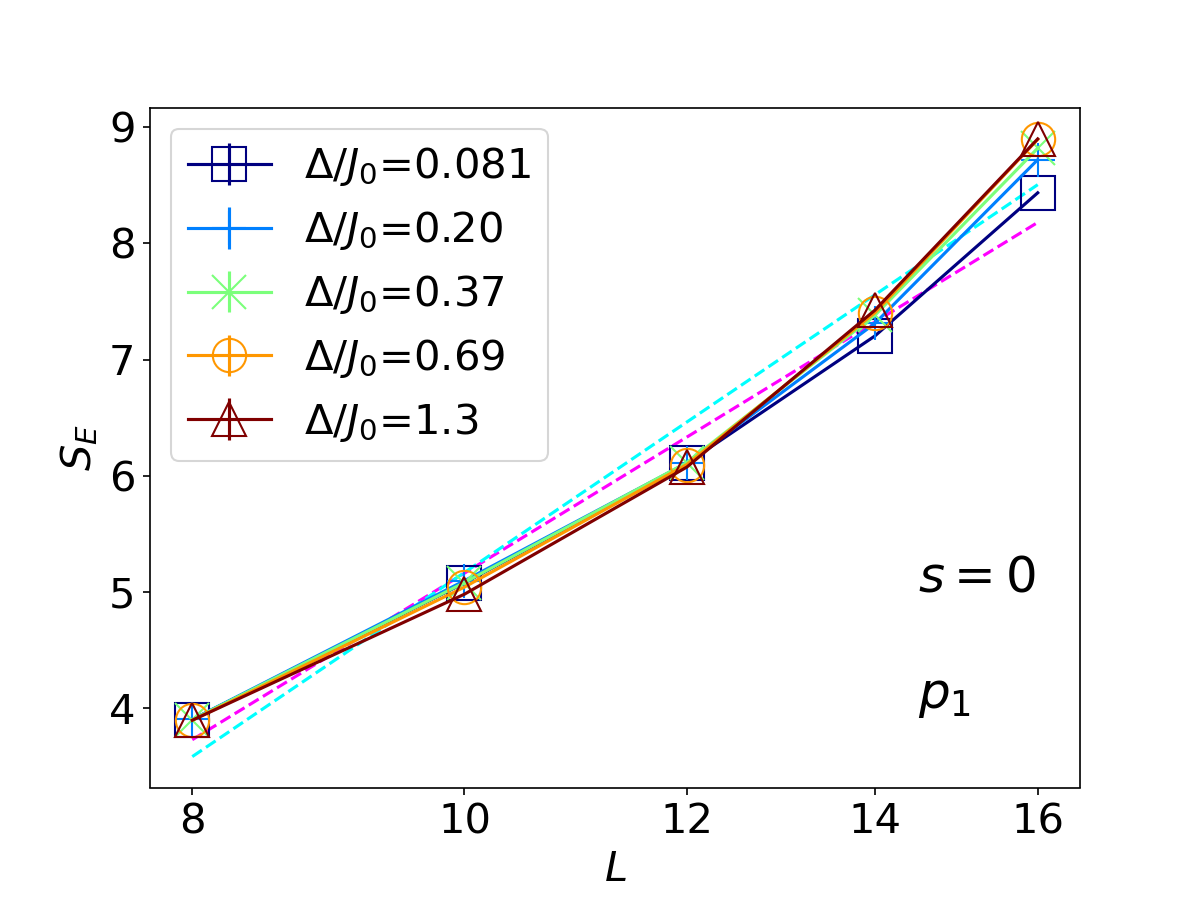}
    (e)\includegraphics[width=0.4\textwidth]{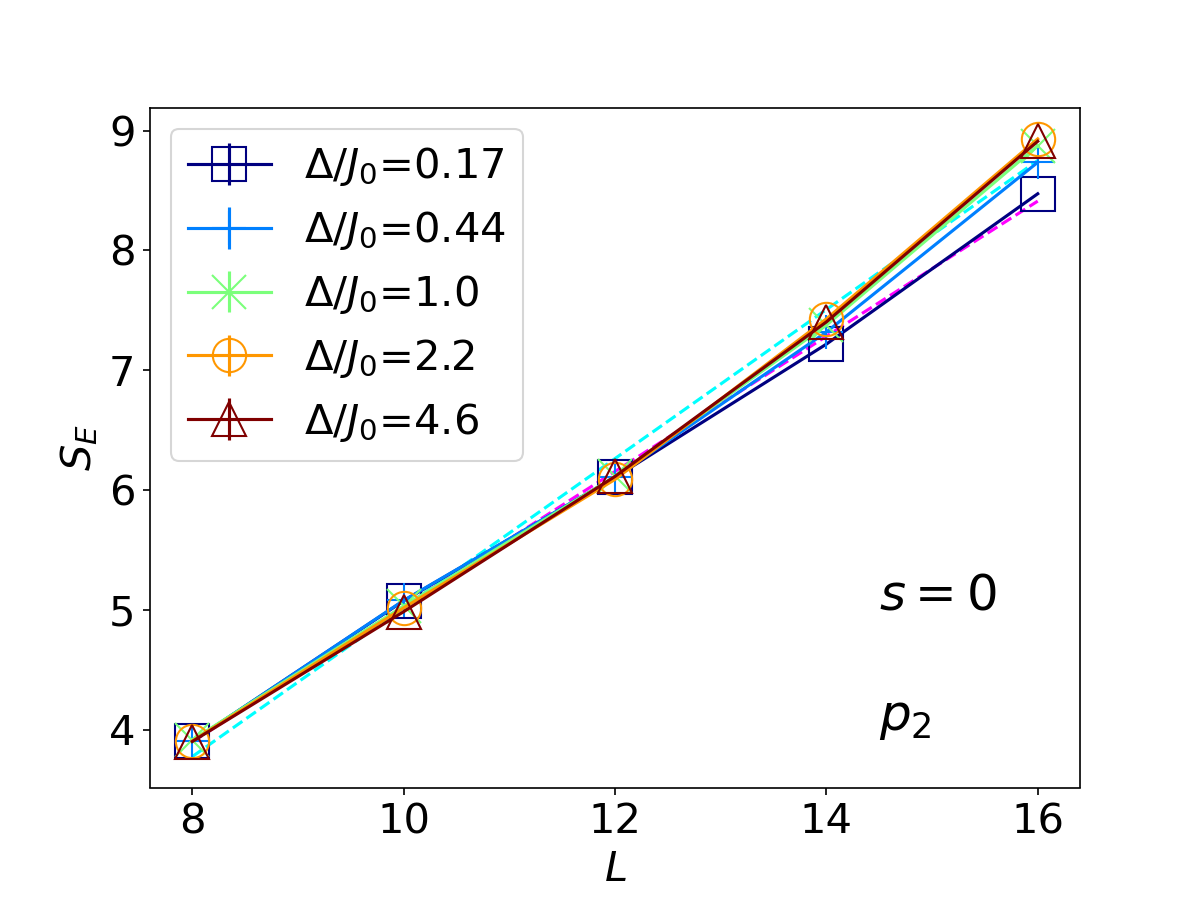}
    (f)\includegraphics[width=0.4\textwidth]{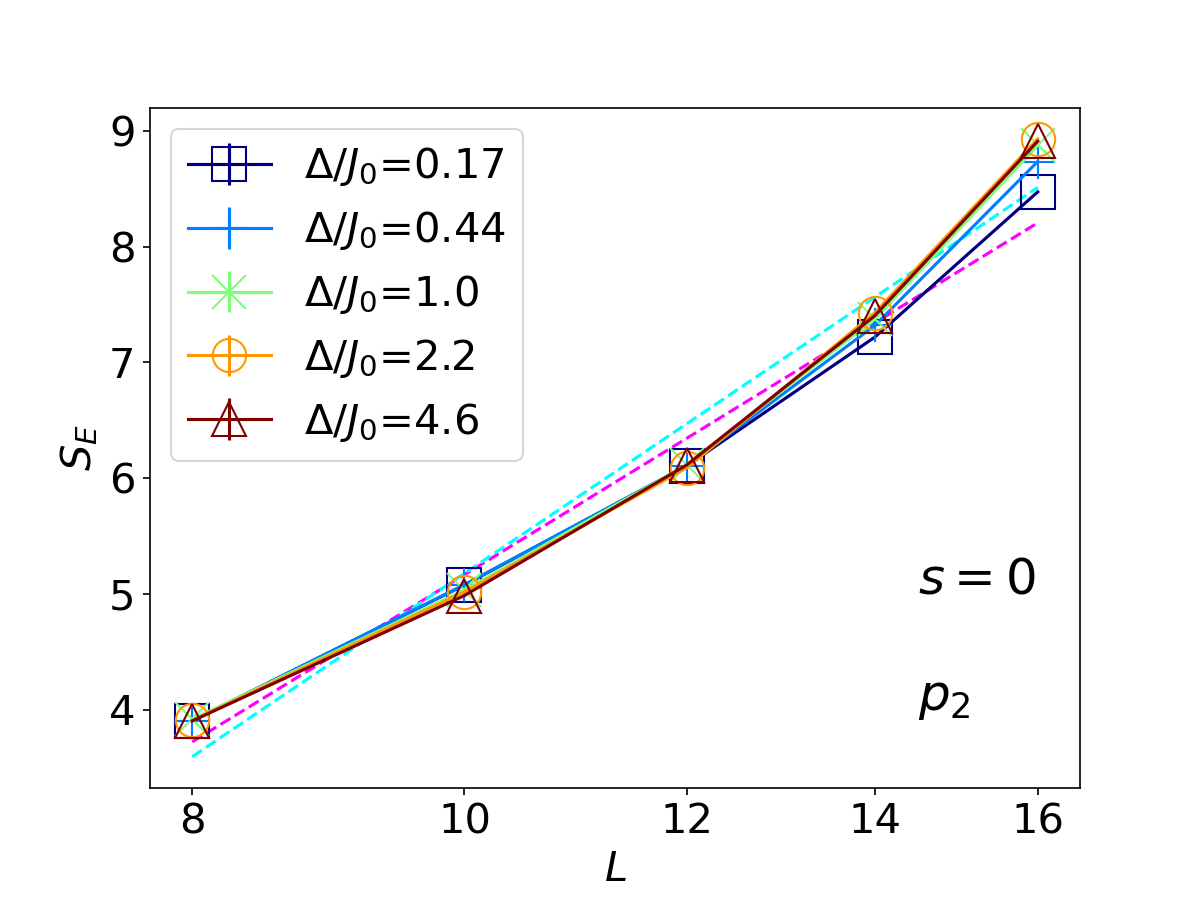}
\caption{Entanglement entropy $S_E$ as a function of system size $L$ for different disorder strengths including $s=0$ sector for states corresponding to $\epsilon \in [0.6,0.7]$. The order of the plots, the meaning of the magenta dashed line and the color of the curves corresponding to different disorder strengths are the same as in Fig.\ \ref{fig-scale-ent-allsz}, Fig.\ \ref{fig-scale-ent-s=2} and Fig.\ \ref{fig-scale-ent-m=0}.} 
\label{fig-scale-ent-s=0}
\end{figure*}

Figs.\ \ref{fig-scale-ent-s=2}, \ref{fig-scale-ent-m=0} and \ref{fig-scale-ent-s=0} show the entanglement entropy $S_E$ as a function of system size $L$ for states $\epsilon \in [0.6,0.7]$ in $s=2$, $m=0$ and $s=0$ respectively. The curves correspond to the weakest and strongest disorders are fitted both linearly and logarithmically. 

\begin{figure*}[tb]

    (a)\includegraphics[width=0.3\textwidth]{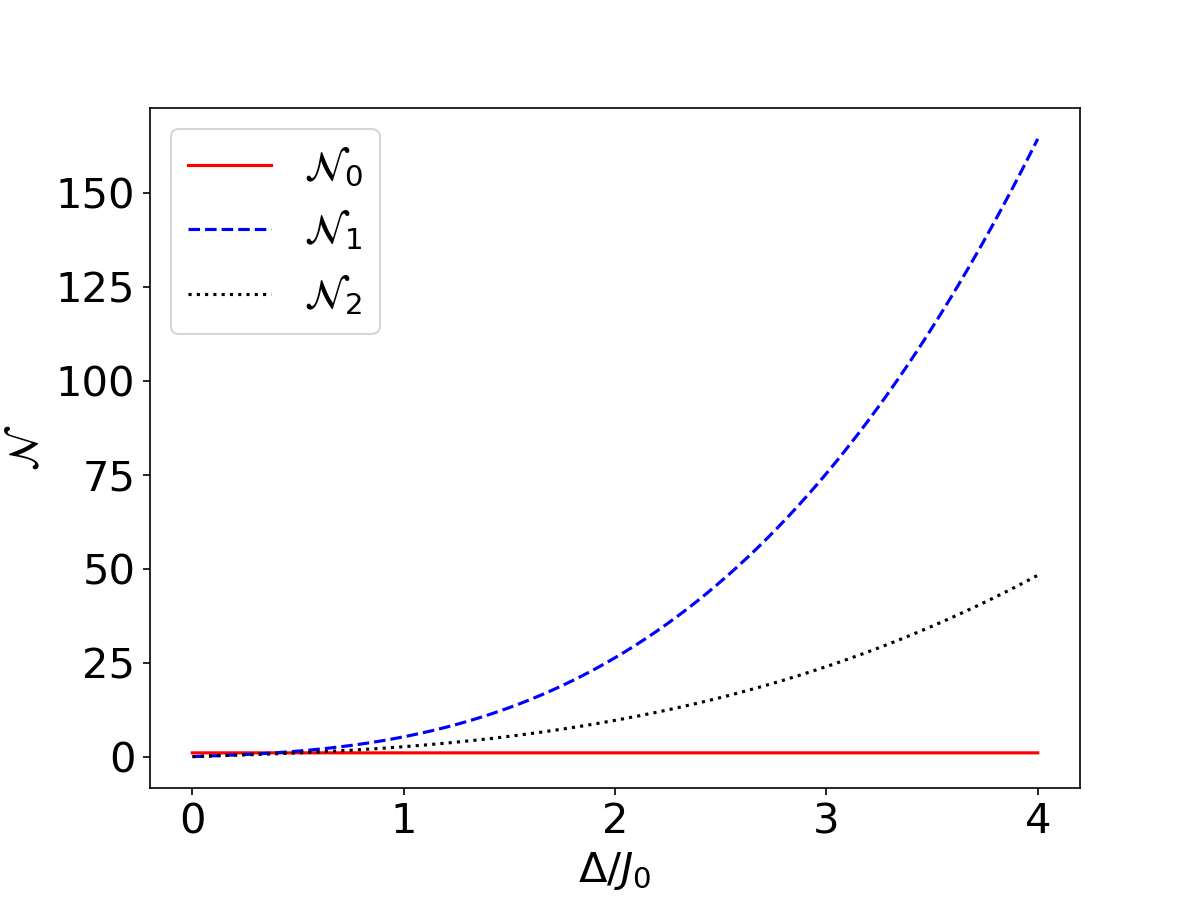}
    (b)\includegraphics[width=0.3\textwidth]{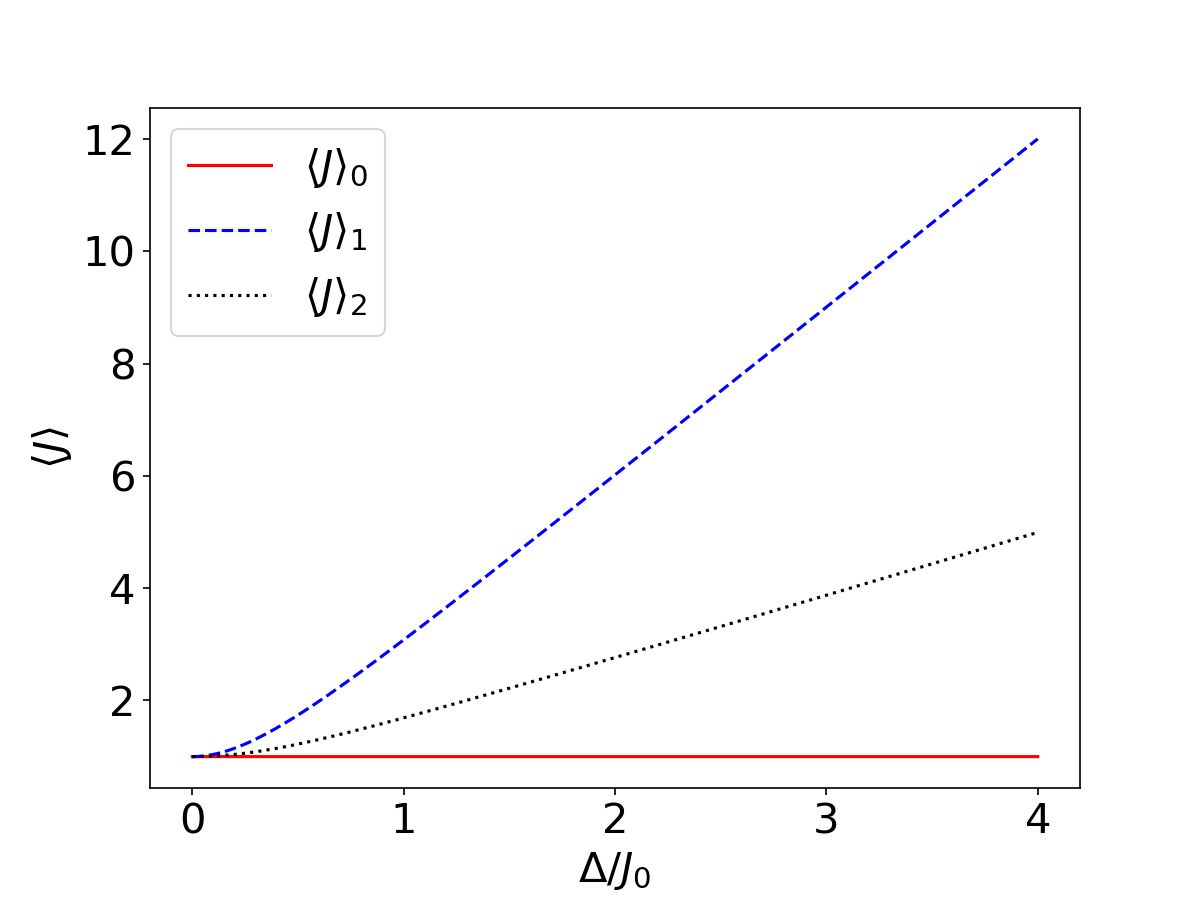}
    (c)\includegraphics[width=0.3\textwidth]{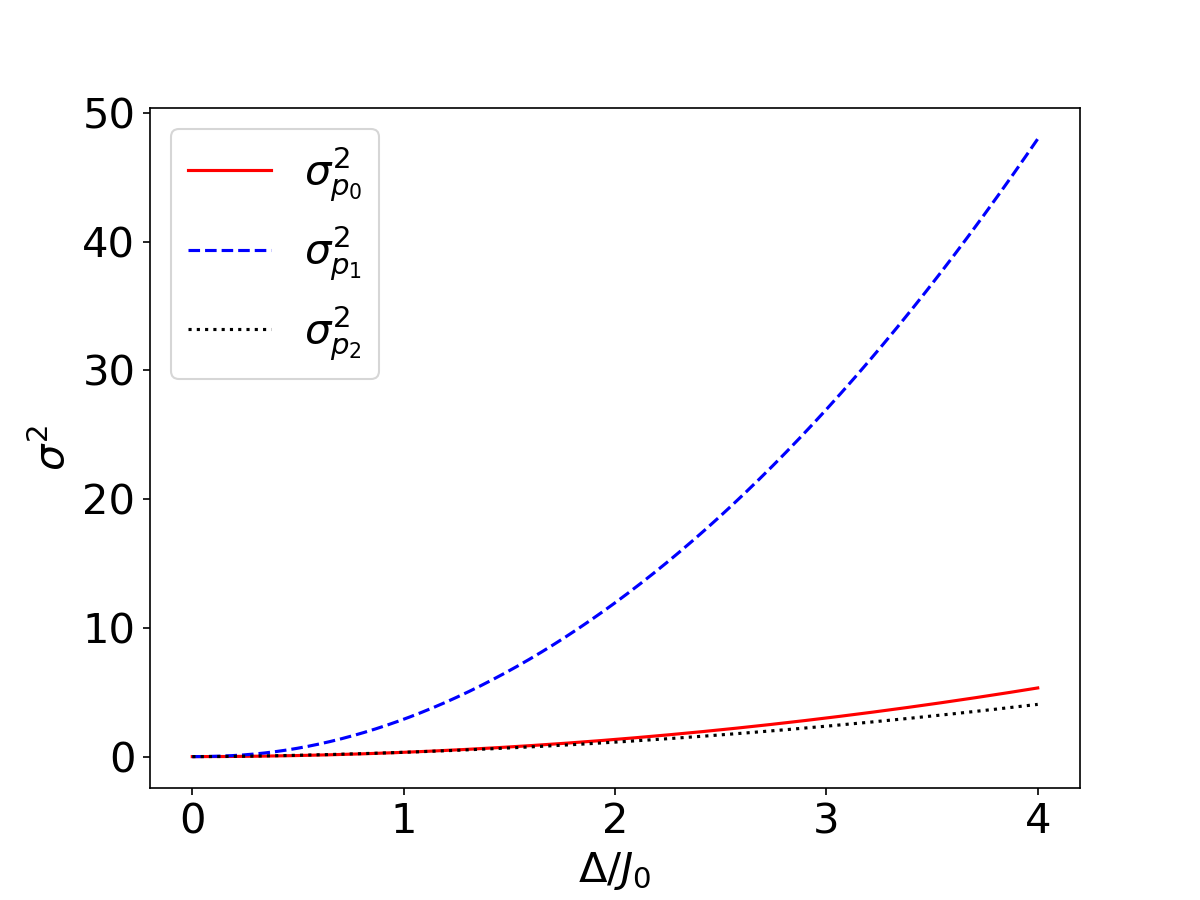}
\caption{(a) Normalization factor (b) $\left<J\right>$ (c) variance $\sigma^2$ all three disorder distributions at $J_0=1$. } 
\label{fig-disorder-distribution-norm-average-variance}
\end{figure*}

Fig.\ \ref{fig-disorder-distribution-norm-average-variance} shows $\mathcal{N}_i(\Delta)$, $\langle J \rangle_{p_i}(\Delta)$ and $\sigma_{p_i}^2(\Delta)$ at $J_0=1$.

%%%%%%%%%%%%%%%%%%%%%%%%%%%%%%%%%%%%%%%%%%%%%%%%%%%%%%%%%%%%%%%%%%%%%%%%%%%%%%
%%%%%%%%%%%%%%%%%%%%%%%%%%%%%%%%%%%%%%%%%%%%%%%%%%%%%%%%%%%%%%%%%%%%%%%%%%%%%%
%%%%%%%%%%%%%%%%%%%%%%%%%%%%%%%%%%%%%%%%%%%%%%%%%%%%%%%%%%%%%%%%%%%%%%%%%%%%%%

%%%%%%%%%%%%%%%%%%%%%%%%%%%%%%%%%%%%%%%%%%%%%%%%%%%%%%%%%%%%%%%%%%%%%%%%%%%%%%
%%%%%%%%%%%%%%%%%%%%%%%%%%%%%%%%%%%%%%%%%%%%%%%%%%%%%%%%%%%%%%%%%%%%%%%%%%%%%%
%%%%%%%%%%%%%%%%%%%%%%%%%%%%%%%%%%%%%%%%%%%%%%%%%%%%%%%%%%%%%%%%%%%%%%%%%%%%%%
%%%%%%%%%%%%%%%%%%%%%%%%%%%%%%%%%%%%%%%%%%%%%%%%%%%%%%%%%%%%%%%%%%%%%%%%%%%%%%
%%%%%%%%%%%%%%%%%%%%%%%%%%%%%%%%%%%%%%%%%%%%%%%%%%%%%%%%%%%%%%%%%%%%%%%%%%%%%%
%%%%%%%%%%%%%%%%%%%%%%%%%%%%%%%%%%%%%%%%%%%%%%%%%%%%%%%%%%%%%%%%%%%%%%%%%%%%%%
%%%%%%%%%%%%%%%%%%%%%%%%%%%%%%%%%%%%%%%%%%%%%%%%%%%%%%%%%%%%%%%%%%%%%%%%%%%%%%
%%%%%%%%%%%%%%%%%%%%%%%%%%%%%%%%%%%%%%%%%%%%%%%%%%%%%%%%%%%%%%%%%%%%%%%%%%%%%%
%%%%%%%%%%%%%%%%%%%%%%%%%%%%%%%%%%%%%%%%%%%%%%%%%%%%%%%%%%%%%%%%%%%%%%%%%%%%%%

%%%%%%%%%%%%%%%%%%%%%%%%%%%%%%%%%%%%%%%%%%%%%%%%%%%%%%%%%%%%%%%%%%%%%%%%%%%%%%
%%%%%%%%%%%%%%%%%%%%%%%%%%%%%%%%%%%%%%%%%%%%%%%%%%%%%%%%%%%%%%%%%%%%%%%%%%%%%%

%%%%%%%%%%%%%%%%%%%%%%%%%%%%%%%%%%%%%%%%%%%%%%%%%%%%%%%%%%%%%%%%%%%%%%%%%%%%%%
%%%%%%%%%%%%%%%%%%%%%%%%%%%%%%%%%%%%%%%%%%%%%%%%%%%%%%%%%%%%%%%%%%%%%%%%%%%%%%

%%%%%%%%%%%%%%%%%%%%%%%%%%%%%%%%%%%%%%%%%%%%%%%%%%%%%%%%%%%%%%%%%%%%%%%%%%%%%%
%%%%%%%%%%%%%%%%%%%%%%%%%%%%%%%%%%%%%%%%%%%%%%%%%%%%%%%%%%%%%%%%%%%%%%%%%%%%%%

%%%%%%%%%%%%%%%%%%%%%%%%%%%%%%%%%%%%%%%%%%%%%%%%%%%%%%%%%%%%%%%%%%%%%%%%%%%%%%
%%%%%%%%%%%%%%%%%%%%%%%%%%%%%%%%%%%%%%%%%%%%%%%%%%%%%%%%%%%%%%%%%%%%%%%%%%%%%%

%%%%%%%%%%%%%%%%%%%%%%%%%%%%%%%%%%%%%%%%%%%%%%%%%%%%%%%%%%%%%%%%%%%%%%%%%%%%%%
\fi\end{document}